\documentclass[aps,rmp,nofootinbib,twocolumn,floatfix]{revtex4}

\usepackage{graphicx}
\usepackage{amsmath}
\usepackage{bm}

\newcommand{\krig}[1]{\stackrel{\circ}{#1}}
\newcommand{\bqa}{\begin{eqnarray}}
\newcommand{\eqa}{\end{eqnarray}}
\newcommand{\be}{\begin{equation}}
\newcommand{\ee}{\end{equation}}
\newcommand{\ba}{\begin{eqnarray}}
\newcommand{\ea}{\end{eqnarray}}
\newcommand{\beq}{\begin{equation}}
\newcommand{\eeq}{\end{equation}}
\newcommand{\beqa}{\begin{eqnarray}}
\newcommand{\eeqa}{\end{eqnarray}}

\newcommand{\barr}[1]{\not\mathrel #1}  
 
\newcommand{\fet}[1]{\mbox{\boldmath $#1$}}
\newcommand{\sing}{^1{\rm S}_0}
\newcommand{\trip}{^3{\rm S}_1}

\newcommand{\mpic}{M_\pi^{crit}}
\newcommand{\Mhi}{M_{high}}
\newcommand{\Mlo}{M_{low}}
\newcommand{\nn}{\nonumber \\ }
\newcommand{\simlt}{\stackrel{<}{{}_\sim}}

\def\palka{\hspace{-6.5pt}/}
\def\palkasmall{\hspace{-5.0pt}/}

\begin{document}

\preprint{HISKP-TH-08/18}

\title{Modern  Theory of Nuclear Forces}

\author{E.~Epelbaum}\email{e.epelbaum@fz-juelich.de}
\affiliation{Forschungszentrum J\"ulich, Institut f\"ur Kernphysik (IKP-3) and
J\"ulich Center for Hadron Physics,
             D-52425 J\"ulich, Germany}
\affiliation{Helmholtz-Institut f\"ur Strahlen- und Kernphysik (Theorie)
and Bethe Center for Theoretical Physics,
 Universit\"at Bonn, D-53115 Bonn, Germany}
\author{H.-W.~Hammer}\email{hammer@itkp.uni-bonn.de}
\affiliation{Helmholtz-Institut f\"ur Strahlen- und Kernphysik (Theorie)
and Bethe Center for Theoretical Physics,
 Universit\"at Bonn, D-53115 Bonn, Germany}
\author{Ulf-G. Mei{\ss}ner}\email{meissner@itkp.uni-bonn.de}
\affiliation{Helmholtz-Institut f\"ur Strahlen- und Kernphysik (Theorie)
and Bethe Center for Theoretical Physics,
 Universit\"at Bonn, D-53115 Bonn, Germany}
\affiliation{Forschungszentrum J\"ulich, Institut f\"ur Kernphysik (IKP-3) and
J\"ulich Center for Hadron Physics,
             D-52425 J\"ulich, Germany}



\begin{abstract}
Effective field theory allows for a systematic and model-independent
derivation of the forces between nucleons in harmony with the symmetries
of Quantum Chromodynamics. We review the foundations of this approach and
discuss its application for light nuclei at various resolution scales. The
extension of this approach to many-body systems is briefly sketched.\\[0.2cm]
\centerline{Commissioned article for {\sl Reviews of Modern Physics}}
\end{abstract}

\maketitle


\tableofcontents

\section{QCD and Nuclear Forces}
\def\theequation{\arabic{section}.\arabic{equation}}
\setcounter{equation}{0}
\setcounter{page}{1}
\label{sec:intro}

Within the Standard Model of particle physics, the strong interactions
are described by Quantum Chromodynmics (QCD).
QCD is a fascinating theory with many intriguing
manifestations. Its structure and interactions are governed by a local
non-abelian gauge symmetry, namely  SU(3)$_{\rm color}$.
Its fundamental degrees of freedom, the quarks (the matter fields) and 
gluons (the force carriers), have never been observed in isolation 
(confinement). The strong coupling
constant $\alpha_S$ exhibits a very pronounced running and is of order one
in the typical energy scales of nuclear physics. The bound states made
from the basic constituents are the hadrons, the strongly interacting
particles. The particle spectrum shows certain regularities that can be
traced back to the flavor symmetries related to the fermions building 
up these states. More precisely, there are six quark flavors. These can 
be grouped into two very different sectors.
While the light quarks ($u,d,s$) are almost massless and thus have to 
be treated relativistically, bound states made from heavy quarks allow for a
precise non-relativistic treatment. In what follows, we will only consider the
light quarks at low energies, where perturbation theory in  $\alpha_S$ is
inapplicable (this regime is frequently called ``strong QCD''). 
A further manifestation  of strong QCD is the appearance of nuclei, 
shallow bound states composed of protons, neutrons, pions or strange 
particles like hyperons. The resulting nuclear forces that are responsible 
for the nuclear binding are residual color  forces, much like the van der 
Waals forces between neutral molecules. It is
the aim of this article to provide the link between QCD and its symmetries,
in particular the spontaneously and explicitely broken chiral symmetry, and
the nuclear forces which will allow to put nuclear physics on  firm
theoretical grounds and also gives rise to a very accurate calculational
scheme for nuclear forces and the properties of nuclei.

This review is organized as follows: In this section, we briefly discuss
some of the concepts underlying the chiral effective field theory of the
nuclear forces and make contact to ab initio lattice simulations of
two-baryon systems as well as to more phenomenological approaches.
Sec.~\ref{sec:EFT} deals with the foundations and applications of
nuclear EFT and should be considered the central piece of this review.
In particular, tests of these forces  in few-nucleon systems are discussed.
Attempts to tackle nuclear matter and finite nuclei are considered in
sec.~\ref{sec:many}. We end with a short summary and outlook.

\subsection{Chiral symmetry}
\label{sec:chisym}
First, we must discuss chiral symmetry in the context of QCD. Chromodynamics is a
non-abelian $SU(3)_{\rm color}$ gauge theory with $N_f = 6$ flavors of quarks,
three of them being light ($u,d,s$) and the other three heavy ($c,b,t$).
Here, light and heavy refers to a typical hadronic scale of about 1~GeV.
In what follows, we consider light quarks only (the heavy quarks are to be 
considered as decoupled). The QCD Lagrangian reads
\beqa\label{LQCD}
{\mathcal L}_{\rm QCD} &=& -\frac{1}{2g^2} {\rm Tr} \left( G_{\mu\nu}G^{\mu\nu}
\right) + \bar q \, i \gamma^\mu D_\mu \, q -  \bar q {\mathcal M} \, q
\nonumber\\
&=&{\mathcal L}_{\rm QCD}^0  -  \bar q {\mathcal M} \, q   ~,
\eeqa
where we have absorbed the gauge coupling in the definition of the gluon
field and color indices are suppressed.  The three-component vector $q$
collects the quark fields, $q^T (x) = \left( u(s), d(x), s(x)\right)$. 
As far as the strong interactions
are concerned, the different quarks $u,d,s$ have identical properties, except
for their masses. The quark masses are free parameters in QCD - the theory
can be formulated for any value of the quark masses. In fact, light quark QCD
can be well approximated by a fictitious world of massless quarks,
denoted ${\mathcal L}_{\rm QCD}^0$ in Eq.~(\ref{LQCD}). Remarkably,
this theory contains no adjustable parameter - the gauge coupling $g$ merely
sets the scale for the renormalization group invariant scale $\Lambda_{\rm
  QCD}$. Furthermore, in the massless world left- and right-handed quarks
are completely decoupled. 
The Lagrangian of massless QCD is invariant under separate unitary
global transformations of the left- and right-hand quark fields, the
so-called {\em chiral rotations},
$q_I \to V_I q_I~,  V_I \in U(3)~,  I = L,R$,
leading to $3^2 =9$ conserved left- and $9$ conserved right-handed 
currents by virtue of Noether's theorem. These can be expressed in terms of 
vector ($V = L + R$) and  axial-vector ($A = L - R$) currents
\beqa\label{conscurr}
V_0^\mu &=& \bar q \, \gamma^\mu \, q~, \quad 
V^\mu_a = \bar q \, \gamma^\mu \frac{\lambda_a}{2} \, q~,\nonumber\\
A_0^\mu &=& \bar q \, \gamma^\mu \gamma_5 \, q~,\quad 
A^\mu_a = \bar q \, \gamma^\mu \gamma_5 \frac{\lambda_a}{2} \, q~,
\eeqa
Here, $a = 1, \ldots 8$, and the $\lambda_a$ are Gell-Mann's $SU(3)$
flavor matrices. The singlet axial current is anomalous,
and thus not conserved. The actual symmetry group of massless QCD
is generated by the charges of the conserved currents, it is
$G_0 = SU(3)_R \times SU(3)_L \times U(1)_V$. The $U(1)_V$ subgroup of 
$G_0$ generates conserved baryon number since the isosinglet vector 
current counts the number of quarks minus antiquarks in a hadron. 
The remaining group $SU(3)_R \times SU(3)_L$ is often
referred to as chiral $SU(3)$. Note that one also considers the light
$u$ and $d$ quarks only (with the strange quark mass fixed at its
physical value), in that case, one speaks of chiral $SU(2)$ and
must replace the generators in Eq.~(\ref{conscurr}) by the Pauli-matrices.
Let us mention that QCD is also invariant under the discrete symmetries of
parity ($P$), charge conjugation ($C$) and time reversal ($T$). 
Although interesting in itself, we do not consider strong $CP$ violation and
the related $\theta$-term in what follows, see e.g.~\cite{Peccei:2006as}.

The chiral symmetry is a symmetry of the Lagrangian of QCD but not of the
ground state or the particle spectrum -- to describe the strong interactions
in nature, it is crucial that chiral symmetry is spontaneously broken. This
can be most easily seen from the fact that hadrons do not appear in parity
doublets. If chiral symmetry were exact, from any hadron one could generate
by virtue of an axial transformation another state of exactly the same 
quantum numbers except of opposite parity. The spontaneous symmetry breaking
leads to  the formation of a quark condensate in the vacuum 
$\langle 0 | \bar q q|0\rangle =\langle 0  | \bar q_L q_R + \bar q_R q_L|0\rangle$, 
thus connecting the left- with the right-handed
quarks. In the absence of quark masses this expectation value
is flavor-independent: $\langle 0 | \bar u u|0\rangle = 
\langle 0 | \bar d d|0\rangle = \langle 0 | \bar q q|0\rangle$. 
More precisely, the vacuum is only invariant under the subgroup of 
vector rotations times the baryon number current, $H_0 = SU(3)_V \times
U(1)_V$. This is the generally accepted picture that is supported by general
arguments \cite{Vafa:1983tf} as well as lattice simulations of QCD (for 
a recent study, see~\cite{Giusti:2007cn}).
In fact, the vacuum expectation value of the quark condensate is only one
of the many possible order parameters characterizing the spontaneous symmetry
violation - all operators that share the invariance properties of the
vacuum 
qualify as order parameters. The quark condensate
nevertheless enjoys a special role, it can be shown to be related to the
density of small eigenvalues of the QCD Dirac operator 
(see \cite{Banks:1979yr} and more recent discussions 
in \cite{Leutwyler:1992yt,Stern:1998dy}),
$\lim_{{\cal M} \to 0} \langle 0 | \bar q q|0\rangle = - \pi \, \rho(0)$.
For free fields, $\rho (\lambda) \sim \lambda^3$ near $\lambda = 0$. Only if
the eigenvalues accumulate near zero, one obtains a non-vanishing condensate.
This scenario is indeed supported by lattice simulations and many model studies
involving topological objects like instantons or monopoles.

Before discussing the implications of spontaneous symmetry breaking for QCD,
we briefly remind the reader of Goldstone's theorem 
\cite{Goldstone:1961eq,Goldstone:1962es}: to every generator of
a spontaneously broken symmetry corresponds a massless excitation of the
vacuum.  These states are the
{\em Goldstone bosons}, collectively denoted as pions $\pi(x)$ in what follows.
Through the corresponding symmetry current the Goldstone bosons couple 
directly to the vacuum,
\beq\label{GBME}
\langle 0 | A^0 (0) | \pi \rangle \neq 0~.
\eeq
In fact, the non-vanishing of this matrix element is a {\it necessary and
sufficient} condition for spontaneous symmetry breaking. In QCD, we have
eight (three) Goldstone bosons for $SU(3)$ ($SU(2)$) with spin zero and 
negative parity -- the latter property is a consequence that these Goldstone 
bosons are generated by applying the axial charges on the vacuum. The
dimensionful scale associated with the matrix element Eq.~(\ref{GBME}) 
is the pion decay constant (in the chiral limit)
\beq\label{GBM}
\langle 0|A^a_\mu(0)|\pi^b(p)\rangle = i \delta^{ab} F p_\mu~,
\eeq
which is a fundamental mass scale of low-energy QCD.
In the world of massless quarks, the value of $F$ differs from the
physical value by terms proportional to the quark masses, to be
introduced later, $F_\pi = F [1 + {\cal O}({\cal M})]$. The physical
value of $F_\pi$ is $92.4\,$MeV, determined from pion decay, $\pi\to \nu\mu$. 

Of course, in QCD the quark masses are not exactly zero. The quark mass term leads
to the so-called {\em explicit chiral symmetry breaking}. Consequently, the
vector and axial-vector currents are no longer conserved (with the exception
of the baryon number current)
\beq\label{div}
\partial_\mu V_a^\mu = \dfrac{1}{2} i \bar q \, [{\cal M},\lambda_a ]\, q~, \quad
\partial_\mu A_a^\mu = \dfrac{1}{2} i \bar q \, \{{\cal M},\lambda_a \} \, \gamma_5\, q~.
\eeq
However, the consequences of the spontaneous symmetry violation  can still be
analyzed systematically because the quark masses are {\em small}. QCD
possesses what is called an approximate chiral symmetry. In that case, the mass spectrum
of the unperturbed Hamiltonian and the one including the quark masses can not be
significantly different. Stated differently, the effects of the explicit symmetry
breaking can be analyzed in perturbation theory. 
As a consequence, QCD has a remarkable mass gap
- the pions (and, to a lesser extent,
the kaons and the eta) are much lighter than all other hadrons.  To be more
specific, consider chiral $SU(2)$. The second formula of Eq.~(\ref{div}) is
nothing but a Ward-identity (WI) that relates the axial current $A^\mu =  \bar d
\gamma^\mu \gamma_5 u$ with the pseudoscalar density $P = \bar d i \gamma_5
u$,
\beq
\partial_\mu A^\mu = (m_u + m_d)\, P~.
\eeq
Taking on-shell pion matrix elements of this WI, one arrives at
\beq\label{pimass}
M_\pi^2 = (m_u + m_d) \frac{G_\pi}{F_\pi}~,
\eeq
where the coupling $G_\pi$ is given by $\langle 0 | P(0)| \pi(p)\rangle =
G_\pi $. This equation leads to some intriguing consequences: In
the chiral limit, the pion mass is exactly zero - in accordance with Goldstone's
theorem. More precisely, the ratio $G_\pi /F_\pi$ is a constant in the chiral
limit and the pion mass grows as $\sqrt{m_u+m_d}$ if the quark masses are
turned on. 

There is even further symmetry related to the quark mass term. It is observed
that hadrons appear in isospin multiplets, characterized by very tiny
splittings of the order of a few MeV. These are generated by the small
quark mass difference $m_u -m_d$ 
and also by electromagnetic effects of the
same size (with the notable exception of the charged to neutral pion mass
difference that is almost entirely of electromagnetic origin). This can be
made more precise: For $m_u = m_d$, QCD is  invariant under $SU(2)$ isospin 
transformations: $q \to q' = U q\,$,
with $U$ a unitary matrix.
In this limit, up and down quarks can not be disentangled as far as the
strong interactions are concerned.  Rewriting of the QCD quark mass term
allows to make the strong isospin violation explicit:
\beqa
{\cal H}_{\rm QCD}^{\rm SB} &=& m_u \,\bar u u + m_d
  \,\bar  d d \\
             &=& \dfrac{m_u+m_d}{2}(\bar u u + \bar d d)
              + \dfrac{m_u-m_d}{2}(\bar u u - \bar d d)~,
\nonumber
\eeqa
where the first (second) term is an isoscalar (isovector). Extending 
these considerations to $SU(3)$, one arrives at the eighfold way of
Gell-Mann and Ne'eman
that played a decisive role in our understanding
of the quark structure of the hadrons. The $SU(3)$ flavor symmetry
is also an approximate one, but the breaking is much stronger than it is
the case for isospin. From this, one can directly infer that the quark mass
difference $m_s - m_d$ must be much bigger than $m_d -m_u$.

The consequences of these broken symmetries can be analyzed systematically
in a suitably tailored effective field theory (EFT), as discussed in more
detail below. At this point, it is important to stress that the chiral symmetry
of QCD plays a crucial role in determining the longest ranged parts of the nuclear
force, which, as we will show, is given by Goldstone boson exchange between
two and more nucleons. This was already stressed long ago, see e.g
\cite{Brown:1970th} (and references therein) but only with the powerful
machinery of chiral effective field theory this connection could be
worked out model-independently, as we will show in what follows.

\subsection{Scales in nuclear physics}
\label{sec:scales}
\begin{figure}[tb]
\includegraphics*[width=5.0cm]{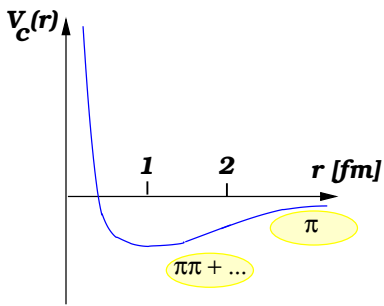}
\vspace{0.1cm}
\begin{center}
\caption{Schematic plot of the central nucleon-nucleon potential.
The longest range contribution is the one-pion-exchange, the intermediate
range attraction is described by two-pion exchanges and other shorter ranged
contributions. At even shorter distances, the NN interaction is strongly
repulsive.
\label{fig:Vc}}
\vspace{-0.5cm}
\end{center}
\end{figure}

\begin{figure}[tb]
\includegraphics*[width=6.0cm]{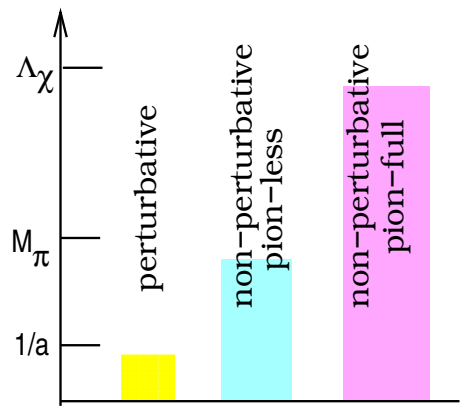}
\vspace{0.1cm}
\begin{center}
\caption{Scales in the two-nucleon problem and the range
of validity of the corresponding EFTs as explained in the
text. Here $\Lambda_\chi$ is the hard scale related to 
spontaneous chiral symmetry breaking, with  $\Lambda_\chi
\simeq M_\rho$, with $M_\rho = 770\,$MeV the mass of the
rho meson. 
\label{fig:nucscales}}
\vspace{-0.5cm}
\end{center}
\end{figure}

To appreciate the complexity related to a theoretical description
of the nuclear forces, it is most instructive to briefly discuss the
pertinent scales arising in this problem. This can most easily
be visualized by looking at the phenomenological central potential
between two nucleons, as it appears e.g. in meson-theoretical approaches
to the nuclear force, see Fig.~\ref{fig:Vc}. The longest range
part of the interaction is the one-pion exchange (OPE) that is firmly rooted
in QCD's chiral symmetry. Thus, the corresponding natural scale 
\noindent of the nuclear force problem is the Compton wavelength of
the pion
\beq
\lambda_\pi = 1 /M_\pi \simeq 1.5\,{\rm fm}~,
\eeq
where $M_\pi = 139.57\,$MeV is the charged pion mass. The central
intermediate range attraction is given by $2\pi$ exchange (and
shorter ranged physics). Finally, the wavefunctions of two nucleons do 
not like to overlap, which is reflected in a short-range repulsion that
can e.g. be modelled by vector meson exchange. From such considerations,
one would naively expect to be able to describe nuclear binding in
terms of energy scales of the order of the pion mass.  However, the true
binding energies of the nuclei are given by much smaller energy scales,
between 1 to 9 MeV per nucleon. Another measure for the shallow
nuclear binding is the
so called binding-momentum $\gamma$. In the deuteron, 
$\gamma = \sqrt{m B_D} \simeq 45\,{\rm MeV} \ll M_\pi$, with
$m = 938.2\,$MeV the nucleon mass and $B_D = 2.224\,$MeV the
deuteron binding energy. The small value of $\gamma$
signals the appearance of energy/momentum scales much below the pion mass. The
most dramatic reflection of the complexity of the nuclear force problem
are the values of  the S-wave neutron-proton scattering lengths,
\beq\label{aSTNN}
|a(^1S_0)| = 23.8\,{\rm fm} \gg 1/M_\pi\, , ~ 
a(^3S_1)   =  5.4\,{\rm fm} \gg 1/M_\pi\, .
\eeq 
Thus, to properly set up an effective field theory for the forces 
between two (or more) nucleons, it is mandatory to deal with these very 
different energy scales. If one were to treat the large S-wave scattering
lengths perturbatively, the range of the corresponding EFT would be
restricted to momenta below $p_{\rm max} \sim 1/|a(^1S_0)|\simeq 8\,$MeV.
To overcome this barrier, one must generate the small binding
energy scales by a non-perturbative resummation. This can e.g. be done
in a theory without explicit pion degrees of freedom, the so-called
pion-less EFT. In such an approach, the limiting hard scale is the
pion mass. To go further, one must include the pions explicitely,
as it is done in the pion-full or chiral nuclear EFT. The relation
between these different approaches is schematically displayed
in Fig.~\ref{fig:nucscales}.
\noindent 
A different and more formal argument that shows the
breakdown of a perturbative treatment of the EFT with 
two or more nucleons is related to the pinch singularities 
in the two-pion exchange diagram in the static limit as will be 
discussed later in the context of the
explicit construction of the chiral nuclear EFT.

In addition, if one extends the considerations to heavier nuclei
or even nuclear matter, the many-body system exhibits yet another
scale, the Fermi momentum $k_F$, with $k_F \simeq 2M_\pi$ at nuclear
matter saturation density. This new scale must be included in a 
properly modified EFT for the nuclear many-body problem which
is not a straightforward exercise as we will show below. It is therefore
not astonishing that the theory for heavier nuclei is still in a much
less developed stage that the one for the few-nucleon problem. These issues
will be taken up in Sec.~\ref{sec:many}.

For more extended discussions of scales in the nuclear force
problem and in nuclei, we refer to 
\cite{Friar:1995dt,Friar:1996zw,Kaiser:2006tu,Delfino:2007zu}.

\subsection{Conventional approaches to the nuclear force problem}
\label{sec:conventional}

Before discussing the application of the  effective field theory approach 
to the nuclear force problem, let us make a few comments on the highly
successful conventional approaches. First, we consider the two-nucleon
case. Historically, meson field theory and dispersion relations have 
laid the foundations for the construction of a two-nucleon potential.
All these approaches incorporate the long-range one-pion exchange
as proposed by Yukawa in 1935~\cite{Yukawa:1935xg} which nowadays is
firmly rooted in QCD. Dispersion relations can be used to construct the
two-pion exchange contribution to the nuclear force as pioneered at 
Paris~\cite{Cottingham:1973wt} and Stony Brook~\cite{Jackson:1975be}.
For a review, see e.g. \cite{Machleidt:2001rw}. In
the 1990ties, the so-called high-precision potentials have been
developed that fit the large basis of $pp$ and $np$ elastic scattering data
with a $\chi^2/{\rm datum} \simeq 1$. One of these is the so-called CD-Bonn
potential~\cite{Machleidt:2000ge} (which was developed at Moscow, Idaho).
Besides one-pion, $\rho$ and $\omega$
vector-meson exchanges, it contains two scalar--isoscalar
mesons in each partial wave up to angular momentum $J=5$ with the mass and
coupling constant of the second $\sigma$ fine-tuned in any partial wave.
The hadronic vertices are regulated with form factors with cut-offs
ranging from 1.3 to 1.7~GeV. Similarly, in  the
Nijmegen~I,II potentials one--pion exchange is supplemented by 
heavy boson exchanges with
adjustable parameters which are fitted for all (low) partial waves
separately~\cite{Stoks:1994wp}.
The Argonne V18 (AV18) potential starts from a very general operator
structure in coordinate space and has fit functions for all these
various operators~\cite{Wiringa:1994wb}. While these various potentials
give an accurate representation of the nucleon-nucleon phase shifts and
of most deuteron properties, the situation becomes much less satisfactory
when it comes to the much smaller but necessary three-nucleon forces.
Such three-body forces are needed to describe the nuclear binding
energies and levels, as most systematically shown by the
Urbana-Argonne group~\cite{Pieper:2001mp}. Systematic studies of 
the dynamics and reactions of systems with three or four-nucleons
further sharpen the case for the necessity of including three-nucleon
forces (3NFs), see e.g.~\cite{Gloeckle:1995jg}.  The archetype of a 3NF
is due to Fujita and Miyazawa (FM) \cite{Fujita:1957zz}, who extended Yukawa's
meson exchange idea by sandwiching the pion-nucleon scattering amplitude 
between nucleon lines, thus generating the 3NF of longest range.
In fact, the work of Fujita and Miyazawa has been the seed for many 
meson-theoretical approaches to the three-nucleon force like the families 
of Tucson-Melbourne \cite{TM1,Coon:1974vc},  Brazilian \cite{Coelho:1984hk} or 
Urbana-Illinois \cite{Pudliner:1997ck,Pieper:2001ap} 3NFs. 

While the conventional approach as briefly outlined here as enjoyed
many successes and is frequently used in e.g. nuclear structure and reaction
calculations, it remains incomplete as there are certain deficiencies
that can only be overcome based on EFT approaches. These are: (i) it is
very difficult - if not impossible - to assign a trustworthy theoretical
error, (ii) gauge and chiral symmetries are difficult to implement, (iii)
none of the three-nucleon forces is consistent with the underlying 
nucleon-nucleon interaction models/approaches and (iv) the connection
to QCD is not at all obvious. Still, as we will show later, there is
a very natural connection between these models and the forces derived 
from EFT by mapping the complicated physics of the short-distance
part of any interaction at length scales $\sim 1/M_\rho$ to the tower 
of multi-fermion contact interactions that naturally arise in the EFT description
(see Sec.~\ref{sec:chiralEFT}).

\subsection{Brief introduction to effective field theory}
\label{sec:EFTintro}

\begin{figure}[tb]
\includegraphics*[width=6.6cm]{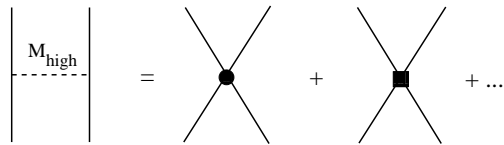}
\vspace{0.1cm}
\begin{center}
\caption{Expansion of a heavy-particle exchange
diagram in terms of local light-particle operators.
The solid and dashed lines denote light and heavy
particles, respectively. The filled circle and square
denote insertions with zero and two derivatives, in order.
The ellipses stands for operators with more derivatives.
\label{fig:resosat}}
\vspace{-0.5cm}
\end{center}
\end{figure}

Effective field theory (EFT) is a general approach to calculate the low-energy
behavior of physical systems by exploiting a separation of scales in
the system (for reviews see e.g. \cite{Georgi:1994qn,Manohar:1996cq,Burgess:2007pt}).
Its roots can be traced to the renormalization group 
\cite{Wilson-83} and the intuitive understanding of ultraviolet divergences
in quantum field theory \cite{Lepage-89}. A succinct formulation of the 
underlying principle was given by Weinberg \cite{Weinberg:1978kz}:
If one starts from the most general Lagrangian consistent with all 
symmetries of the underlying interaction, one will get the most general
S-matrix consistent with these symmetries. Together with a power counting 
scheme that specifies which terms are required at a desired accuracy
leads to a predictive paradigm for a low-energy theory.
The expansion is typically in powers of a low-momentum scale $\Mlo$
which can be the typical external momentum
over a high-momentum scale $\Mhi$. However, what physical scales 
$\Mhi$ and $\Mlo$ are identified with depends on the considered
system.  In its most simple setting, consider a theory that is made of two
particle species, the light and the heavy
ones with $\Mlo \ll \Mhi$. Consider now soft processes
in which the energies and momenta are of the order of the 
light particle mass (the so-called soft scale). 
Under such conditions, the short-distance
physics related to the heavy particles can never be resolved.
However, it can be represented by light-particle contact interactions 
with increasing dimension (number of derivatives). Consider
e.g. heavy particle exchange between light ones in the limit
that $\Mhi \to \infty$ while keeping the ratio $g/\Mhi$ fixed, with
$g$ the light-heavy coupling constant. As depicted in
Fig.~\ref{fig:resosat}, one can represent such exchange diagrams by a 
sum of local operators of the light fields with increasing
number of derivatives. In a highly symbolic notation
\beq
\frac{g^2}{\Mhi^2 - t} = \frac{g^2}{\Mhi^2} + 
\frac{g^2 \, t}{\Mhi^4} + \ldots~,
\eeq
with $t$ the squared invariant momentum transfer.
\noindent In many cases, the corresponding high-energy theory
is not known. Still, the framework of EFT offers a predictive
and systematic framework for performing calculations in the
light particle sector. Denote by ${\cal Q}$ a typical energy
or momentum of the order of $\Mlo$ and by $\Lambda$ the
hard scale where the EFT will break down. In many cases, this
scale is set by the masses of the heavy particles not considered
explicitely. In such a setting, any matrix element or Greens
function admits an expansion in the small parameter ${\cal Q}/
\Lambda$ \cite{Weinberg:1978kz} 
\beq
{\cal M} = \sum\limits_{\nu} \left(\frac{\cal Q}{\Lambda}\right)^{\nu}  
\,{\cal F} \left(\frac{{\cal Q}}{\mu}, g_i\right)
\eeq
where ${\cal F}$ is a function of order one (naturalness), $\mu$
a regularization scale (related to the UV divergences appearing
in the loop graphs) and the $g_i$ denotes a collection of
coupling constants, often called low-energy constants (LECs).
These parameterize (encode) the unknown high-energy (short-distance)
physics and must be determined by a fit to  data (or
can be directly calculated if the corresponding high-energy theory
is known/can be solved). The counting index $\nu$ in general depends 
on the fields in the effective theory, the number of derivatives and
the number of loops. This defines the so-called power counting 
which allows to categorize all contributions to any matrix element
at a given order.  It is important to stress that $\nu$ must be
bounded from below to define a sensible EFT. In QCD e.g. this is 
a consequence of the spontaneous breaking of its chiral symmetry. 
The contributions with the lowest possible value of $\nu$ 
define the so-called leading order (LO) contribution, the first
corrections with the second smallest allowed value of $\nu$ the
next-to-leading order (NLO) terms and so on. In contrast to more
conventional perturbation theory, the small parameter is not a 
coupling constant (like, e.g., in Quantum Electrodynamcis) but 
rather one expands in small energies or momentum, where small 
refers to  the hard scale $\Lambda$. The archetype of such
a perturbative EFT is chiral perturbation theory that exploits the
strictures of the spontaneous and explicit chiral symmetry breaking
in QCD \cite{Gasser:1983yg,Gasser:1984gg}.  Here, the light degrees
of freedom are the pions, that are generated through the symmetry violation.
Heavier particles like e.g. vector mesons only appear indirectly
as they generate local four-pion interactions with four, six, $\ldots$
derivatives. For a recent review, see Ref.~\cite{Bernard:2006gx}.  
Of course, the pions also couple to heavy matter fields like e.g. 
nucleons, that can also be included in CHPT, as reviewed by Bernard
\cite{Bernard:2007zu}.

So far, we have made the implicit assumption of naturalness,
which implies e.g. that the scattering length $a$  
is of natural size as e.g. in CHPT, where the
scale is set by $1/\Lambda_\chi \simeq 1\, {\rm GeV}^{-1}\simeq 0.2\,$fm.
This also implies that there are no bound states close to the 
scattering thresholds. In many physical systems and of particular 
interest here, especially in the two-nucleon system, this is not the case, 
but one rather has to deal with unnaturally large scattering lengths (and
also shallow bound states). To be specific, let us consider nucleon-nucleon
scattering at very low energies in the $^1S_0$ channel, cf. Eq.~(\ref{aSTNN}).
For such low energies, even the pions can be considered heavy and are
thus integrated out. To construct an EFT that is applicable for momenta 
$p > 1/a$, one must retain all terms $a p \sim 1$ in the scattering matrix.  
This requires  a non-perturbative resummation and is most elegantly done
in the power divergence scheme of Kaplan, Savage and 
Wise~\cite{Kaplan:1998tg,Kaplan:1998we}. 
This amounts to summing the leading four-nucleon 
contact term $\sim C_0 (\psi^\dagger \psi)^2$
to all orders in $C_0$ and matching the scale-dependent LEC
$C_0$ to the scattering length. This leads to the T-matrix
\beq
T = \frac{4\pi}{m} \frac{1}{1/a + ip} \left[ 1 + {\cal O}(p^2)\right]~,
\eeq
where the expansion around the large scattering length is made explicit.
All other effects, like e.g. effective range corrections, are treated
perturbatively. This compact and elegant  scheme is, however, not
sufficient for discussing nuclear processes with momenta $p \ge M_\pi$.
We will come back to this topic when we give the explicit construction
of the chiral nuclear EFT in sec.~\ref{sec:chiralEFT}. It is important to 
stress that such EFTs with unnaturally large scattering length can exhibit
universal phenomena that can be observed in physical systems which differ
in their typical energy scale by many orders of magnitude, for a review
see Braaten and Hammer~\cite{Braaten:2004rn}.
We also remark that there are many subtleties in constructing a proper EFT, 
but space forbids to discuss these here. Whenever appropriate and/or
neccessary, we will mention these in the following sections and provide
explicit references.

\subsection{First results from lattice QCD}
\label{sec:lattice}

Lattice QCD (LQCD) is a promising tool to calculate hadron properties ab
initio from the QCD Lagrangian on a discretized Euclidean space-time.
This requires state-of-the-art high performance computers and refined
algorithms to analyse the QCD partition function by Monte Carlo methods.
Only recently soft- and hardware developments have become available
that allow for full QCD simulations at small enough quark masses
(corresponding to pion masses below 300~MeV), large enough volumes
(corresponding to spatial dimensions larger than 2.5~fm) and
sufficiently fine lattice spacing ($a \simeq 0.05\,$fm) so that the
results are not heavily polluted by computational artefacts and
can really be connected to the physical quark masses by sensible
chiral extrapolations.

For the nuclear force problem, there are two main developments
in LQCD to be reviewed here. These concern the extraction of hadron-hadron
scattering lengths from unquenched simulations and the first attempts
to construct a nuclear potential. These are groundbreaking studies,
but clearly at present one has not yet achieved an accuracy to obtain
high-precision predictions for nuclear properties. We look very much 
forward to the development of these approaches in the years to come.

The first exploratory study of the nucleon-nucleon scattering lengths
goes back to Fukugita et al.~\cite{Fukugita:1994na,Fukugita:1994ve}
in the quenched approximation. They make use of an elegant formuala,
frequently called the ``L\"uscher formula'',
that relates the S-wave scattering length $a_0$ between two hadrons $h_1$
and $h_2$ to the energy shift $\delta E = E_{h_1h_2} - (m_1+m_2)$ of the
two-hadron state at zero relative momentum  confined in spatial box of
size $L^3$. It is given by \cite{Hamber:1983vu,Luscher:1986pf,Luscher:1990ux}
\beq\label{eq:luescher}
\delta E = -\frac{2\pi a_0}{\mu L^3} \left[ 1 + c_1 \frac{a_0}{L}
 + c_2 \frac{a_0^2}{L^2}\right] + {\cal O}(L^{-6})~,
\eeq
with $\mu = m_1 m_2/(m_1+m_2)$ the reduced mass and 
$c_1 = 2.837297$ and $c_2 = 6.375183$.
A generalization of this formalism was given by Beane et al.\cite{Beane:2003da}
utilizing methods developed for the so-called pionless nuclear EFT (EFT
with contact interactions, for a review see e.g. \cite{Bedaque:2002mn}).
It reads
\beqa\label{eq:Sbeane}
p \, \cot\delta_0(p) &=& \frac{1}{\pi L} {\cal S}\left((Lp/2\pi)^2\right)~,
\nonumber\\
{\cal S}(\eta) &=& \sum_{\vec j}^{\Lambda_j} \, \frac{1}{|\vec j\,|^2 - \eta}
- 4\pi \Lambda_j~,
\eeqa
which gives the location of all energy eigenstates in the box. Here,
$\delta_0$ is the S-wave phase shift. The sum over
all three-vectors of integers ${\vec j}$ is such that $|{\vec j}\,| < \Lambda_j$
and the limit $\Lambda_j \to \infty$ is implicit. In the limit $L \gg |a_0|$,
Eq.~(\ref{eq:Sbeane}) reduces to the L\"uscher formula,
Eq.~(\ref{eq:luescher}). On the other hand, for large scattering length,
$|(p \cot\delta_0)^{-1} | \gg L$, the energy of the lowest state is given 
by 
\beq
E_0 = \frac{4\pi^2}{\mu L^2} \left[ d_1 + d_2 Lp\cot\delta_0+ \ldots\right] ~,
\eeq
with $d_1 = 0.472895$, $d_2 = 0.0790234$ and $p \cot\delta_0$ is evaluated at
the energy $E = 4\pi^2 d_1/(\mu L^2)$. Within this framework, in 
Ref.~\cite{Beane:2006mx} the first fully dynamical simulation of the
neutron-proton scattering lengths was performed, with a lowest pion mass of 354~MeV.
This mass is still too high to perform a precise chiral extrapolation
to the physical pion mass, but this calculation clearly demonstrates the
feasilbilty of this approach (see also sec.~\ref{sec:chiralextrap}). 
This scheme can also be extended to hyperon-nucleon interactions, 
see \cite{Beane:2003yx}. A first signal  for $\pi \Lambda$  and 
$n \Sigma^-$ scattering was reported in Ref.~\cite{Beane:2006gf}.
For a recent review on these activities of the NPLQCD collaboration, 
we refer the reader to~\cite{Beane:2008dv}.

\begin{figure}[tb]
\includegraphics*[width=7.5cm]{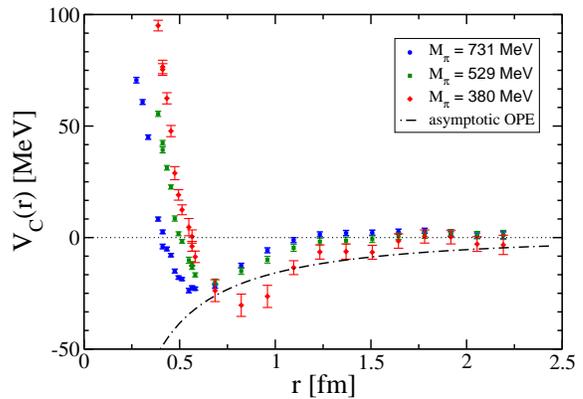}
\begin{center}
\caption{Effective potential in the $^1S_0$ channel for three different
quark masses in quenched LQCD after Ref.~\protect\cite{Ishii:2007xz}.
The dashed line is the asymptotic OPEP for $M_\pi =380\,$MeV, $m =
1.2\,$GeV and $g_{þi N}^2/(4\pi) = 14.0$.
We are grateful to Dr. N.~Ishii for providing us with the data.
\label{fig:Vlatt}}
\vspace{-0.5cm}
\end{center}
\end{figure}

Another interesting development was initiated and carried out 
by Aoki, Hatsuda and Ishii \cite{Ishii:2006ec}. They have 
generalized the two-center Bethe-Salpeter wavefunction approach 
of the CP-PACS collaboration~\cite{Aoki:2005uf}, which
offers an alternative to L\"uscher's formula, to the 
two-nucleon (NN) system.
Given an interpolating field for the neutron and for the proton, the
NN potential can be defined from the properly reduced 6-quark Bethe-Salpeter
amplitude $\phi (\vec r\,)$. The resulting Lippmann-Schwinger equation 
defines a non-local  potential for a given, 
fixed separation $r = |{\vec r}|$. Performing a derivative expansion, the
central potential $V_c (r) $ at a given energy $E$ is extracted from
\beq\label{Vlatt}
V_C(r) = E + \frac{1}{m}\frac{\vec\nabla^2 \phi(r)}{\phi (r)}~.
\eeq
Monte Carlo Simulations are then performed to generate the 6-quark 
Bethe-Salpeter
amplitude in a given spin and isospin state of the two-nucleon system on
a large lattice, $V = (4.4\,{\rm fm})^4$ in the quenched approximation for
pion masses between 380 and 730~MeV. Despite these approximations, the 
resulting effective potential extracted using Eq.~(\ref{Vlatt}) shares the
features of the phenomenological NN potentials - a hard core (repulsion)
at small separation surrounded by an attractive well at intermediate and
larger distances, see Fig.~\ref{fig:Vlatt}.
\noindent  Furthermore, the asymptotic form of this potential has
exactly the form of the OPE, provided one rescales the formula
\beq
V_C^{\rm OPE} (r) = \frac{g_{\pi N}^2}{4\pi}\,\fet\tau_1\cdot\fet\tau_2
\frac{\vec\sigma_1 \cdot \vec\sigma_2}{3}\left( \frac{M_\pi}{2m}\right)^2
\frac{{\rm e}^{-M_\pi r}}{r}
\eeq   
with the pion and nucleon masses used in the simulations but keeping
the pion-nucleon coupling at its physical value, $g_{\pi N}^2/(4\pi) \simeq
14.0$. These interesting results 
have led to some enthusiastic appraisal,
see e.g. \cite{Wilczek:2007}. However, it is important to stress that 
the so-calculated potential is not unique, especially its properties at
short distances, since it depends on the definition of the interpolating
nucleon fields. Furthermore, the quenched approximation is known to have
uncontrolled systematic uncertainties as it does not even define a quantum
field theory. In this context, the authors of Ref.~\cite{Aoki:2005uf} 
report on the numerical absence of the large-distance-dominating 
$\eta$-exchange from the  flavor-singlet hair-pin diagram. Still, 
one would like to see this promising calculation repeated with dynamical 
quarks of sufficiently small masses. 
In Ref.~\cite{Nemura:2008sp} this framework was used to study the $\Xi N$
interaction. Interestingly, the central potential of the $p\Xi^0$ interaction
looks very similar to the central $np$ potential. It would be interesting 
to extend these calculations to other hyperon-nucleon channels and also
study the effects of SU(3) symmetry breaking. We will come back to these
issues in the context of an three-flavor chiral EFT in sec.~\ref{sec:YN}.
For recent developments in this scheme concerning the calculation of the 
tensor force, the energy dependence of the NN potential and  preliminary
results for full QCD (2+1 flavors), see the talks by  Aoki, Ishii, and
Nemura at the Lattice 2008 conference \cite{Latt08conf}.

\subsection{Observables and not-so observable quantities}

There is an extensive literature, primarily from the sixties and seventies,
on the role of off-shell physics in nuclear phenomena
(see, e.g., Ref.~\cite{SRIVASTAVA75} and references therein).
This includes
not only few-body systems (e.g., the triton) and nuclear matter,
but interactions of two-body systems with external probes,
such as nucleon-nucleon bremstrahlung and the electromagnetic form
factors of the deuteron.
The implicit premise was that there is
a true underlying potential governing the nucleon-nucleon force,
so that its off-shell properties can be determined.
Indeed,
the nuclear many-body problem has traditionally been posed as
finding approximate solutions to the many-particle Schr\"odinger
equation, given a fundamental two-body interaction that reproduces
two-nucleon observables.

In contrast, effective field theories {\em are\/} determined
completely by on-energy-shell information, up to a well-defined
truncation error. In writing down the most general Lagrangian
consistent with the symmetries of the underlying theory,
many-body forces arise naturally. Even though they are
usually suppressed at low energies, they enter at some order
in the EFT expansion. These many-body forces have to be determined
from many-body data. The key point is, however, that no
off-energy-shell information is needed or experimentally accessible.
A fundamental theorem of quantum field theory states that
physical observables (or more precisely, S-matrix elements) are
independent of the choice of interpolating fields that appear in
a Lagrangian \cite{Haag58,COLEMAN69}.
Equivalently, observables are invariant
under a change
of field variables in a field theory Lagrangian (or Hamiltonian):
\beq
\psi(x) \to \psi(x) + \eta P[\psi]\,,
\eeq
where $P[\psi]$ is a local polynomial of the field $\psi$ and its derivatives
and $\eta$ is an arbitray counting parameter.
Newly generated contributions to observables have to cancel separately at
each order in $\eta$. 
This ``equivalence theorem'' holds for renormalized field theories.
In an EFT, one exploits the invariance under field redefinitions
to eliminate redundant terms in the effective Lagrangian and to choose
the most convenient or efficient form for practical
calculations \cite{POLITZER80,GEORGI91,KILIAN94,SCHERER95a,ARZT95}.
Since off-shell Green's functions and the corresponding off-shell amplitudes
{\it do\/} change under field redefinitions, one must conclude that off-shell
properties are unobservable.

Several recent works have emphasized from a field theory
point of view the impossibility of observing off-shell effects.
In Refs.~\cite{FEARING98,FEARING99},
model calculations were used to illustrate how apparent determinations of
the two-nucleon off-shell T-matrix in
nucleon-nucleon bremstrahlung are illusory, since field redefinitions
shift contributions between off-shell contributions and contact interactions.
Similarly, it was shown in Ref. \cite{SCHERER95b} that Compton scattering
on a pion cannot be used to extract information on the off-shell
behavior of the pion form factor.
The authors of Refs.~\cite{FHK99,CFM96} emphasized the nonuniqueness
of chiral Lagrangians for three-nucleon forces and pion production.
Field redefinitions lead to different off-shell forms that yield the same
observables within a consistent power counting.
In Ref.~\cite{KSW99}, an interaction proportional to the equation of
motion is shown to have no observable consequence
for the deuteron electromagnetic form factor, even though it contributes
to the off-shell T-matrix.

In systems with more than two nucleons, one can trade off-shell, two-body
interactions for many-body forces.
This explains how two-body interactions related by unitary transformations
can predict different binding energies for the triton \cite{ARNAN73}
if many-body forces are not consistently included.
These issues were discussed from the viewpoint of unitary transformations
in Refs.~\cite{POLYZOU90} and \cite{AMGHAR95}.
The extension to many-fermion systems in the thermodynamic limit was
considered in \cite{Furnstahl:2000we}. The effects of field redefinitions 
were
illustrated using the EFT for the dilute Fermi gas \cite{Hammer:2000xg}. 
If many-body interactions generated by the field redefinitions are neglected,
a Coester line similar to the one observed for
nuclear matter \cite{COESTER70} is generated. 
Moreover, the connection to more traditional treatments using unitary 
transformations was elucidated.
The question of whether occupation numbers and momentum distributions of 
nucleons in nuclei are observables 
was investigated in Ref.~\cite{Furnstahl:2001xq}. 
Field redefinitions lead to variations in the occupation numbers
and momentum distributions that imply the answer is negative.
The natural size of the inherent ambiguity (or scheme dependence)
in these quantities is determined by the applicability of the 
impulse approximation. Only if the impulse approximation is well justified,
the ambiguity is small and these quantities are approximately scheme
independent. This has important  implications for the interpretation 
of ($e$,$e'p$) experiments with nuclei. Whether the stark difference in 
occupation numbers between nonrelativistic and relativistic Brueckner
calculations can be explained by this ambiguity is another interesting
question \cite{Jaminon:1990zz}.

\section{EFT for Few-Nucleon Systems: Foundations and Applications}
\def\theequation{\arabic{section}.\arabic{equation}}
\setcounter{equation}{0}
\label{sec:EFT}

\subsection{EFT with contact interactions and universal aspects}
\label{sec:EFTfoundations}

In nuclear physics, there are a number of EFTs which are all useful 
for a certain range of systems (cf. Fig.~\ref{fig:nucscales}). 
The simplest theories
include only short range interactions and even integrate out the pions.
At extremely low energies,
$\Mhi$ is given by the NN scattering lengths and one can formulate
a perturbative EFT in powers of of the typical momentum $k$ divided
by $\Mhi$. Since the NN scattering lengths are large this theory has
a very limited range of applicability. 
It is therefore useful to construct another EFT with short-range 
interactions that resums the interactions generating the large 
scattering length. This so-called pionless EFT 
can be understood as an expansion
around the limit of infinite scattering length or equivalently 
around threshold bound states. 
Its breakdown scale is set by one-pion exchange, $\Mhi\sim M_\pi$, while
$\Mlo \sim 1/a \sim k$.
For momenta $k$ of the order of the pion mass $M_\pi$, pion exchange becomes
a long-range interaction and has to be treated explicitly. This 
leads to the chiral EFT whose breakdown scale $\Mhi$ is set by the chiral
symmetry breaking scale $\Lambda_\chi$ and will be discussed in detail below. 

The pionless theory relies only on the large scattering length and 
is independent of the mechanism responsible for it. 
It is very general and can be applied in systems ranging from
ultracold atoms to nuclear and particle physics. It is therefore 
ideally suited to unravel universal phenomena driven by the 
large scattering length such as limit cycle physics 
\cite{Mohr:2005pv,Braaten:2003eu} and the Efimov effect \cite{Efimov-70}.
For recent reviews of applications to the physics of ultracold 
atoms, see Refs.~\cite{Braaten:2004rn,Braaten:2006vd}. 
Here we consider applications of this theory in nuclear physics.  

The pionless EFT is designed to reproduce the well known effective range 
expansion. The leading order Lagrangian can be written as:
\beqa
{\cal L}&=&N^\dagger \left(i\partial_0 +\frac{\vec{\nabla}^2}{2m}\right)N
\nonumber \\
&-&C_0^t\left(N^T \tau_2 \sigma_i \sigma_2 N\right)^\dagger\left(
N^T \tau_2 \sigma_i \sigma_2 N\right) 
\nonumber \\
&- &C_0^s\left(N^T \sigma_2 \tau_a \tau_2 N\right)^\dagger\left(
N^T \sigma_2 \tau_a \tau_2 N\right) +...\,,
\label{lagN}
\eeqa
where the dots represent higher-order terms suppressed by derivatives
and more nucleon fields. The Pauli matrices
$\sigma_i$ $(\tau_a)$ operate in spin (isospin) space, respectively.
The contact terms proportional to $C_0^t$ ($C_0^s$) correspond to 
two-nucleon interactions in the $\trip$ ($\sing$) NN channels.
Their renormalized values are related to the 
corresponding large scattering lengths $a_t$ and $a_s$
in the spin-triplet and spin-singlet channels, respectively.
The exact relation, of course, depends on the renormalization scheme.
Various schemes can be used, such as a momentum cutoff or dimensional 
regularization. Convenient schemes that have a manifest power counting
at the level of individual diagrams are dimensional regularization
with PDS subtraction, where poles in 2 and 3 spatial dimensions are subtracted
\cite{Kaplan:1998we}, or momentum subtractions schemes as in 
\cite{Gegelia:1998xr}. However, a simple momentum cutoff can be used as
well.

\begin{figure}[tb]
\bigskip
\centerline{\includegraphics*[width=8cm,angle=0]{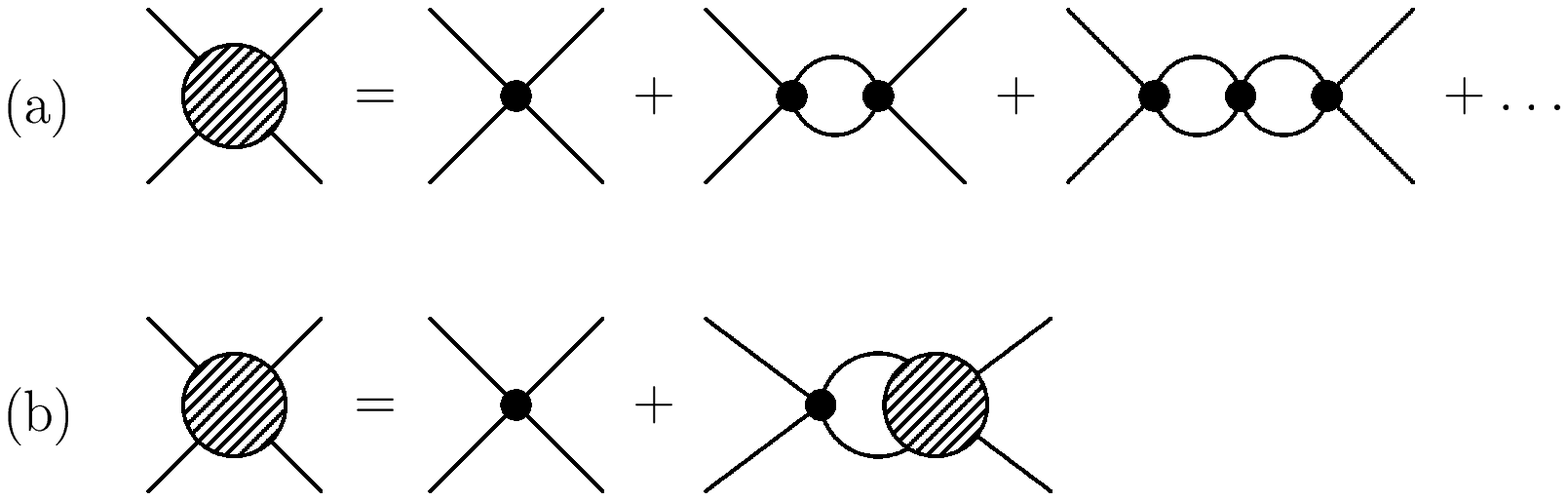}}
\caption{The bubble diagrams with the contact interaction 
$C_0^t$ or $C_0^s$ contributing to the two-nucleon scattering 
amplitude.}
\label{fig:bubble}
\end{figure}

\begin{figure}[b]
\bigskip
\centerline{\includegraphics*[width=8cm,angle=0]{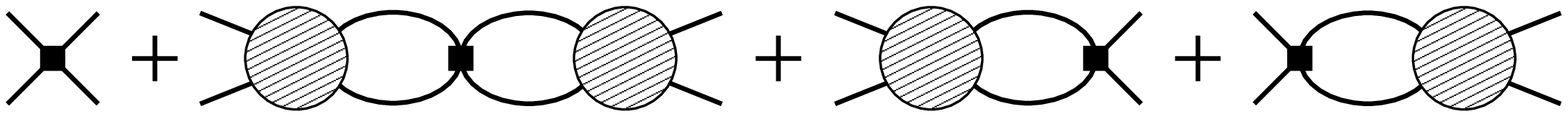}}
\caption{Diagrams for the inclusion of higher order contact interactions.}
\label{fig:bubble2}
\end{figure}

\begin{figure}[tb]
\centerline{\includegraphics*[width=8cm,angle=0,clip=true]{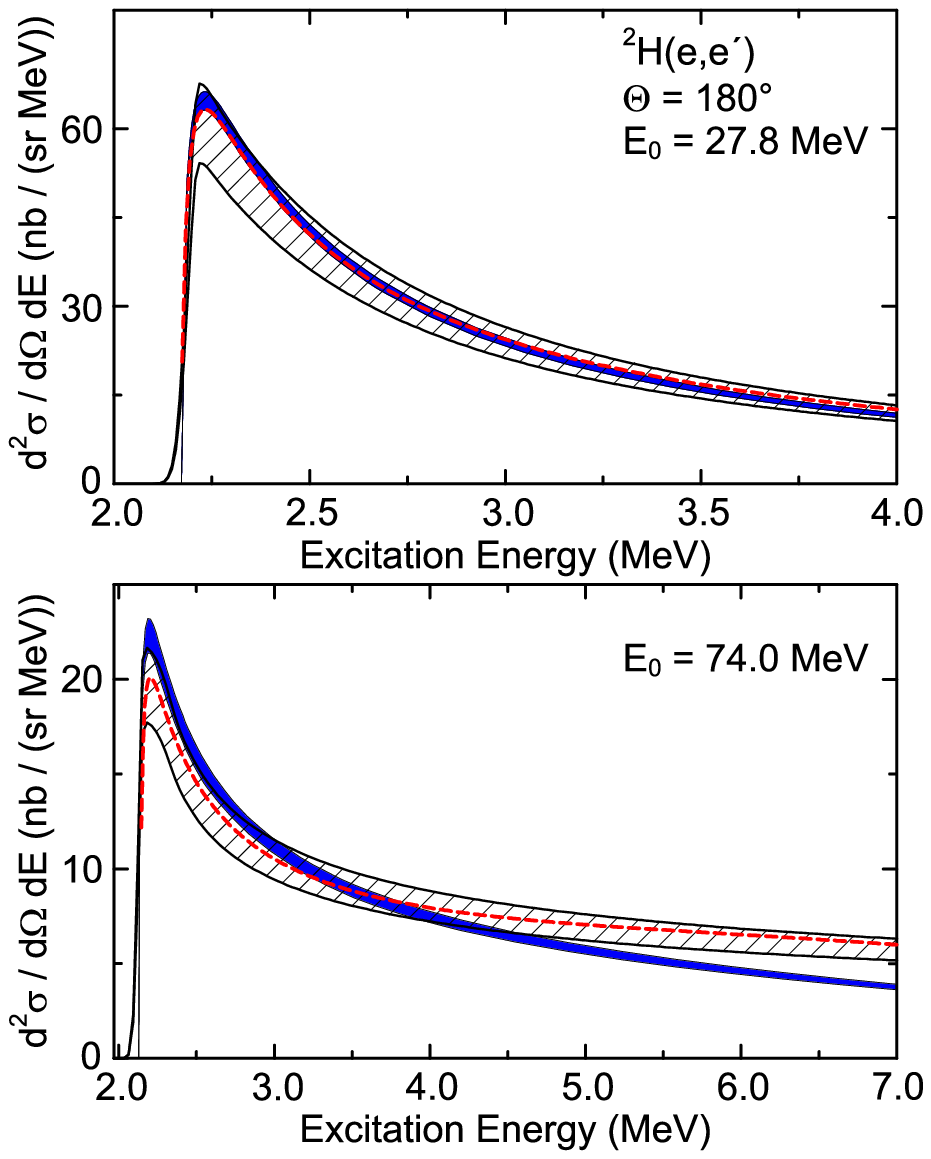}}
\caption{Double-differential cross sections of the $^2$H$(e,e')$ reaction
with errors (hatched bands) extracted from the experiment. The gray
bands and dashed lines are calculations in pionless EFT and a potential 
model. Figure courtesy of H.W.~Grie\ss hammer. 
}
\label{fig:dbreakup}
\end{figure}

Since the scattering lengths are set by the low-momentum scale $a\sim 1/\Mlo$,
the leading contact interactions
have to be resummed to all orders \cite{Kaplan:1998we,vanKolck:1998bw}.
The nucleon-nucleon scattering amplitude in the $\trip$ ($\sing$)
channels is obtained by summing the so-called bubble diagrams with the 
$C_0^t$ ($C_0^s$) interactions shown in Fig.~\ref{fig:bubble}.
This summation gives the exact solution of the Lippmann-Schwinger equation
for the $C_0^t$ or $C_0^s$ interactions.
Higher order derivative terms which are not shown explicitly
in Eq.~(\ref{lagN})
reproduce higher order terms in the effective range expansion. Since
these terms are natural and their size is set by $\Mhi$, their
contribution at low energies is suppressed by powers of $\Mlo /\Mhi$
and can be treated in perturbation theory. The subleading correction
is given by the effective range $r_0 \sim 1/\Mhi$ and the corresponding
diagrams are illustrated in Fig.~\ref{fig:bubble2}.
The renormalized S-wave scattering amplitude to next-to-leading 
order in a given channel then takes the form
\beqa
T_2 (k) &=& \frac{4\pi}{m} \frac{1}{-1/a-ik} \left[
1-\frac{r_0 k^2/2}{-1/a-ik}+\ldots \right]\,,
\eeqa
where $k$ is the relative momentum of the nucleons and the dots 
indicate corrections of order $(\Mlo /\Mhi)^2$ for typical
momenta $k\sim\Mlo$.
The pionless EFT becomes very useful in the two-nucleon sector when 
external currents are considered and
has been applied to  a variety of electroweak processes.
These calculations are reviewed in detail in
Refs.~\cite{Beane:2000fx,Bedaque:2002mn}. 
More recently Christlmeier and Grie\ss hammer have 
calculated  low-energy deuteron 
electrodisintegration in the framework of the pionless EFT
\cite{Christlmeier:2008ye}. For the double differential cross
sections of the $d(e,e')$ reaction at $\theta=180^\circ$ excellent 
agreement was found with a recent experiment at S-DALINAC
\cite{Ryezayeva:2008zz}.\footnote{However, there is a 
disagreement between theory and data for the small longitudinal-transverse 
interference contribution $\sigma_{LT}$ that is currently not understood
\cite{Christlmeier:2008ye}.}
The double-differential cross section for
an incident electron energy  $E_0 = 27.8$ MeV  and $\theta=180^\circ$
is shown in Fig.~\ref{fig:dbreakup}.
The data were used to precisely map the $M1$ response which governs the 
reaction $np\to d\gamma$ relevant to big-bang nucleosythesis.
Finally, the reaction $pp \to pp\pi^0$ near threshold
was studied by Ando \cite{Ando:2007in}.

We now proceed to the three-nucleon system.
Here it is convenient (but not mandatory) to rewrite the 
theory using so-called \lq\lq dimeron'' 
auxilliary fields \cite{Kaplan:1996nv}.
We need two dimeron fields, one for each S-wave channel:
(i) a field $t_i$ with
spin (isospin) 1 (0) representing two nucleons interacting in the $^3 S_1$
channel (the deuteron) and
(ii) a field $s_a$ with
spin (isospin) 0 (1) representing two nucleons interacting in the $^1 S_0$
channel \cite{Bedaque:1999ve}:
\beqa
{\cal L}&=&N^\dagger \Big(i\partial_t +\frac{\vec{\nabla}^2}
{2m}\Big)N - t^\dagger_i \Big(i\partial_t -\frac{\vec{\nabla}^2}
{4m} -\Delta_t \Big) t_i \nonumber\\
&-& s^\dagger_a \Big(i\partial_t -\frac{\vec{\nabla}^2}
{4m} -\Delta_s \Big) s_a
- \frac{g_t}{2}\Big( t^\dagger_i N^T \tau_2 \sigma_i
\sigma_2 N +h.c.\Big) 
\nonumber \\
&-&\frac{g_s}{2}\Big(s^\dagger_a N^T \sigma_2 \tau_a \tau_2 N +h.c.\Big) 
-G_3 N^\dagger \Big[ g_t^2
(t_i \sigma_i)^\dagger (t_j\sigma_j) \nonumber \\
&+& \frac{g_t g_s}{3} \left( (t_i\sigma_i)^\dagger
(s_a \tau_a) + h.c. \right) 
+ g_s^2 (s_a \tau_a)^\dagger (s_b \tau_b) \Big] N\nonumber \\
&+&\ldots\,,
\label{lagd}
\eeqa
where $i,j$ are spin and $a,b$ are isospin indices
while  $g_t$, $g_s$, $\Delta_t$, $\Delta_s$
and $G_3$ are the bare coupling constants.
This Lagrangian goes beyond leading order and
already includes the effective range terms. The coupling
constants $g_t$, $\Delta_t$, $g_s$, $\Delta_s$ are matched to the
scattering lengths $a_\alpha$ and effective ranges $r_{0\alpha}$
in the two channels ($\alpha=s,t$). Alternatively, one can 
match to the position of the 
bound state/virtual state pole $\gamma_\alpha$ 
in the $T$-matrix instead of the 
scattering length which often improves convergence \cite{Phillips:1999hh}.
The two quantities are related through:
\beq
\gamma_\alpha=\frac{1}{r_{0\alpha}}\left(1-\sqrt{1-2r_{0\alpha}/a_\alpha}
\right)\,,
\eeq
where $\alpha=s,t$. 
The term proportional to $G_3$ constitutes a Wigner-$SU(4)$
symmetric three-body interaction. It only contributes in the 
spin-doublet S-wave channel.
When the auxilliary dimeron fields $t_i$
and $s_a$ are integrated out, an equivalent form containing only nucleon 
fields is obtained.
At leading order when the effective range corrections are neglected, 
the spatial and time derivatives acting on the dimeron
fields are omitted and the field is static. 
The coupling constants $g_\alpha$ and $\Delta_\alpha$,
$\alpha=s,t$ are then not
independent and only the combination $g_\alpha^2/\Delta_\alpha$ enters in 
observables. This combination can then be matched to the scattering length
or pole position.

\begin{figure}[tb]
\bigskip
\centerline{\includegraphics*[width=7cm,angle=0]{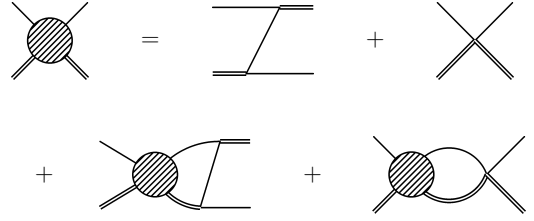}}
\caption{The integral equation for the boson-dimeron scattering amplitude.
The single (double) line indicates the boson (dimeron)
propagator.}
\label{fig:ineq}
\end{figure}

The simplest three-body process to consider is neutron-deuteron scattering
below the breakup threshold.
In order to focus on the main aspects of renormalization, we
suppress all spin-isospin  indices and complications from coupled channels
in the three-nucleon problem.  This corresponds to
a system of three spinless bosons with large scattering length.
If the scattering length is positive, the bosons form a two-body
bound state analog to the deuteron which we call dimeron. 
The leading order integral equation for boson-dimeron scattering is shown
schematically in Fig.~\ref{fig:ineq}. 
For total orbital angular momentum $L=0$, it takes the following form:
\beqa
T_3(k, p; \, E) &=& \frac{16}{3 a} M (k, p; \, E) + \frac{4}{\pi} 
\int_0^\Lambda dq\, q^2\, 
T_3(k, q; \, E) \nonumber \\
&&\times \, \frac{ M (q, p; \, E)}{- 1/a + \sqrt{3 q^2/4 - m E - i \epsilon}} 
\,, 
\label{STM}
\eeqa
where the inhomogeneous term reads
\beq
M (k, p; \, E) = \frac{1}{2 k p} \ln 
\left( \frac{k^2 + k p + p^2 - m E}{k^2 - k p + p^2 - m E} \right)
+ \frac{H (\Lambda )}{\Lambda^2}\,.
\eeq
Here, $H$ determines  the strength of the three-body force 
$G_3(\Lambda)=2mH(\Lambda)/\Lambda^2$ which enters already at 
leading order and $\Lambda$ is 
a UV cutoff introduced to regularize the integral equation.
The magnitude of the incoming (outgoing) relative momenta is
$k$ ($p$) and $E = 3 k^2/(4 m) - 1/(ma^2)$. 
The on-shell point corresponds to $k = p$ and 
the phase shift can be obtained via
$k \cot \delta = 1/T_3(k, k; \, E)+ik$.
For $H=0$ and $\Lambda \to \infty$,  Eq.~(\ref{STM}) reduces to the
STM equation first derived by Skorniakov and Ter-Martirosian
\cite{Skorniakov:1957aa}. It is well known that the STM equation has 
no unique solution \cite{Danilov:1961aa}.
The regularized equation has a unique solution for any given (finite) value 
of the ultraviolet cutoff $\Lambda$ but 
the amplitude in the absence of the three-body force shows an
oscillatory behavior on $\ln\Lambda$. 
Cutoff independence of the amplitude is restored by an appropriate 
``running'' of $H (\Lambda )$ which turns out to be 
a limit cycle \cite{Bedaque:1998kg,Bedaque:1998km}:
\beq
H(\Lambda)=\frac{\cos[s_0 \ln(\Lambda/\Lambda_*)+\arctan s_0]}
{\cos[s_0 \ln(\Lambda/\Lambda_*)-\arctan s_0]}\,,
\label{eq:Heq}
\eeq
where $\Lambda_*$ is a dimensionful three-body parameter generated
by dimensional transmutation. Adjusting 
$\Lambda_*$ to a single three-body observable allows to determine all 
other low-energy properties of the three-body system.
Note that the choice of the three-body parameter $\Lambda_*$ is not unique
and there are other definitions more directly related to 
experiment \cite{Braaten:2004rn}.
Because $H(\Lambda)$ in Eq.~(\ref{eq:Heq}) vanishes for certain values of the 
cutoff $\Lambda$ it is possible to eliminate the explicit three-body force
from the equations by working with a fixed cutoff that encodes the
dependence on $\Lambda_*$. This justifies 
tuning the cutoff $\Lambda$ in the STM equation
to reproduce a three-body datum and using the same cutoff to calculate 
other observables as suggested by Kharchenko \cite{Kharchenko:1973aa}. 
Equivalently, a subtraction can  be
performed in the integral equation~\cite{Hammer:2000nf,Afnan:2003bs}.  
In any case, one
three-body input parameter is needed for the calculation of
observables.
A comprehensive study of the range corrections to the three-boson spectrum
was carried out in Ref.~\cite{Platter:2008cx}.
The authors showed that all range corrections vanish in the 
unitary limit due to the discrete scale invariance.
While the corrections proportional to $r_0 /a$ vanish trivially, 
this includes also the corrections proportional to 
$\kappa_* r_0$ where $\kappa_*=\sqrt{mB_3^*}$ is the binding momentum of the 
Efimov state fixed by the chosen renormalization condition.
Moreover, they have calculated the corrections to the Efimov spectrum
for finite scattering length. 
The range corrections are negligible for the 
shallow states but become important for the deeper bound states.

The integral equations for the three-nucleon problem derived from the
Lagrangian (\ref{lagd}) are a generalization of Eq.~(\ref{STM}).
(For their explicit form and derivation, 
see e.g. Ref.~\cite{Bedaque:2002yg}.)
For S-wave nucleon-deuteron scattering in the spin-quartet channel
only the spin-1 dimeron field contributes. 
This integral equation has a unique solution for $\Lambda \to \infty$ 
and there is no three-body force in the first few orders. The spin-quartet
scattering phases can therefore be predicted to high precision
from two-body data \cite{Bedaque:1997qi,Bedaque:1998mb}.
In the spin-doublet  channel both dimeron fields as well
as the the three-body force in the Lagrangian (\ref{lagd})
contribute \cite{Bedaque:1999ve}. This leads to 
a pair of coupled integral equations for the T-matrix.
Thus, one needs a new parameter which is not determined in the 2N system 
in order to fix the (leading) low-energy behavior of the 3N system in 
this channel. The three-body parameter gives a natural explanation of
universal correlations between different three-body observables such as
the Phillips line, a correlation between the triton binding energy and
the spin-doublet neutron-deuteron scattering length \cite{Phillips68}.
These correlations are purely driven by the large scattering length
independent of the mechanism responsible for it.
As a consequence, they occur in atomic systems such as $^4$He atoms
as well \cite{Braaten:2004rn}.

\begin{figure}[tb]
\centerline{\includegraphics*[width=7cm,angle=0,clip=true]{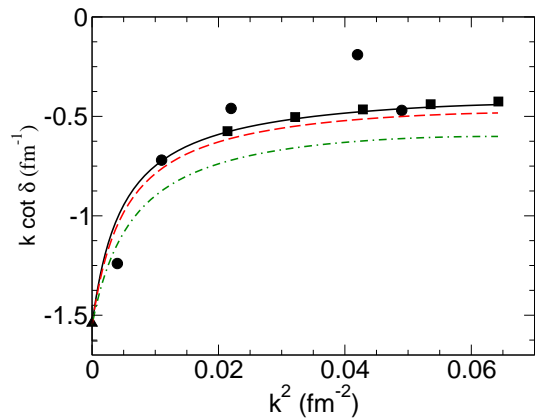}}
\caption{
Phase shifts for neutron-deuteron scattering below the deuteron
breakup at LO (dash-dotted line), NLO (dashed line), and N$^2$LO
(solid line). The filled squares and circles give the results of
a phase shift analysis and a calculation using AV18 and the 
Urbana IX three-body force, respectively. Figure courtesy of
L.~Platter.
}
\label{fig:nddoublet}
\end{figure}

Higher-order corrections to
the amplitude including the ones due to 
2N effective range terms can be included perturbatively.
This was first done at NLO for the scattering length and triton 
binding energy in \cite{Efimov:1991aa}
and for the energy dependence of the phase shifts in \cite{Hammer:2000nf}.
In Refs.~\cite{Bedaque:2002yg,Griesshammer:2004pe}, it was demonstrated that
it is convenient to iterate certain higher order range terms in order 
to extend the calculation to N$^2$LO. Here, also a subleading three-body
force was included as required by dimensional analysis. More recently, 
Platter and Phillips showed using
the subtractive renormalization that the leading three-body 
force is sufficient to achieve cutoff independence up to N$^2$LO in the expansion
in $\Mlo/\Mhi$ \cite{Platter:2006ev}. The  results
for the spin-doublet neutron-deuteron scattering phase shift 
at LO \cite{Bedaque:1999ve}, NLO \cite{Hammer:2000nf}, and 
N$^2$LO \cite{Platter:2006ad} are shown in Fig.~\ref{fig:nddoublet}.
There is excellent agreement with the available phase shift analysis and
a calculation using a phenomenological NN interaction.
Whether there is
a suppression of the subleading three-body force or simply a correlation
between the leading and subleading contributions is not fully understood.
The extension to 3N channels with higher orbital angular momentum
is straightforward \cite{Gabbiani:1999yv}
and three-body forces do not appear until very high orders.
A general counting scheme for three-body forces
based on the asymptotic behavior of the solutions of the leading
order STM equation was proposed in \cite{Griesshammer:2005ga}. 
A complementary approach to the few-nucleon problem is given by the
renormalization group where the power counting is determined
from the scaling of operators under the renormalization group
transformation \cite{Wilson-83}. This method leads 
 to consistent results for the power counting
\cite{Barford:2004fz,Birse:2008wt,Ando:2008jb}.
Universal low-energy properties of few-body systems with short-range 
interactions and large two-body scattering length were reviewed in 
\cite{Braaten:2004rn}. (See also \cite{Efimov:1981aa} for an early work on 
this subject.) 
Three-body calculations with external currents are still in their
infancy. However, a few exploratory calculations have been carried out.
Universal properties of the triton charge form factor were investigated
in Ref.~\cite{Platter:2005sj} and neutron-deuteron radiative capture
was calculated in Refs.~\cite{Sadeghi:2005aa,Sadeghi:2006aa}.
This opens the possibility to carry out accurate calculations of
electroweak reactions at very low energies for astrophysical processes.

The pionless approach has also been extended to the four-body sector
\cite{Platter:2004qn,Platter:2004zs}. In order to be able to apply
the Yakubovsky equations, an equivalent
effective quantum mechanics formulation was used. The study of the 
cutoff dependence of the four-body binding energies revealed that
no four-body parameter is required for renormalization at leading order.
As a consequence, there are universal correlations in the 
four-body sector which are also driven  by the large scattering length. 
The best known example is the Tjon line:
a correlation between the triton and alpha particle binding energies,
$B_t$ and $B_\alpha$, respectively. Of course, higher
order corrections break the exact correlation and generate a band.
\begin{figure}[tb]
\centerline{\includegraphics*[width=7cm,angle=0]{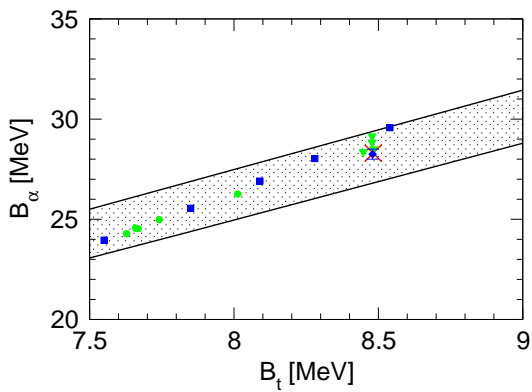}}
\caption{\label{fig:tjon}
The Tjon line correlation as predicted by the
pionless theory. The grey circles and triangles show
various calculations using phenomenological potentials \cite{Nogga:2000uu}.
The squares show the results of chiral EFT at NLO for different cutoffs
while the diamond gives the N$^2$LO result 
\cite{Epelbaum:2000mx,Epelbaum:2002vt}.
The cross shows the experimental point.
}
\end{figure}
In Fig.~\ref{fig:tjon}, we show this band together with some
some calculations using phenomenological 
potentials \cite{Nogga:2000uu} and a chiral EFT potential with explicit
pions \cite{Epelbaum:2000mx,Epelbaum:2002vt}.
All calculations with interactions that give a
large scattering length must lie within the band. Different
short-distance physics and/or cutoff dependence should only move 
the results along the band. This can for example be observed in
the NLO results with the chiral potential indicated by the squares
in Fig.~\ref{fig:tjon} or in
the few-body calculations with the low-momentum NN potential 
$V_{\rm low\: k}$ carried out in Ref.~\cite{Nogga:2004ab}. The 
$V_{\rm low\: k}$ potential
is obtained from phenomenological NN interactions by intergrating out 
high-momentum modes above a cutoff $\Lambda$
but leaving two-body observables (such as the large
scattering lengths) unchanged. The results of few-body calculations 
with $V_{\rm low\: k}$ are not independent of $\Lambda$ but lie all
close to the Tjon line (cf. Fig.~2 in Ref.~\cite{Nogga:2004ab}).
The studies of the four-body system in the pionless theory
were extendend further in Ref.~\cite{Hammer:2006ct}. Here the 
dependence of the four-body bound state spectrum on the two-body
scattering length was investigated in detail and summarized in a
generalized Efimov plot for the four-body spectrum.

The question of whether a four-body parameter has to enter at leading order
was reanalyzed by Yamashita et al.~\cite{Yama06}.
Within the renormalized zero-range model, they found a strong sensitivity
of the deepest four-body energy
to a four-body subtraction constant in their equations.
They motivated this observation from a general model-space reduction
of a realistic two-body interaction close to a Feshbach resonance.
The results of Ref.~\cite{Platter:2004qn} for the $^4$He tetramer
that include a four-body parameter were also reproduced. 
Yamashita et al. concluded that a four-body
parameter should generally enter at leading order. They argued that
four-body systems of $^4$He atoms and nucleons (where this sensitivity
is absent \cite{Platter:2004qn,Platter:2004zs,Nogga:2004ab})
are special because repulsive interactions strongly reduce
the probability to have four particles close together.
However, the renormalization of the four-body problem was not explicitly
verified in their calculation.
Another drawback of their analysis is the focus on the deepest four-body
state only. Therefore, their findings could be an artefact of
their particular regularization scheme. 
Another recent study by von Stecher and collaborators 
\cite{vStech08} confirmed the absence of a four-body parameter 
for shallow states while some sensitivity was found for the 
deepest four-body state.

The pionless theory has also been extended to more than four particles
using it within the no-core shell model approach. Here the expansion
in a truncated harmonic oscillator basis is used as the 
ultraviolet regulator of the EFT. The effective interaction is determined
directly in the model space, where an exact diagonalization in a complete 
many-body basis is performed. In Ref.~\cite{Stetcu:2006ey},
the $0^+$ excited state of $^4$He and the $^6$Li ground state were 
calculated using the deuteron, triton, alpha particle ground states 
as input. The first $(0^+;0)$ excited state in $^4$He is calculated within
10\% of the experimental value, while the $^6$Li ground state comes out
at about 70\% of the experimental value in agreement with the 
30 \% error expected for the leading order approximation. These results
are promising and should be improved if range corrections are
included.
Finally, the spectrum of trapped three- and four-fermion systems
was calculated using the same method \cite{Stetcu:2007ms}. In this 
case the harmonic potential is physical and not simply used as an 
ultraviolet regulator.

\subsection{Chiral EFT for few nucleons: foundations}
\label{sec:chiralEFT}

\begin{figure}[tb]
\includegraphics*[width=0.47\textwidth]{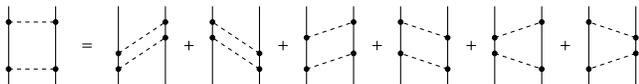}
\vspace{0.1cm}
\begin{center}
\caption{Representation of the two-pion exchange Feynman diagram in terms
  of time-ordered graphs. Solid and dashed lines
         represent nucleons and pions,
         respectively. 
\label{fig:TPEtime}}
\vspace{-0.5cm}
\end{center}
\end{figure}

The extension of the previously discussed EFT with contact interactions to
higher energies requires the inclusion of pions as explicit degrees of freedom.
The interaction between pions and nucleons can be described in a systematic
way using chiral perturbation theory. In contrast, the interaction
between the nucleons is strong and leads to nonperturbative phenomena at low
energy such as e.g.~shallow-lying bound states. This breakdown of perturbation
theory can be linked to the fact that the interaction between the nucleons
is \emph{not} suppressed in the chiral limit contrary to the pion and
pion-nucleon interactions. Moreover, an additional enhancement occurs for
Feynman diagrams involving two and more nucleons due to the appearance of the
so-called pinch singularities in the limit of the infinite nucleon
mass. Although such infrared singularities disappear if one keeps the nucleon
mass at its physical value, they do generate large enhancement factors which
destroy the chiral power counting. This can be more easily understood
utilizing the language of  time-ordered perturbation theory. 
Consider, for example, the two-pion exchange box diagram shown in Fig.~\ref{fig:TPEtime}.
While all intermediate states in the first two time-ordered graphs, often
referred to as irreducible, involve at
least one virtual pion and thus lead to energy denimonators of the expected
size, $E - E_i \sim M_\pi$, the remaining reducible diagrams involve an intermediate
state with nucleons only which produces unnaturally small energy
denominators of the order $E - E_i \sim M_\pi^2/m \ll M_\pi$. Clearly, the
enhanced reducible time-ordered diagrams are nothing but the iterations of the
Lippmann-Schwinger equation with the kernel which  contains all possible
irreducible diagrams and defines the nuclear Hamiltonian. 
It is free from infrared enhancement factors and can be
worked out systematically using the machinery of chiral perturbation theory as
suggested in Weinberg's seminal work
\cite{Weinberg:1990rz,Weinberg:1991um}.\footnote{An alternative
  framework based on the perturbative treatment of the pion exchange
  contributions has been introduced by Kaplan, Savage and Wise
  \cite{Kaplan:1998tg,Kaplan:1998we}, see 
  \cite{Beane:2000fx,Bedaque:2002mn} for review articles. As shown in
  Refs.~\cite{Cohen:1998jr,Cohen:1999iaa,Fleming:1999ee},  the perturbative
  inclusion of pions does not allow to significantly increase the applicability
  range of the theory as compared to pionless EFT. For yet different proposals
  to include pions in EFT for the nucleon-nucleon system 
  see Refs.~\cite{Lutz:1999yr,Oller:2003px,Soto:2007pg}.} 
This natural reduction to the quantum mechanical $A$-body
problem is a welcome feature for practical calculations as it allows to apply various existing few-body
techniques such as e.g.~the Faddeev-Yakubovsky scheme, the no-core shell
model, Green's function Monte Carlo and hyperspherical harmonics methods. On
the other hand, the framework offers a systematic and perturbative scheme to
derive nuclear forces and current operators in harmony with the chiral symmetry of QCD. 
The expansion parameter is given by the ratio $Q/\Lambda$ where $Q$ is the
soft scale associated with the pion mass and/or external nucleon momenta and
$\Lambda$ is the pertinent hard scale.  
For a given connected irreducible diagram with
$N$ nucleons, $L$ pion loops and $V_i$ vertices of 
type $i$, the  power $\nu$ of the soft scale $Q$  which
determines its importance can be obtained based on naive dimensional
analysis (i.e.~assuming classical scaling dimensions for various operators
in the effective Lagrangian):
\beq
\label{powc}
\nu = -4 + 2 N + 2 L + \sum_i V_i \Delta_i\,, 
 \quad \Delta_i = d_i + \frac{1}{2} n_i - 2\,.
\eeq
Here, $n_i$ is the number of nucleon field operators and $d_i$ the  
number of derivatives and/or insertions of $M_\pi$. The spontaneously broken
chiral symmetry of QCD guarantees $\Delta_i \geq 0$. As a consequence, the
chiral dimension $\nu$ is bounded from below, and only a finite number of diagrams
contribute at a given order. In addition, Eq.~(\ref{powc}) provides a natural
explanation to 
the dominance of the two-nucleon interactions and the hierarchy of nuclear
forces observed in nuclear physics. In particular, it implies that two-,
three-, and four-nucleon forces start to contribute at orders $\nu = 0$, $2$
and $4$, respectively. Notice that as argued in Ref.~\cite{Weinberg:1991um},
the nucleon mass $m$ should be counted as $Q/m \sim Q^2/\Lambda^2$ 
(which implies that $m \gg \Lambda$) 
in order to maintain consistency with the appearance of shallow-lying
bound states.\footnote{This statement only applies for the power 
counting based on
naive dimensional analysis.} Notice further that
according to this counting rule, the momentum
scale associated with the real pion production is treated as
the hard scale, $\sqrt{m M_\pi} \sim \Lambda$, and
needs not be explicitly kept track of, see also \cite{Mondejar:2006yu}
and section \ref{sec:further_scales} for a related discussion. Clearly, such a framework is
only applicable at energies well below the pion
production threshold. We also emphasize that the validity of the naive
dimensional scaling rules for few-nucleon contact operators has been
questioned in Refs.~\cite{Nogga:2005hy,Birse:2005um}. We will come back
to this issue in section \ref{sec:chiralEFTappl}.  

Before discussing the chiral expansion of the nuclear forces it is
important to clarify the relation between the underlying
chiral Lagrangian for pions and nucleons and the nuclear Hamiltonian we are
finally interested in. The derivation of the nuclear potentials from field theory is an old
and extensively studied problem in nuclear physics. Different approaches have
been developed in the fifties of the last century in the context of the
so-called meson theory of nuclear forces, see e.g.~the review article 
\cite{Phillips:1959aa}. In the modern framework of chiral EFT, the most
frequently used methods besides the already mentioned time-ordered perturbation theory
are the ones based on S-matrix and the unitary transformation. In the former
scheme, the nuclear potential is defined through matching the amplitude to the
iterated Lippmann-Schwinger equation \cite{Kaiser:1997mw}. In the
second approach, the 
potential is obtained by applying an appropriately choosen unitary transformation 
to the underlying pion-nucleon Hamiltonian which eliminates the coupling
between the purely nucleonic Fock space states and the ones which contain pions, see
\cite{Epelbaum:1998ka} for more details. We stress that both methods lead
to \emph{energy-independent} interactions as opposed by the ones obtained in
time-ordered perturbation theory. The energy independence of the potential is
a welcome feature which enables applications to three- and
more-nucleon systems.   

\begin{figure}[tb]
\includegraphics*[width=0.44\textwidth]{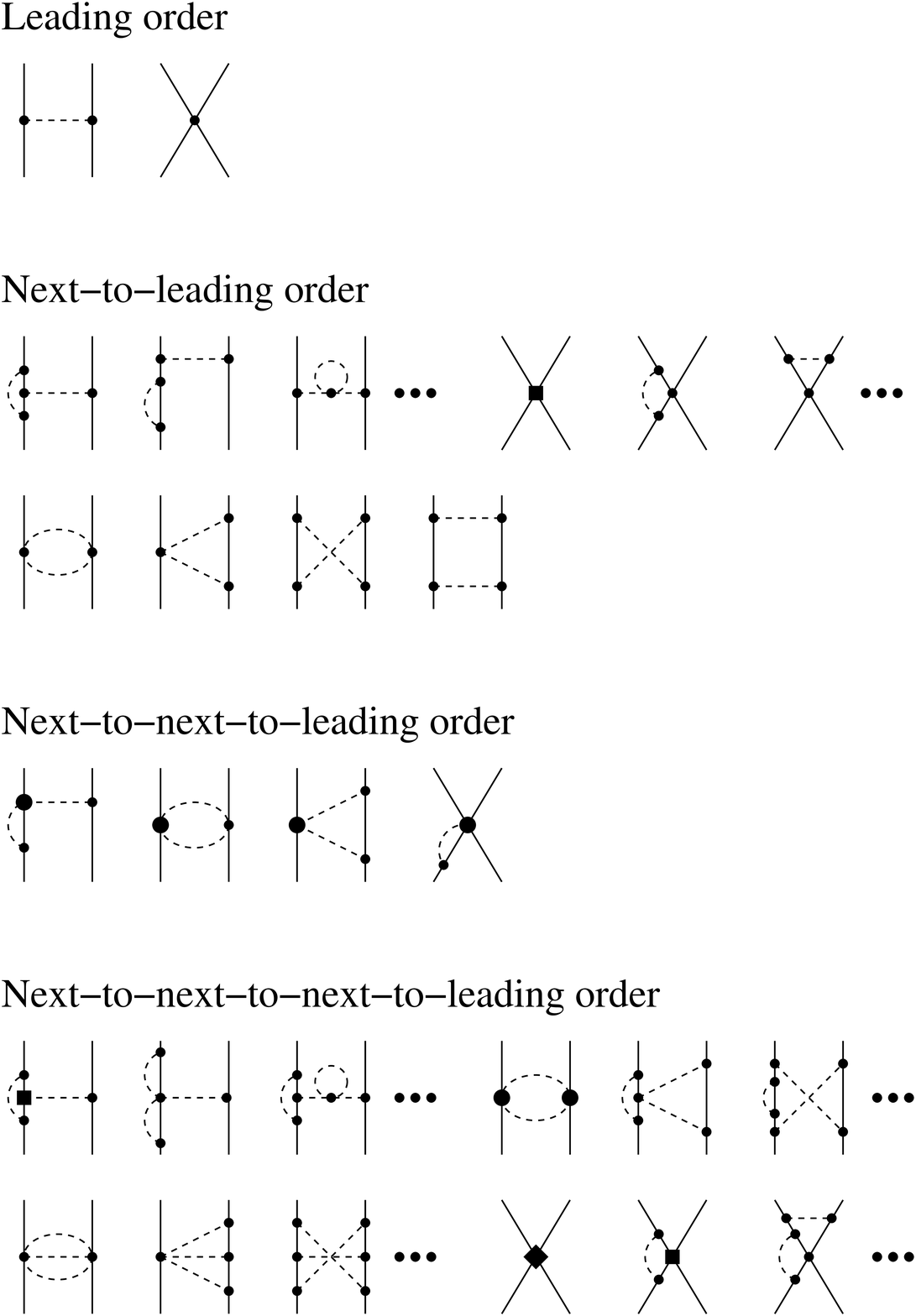}
\vspace{0.2cm}
\begin{center}
\caption{Chiral expansion of the two-nucleon force up to N$^3$LO. Solid dots,
  filled circles, squares and diamonds denote vertices with $\Delta_i = 0$,
  $1$, $2$ and $4$, respectively. Only irreducible contributions of the
  diagrams are taken in to account as explained in the text. 
\label{fig:2NF}}
\vspace{-0.5cm}
\end{center}
\end{figure}

We are now in the position to discuss the structure of the nuclear force
at lowest orders of the chiral expansion. 
The leading-order (LO) contribution results, according to Eq.~(\ref{powc}),
from two-nucleon tree diagrams constructed from the Lagrangian of lowest dimension
$\Delta_i=0$, $\mathcal{L}^{(0)}$, which has the following form in the heavy-baryon
formulation \cite{Jenkins:1990jv,Bernard:1992qa}:
\beqa
\label{lagr0}
\mathcal{L}_{\pi}^{(0)} &=& \frac{F^2}{4} \langle \nabla^\mu U \nabla_\mu 
U^\dagger + \chi_+ \rangle \,, \\
\mathcal{L}_{\pi N}^{(0)} &=& \bar{N} \left(
 i \, v\cdot D + \krig{g}_A \, u \cdot S \right) N\,,\nn
\mathcal{L}_{NN}^{(0)} &=& -\frac{1}{2} C_S ( \bar N N)   ( \bar N N )  +
2 C_T  ( \bar N S N ) 
\cdot ( \bar N S N ) \,, 
\nonumber
\eeqa
where $N$, $v_\mu$  and  $S_\mu \equiv ( 1/2)  i \gamma_5 \sigma_{\mu \nu} v^\nu$
denote the large component of the nucleon
field, the nucleon four-velocity and the covariant spin vector, respectively.  
The brackets $\langle \ldots \rangle$ denote traces in the flavor space
while $F$ and $\krig{g}_A$ refer to the chiral-limit values of the pion decay
and the nucleon axial vector coupling constants. 
The low-energy constants (LECs)
$C_S$ and $C_T$ determine the strength of the leading NN short-range
interaction. 
Further, the unitary $2 \times 2$ matrix $U( \fet \pi ) = u^2( \fet \pi )$ in
the flavor space collects the pion fields, 
\beq
U (\fet \pi ) = 1 + \frac{i}{F} \fet \tau \cdot \fet \pi - 
\frac{1}{2 F^2} \fet \pi^2 + \mathcal{O} (\pi^3)\,,
\eeq
where $\tau_i$ denotes the isospin Pauli matrix. 
The covariant derivatives of the nucleon and pion fields are defined via
$D_\mu = \partial_\mu + [ u^\dagger , \, \partial_\mu u ]/2$ and 
$u_\mu = i(u^\dagger \partial_\mu u
- u \partial_\mu u^\dagger)$. The quantity $\chi_+ = u^\dagger \chi u^\dagger + u
\chi^\dagger u$ with $\chi =2 B \mathcal{M}$ involves the explicit chiral
symmetry breaking due to the finite light quark masses, $\mathcal{M} =
\mbox{diag} (m_u ,\, m_d )$. The constant $B$ is related to the value of the
scalar quark condensate in the chiral limit, $\langle 0 | \bar u u | 0 \rangle
= -F^2 B$, and relates the pion mass $M_\pi$ to the quark mass $m_q$ via
$M_\pi^2 = 2 B m_q + \mathcal{O} (m_q^2 )$.  For more details on the notation
and the complete expressions for the pion-nucleon Lagrangian including up to
four derivatives/$M_\pi$-insertions the reader is referred to \cite{Fettes:2000gb}.
Expanding the effective
Lagrangian in Eqs.~(\ref{lagr0}) in powers of the pion fields one can easily
verify that the only possible connected two-nucleon tree diagrams
are the one-pion exchange
and the contact one, see the first line in Fig.~\ref{fig:2NF},
yielding the following potential in the two-nucleon center-of-mass system
(CMS):  
\beq 
\label{VLO}
V_{\rm 2N}^{(0)} = -\frac{g_A^2}{4 F_\pi^2} \frac{\vec
  \sigma_1 \cdot \vec q \vec \sigma_2 \cdot \vec q}{\vec q {}^2 +M_\pi^2} \fet \tau_1
\cdot \fet \tau_2  + C_S + C_T
\vec \sigma_1 \cdot \vec \sigma_2\,,
\eeq
where the superscript of $V_{\rm 2N}$ denotes the chiral order $\nu$, $\sigma_i$
are the Pauli spin matrices, $\vec q
= \vec p \, ' - \vec p$ is the nucleon momentum transfer and $\vec{p}$
($\vec{p}~'$) refers to initial (final) nucleon momenta in the CMS. Further,
$F_\pi = 92.4$ MeV  and $g_A = 1.267$ denote the pion decay and the nucleon
axial coupling constants, respectively.  

The first corrections to the LO result are suppressed by two powers of the
low-momentum scale.  The absence of the
contributions at order $\nu =1$ can be traced back to parity conservation  which forbids 
$( \bar NN)( \bar NN)$ vertices with one spatial derivative and $\pi NN$ vertices with two derivatives
(i.~e.~$\Delta_i =1$). The next-to-leading-order (NLO) contributions to the
2NF therefore
result from tree diagrams with one insertion of the $\Delta_i=2$-interaction
and one-loop diagrams constructed from the lowest-order vertices, see
Fig.~\ref{fig:2NF}. The relevant terms in the effective Lagrangian read \cite{Gasser:1987rb}
\beqa
\mathcal{L}_{\pi}^{(2)} &=&\frac{l_3}{16} \langle \chi_+ \rangle^2 
+ \frac{l_4}{16} \Big( 2 \langle \nabla_\mu U \nabla^\mu U^\dagger 
\rangle \langle \chi_+ \rangle \nn
&+& 2 \langle \chi^\dagger U \chi^\dagger U + \chi U^\dagger \chi U^\dagger 
\rangle - 4 \langle \chi^\dagger
\chi \rangle - \langle \chi_- \rangle^2 \Big) \nn
&+& \ldots\,, \nn
\mathcal{L}_{\pi N}^{(2)} &=& \bar N \bigg( \frac{1}{2 \krig{m}} (v \cdot D)^2 
- \frac{1}{ 2 \krig{m}}
D \cdot D + d_{16} S \cdot u \langle \chi_+ \rangle \nn
&+& i d_{18} S^\mu [ D_\mu , \, \chi_-]
+ \ldots
 \bigg) N \,,\nonumber
\eeqa
\beqa
\mathcal{L}_{NN}^{(2)} &=& - \tilde C_1 \big( ( \bar N D N) \cdot  ( \bar N D N) 
+ ( ( D \bar N) N) \cdot  ((D  \bar N) N) \big) \nn
&-&  2 (\tilde C_1  + \tilde C_2 )  ( \bar N D N) \cdot  ( (D \bar N ) N) \nn
&-&  \tilde C_2   ( \bar N N) \cdot  ( (D^2 \bar N ) N + \bar N D^2 N) + \ldots
\,, 
\label{lagr2}
\eeqa
where $l_i$, $d_i$ and $\tilde C_i$ denote further LECs  and $\krig m$ is the
nucleon mass in the chiral limit. The ellipses in the pion and pion-nucleon
Lagrangians refer to terms which do not contribute to the nuclear force at
NLO. In the case of the nucleon-nucleon Lagrangian
$\mathcal{L}_{NN}^{(2)}$ only a few terms are given explicitly.  The complete
reparametrization-invariant set of terms can be found in \cite{Epelbaum:2000kv}. 
The NLO contributions to the two-nucleon potential have been first  
considered in \cite{Ordonez:1993tn,Ordonez:1995rz} utilizing the
framework of time-ordered perturbation theory. The corresponding
energy-independent expressions have been worked out in \cite{Friar:1994zz}
using the method described in \cite{Friar:1977xh} and then re-derived in
\cite{Kaiser:1997mw} using an S-matrix-based approach and, independently,  in
\cite{Epelbaum:1998ka,Epelbaum:1999dj} based on the method of unitary transformation.  
The one-pion ($1\pi$) exchange diagrams at NLO do not produce any
new momentum dependence. Apart from renormalization of various LECs in
Eq.~(\ref{VLO}), one obtains the leading contribution to the Goldberger-Treiman 
discrepancy \cite{Epelbaum:2002gb}
\beq
\label{GTD}
\frac{g_{\pi N}}{m} = \frac{g_A}{F_\pi} 
 - \frac{ 2 M_\pi^2}{F_\pi} \, d_{18} + \ldots
\eeq
where the ellipses refer to higher-order terms. Similarly, loop diagrams
involving NN short-range interactions only lead to ($M_\pi$-dependent) shifts
in the LO contact terms. The remaining contributions to the 2NF due to 
higher-order contact interactions and two-pion exchange have the form: 
\beqa
\label{VNLO}
V^{(2)}_{\rm 2N} &=&  - \frac{ \fet{\tau}_1 \cdot \fet{\tau}_2 }{384 \pi^2 F_\pi^4}\,
L^{\tilde \Lambda} (q) \, \bigg( 4M_\pi^2 (5g_A^4 - 4g_A^2 -1)\nonumber\\
&+& \vec{q}\, ^2 (23g_A^4 - 10g_A^2 -1)
+ \frac{48 g_A^4 M_\pi^4}{4 M_\pi^2 + \vec{q} \, ^2} \bigg) \nn
&-&  \frac{3 g_A^4}{64 \pi^2 F_\pi^4} \,L^{\tilde \Lambda} (q)  \, 
\left(  \vec \sigma_1 \cdot \vec q  \, \vec \sigma_2 \cdot \vec q   -  \vec \sigma_1 \cdot\vec
\sigma_2  \, \vec{q} \, ^2 \right) \nn 
&+& C_1 \, \vec{q}\,^2 + C_2 \, \vec{k}^2 +
( C_3 \, \vec{q}\,^2 + C_4 \, \vec{k}^2 ) \,  \vec{\sigma}_1 \cdot
\vec{\sigma}_2 \nn
&+& iC_5\, \frac{1}{2} \, ( \vec{\sigma}_1 + \vec{\sigma}_2) \cdot  \vec{q} \times
\vec{k}
 + C_6 \, \vec{q}\cdot \vec{\sigma}_1 \, \vec{q}\cdot \vec{\sigma}_2  
\nn
&+& C_7 \, \vec{k}\cdot \vec{\sigma}_1 \, \vec{k}\cdot \vec{\sigma}_2 \,,
\eeqa
where $q \equiv | \vec q \, |$ and the LECs $C_i$ can be written as linear
combinations of $\tilde C_i$ in Eq.~(\ref{lagr2}). The loop function
$L^{\tilde \Lambda} (q)$ is defined in the spectral function regularization (SFR)
\cite{Epelbaum:2003gr,Epelbaum:2003xx} as
\beq
\label{def_LA}
L^{\tilde \Lambda} (q) = \theta (\tilde \Lambda - 2 M_\pi ) \, \frac{\omega}{2 q} \, 
\ln \frac{\tilde \Lambda^2 \omega^2 + q^2 s^2 + 2 \tilde \Lambda q 
\omega s}{4 M_\pi^2 ( \tilde \Lambda^2 + q^2)} \,,
\eeq
where we have introduced the following abbreviations: 
$\omega = \sqrt{4 M_\pi^2 + \vec{q}\, ^2}$ and 
$s = \sqrt{\tilde \Lambda^2 - 4 M_\pi^2}$. Here, $\tilde \Lambda$ denotes the
ultraviolet cutoff in the mass spectrum of the two-pion-exchange potential. If dimensional regularization
(DR) is employed, the expression for the loop function simplifies to 
\beq
L (q) = \lim_{\tilde \Lambda \to \infty} L^{\tilde \Lambda} (q) = 
\frac{\omega}{q} \, \ln \frac{\omega + q}{2 M_\pi}\,.
\eeq

\begin{figure}[tb]
\includegraphics*[width=0.47\textwidth]{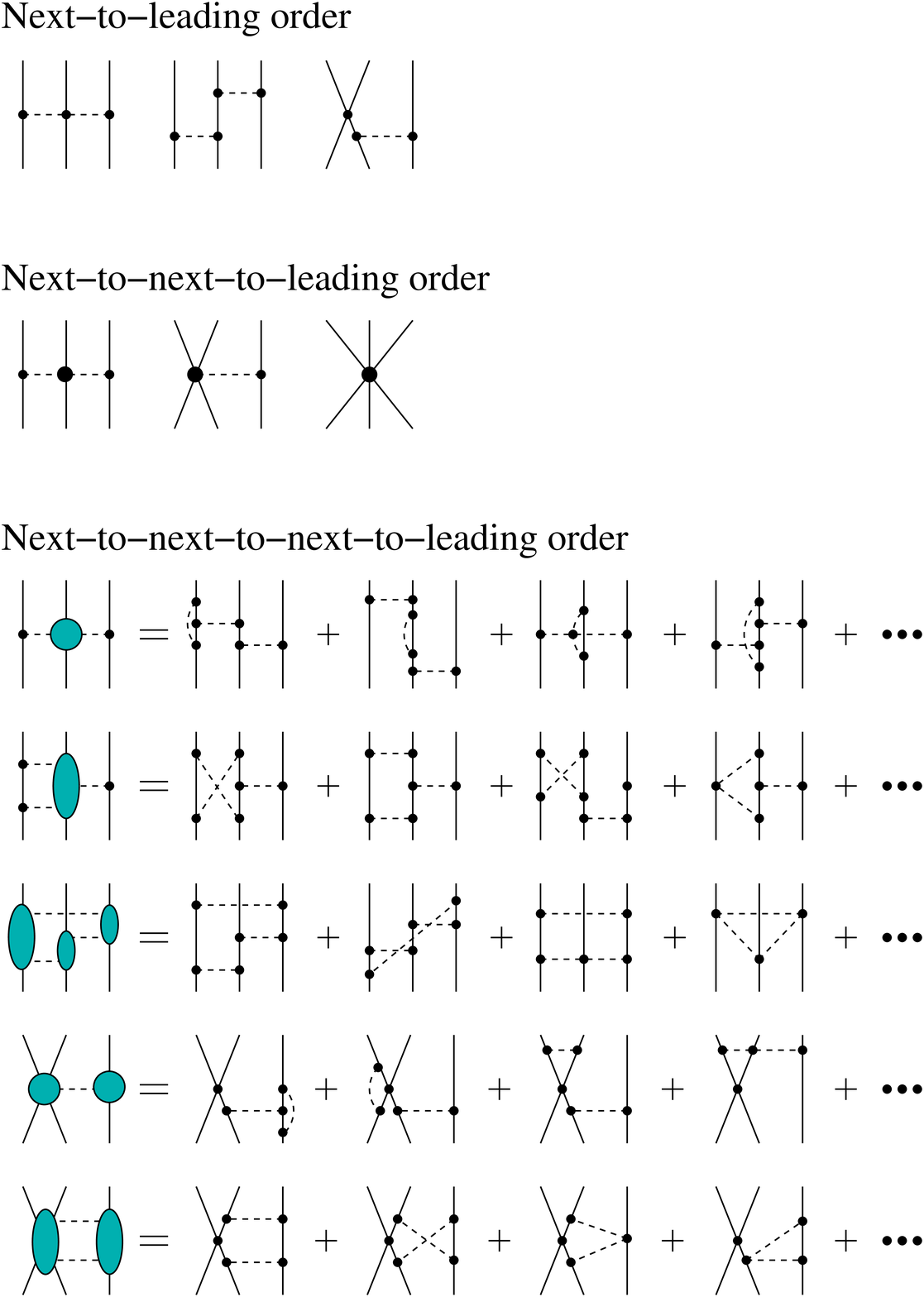}
\vspace{0.2cm}
\begin{center}
\caption{Chiral expansion of the three-nucleon force up to N$^3$LO. Diagrams
  in the first line (NLO) yield vanishing contributions to the 3NF if
  one uses energy-independent formulations as explained in the text. The five
  topologies at N$^3$LO involve the two-pion exchange,
  one-pion-two-pion-exchange, ring, contact-one-pion exchange and
  contact-two-pion-exchange diagrams in order. Shaded blobs represent the
  corresponding amplitudes.  For remaining 
  notation see Fig.~\ref{fig:2NF}. 
\label{fig:3NF}}
\vspace{-0.5cm}
\end{center}
\end{figure}

In addition to the two-nucleon contributions, at NLO one also needs to
consider three-nucleon diagrams shown in the first line of Fig.~\ref{fig:3NF}.  
The first diagram does not involve reducible topologies and,
therefore, can be dealt with using the Feynman graph technique. It is then easy to
verify that its contribution is shifted to higher orders due to the
additional suppression by the factor of $1/m$ caused by the appearance of time
derivative at the leading-order $\pi \pi \bar N N$ vertex, the so-called  Weinberg-Tomozawa vertex.  
The two remeining diagrams have been considered by Weinberg
\cite{Weinberg:1990rz,Weinberg:1991um} and later by Ordonez and van Kolck
\cite{Ordonez:1992xp} using the energy-dependent formulation based on
time-ordered perturbation theory. In this approach, it was shown that the
resulting 3NF cancels exactly (at the order one is working) against the
recoil correction to the 2NF when the latter is iterated in
the dynamical equation. In energy-independent approaches such as e.g.~the
method of unitary transformation which are employed in most of the existing
few-nucleon calculations one observes that the irreducible contributions from
the last two diagrams in the first line of Fig.~\ref{fig:3NF} are suppressed
by the factor $1/m$ and thus occur at higher orders \cite{Epelbaum:2000kv},
see also \cite{Coon:1986kq,Eden:1996ey}. Consequently, there is no
3NF at NLO in the chiral expansion. 

The contributions at next-to-next-to-leading order (N$^2$LO) involve one-loop
diagrams with one insertion of the subleading vertices of
dimension $\Delta_i = 1$, see Fig.~\ref{fig:2NF}. The corresponding Lagrangians read:
\beqa
\label{lagr1}
{\cal L}^{(1)}_{\pi N} & = & \bar{N} \big(  c_1 \, \langle \chi_+ \rangle + c_2 \, (v \cdot
       u)^2 + c_3 \, u \cdot u \nn
&+& c_4 \, [ S^\mu, S^\nu ] u_\mu u_\nu    + c_5 \langle \hat \chi_+ \rangle \big)
  N \,, \nn
{\cal L}^{(1)}_{\pi NN} & = & \frac{D}{2} (\bar{N} N) (\bar{N} S \cdot u N)\,,
\eeqa
where $\hat \chi_+ \equiv \chi_+ - \langle \chi_+ \rangle /2$ and $D$, $c_i$ are
the LECs. The $1\pi$-exchange loop diagram again only lead to renormalization
of the corresponding LECs. Similarly, the contribution from the last diagram
which involves the two-nucleon contact interaction can be absorbed into
a re-definition of the LECs $C_{S,T}$ and $C_i$ in Eqs.~(\ref{VLO}), (\ref{VNLO})
(provided one is not interested in the quark mass dependence of the nuclear
force). Further, the football diagram yields vanishing
contribution due to the antisymmetric (with respect to pion isospin quantum numbers)
nature of the Weinberg-Tomozawa vertex.   
Thus, the only nonvanishing contribution at this order results from the
$2\pi$-exchange triangle diagram:
\beqa
\label{VNNLO}
V^{(3)}_{\rm 2N} &=&  -\frac{3g_A^2}{16\pi F_\pi^4}  \big( 2M_\pi^2(2c_1 -c_3)
-c_3 \vec{q} \, ^2 \big) 
 (2M_\pi^2+\vec{q} \, ^2) \nn 
& \times & A^{\tilde \Lambda} (q)   - \frac{g_A^2 c_4 }{32\pi F_\pi^4} \, \fet{ \tau}_1 \cdot \fet{
  \tau}_2 \,  (4M_\pi^2 + q^2) A^{\tilde \Lambda}(q)\, \nn
& \times &
\big( \vec \sigma_1 \cdot \vec q\, \vec \sigma_2 \cdot \vec q\, 
-\vec{q} \, ^2 \, \vec \sigma_1 \cdot\vec \sigma_2 \big)\,,
\eeqa
where the loop function $A^{\tilde \Lambda} (q)$ is given by
\beq
A^{\tilde \Lambda} (q) = \theta (\tilde \Lambda - 2 M_\pi ) \, \frac{1}{2 q} \, 
\arctan \frac{q ( \tilde \Lambda - 2 M_\pi )}{q^2 + 2 \tilde \Lambda M_\pi}\,.
\eeq
In DR, the expression for $A(q)$ takes the following simple form:
\beq
A (q)  \equiv \lim_{\tilde \Lambda \to \infty} A^{\tilde \Lambda} (q) =
\frac{1}{2q}\, \arctan \frac{q}{2 M_\pi}\,.
\eeq
Notice that the triangle diagram also generates short-range contributions
which may be absorbed into re-definition of contact interactions. The
isoscalar central contribution proportional to the LEC $c_3$ is attractive and
very strong. It is by far the strongest two-pion exchange contribution and
reaches a few 10s of MeV (depending on the choice of
regularization) at internucleon distances of the order $r \sim M_\pi^{-1}$. 
The origin of the unnaturally strong subleading $2\pi$-exchange contributions
can be traced back to the (numerically) large values of the LECs $c_{3,4}$ and
is well understood in terms of resonance exchange related to $\Delta$ excitation
\cite{Bernard:1996gq}.  We will come back to this issue in section \ref{sec:chiralEFTdelta}
where the chiral EFT formulation with explicit $\Delta$ degrees of freedom will
be discussed.  The central $2\pi$-exchange potential was also
calculated by Robilotta \cite{Robilotta:2000py} using the infrared-reglarized version of
chiral EFT which enables to sum up a certain class of relativistic
corrections \cite{Becher:1999he}. He found that the results in the  heavy-baryon
limit overestimate the ones obtained using infrared regularization by
about $25 \%$, see also \cite{Epelbaum:2005pn} for a related discussion. 
Last but not least, the chiral $2\pi$-exchange potential up to N$^2$LO has been
tested in the Nijmegen partial wave analysis (PWA) of both proton-proton and
neutron-proton data \cite{Rentmeester:1999vw,Rentmeester:2003mf} where also an
attempt has been done to determine the values of the LECs $c_{3,4}$. As
demonstrated in these studies, the representation of the (strong) long-range
interaction based on the combination of the $1\pi$- and the chiral
$2\pi$-exchange potentials rather than on the pure $1\pi$-exchnge potential
allows to considerably reduce the number of 
phenomenological parameters entering the energy-dependent boundary conditions
which are needed to parametrize the missing short- and medium-range
interactions. Also the extracted values of the LECs $c_{3,4}$ agree reasonably
well with various determinations in the pion-nucleon system. These studies
provide a beautiful confirmation of the important role of the
$2\pi$-exchange potential in nucleon-nucleon scattering observables, see,
however, Ref.~\cite{Entem:2003cs} for a criticism. For a similar work
utilizing the distorted-wave methods the reader is referred to
\cite{Birse:2003nz,Birse:2007sx}.

The first nonvanishing contributions to the 3NF also show up at N$^2$LO and
arise from tree diagrams shown in Fig.~\ref{fig:3NF} which involve a single
insertion of the subleading vertices $\mathcal{L}^{(1)}$ in Eq.~(\ref{lagr1})
and 
\beq
\mathcal{L}_{NNN}^{(1)} = - \frac{1}{2}  (\bar{N} N) (\bar{N} \fet \tau N)
\cdot (\bar{N} \fet \tau N)\,.
\eeq
where $E$ is a low-energy constant. 
The corresponding 3NF expression read: 
\beqa
\label{leading}
V_{3N}^{(3)} &=& \frac{g_A^2}{8 F_\pi^4}\; 
\frac{\vec \sigma_1 \cdot \vec q_1  \; \vec \sigma_3 \cdot \vec q_3 }{[
  q_1^2 + M_\pi^2] \, [q_3^2 + M_\pi^2]} \;\Big[ \fet \tau_1
  \cdot \fet \tau_3  \, \big( - 4 c_1 M_\pi^2 \nn
&& {} + 2 c_3 \, \vec q_1 \cdot \vec
    q_3 \big)  
+  c_4 \fet \tau_1
  \times \fet \tau_3  \cdot \fet \tau_2  \; \vec q_1 \times \vec q_3 
\cdot \vec \sigma_2  \Big]   \nn
&-& \frac{g_A \, D}{8 F_\pi^2}\;  
\frac{\vec \sigma_3 \cdot \vec q_3 }{q_3^2 + M_\pi^2} \; 
\fet \tau_1 \cdot \fet \tau_3 \; \vec \sigma_1 \cdot \vec q_3 \nn
&+& \frac{1}{2} E \, \fet \tau_2 \cdot \fet \tau_3 \,,
\eeqa
where the subscripts refer to the nucleon labels and $\vec q_{i} = \vec p_i \,
' - \vec p_i$,  with $\vec p_i \, '$
and $\vec p_i$ being the final and initial momenta of the nucleon $i$. 
The expressions in Eq.~(\ref{leading}) correspond to a particular 
choice of nucleon labels. The full expression for
the 3NF results by taking into account all possible permutations of the
nucleons (for 
three nucleons there are altogether six permutations), i.e.:
\beq
V_{\rm 3N}^{\rm full} = V_{\rm 3N} + \mbox{all permutations}\,.
\eeq
We further emphasize that the expressions for the 3NF given in Refs.~\cite{Ordonez:1992xp,vanKolck:1994yi}
contain one redundant $1\pi$-exchange and two redundant contact
interactions. As shown in \cite{Epelbaum:2002vt}, only one independent linear combination
contributes in each case if one considers matrix elements between
antisymmetrized few-nucleon states, see also \cite{Bedaque:1999ve} for a related discussion. 

We now turn to next-to-next-to-next-to-leading order (N$^3$LO) and discuss
first the corrections to the 2NF. As follows from Eq.~(\ref{powc}), one has to 
account for contributions from tree diagrams with one insertion from
$\mathcal{L}^{(4)}$ or two insertions from $\mathcal{L}^{(2)}$, one-loop
diagrams with one insertion from  $\mathcal{L}^{(2)}$ or two insertions from $\mathcal{L}^{(1)}$
as well as two-loop graphs constructed from the lowest-order vertices, see
Fig.~\ref{fig:2NF}.  Apart from renormalization of various LECs, the
$1\pi$-exchange potential receives at this order (in the scheme based on the
counting $m \sim \mathcal{O} (\Lambda^2/M_\pi)$) the first relativistic
corrections proportional to $m^{-2}$. These are scheme-dependent and have to
be chosen consistently with the $1/m$-corrections to the $2\pi$-exchange
potential and the relativistic
extension of the dynamical equation, see \cite{Friar:1999sj} for a comprehensive
discussion. The two-pion exchange contributions at N$^3$LO were worked out by
Kaiser \cite{Kaiser:2001pc} based on the one-loop representation of the $\pi
N$ scattering amplitude. We refrain from giving here the rather involved
expressions for the sub-subleading $2\pi$-exchange potential and refer to
the original work \cite{Kaiser:2001pc} where the results are given in terms of
the corresponding spectral functions. For certain classes of contributions,
the integrals over the two-pion exchange spectrum could be performed
analytically and are given in Ref.~\cite{Entem:2002sf}.  Notice further that
the subleading (i.e.~the ones proportional to $m^{-2}$) relativistic
corrections of the $2\pi$-exchange range have also been worked out by Kaiser
\cite{Kaiser:2001at}. In the counting scheme with $m \sim \mathcal{O}
(\Lambda^2/M_\pi)$, these terms, however, would only appear at N$^5$LO.  
It should also be emphasized that the N$^3$LO contributions to the
$2\pi$-exchange potential  were worked out in the covariant version of
chiral EFT (more precisely, using the formulation by Becher and Leutwyler
\cite{Becher:1999he}) by Higa et al.~\cite{Higa:2003jk,Higa:2003sz,Higa:2005ip}.

$3\pi$-exchange contributions also appear at this
order in the chiral expansion and have been worked out in
Refs.~\cite{Kaiser:1999ff,Kaiser:1999jg}, see also Ref.~\cite{Pupin:1999ba}
for a related work. The resulting potentials turn out to be rather weak.
For example, the strongest contribution is of the isoscalar spin-spin type
(i.e.~proportional to $\vec \sigma_1 \cdot \vec \sigma_2$) and about $10$ times
weaker than the corresponding $2\pi$-exchange contribution at the same order at
relative distances $r \sim M_\pi^{-1}$. It should, however, be emphasized that
the subleading $3\pi$-exchange contributions at N$^4$LO are
larger in size \cite{Kaiser:2001dm} which, again, can be traced back to the large values of the
LECs $c_i$. Finally, the last type of the 2NF corrections at this order results from
diagrams involving contact interactions. The most general polynomial (in
momenta) representation of the short-range part of the potential involves,
apart from the two leading and seven subleading terms given in
Eqs.~(\ref{VLO}) and (\ref{VNLO}) fifteen new contact interactions (in the
isospin invariant sector) yielding in total 24 LECs to be determined from
nucleon-nucleon data.  

The 3NF contributions at N$^3$LO feed into five different topologies, see
Fig.~\ref{fig:3NF}, and  are currently being worked out. Currently, the
expressions for the  first three topologies which do not involve short-range
contact interactions are available. The one-loop corrections to the
$2\pi$-exchange diagrams can, to a large extent, be accounted for by a finite
shift $c_i \to \bar c_i = c_i + \delta c_i$ of the LECs $c_i$
\cite{Bernard:2007sp,Ishikawa:2007zz}  
\beq
\label{corrections_c}
\delta c_1 = - \frac{g_A^2 \,M_\pi}{64 \pi F_\pi^2}\,, \quad  \quad 
\delta c_3 =  - \delta c_4 = \frac{g_A^4 \,M_\pi}{16 \pi F_\pi^2}\,.  
\eeq
Numerically, these corrections are of the order of 20\% of the corresponding
LECs and are consistent with the difference in values of $c_i$ between the
order-$Q^2$ and $Q^3$ determinations from the pion-nucleon system, see 
\cite{Bernard:1995gx,Bernard:1996gq,Fettes:1998ud,Buettiker:1999ap}. 
The only $2\pi$-exchange contribution that cannot be cast into redefinition of
the LECs $c_i$ arises from the diagram which involves pions interacting in
flight, see \cite{Bernard:2007sp,Ishikawa:2007zz} for the explicit
expression. We also emphasize that there are no $2\pi$-exchange contributions
from tree diagrams with one insertion from $\mathcal{L}_{\pi N}^{(2)}$ in
Eq.~(\ref{lagr2}) (except for the relativistic corrections). This is because diagrams
involving subleading $\pi NN$ interaction do not yield any irreducible
contributions while the ones with the $\pi \pi NN$-vertices of dimension $\nu
= 2$  involve at least
one time derivative and are, therefore, suppressed by a factor of $1/m$.  
This observation is consistent with the absence of logarithmic ultraviolet
divergences in the loop diagrams. In this context, it should be emphasized that
the requirement of renormalizability at N$^3$LO (and, presumably, also at
higher orders) was found to impose strong constraints on the unitary ambiguity
in the form of the resulting nuclear potentials. This issue is discussed
extensively in \cite{Epelbaum:2007us} and may remind one of the recent 
findings in the context of large-$N_c$ QCD 
\cite{Belitsky:2002ni,Cohen:2002qn,Cohen:2002im} where it was
shown that the multiple-meson-exchange potential derived in the energy-dependent
formulation is inconsistent with large-$N_C$ counting rules. The consistency
could be maintained using a different (but equivalent up to the considered order)
form of the potential based on the energy-independent formalism, 
see Ref.~\cite{Cohen:2002im} for more details. 
The contributions from the two-pion-one-pion exchange and
ring diagrams are given explicitly in Ref.~\cite{Bernard:2007sp} where expressions
 are shown  both in momentum and coordinate spaces. 
Especially in the case of
ring diagrams where loop integrals involve two independent external momenta
and, therefore, yield rather involved expressions in momentum space.  It is
advantageous to switch to coordinate space where a much more compact
representation emerges.  Notice further that ring
diagrams were already studied in the pioneering work \cite{Fujita:1962}. 
The  calculation  of the last two
topologies involving the leading contact interactions is in progress. 
Last but not least, one should also take into account the leading relativistic
$1/m$-corrections to the NLO three-nucleon diagrams, see the first line in
Fig.~\ref{fig:3NF}. Again, these contributions are scheme-dependent and should
be chosen consistently with the relativistic corrections to the 2NF and the
form of the dynamical equation. The $1/m$-corrections to the $2\pi$-exchange
3NF have already been 
worked out long time ago by Coon and Friar and are given in the most general
form in Ref.~\cite{Coon:1986kq}. Notice further that at this order, one 
needs to
account for the dependence of the 2NF on the total momentum of the 2N system
(effects due to drift of the CMS of a two-body subsystem). Such boosted 2N
operators may, in fact, also be viewed as
3N operators. In the context of chiral EFT, this kind of corrections is
discussed in \cite{Robilotta:2006xq}. 

\begin{figure}[tb]
\includegraphics*[width=0.38\textwidth]{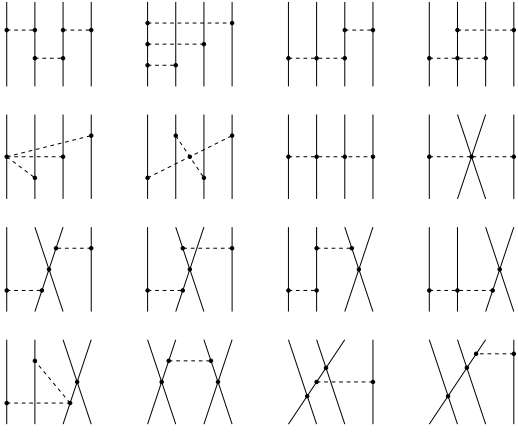}
\vspace{0.2cm}
\begin{center}
\caption{Diagrams contributing to the four-nucleon force at N$^3$LO.  For
  notation see Fig.~\ref{fig:2NF}. 
\label{fig:4NF}}
\vspace{-0.5cm}
\end{center}
\end{figure}

The last type of N$^3$LO contributions arises from four-nucleon tree diagrams
constructed from the lowest-order vertices, see
in Fig.~\ref{fig:4NF}, which have been evaluated recently using the method of
unitary transformation \cite{Epelbaum:2006eu,Epelbaum:2007us}.  
Notice that the first two diagrams in the second line were already discussed 
long ago, see e.g.~\cite{Mcmanus:1980ep,Robilotta:1985gv}. Furthermore, it
should be emphasized that disconnected diagrams calculated e.g.~in
\cite{vanKolck:1994yi}
using time-ordered perturbation theory do not contribute to the nuclear
force in the method of unitary transformation.  
It has been 
conjectured in Ref.~\cite{Robilotta:1985gv} that 4N diagrams which involve
reducible topologies do not generate irreducible pieces in the
amplitude and thus lead to vanishing 4NFs. While this is indeed the case for
the leading 3N diagrams at NLO, it is explicitly shown in
Ref.~\cite{Epelbaum:2006eu,Epelbaum:2007us} that many of the reducible-like
diagrams in Fig.~\ref{fig:4NF} do generate nonvanishing 4NFs which are not
suppressed by inverse powers of the nucleon mass. As a representative example,
we give here the $g_A^6$-contribution which results entirely from the first diagram in 
Fig.~\ref{fig:4NF} (the second graph appears to be truly reducible and does
not produce any contribution to the 4NF):
\beqa
\label{4nf_class1}
V_{4N}^{(4)} &=& - \frac{2 g_A^6}{( 2 F_\pi )^6} 
\frac{\vec \sigma_1 \cdot \vec q_1 \;\vec \sigma_4 \cdot \vec q_4}{[\vec q_1^{\;2}  + M_\pi^2]\,
[\vec q_{12}^{\;2}  + M_\pi^2]^2 \, [\vec q_4^{\;2}  + M_\pi^2]} \nn
&\times & {} \Big[ ( \fet \tau_1 \cdot \fet \tau_4 \,  \fet \tau_2 \cdot \fet \tau_3
-  \fet \tau_1 \cdot \fet \tau_3 \,  \fet \tau_2 \cdot \fet \tau_4 ) \,\vec q_1 \cdot \vec q_{12}
\, \vec q_4 \cdot \vec q_{12} \nn
&&{} + \fet \tau_1 \times \fet \tau_2 \cdot \fet \tau_4  \; \vec q_1 \cdot \vec q_{12}  \; 
\vec q_{12} \times \vec q_4 \cdot \vec \sigma_3 \nn
&& {}  + \fet \tau_1 \times \fet \tau_3 \cdot \fet \tau_4  \; \vec q_4 \cdot \vec q_{12}  \; 
\vec q_{1} \times \vec q_{12} \cdot \vec \sigma_2 \\
&& {} + \fet \tau_1 \cdot \fet \tau_4 \; \vec q_{12} \times \vec q_{1} \cdot \vec \sigma_2 \;
\vec q_{12} \times \vec q_4 \cdot \vec \sigma_3  \Big] +\mbox{all perm.},
\nonumber
\eeqa
where $\vec q_{12} = \vec q_1 + \vec q_2 = - \vec q_3 - \vec q_4 = -\vec
q_{34}$ is the momentum transfer between the 
nucleon pairs 12 and 34. The complete expression for the leading 4NF both in
momentum and coordinate space can be found in Ref.~\cite{Epelbaum:2007us}. 
A rough estimation of the 4NF contributions to e.g.~the $\alpha$-particle
binding energy is provided by the strength of the corresponding $r$-space
potentials expressed in terms of dimensionless variables $r_{ij} M_\pi$. One
then finds e.g.~for the $g_A^6$-terms $g_A^6 M_\pi^7 (16 \pi F_\pi^2 )^{-3}
\sim 50$ keV. This agrees qualitatively with a more accurate numerical
estimation carried out in Ref.~\cite{Rozpedzik:2006yi} which, however, still
involved severe approximations to simplify the calculations. 

So far we only discussed isospin-invariant contributions to the nuclear
forces. It is well established that nuclear forces
are charge dependent (for reviews see e.g.~\cite{Miller:1990iz,Miller:1994zh}).
For example, in 
the nucleon-nucleon $^1S_0$ channel one has for the scattering lengths $a$ and
the effective ranges $r$ (after removing electromagnetic effects)
\beqa\label{CIBval}
a_{\rm CIB} &=& \frac{1}{2} \left( a_{nn} + a_{pp} \right) - a_{np}
= 5.64 \pm 0.40~{\rm fm}~,\nn
r_{\rm CIB} &=& \frac{1}{2} \left( r_{nn} + r_{pp} \right) - r_{np}
= 0.03 \pm 0.06~{\rm fm}~.
\eeqa
These numbers for charge independence breaking (CIB)
are taken from the recent compilation of Machleidt~\cite{Machleidt:2000ge}.
The charge independence breaking in the scattering lengths is large, of
the order of 25\%, since $a_{np} = (-23.740 \pm 0.020)\,$fm. Of course,
it is magnified at threshold due to kinematic factors (as witnessed
by the disappearance of the effect in the effective
range). In addition, there are
charge symmetry breaking (CSB) effects leading to different values for
the $pp$ and $nn$ phase shifts/threshold parameters, 
\beqa\label{CSBval}
a_{\rm CSB} &=&  a_{pp} - a_{nn} = 1.6 \pm 0.6~{\rm fm}~, \nn
r_{\rm CSB} &=&  r_{pp} - r_{nn} = 0.10 \pm 0.12~{\rm fm}~.
\eeqa
Combining these numbers gives as central values $a_{nn} = -18.9\,$fm
and $a_{pp} = -17.3\,$fm. Notice that this value for $a_{nn}$ is in agreement
with the recent experimental determinations from the reaction $\pi^- d 
\to  n n \gamma$, $a_{nn} = -18.5 \pm 0.5\,$fm \cite{Howell:1998aa}, and
the kinematically complete deuteron breakup reaction $n d  \to  n n p$
at $E_{\rm lab } = 13$~MeV, $a_{nn} = -18.7 \pm 0.6\,$fm
\cite{GonzalezTrotter:1999zz}.  However, another recent experiment also based on the
deuteron breakup reaction at  $E_{\rm lab } = 25.2$~MeV yielded a considerably
smaller value, $a_{nn} = -16.3 \pm 0.4\,$fm \cite{Huhn:2000aa}. 
For a review of indirect methods to measure the $^1S_0$ $nn$ scattering
length and the current experimental status for this observable see \cite{Howell:2008dt}. 

Within the Standard Model,
isospin violation has its origin in the different masses of the up and down
quarks and the electromagnetic interactions. Chiral EFT is well suited to
explore the consequences of these two effects for low-energy dynamics of few-
and many-nucleon systems. Consider first the strong isospin-violating
effects. The QCD quark mass term can be written in the two-flavor case as
\beq
\label{qcd_strong_iso}
\mathcal{L}_{\rm mass}^{\rm QCD} = - \frac{1}{2} \bar q (m_u + m_d ) (1 +
\epsilon \tau^3 ) q\,,
\eeq
where the superscript of the Pauli isospin matrix denotes the corresponding
Cartesian component and 
\beq
\epsilon \equiv \frac{m_u - m_d}{m_u + m_d} \sim - \frac{1}{3}\,.
\eeq
Here, the numerical estimation corresponds to the modified  $\overline{\rm
  MS}$ subtraction scheme 
at a renormalization scale of 1~GeV \cite{Leutwyler:1996qg}. In
Eq.~(\ref{qcd_strong_iso}), the isoscalar term breaks chiral but preserves
isospin symmetry. It is responsible e.g.~for the nonvanishing pion mass
$M^2 = (m_u +m_d) B$, $M_\pi^2 = M^2 + \mathcal{O} (m_{u,d}^2)$,  and
generates a string of 
chiral-symmetry-breaking terms in the effective hadronic Lagrangian 
proportional to $M^{2n}$ with $n=1, 2, \ldots$. The
isovector term gives rise to the strong isospin breaking and leads to hadronic
effective interactions $\propto (\epsilon M^2)^n$. Consequently, the typical
size of the strong isospin violation in hadronic observables is given by 
$\epsilon M_\pi^2 / \Lambda^2 \sim 1\%$ if one takes $\Lambda = M_\rho$ (this,
however, does not apply e.g.~to the
pion masses). The leading and subleading strong isospin-violating
contributions are already incorporated in the Lagrangians $\mathcal{L}_{\pi}^{(2)}$
and $\mathcal{L}_{\pi N}^{(1,2)}$ in Eqs.~(\ref{lagr0}), (\ref{lagr2}),
(\ref{lagr1})  and correspond to terms involving $\chi$, $\chi_\pm$. Notice
that the strong isospin violation is additionally suppressed in the meson
sector (due to G-parity). In particular, the charged-to-neutral pion mass
difference is almost entirely of electromagnetic origin.  
Electromagnetic terms in the effective Lagrangian resulting from
exchange of hard virtual photons can be generated using the method of external
sources \cite{Gasser:1983yg}. 
All such terms are proportional to positive powers of the
nucleon charge matrix $Q = e (1 + \tau^3 )/2$ where $e$ denotes the electric
charge.  In addition, soft photons have
to be included explicitly. For more details on the inclusion of virtual
photons in chiral EFT the reader is referred to 
\cite{Urech:1994hd,Neufeld:1995mu,Meissner:1997fa,Knecht:1997jw,Meissner:1997ii,
Muller:1999ww,Gasser:2002am}.    

To explore isospin-breaking (IB) effects in nuclear forces/few-nucleon
observables it is useful to relate the corresponding small parameters
$\epsilon$ and $e$ to the chiral expansion parameter $M_\pi /
\Lambda$.  Clearly, 
this can be done in various ways. For example, in
Refs.~\cite{Walzl:2000cx,Epelbaum:2004xf,Epelbaum:2005fd} the following rules
have been adopted:
 \beq\label{CountRulesIso}
\epsilon \sim e \sim \frac{M_\pi}{\Lambda}; \quad \quad
\frac{e^2}{(4 \pi )^2}  \sim \frac{M_\pi^4}{\Lambda^4}\,.
\eeq
Similar counting rules were also used in
Refs.~\cite{VanKolck:1993ee,vanKolck:1996rm,Friar:1999zr,Friar:2003yv,Friar:2004ca,Friar:2004rg}. Notice,
however, that in the meson and single-baryon sectors one usually counts
$\epsilon \sim 1$ but  $e \sim M_\pi/\Lambda$.
Utilizing the counting rules in Eq.~(\ref{CountRulesIso}),
the leading and subleading IB contributions from the hard virtual photons have
the form \cite{Muller:1999ww}:  
\beqa
\label{lagr_em}
\mathcal{L}_{\pi, \; \rm em}^{(2)} &=& C \langle Q U Q U^\dagger \rangle \,, \nn
\mathcal{L}_{\pi N , \; \rm em}^{(3)} &=& F^2 \bar{N} \big( f_1  \langle \tilde
Q_+^2 - \tilde Q_-^2 \rangle  +  f_2  \langle 
Q_+ \rangle \tilde  Q_+  \nn
&+&   f_3  \langle \tilde
Q_+^2 + \tilde Q_-^2 \rangle \big) N \,,
\eeqa
where $\tilde Q_\pm \equiv  Q_\pm - 1/2 \langle  Q_\pm \rangle$ and 
$f_i$ refer to the corresponding LECs. The leading Lagrangians for the 
strong and electromagnetic IB contact interactions $\mathcal{L}_{NN, \; \rm
  str}^{(3)}$ and $\mathcal{L}_{NN, \; \rm
  em}^{(4)}$ are given explicitly in Ref.~\cite{Walzl:2000cx}. 

Let us first discuss IB contributions to the pion and nucleon masses. As
already pointed out, the leading contribution to the charged--to--neutral pion
mass difference is entirely of electromagnetic origin, 
\beq
\delta  M_\pi^2 \equiv M_{\pi^\pm}^2 - M_{\pi^0}^2 \simeq  \frac{2}{F^2} e^2 C\,.
\eeq
The experimentally known pion mass difference $M_{\pi^\pm} - M_{\pi^0} = 4.6$~MeV allows  
to fix the value of the LEC $C$, $C = 5.9 \cdot 10^{-5}$ GeV$^4$. Notice that
the natural scale  
for this LEC is $F_\pi^2 \Lambda^2 /(4 \pi)^2 \sim 3 \cdot 10^{-5}$ GeV$^4$
if one adopts $\Lambda \sim M_\rho$. Writing the nucleon mass $m$ 
as 
\beq
m \equiv \mbox{diag} (m_p , \; m_n ) =  m + \frac{1}{2} \delta m \, \tau^3 \,,
\eeq
one obtains for the proton-to-neutron mass difference $\delta m$:
\beq
\label{m_nucl}
\delta m =  -4 c_5 \epsilon M_\pi^2   - f_2 \, e^2 F_\pi^2  + \ldots \,, 
\eeq
where the ellipses refer to higher-order corrections. 
Notice that the $f_3$-term in Eq.~(\ref{lagr_em}) is isospin-invariant. On the
other hand, the $f_1$-term does produce IB vertices with two and more pions
but  does not contribute to the leading electromagnetic
nucleon mass shift. 
The LECs $c_5$ and $f_2$ can be determined from the strong and electromagnetic
nucleon mass shifts:
\beqa
\label{m_shift2}
( m_p - m_n )^{\rm str} &=& (\delta m )^{\rm str} = -2.05 \pm 0.3 \mbox{
  MeV ,}  \nn 
( m_p - m_n  )^{\rm em} &=& (\delta  m )^{\rm em} = 0.76 \pm 0.3  \mbox{ MeV ,} 
\eeqa
which leads to \cite{Meissner:1997ii}
\beq
c_5 = -0.09 \pm 0.01  \mbox{ GeV}^{-1}\,, \quad  
f_2 = -0.45 \pm 0.19    \mbox{ GeV}^{-1}\,.
\nonumber
\eeq
The values for the strong and electromagnetic nucleon mass shifts 
are taken from \cite{Gasser:1982ap}. Notice that the recent lattice QCD 
result \cite{Beane:2006fk} for  the strong nucleon mass shift 
$(\delta m )^{\rm str} = - 2.26 \pm 0.57 \pm 0.42 \pm 0.10$~MeV is in
agreement with the one of Ref.~\cite{Gasser:1982ap}.  
We further emphasize that according to
the counting rules in Eq.~(\ref{CountRulesIso}), the electromagnetic
contribution to the nucleon mass shift is
formally of higher order than the strong one. Based on naive dimensional
analysis,  these contributions are expected to be of the size
$| (\delta m )^{\rm str}  | \sim \epsilon M_\pi^2 / M_\rho \sim 8$~MeV
and  $| (\delta  m )^{\rm em}  | \sim e^2 M_\rho /(4 \pi)^2 \sim 0.5$~MeV.  

We are now in the position to overview the structure of the IB nuclear
forces. The general isospin structure of the two-nucleon force feeds,
according to the classification of Henley and Miller \cite{Henley:1979ig}, into the
four classes:  $V_{2N}^I = \alpha + \beta \, \fet \tau_1 \cdot \fet \tau_2$
(isospin-invariant), $V_{2N}^{II} = \alpha \, \tau_1^3 \tau_2^3$
(charge-independence-breaking), 
 $V_{2N}^{III} = \alpha \, (\tau_1^3  + \tau_2^3 )$
(charge-symmetry-breaking) and 
$V_{2N}^{IV} = \alpha \, (\tau_1^3  - \tau_2^3 ) + \beta \, [ \fet \tau_1 \times
\fet \tau_2 ]^3$ (issospin-mixing). Here, $\alpha$ and $\beta$ denote
the corresponding space and spin operators. Notice that for the class-IV
terms, $\beta$ has to be odd under a time reversal transformation. The
most general isospin structure of the 3NF is worked out in
Ref.~\cite{Epelbaum:2004xf}. While the distinction between the class-I, II and
III forces based on the conservation of the total isospin operator $\fet T =
(\sum_i \fet \tau_i )/2$ and charge-symmetry operator $P_{cs} = \exp(i \pi
T_2)$ can be straightforwardly generalized to any number of nucleons, the
conservation of the operator $\fet T^2$ responsible for the
distinction between the class-III and IV 2NFs depends, in general, on the
number of nucleons. In particular, the class-II and III 2NFs commute with the
operator $\fet T_{2N}^2$ (i.e.~do not mix isospin in the 2N system) but do not
commute with $\fet T_{>2N}^2$. For this reason, the general isospin structure
of the 3NF was classified in Ref.~\cite{Epelbaum:2004xf} in terms of class-I,
II and III contributions. 

The dominant IB  
contribution to the 2NF occurs at NLO$\palka$ (the slash indicates
that we now use the power counting rules extended as in
Eq.~(\ref{CountRulesIso})) due to the charge-to-neutral pion mass difference
in the $1\pi$-exchange potential. It can be accounted for by taking the
proper pion masses in the $1\pi$-exchange potential for various physical
channels: 
\beqa
\label{ope_iso}
V_{1\pi}^{pp} &=& V_{1\pi}^{nn} = V_{1\pi} (M_{\pi^0}) \,, \nn
V_{1\pi}^{np} &=& -V_{1\pi} (M_{\pi^0}) + 2 (-1)^{I+1} V_{1\pi} (M_{\pi^\pm}) \,,
\eeqa
where $I$ denotes the total isospin of the two-nucleon system and 
\beq
V_{1\pi} (M_{\pi^\pm}) = - \frac{g_A^2}{4 F_\pi^2} \frac{\vec \sigma_1 \cdot
  \vec q \, \vec \sigma_2 \cdot \vec q}{\vec q \, ^2 + M_\pi^2}\,.
\eeq
Notice that the resulting IB interaction conserves charge symmetry
(i.e.~class-II) and reaches about $\delta M_\pi^2/M_\pi^2 \sim 7\%$ of the
strength of the isospin-invariant $1\pi$-exchange potential. It is known
to yield a sizeable contribution to the CIB in the $^1S_0$ NN scattering
length, see e.g.~\cite{Walzl:2000cx}. Another IB effect at the same
order comes from the Coulomb interaction between the protons 
(class-II and III). We emphasize that effects of
the purely electromagnetic interactions in two-nucleon scattering observables
get enhanced under certain kinematical conditions (low energies and/or
forward angles) due to the long-range nature of these interactions. Clearly,
such an enhancement goes beyond the simple power counting rules in
Eq.~(\ref{CountRulesIso}). Consequently, despite the fact that the first
corrections to the point-like static one-photon exchange (Coulomb interaction) due to
recoil and two-photon exchange \cite{Austen:1983te}, pion loop contributions
to the nucleon form factors \cite{Kaiser:2006ws}, vacuum polarization
\cite{Ueling:1935aa,Durand:1957zz} and magnetic moment interaction
\cite{Stoks:1990bb} are suppressed by the factor $1/m^2$ and, according to the
power counting, contribute at rather high orders, sizeable
effects may show up in certain observables. For example, the magnetic moment
interaction strongly affects the nucleon analyzing power $A_y$  at low energy
and forward angles. Effects of subleading electromagnetic interactions were also
investigated in 3N continuum using phenomenological nuclear forces
\cite{Kievsky:2003eu,Witaa:2003gs}, see \cite{Rupak:2001ci} for a formulation 
based on pionless EFT. 

The corrections to the IB 2NF at N$^2$LO$\palka$ are CSB and arise from 
charge dependence of the pion-nucleon coupling constant in the $1\pi$-exchange
potential and the derivative-less NN contact interaction $\propto m_u - m_d$
\cite{vanKolck:1996rm,Epelbaum:2005fd}. Notice, however, that the
energy-dependent Nijmegen PWA does not
yield any evidence for charge dependence of the pion-nucleon coupling constant
\cite{deSwart:1997ep}.  The leading
CIB contact interactions are of electromagnetic
origin and (formally) start to contribute at N$^3$LO$\palka$. At this order,
one also has to take into account further IB contributions to the
$1\pi$-exchange potential at the one-loop level which, to a large
extent, can be accounted for by a further (charge-dependent) renormalization
of the $\pi N$ coupling constants in Eq.~(\ref{ope_iso}). The only
contributions which have a different momentum dependence and, therefore,
cannot be cast into the form of Eq.~(\ref{ope_iso}) are 
the ones $\propto (\delta M_\pi )^2$ and  the 
proton-to-neutron mass difference which involve class-IV operators 
\cite{Epelbaum:2005fd,Friar:2004ca}, see \cite{Cheung:1979ma} for a much earlier derivation
of these terms. Notice that the power counting rules in
Eq.~(\ref{CountRulesIso})  suggests the following
hierarchy of the 2NF \cite{VanKolck:1993ee}: $V_{2N}^I > V_{2N}^{II} >
V_{2N}^{III} >  V_{2N}^{IV} $ which is consistent with the observations. 
Next,  $\pi \gamma$-exchange also
contributes at this order. The resulting CIB potential has been worked out by
van Kolck at al.~\cite{vanKolck:1997fu} and re-derived recently by Kaiser
\cite{Kaiser:2006ws}. It can be written in a rather compact way and leads to
negligibly small effects in NN scattering. 
Kaiser also calculated subleading contributions to the
$\pi \gamma$-exchange 
potential proportional to the large isovector magnetic moment $\kappa_v = 4.7$
of the nucleon and found that the resulting potentials, which
are also CIB, have a similar strength as the leading-order one
\cite{Kaiser:2006ws}.\footnote{Notice, however, that these corrections are
  suppressed by the factor $1/m$ relative to the leading-order contributions
  and, therefore, appear formally at N$^5$LO$\palkasmall$.} IB
$2\pi$-exchange also starts to contribute at N$^3$LO$\palka$ and is driven by
the neutral-to-charged pion mass difference (CIB) \cite{Friar:1999zr} and the strong 
contribution to the nucleon mass shift (CSB) 
\cite{Coon:1995qh,Niskanen:2001aj,Friar:2003yv,Epelbaum:2005fd}, see \cite{Walzl:2000cx}
for the application to NN phase shifts. Finally, there are also the first IB
3NFs. While the dominant CIB $2\pi$-exchange 2NF is generated by the pion mass
difference, the 3N diagrams with one insertion of $\delta M_\pi^2$ at
N$^3$LO$\palka$ are additionaly suppressed by the factor $1/m$ (if one uses an
energy-independent formulation), see the
discussion about the 3NFs at NLO earlier in the text.  The
nonvanishing 3NFs at N$^3$LO$\palka$ result from $1\pi$- and $2\pi$-exchange
3N diagrams constructed with the leading-order isospin-invariant vertices and
a single insertion of $\delta m$ as well as $2\pi$-exchange diagram with the
leading IB $\pi \pi NN$ interactions $\propto f_{1,2}$
\cite{Epelbaum:2004xf,Friar:2004rg}. One finds that all these
contributions are CSB except the one which is proportional to LEC $f_1$ and is
CIB. We further emphasize that while the value of the LEC $f_2$ is determined
by the electromagnetic nucleon mass shift, the value of the LEC $f_1$ is
unknown. However, see \cite{Gasser:2002am} for an estimation of $f_1$
based on dimensional analysis and \cite{Meissner:2005ne}
for an attempt to determine $f_1$ from data.

Remarkably, even the N$^4$LO$\palka$ contributions to the two- and
three-nucleon forces have been worked out. At this order, no new
structures appear in the $1\pi$-exchange potentials. 
The corrections to the leading IB $2\pi$-exchange potential result from
a single insertion of 
either the subleading isospin-conserving $\pi \pi NN$
vertices proportional to the LECs $c_i$, see Eq.~(\ref{lagr1}), the leading
electromagnetic vertex proportional to the LEC $f_{2}$, see Eq.~(\ref{lagr_em}),
\footnote{The two other LECs $f_{1,3}$ do not contribute to the $2\pi$-exchange 2NF at
this order.} or the (poorly known) leading charge-dependance of the pion-nucleon coupling
constant \cite{Epelbaum:2005fd}. The resulting IB potentials involve the class-II
and III central, tensor and spin-spin components. The CIB potentials typically
have the strength of a few 10 keV at relative distances $r \sim M_\pi^{-1}$. The
CSB tensor and spin-spin potentials are weaker ($< 10$ keV) while the CSB
central potential is comparable in size to the CIB contributions. Similarly to
the isospin-conserving $2\pi$-exchange potential, the subleading contributions
turn out to be numerically large in comparison to the leading-order ones. In
particular, for the class-III central $2\pi$-exchange potential one obtains
$V_{2N}^{2\pi , \,  (5)}/V_{2N}^{2\pi , \, (4)} \simeq 3$ for $r \sim
M_\pi^{-1}$. The main reason for this unpleasant convergence pattern
is the same as in the isospin-conserving case and
can be traced back to the (large) $\Delta$-isobar contributions to the LECs
$c_{3,4}$. We will discuss this issue in more detail in section \ref{sec:chiralEFTdelta}. 
Last but not least, there are also numerous IB contact interactions with up to
two derivatives involving class-II, III and IV terms, see also
\cite{Friar:2004ca}. The corrections to the 3NF at N$^4$LO$\palka$ are worked
out in Refs.~\cite{Epelbaum:2004xf,Friar:2004rg}. At this order, the first IB
but charge symmerty conserving 3NFs show up which result from the
neutral-to-charged pion mass difference in the N$^2$LO $2\pi$- and
$1\pi$-exchange diagrams in the second line of Fig.~\ref{fig:3NF} and the 
$2\pi$-exchange diagram involving the $\pi \pi NN$ vertex $\propto f_1$. 
In addition, there are CSB 3NFs of the $2\pi$- and $1\pi$-exchange types
driven by the electromagnetic nucleon mass shift. 
Again, the strongest 3NFs turn out to be the ones which are proportional to
the LECs $c_{3,4}$. They are charge-symmetry conserving and arise from a
single insertion of $\delta M_\pi^2$ into the pion propagators of
the $2\pi$-exchange 3N graph. The expected strength of such IB potentials is $\sim
2\delta M_\pi^2/M_\pi^2 \sim 13 \, \%$ as compared to the isospin-invariant
ones given in Eq.~(\ref{leading}) which are known to yield about $\sim 500 \ldots
1000$ keV to the triton binding energy (the precise numbers are renormalization
scheme dependent). Also the strength of the corresponding coordinate-space
potentials, e.g.~ $| g_A^2 \delta M_\pi^2  M_\pi^4 c_3 /(64 \pi^2 F_\pi^4)|
\sim 70$ keV (here we picked out one particular term), indicates that the
resulting IB effects in few-nucleon observables might be 
sizeable. The CSB 3NFs, on the other hand, do show a more natural convergence
pattern and are considerably waeker. Their contribution to e.g.~the $^3$H-$^3$He
binding energy difference is expected to be of the order $\sim 10$ keV. 

Recently, a certain classes of even higher-order contributions have
been worked out by Kaiser. In particular, he calculated the subleading  $\pi \pi
\gamma$-exchange potentials proportional to the LECs $c_{3,4}$ at the 2-loop
level which (formally) contribute at order N$^6$LO$\palka$  $\,$
\cite{Kaiser:2006ck,Kaiser:2006na}. Especially the contributions driven by the
LEC $c_3$ were found to generate astonishingly strong CSB and CIB potentials
which amount to $\sim 1\%$ of the strongly attractive isoscalar central
potential at N$^2$LO and reach a few 100s of keV at $r \sim M_\pi^{-1}$.
Notice, however, that effects of these very strong potentials in S-, P-
and D-waves may, to some extent, be compensated by the corresponding 
IB contact interactions.
The effects in higher partial waves are presumably suppressed due to the
shorter range of the $2\pi$-exchange potential compared with the
$1\pi$-exchange one.

\subsection{Chiral EFT for few nucleons: applications}
\label{sec:chiralEFTappl}

We now turn our attention to applications. As discussed in the previous
section, the two-nucleon chiral potential involves the long-range
contributions due to the multiple pion exchanges and short-range ones 
parametrized by contact interactions. Both kinds of terms typically grow with
increasing nucleon momenta and become meaningless in the large-momentum region
as follows from the very nature of EFT being the low-momentum expansion. 
As a consequence, the Schr\"odinger equation is ultraviolet divergent 
and needs to be regularized (and renormalized). The problem of renormalization
in the nonperturbative regime in the context of both pion-less 
\cite{Beane:1997pk,Phillips:1997xu,Birse:1998dk,
Phillips:1998uy,Gegelia:1998iu,Yang:2004ss,Harada:2006cw,Harada:2007ua,
Braaten:2004rn}
and pion-full EFT 
\cite{Lepage:1997,Cohen:1998bv,Frederico:1999ps,
Phillips:1999bf,Lepage:2000,Gegelia:2001ev,Beane:2001bc,PavonValderrama:2003np,
PavonValderrama:2004nb,Gegelia:2004pz,PavonValderrama:2005gu,Nogga:2005hy,
PavonValderrama:2005wv,PavonValderrama:2005uj,Birse:2005um,
Epelbaum:2006pt,Djukanovic:2006mc,Higa:2007gz,Yang:2007hb,
Long:2007vp,Entem:2007jg,Birse:2007sx,Valderrama:2008kj,Shukla:2008sp}
has attracted a lot of interest in the past years. 
The standard procedure to renormalize the Lippmann-Schwinger (LS) 
equation is based on Wilson's
method and implies the following two steps \cite{Lepage:1997}. First, 
one solves the LS equation regularized with the finite
momentum (or coordinate-space) cutoff and using the
potential truncated at a given order in the chiral expansion as the kernel.    
Secondly, the LECs accompanying the contact terms in the potential are 
determined by matching the resulting phase shifts to experimental data which,
in this framework, can be viewed as renormalization. Notice that iterating the
\emph{truncated} expression for the chiral potential in the LS equation
necessarily generates
ultraviolet divergencies in the Neumann series which require counterterms
beyond the given approximation for the potential. 
As a consequence, taking the limit of the infinite cutoff
in such a manifestly nonrenormalizable (in the above mentioned sense) approach
might result e.g.~in impossibility to resolve the (nonlinear) matching
conditions for the corresponding LECs. A detailed discussion on the 
choice of ultraviolet cutoff and its role in renormalization of the
Schr\"odinger equation is 
given by Lepage \cite{Lepage:1997,Lepage:2000}. He argued that the
coordinate-space (momentum-space) cutoff should not be decreased (increased)
beyond the separation scale, after which the description of  
the data stops to improve. Taking the cutoff near this separation scale is the
most efficient choice. This strategy has been followed by the currently
most advanced N$^3$LO analyses of the 2N system of
Refs.~\cite{Entem:2003ft,Epelbaum:2004fk} where the cutoffs $\Lambda = 450
\ldots 600$ MeV have been employed. These studies were critisized by Nogga et
al.~\cite{Nogga:2005hy} who considered low NN partial waves based on the 
$1\pi$-exchange
potential and contact interactions using a much bigger cutoff variation with
$\Lambda < 4$~GeV. They found that higher-order counterterms have to be
promoted to LO in the $^3P_0$, $^3P_2$-$^3F_2$ and $^3D_2$ channels in order
to stabilize the amplitude. On the other hand, the efficiency of such a
modified power counting framework was questioned in 
Ref.~\cite{Epelbaum:2006pt}, where it has
been demonstrated that increasing the cutoff and promoting counterterms as
suggested in Ref.~\cite{Nogga:2005hy}
does not improve the overall description of the scattering observables. 
For more discussions on the conceptual issues related to the
power counting in the NN system the reader is referred to
\cite{Lepage:1997,Lepage:2000,Gegelia:2004pz,Nogga:2005hy,Epelbaum:2006pt,
Birse:2005um,Long:2007vp,Birse:2007sx}. More work is
needed in the future in order to clarify the relation between 
the well-established chiral expansion of the nuclear potential and 
the scattering amplitude.  

We further emphasize that it is possible to \emph{nonperturbatively}
renormalize the partial-wave-projected LS equation with
singular $1/r^n$-potentials
\cite{Beane:2000wh,Bawin:2003dm,Braaten:2004pg,Barford:2004fz,
Hammer:2005sa,Long:2007vp}.
This program was applied to different NN channels
based on the $1\pi$- and $2\pi$-exchange potentials
at various orders in the chiral expansion by the Granada group
\cite{PavonValderrama:2003np,PavonValderrama:2004nb,PavonValderrama:2005gu,
PavonValderrama:2005wv,PavonValderrama:2005uj,Higa:2007gz,Entem:2007jg,Valderrama:2008kj}. 
In these studies, the short-range counter
terms are replaced by adjustable parameters entering the short-distance boundary
conditions. The number of such parameters in each channel is uniquely
determined by the sign (attractive vs.~repulsive) of the strongest
singularity which raises concerns about a systematic improvability (in the EFT
sense) of such a framework. Nevertheless, the findings 
of these studies in attractive channels provide an impressive demonstration of
the existence of the long-range correlations in the NN scattering
observables. 

\begin{figure}[tb]
\includegraphics*[width=0.47\textwidth]{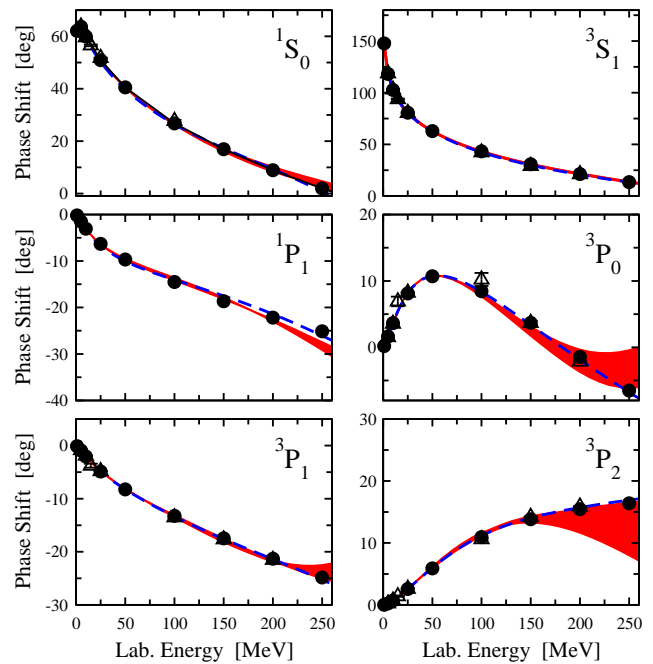}
\vspace{0.1cm}
\begin{center}
\caption{Neutron-proton phase shifts in S- and P-waves at N$^3$LO in
  comparison with the Nijmegen \cite{Stoks:1993tb,NNonline} (filled circles) and Virginia Tech
  \cite{SAID} (open triangles) PWA. Shaded bands (dashed lines) refer to the calculations by
  EGM \cite{Epelbaum:2004fk} (EM \cite{Entem:2003ft}). 
\label{fig:SP}}
\vspace{-0.9cm}
\end{center}
\end{figure}

\begin{figure}[tb]
\includegraphics*[width=0.47\textwidth]{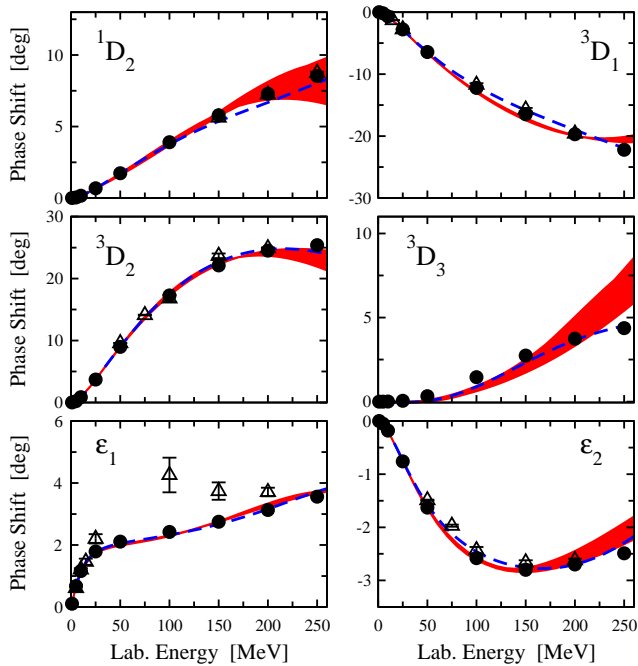}
\vspace{0.1cm}
\begin{center}
\caption{Neutron-proton phase shifts in D-waves and the mixing angles
  $\epsilon_{1,2}$ at N$^3$LO. For notation see Fig.~\ref{fig:SP}.
\label{fig:D}}
\vspace{-0.9cm}
\end{center}
\end{figure}

The most advanced analyses of the two-nucleon system based on the Weinberg
power counting take into account the 2NF contributions up to N$^3$LO
\cite{Entem:2003ft,Epelbaum:2004fk}. Most of the LECs $c_i$, $d_i$ entering
the long-range part of the potential are sufficiently well determined in
the pion-nucleon system \cite{Fettes:1998ud}.\footnote{Notice, however, that
  the value of the LEC $c_4$ adopted in \cite{Entem:2003ft}, $c_4=5.4$
  GeV$^{-1}$ is not compatible with pion-nucleon scattering where one finds at
order $Q^3$ \cite{Buettiker:1999ap}: $c_4 = 3.40 \pm 0.04$ GeV$^{-1}$.} The
24\footnote{This  number
  refers to isospin-invariant contact interactions.} 
unknown LECs entering the short-range  part of the 2NF at N$^3$LO have
been extracted from the low-energy NN data for several choices of the cutoff
in the Schr\"odinger equation. 
Both N$^3$LO potentials of Entem and Machleidt \cite{Entem:2003ft} (EM) and
Epelbaum, Gl\"ockle and Mei{\ss}ner \cite{Epelbaum:2004fk} (EGM) yield accurate results for the
neutron-proton phase shifts up to $E_{\rm lab} \sim 200$~MeV and the deuteron observables.  
This is exemplified in Figs.~\ref{fig:SP}, \ref{fig:D} where the EGM and EM results for
the neutron-proton S- P- and D-waves and the corresponding mixing angles are
shown in comparison with PWA results from
Refs.~\cite{Stoks:1993tb,NNonline,SAID}. The bands in the EGM analysis result
from the variation of the cutoff in the LS equation (spectral function
regularization)  in the range $\Lambda = 450 \ldots 600$ MeV ($\tilde \Lambda
= 500 \ldots 700$ MeV). 
It is comforting to see that in most cases the results of both analyses agree
with each other within the estimated theoretical uncertainty.
Notice, however, that the EM and EGM analyses differ from each other in
several important aspects. For example, the so-called spectral function
regularization \cite{Epelbaum:2003gr,Epelbaum:2003xx} 
of the $2\pi$-exchange contributions has been adopted by EGM
while the analysis by EM is based upon dimensionally regularized expressions. 
Further differences can be attributed to the implementation of the
momentum-space cutoff in the Schr\"odinger equation and the treatment of
relativistic effects. More precisely, the work by EGM is based on the
``relativistic'' Schr\"odinger/Lippmann-Schwinger equation, a natural
extension of the usual nonrelativistic equations utilizing the
relativistic relation between the CMS energy and momentum, see \cite{Friar:1999sj} for
more details. This equation can be straightforwardly generalized to the case
of several nucleons, see \cite{Witala:2004pv,Lin:2007kg,Lin:2008sy} for recent studies of relativistic
effects in 3N observables, and can also be cast into equivalent 
nonrelativistic-like forms \cite{Kamada:1999wy,Friar:1999sj} 
(provided the potential is appropriately modified). On the other hand, the
analysis of Ref.~\cite{Entem:2003ft} is based on the  
nonrelativistic Schr\"odinger equation and uses the static $1\pi$-exchange
potential and 
the $1/m$- and $1/m^2$-corrections to the  
$2\pi$-exchange potential from Refs.~\cite{Kaiser:1997mw,Kaiser:2001at}, where no
particular dynamical equation is specified.  
It is unclear whether the relativistic corrections to the
potential used in \cite{Entem:2003ft} 
are consistent with the nonrelativistic Schr\"odinger equation. Further
differences between the EGM and EM analyses result from the fitting
procedure: the LECs accompanying the contact interactions were determined by
EM/EGM by fitting directly to the scattering data/to the Nijmegen PWA
\cite{Stoks:1993tb,NNonline}. For this reason, EGM adopted the same treatment
of IB effects as followed by the Nijmegen group and did not include e.g.~the
leading IB contributions to the $2\pi$-exchange potential. Perhaps, the most
important difference between the two studies is related to the estimation of
the theoretical uncertainty. In the work by EGM, the theoretical uncertainty
was estimated by varying the cutoffs in the Schr\"odinger equation and
the spectral function representation for the $2\pi$-exchange potential 
restricted by the condition that the resulting
LECs  are of a natural size
which might be viewed as a self-consistency check for calculations carried out
within the power counting scheme based on naive dimensional analysis, see
\cite{Beane:2000fx,Epelbaum:2005pn}. Notice further that at NLO and N$^2$LO,
the strengths of various contact interactions are well understood in terms of
resonance saturation on the basis of phenomenological one-boson exchange
models \cite{Epelbaum:2001fm}. No serious attempt to
provide a realistic error estimation was done in the analysis of EM.  
On the other hand,
their work clearly demonstrates that for a particularly chosen regularization
prescription it is even possible to accuratly describe two-nucleon scattering data for
$E_{\rm lab} > 200$~MeV.  For further technical 
details, results for various scattering observables and the properties of the
deuteron the reader is referred to the original publications \cite{Entem:2003ft,Epelbaum:2004fk}
and the review article \cite{Epelbaum:2005pn}. 

\begin{figure}[tb]
\includegraphics*[width=0.35\textwidth]{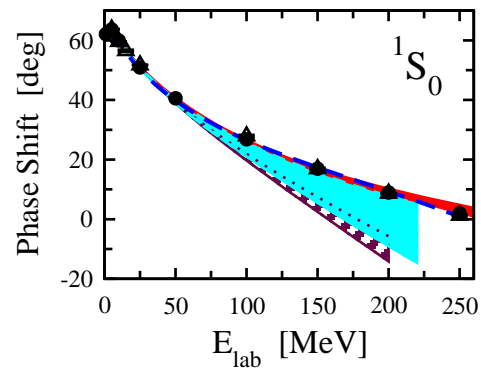}
\vspace{0.1cm}
\begin{center}
\caption{Neutron-proton $^1$S$_0$ partial wave at NLO (dashed band), N$^2$LO
  (light-shaded band) and N$^3$LO (dark-shaded band) 
  in comparison with the Nijmegen \cite{Stoks:1993tb,NNonline} (filled circles) and Virginia Tech
  \cite{SAID} (open triangles) PWA.
\label{fig:1S0}}
\vspace{-0.5cm}
\end{center}
\end{figure}

To illustrate the
convergence of the chiral expansion for NN phase shifts, we show in
Fig.~\ref{fig:1S0} the results for the $^1S_0$ partial wave at NLO, N$^2$LO
\cite{Epelbaum:2003xx} and N$^3$LO \cite{Epelbaum:2004fk}.  We emphasize that
the variation of the cutoff at both NLO and N$^2$LO only shows the effects of
missing N$^3$LO contact interactions. It, therefore, does not provide a
realistic estimation of the
theoretical uncertainty at NLO, see \cite{Epelbaum:2005pn} for an extended
discussion.   

Applications to the three-nucleon system have so far been carried out
up to N$^2$LO. At NLO, no 3NF needs to be taken into account. This allowed for
a parameter-free predictions of various 3N scattering observables at low
energies as well as for the triton and $\alpha$-particle binding energies 
\cite{Epelbaum:2000mx}. Using the most recent version of the NLO potential
based on the spectral functions regularization, one finds
at NLO \cite{Epelbaum:2005pn} $B_{^3\rm H}= 7.71 \ldots 8.46$~MeV and
$B_{^3\rm He}= 24.38 \ldots 28.77$~MeV to be compared with the experimental
values $B_{^3\rm H}= 8.482$~MeV and 
$B_{^3\rm He}= 28.30$~MeV. These numbers are similar to the ones obtained in
Ref.~\cite{Epelbaum:2000mx} within the framework based on dimensional
regularization. 

At N$^2$LO one, for the first time, has to take into account the crresponding
3NFs.  The two LECs $D$ and $E$ entering the expressions for the 3NF in
Eq.~(\ref{leading}) have been determined by fitting the $^3$H binding energy   
and either the {\it nd} doublet scattering length \cite{Epelbaum:2002vt}, 
the $^4$He binding energy \cite{Nogga:2005hp} or the properties of light nuclei
\cite{Navratil:2007we}.  
Notice that the $\pi NNNN$ vertex entering the $1\pi$-exchange-contact 3NF 
also plays an important role in processes with a completely different
kinematics such as e.g.~the pion production in the NN collisions
\cite{Hanhart:2000gp}, see section \ref{sec:further_scales},  or weak
reactions like $pp\to d e^+ \nu_e$,  
see \cite{Nakamura:2007vi} and references therein. This offers the
possibility to extract the corresponding LEC from these processes, see
\cite{Nakamura:2007vi} for a recent attempt. With the LECs being determined as
described above, the resulting nuclear Hamiltonian can be used to describe the
dynamics of few-nucleon systems. In particular,  3N
continuum observables offer a natural and rich 
testing ground for the chiral forces. In
Refs.~\cite{Epelbaum:2000mx,Epelbaum:2002ji,Epelbaum:2002vt,Ermisch:2003zq,
Duweke:2004xv,Ermisch:2005kf,Kistryn:2005fi,Witala:2006nn,Biegun:2006zc,
Ley:2006hu,Stephan:2007zza}
various 3N scattering
observables have been explored by solving the momentum-space Faddeev
equations with chiral two- and three-nuclein forces as input. 
In the formulation 
of Ref.~\cite{Gloeckle:1995jg}, one first computes the $T$-matrix
by solving the Faddeev-like integral equation 
\beqa
\label{eq2}
T &=& t \,  P  \, \phi + (1 + t  \, G_0)  \, V_{3N}^{1}  \, (1 +
P)  \, \phi + t  \, P  \, G_0  \, T  \nn
&+&  (1 + t  \, G_0)   
\, V_{3N}^{1} \,  (1 + P)  \, G_0  \, T\,,
\eeqa
where the initial state $\phi$ is composed of a deuteron and a
momentum eigenstate of the projectile nucleon.  
Here $V_{3N}^{i}$ is that part of the 3N force
which singles out the particle $i$ and which is symmetric under the
interchange of the two other particles. The complete 3NF is decomposed as
$V_{3N} = V_{3N}^{1}+ V_{3N}^{2}+V_{3N}^{3}$.
Further, $G_0=1/(E-H_0)$ is the free propagator of the nucleons,  
$P$ is a sum of a cyclical and anti-cyclical permutation of the three particles
and $t$ denotes the two-body $t$-matrix.
Once $T$ is calculated, the transition operators
$U_{\rm el}$ and $U_{\rm br}$ for the elastic and break-up channels can
be obtained via
\beqa
U_{\rm el} &=&P  \, G_0^{-1} + P  \, T + V_{3N}^{1} \,  (1 + P) \,  (1 +
G_0  \, T)\,, \nn
U_{\rm br} &=& (1 + P)  \, T\,.
\eeqa
For details on solving these equations in momentum space using a
partial wave decomposition the reader is referred to 
\cite{Huber:1996td}. The partial wave decomposition of the $1\pi$-exchange and
contact 3NF at N$^2$LO and the one-pion-two-pion-exchange topology at N$^3$LO
is detailed in Refs.~\cite{Epelbaum:2002vt} and \cite{Kamada:2008zz},
respectively. The expressions
for various observables in terms of the transition operators are given in
\cite{Gloeckle:1995jg}. 
The inclusion of the long-range 
electromagnetic interaction requires a non-trivial generalization of the formalism,
see \cite{Deltuva:2005cc,Deltuva:2005zz} for recent progress along this line.  
 
\begin{figure}[tb]
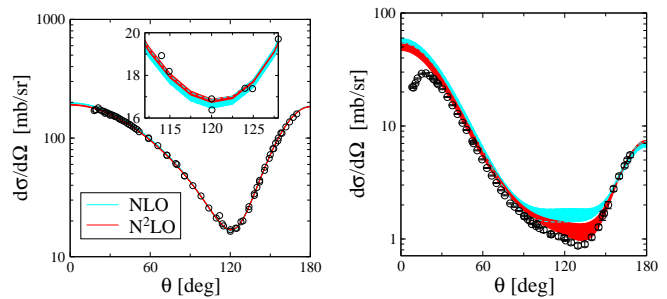

\begin{center}
\begin{minipage}{0.48\textwidth}
\includegraphics[width=0.48\textwidth]{ds10_new.eps}
\hfill
\includegraphics[width=0.48\textwidth]{ds65_new.eps}
\end{minipage}
\caption{Differential cross section for elastic {\it nd}
  scattering at $E_{\rm lab} = 10$~MeV (left panel) and 65~MeV (right
  panel). Light (dark) shaded bands depict the results at NLO (N$^2$LO). The
  neutron-deuteron data at 10~MeV are from Ref.~\cite{Howell:1987aa}. The 
  remaining data at 10~MeV are the Coulomb/IB-corrected proton-deuteron 
  data from Refs.~\cite{Sagara:1994zz,Rauprich:1988aa,Sperisen:1984aa}. The
  data at 65~MeV are proton-deuteron data from Ref.~\cite{Witala:1993aa}.}  
\label{ds_10_65} 
\end{center}
\end{figure}

\begin{figure}[tb]
\begin{center}
\vskip -0.2 true cm
\begin{minipage}{0.49\textwidth}
\includegraphics[width=0.49\textwidth]{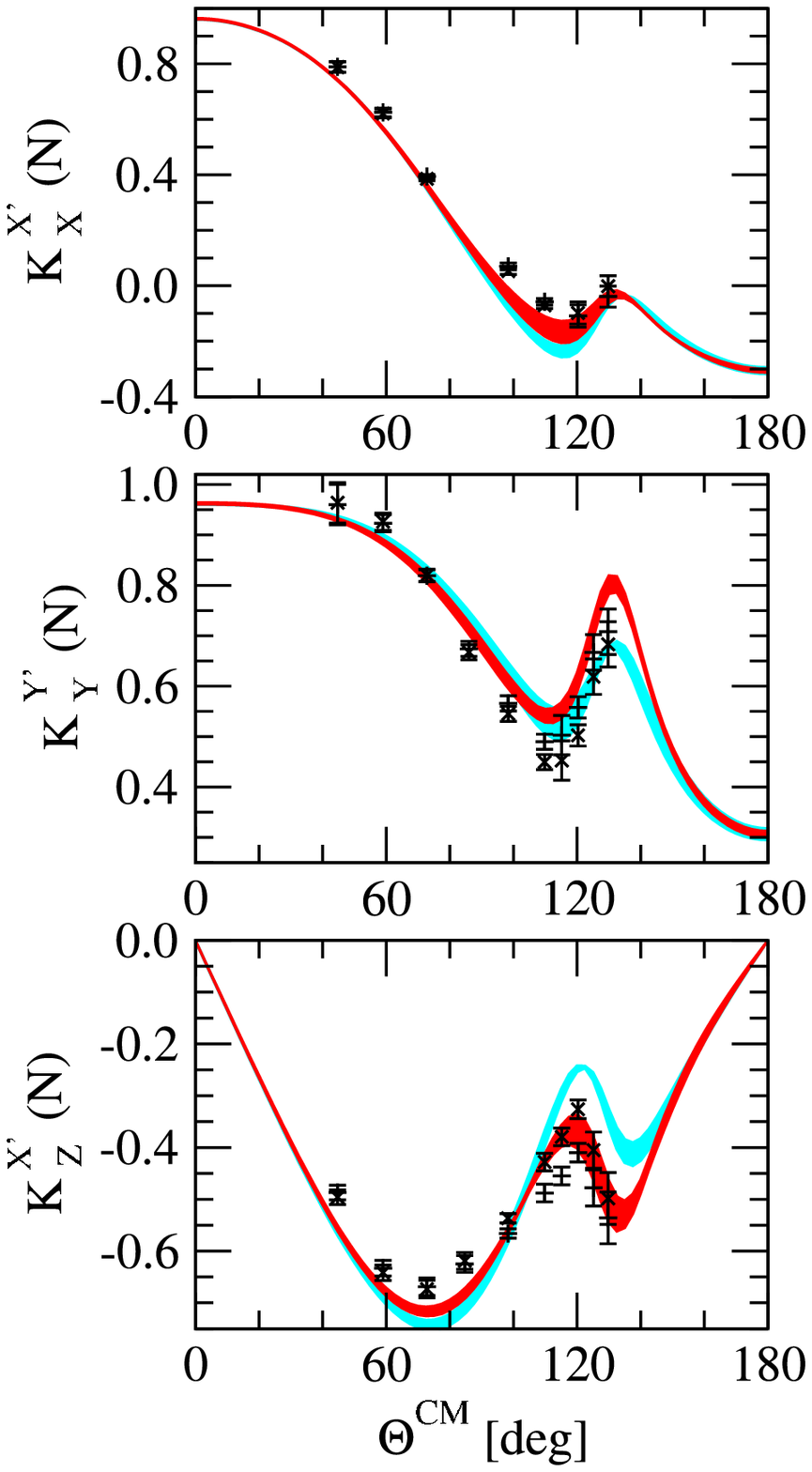}
\hfill
\includegraphics[width=0.49\textwidth]{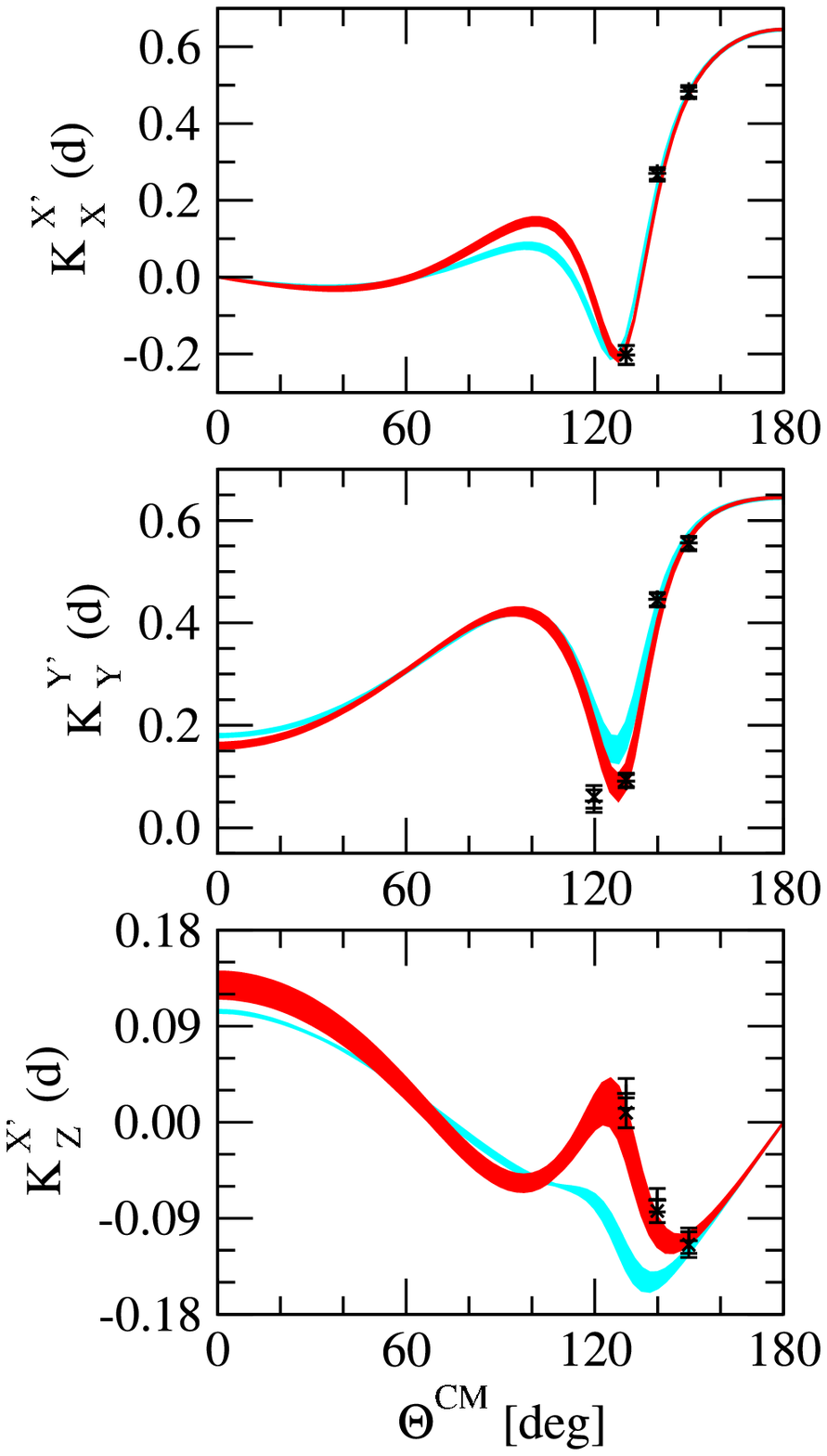}
\end{minipage}
\caption{The 
proton-to-proton (left panel) and proton-to-deuteron (right panel) polarization transfer
coefficients in $d (\vec p 
, \, \vec p \,) d$ and $d (\vec p , \, \vec d \,) p$  
reactions at $E_p^{\rm lab} = 22.7$.  Light (dark) shaded bands depict the
results at NLO (N$^2$LO).  Data are from
Refs.~\cite{Glombik:1995aa,Kretschmer:1995aa}. }  
\label{fig_spin_transf} 
\end{center}
\end{figure}

The results for the
differential cross section in elastic {\it nd} scattering are in a
good agreement with the data, see Fig.~\ref{ds_10_65} for two representative
examples.  
Notice, however, that the theoretical uncertainty
becomes significant already at intermediate energies. Qualitatively, this
behavior is consistent with the one observed in the two-nucleon system
\cite{Epelbaum:2004fk}. Notice further that the description of the
data improves significantly when going from NLO to N$^2$LO.   
The situation is similar for vector and tensor analyzing
powers, see Ref.~\cite{Epelbaum:2005pn} for a recent review article. 
More  complicated spin observables have also been studied. As a representative
example, we show in Fig.~\ref{fig_spin_transf} a selection of the 
proton-to-proton and proton-to-deuteron polarization transfer coefficients 
measured in $d (\vec p , \, \vec p \,) d$ and $d (\vec p , \, \vec d \,) p$ 
reactions at $E_p^{\rm lab} = 22.7$~MeV \cite{Glombik:1995aa,Kretschmer:1995aa}. 
The results at N$^2$LO are in a reasonable agreement with the data, see
\cite{Witala:1993aa} for more examples. One further observes that the
theoretical uncertainty obtained by the cutoff variation is underestimated at
NLO, see the discussion earlier in the text. It is, however, comforting to see
that the description of the
data improves significantly when going from NLO to N$^2$LO. 

\begin{figure}[tb]
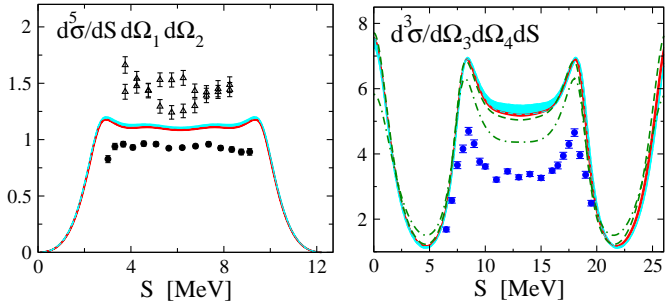

\begin{minipage}{0.49\textwidth}
\includegraphics*[width=0.50\textwidth]{sst.eps}
\hfill
\includegraphics*[width=0.47\textwidth]{3N_scatt.eps}
\end{minipage}
\vspace{-0.3cm}
\begin{center}
\caption{Chiral EFT predictions for neutron-deuteron breakup cross section (in mb
         MeV$^{-1}$ sr$^{-2}$) along the kinematical locus $S$. Light-shaded
         (dark-shaded) bands refer to the results at NLO (N$^2$LO). Left
         panel: The SST configuration at $E_{N}=13$~MeV. Neutron-deuteron data
         (open triangles) are from \cite{Strate:1989aa,Setze:1996aa},
         proton-deuteron data (filled 
         circles) are from \cite{Rauprich:1991aa}. 
         Right panel: The SCRE configuration with $\alpha = 56^\circ$ at
         $E_{N}=19$~MeV  \cite{Ley:2006hu}. Dashed and 
         dashed-dotted lines are results based on the CD
         Bonn 2000 2NF \cite{Machleidt:2000ge} combined with the TM99 3NF
         \cite{Coon:2001pv} and the coupled  channel
         calculation including the explicit $\Delta$ and the Coulomb
         interaction \cite{Deltuva:2005cc}, respectively.
\label{fig:breakup}}
\vspace{-0.5cm}
\end{center}
\end{figure}

The nucleon-deuteron breakup reaction offers even more
possibilities than the elastic channel due to the much richer kinematics
corresponding to three nucleons in the final state. It 
has also been studied extensively over
the last years, both theoretically and experimentally, 
leaving one with mixed conclusions. While the differential cross section in
some configurations such as e.g.~the recently measured {\it np} final-state
interaction, co-planar star and an intermediate-star geometries at low
energies are in a very
good agreement with the data \cite{Duweke:2004xv}, large deviations are
observed in certain other configurations. In particular, the so-called
symmetric space-star configuration (SST) 
appears rather puzzling. 
In this configuration, the plane in the CMS spanned by the 
outgoing nucleons is perpendicular to the beam axis, and the angles between the
nucleons are $120^\circ$. 
At $E_{\rm lab} = 13$~MeV, the
proton-deuteron and neutron-deuteron ({\it nd}) cross section data deviate
significantly from each other. Theoretical calculations based on both
phenomenological and 
chiral nuclear forces have been carried out for the {\it nd} case and 
are unable to describe the data, see Fig.~\ref{fig:breakup}.  
Moreover, the Coulomb effect was 
found to be far too small to explain the difference between the {\it pd} 
and {\it nd} data sets. Recently, proton-deuteron data for a similar symmetric 
constant relative-energy (SCRE) configuration have been measured in
Cologne \cite{Ley:2006hu}. This geometry is characterized by the angle $\alpha$
between the beam axis and the plane in the CMS spanned by the 
outgoing nucleons. Similar to the SST
geometry, one observes large deviations between the theory and the data, in
particular for $\alpha = 56^\circ$, see Fig.~\ref{fig:breakup}. The included
3NFs have 
little effect on the cross section while the effect of the Coulomb interaction 
is significant and removes a part of the discrepancy. Notice that all above
cases correspond to rather 
low energies where one expects good convergence of the chiral
expansion. Furthermore, contrary to the $A_y$-puzzle, the cross sections
discussed above are mainly sensitive to the two-nucleon S-waves without any 
known fine tuning between partial waves. 
First attempts have been made in the past few years to perform deuteron breakup
experiments at intermediate energies, in particular at $E_{N}=65$~MeV \cite{Kistryn:2005fi}, 
in which a large part of the phase space is covered at once. Chiral EFT
results at N$^2$LO for more than 155 data points were shown to be of a comparable 
quality to the ones based on modern phenomenological nuclear forces. 

Recently, first results for the 4N continuum based on both phenomenological and
chiral nuclear forces and including the Coulomb interactions have become
available, see \cite{Fisher:2006pm,Deltuva:2007xv} for $p -  ^3$He
scattering, \cite{Deltuva:2007xw} for the $n - ^3$He,  $p - ^3$H and $d-d$
scattering, and \cite{Lazauskas:2004uq} for the related earlier work. These
studies do not yet 
include effects of 3NFs but clearly indicate that at least 
some of the puzzles observed in the 3N continuum also persist in the 4N
continuum  (such
as e.g.~the $A_y$-puzzle in $p - ^3$He scattering \cite{Deltuva:2007xv}).  
For a promising new approach to describe scattering states in even heavier
systems the reader is referred to \cite{Quaglioni:2008sm}. 

\begin{figure}[tb]
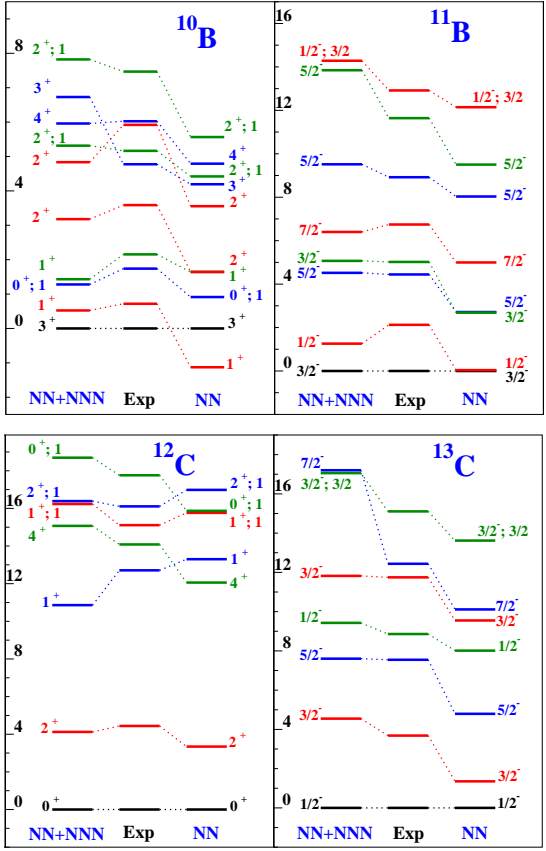

\includegraphics*[width=0.4\textwidth]{temp1.eps}
\vskip 0.2 true cm 
\hskip 0.1 true cm
\includegraphics*[width=0.4\textwidth]{temp2.eps}
\vspace{0.13cm}
\begin{center}
\caption{States dominated by P-shell configurations for $^{10}$B, $^{11}$B,
  $^{12}$C and $^{13}$C. The excitation energy scales are in MeV. Calculation
  is carried out in the framework of NCSM based on chiral N$^3$LO 2NF of
  Ref.~\cite{Entem:2002sf} and N$^2$LO 3NF.  For more details on the
  calculation see \cite{Navratil:2007we}. Figure courtesy of Petr Navratil. 
\label{fig:NCSM}}
\vspace{-0.5cm}
\end{center}
\end{figure}

The properties of certain S-shell and P-shell nuclei with $A \leq 13$ have
been analyzed recently based on the no-core shell model (NCSM), see
\cite{Nogga:2005hp,Navratil:2007we} and \cite{Navratil:2007aj} for an overview.   
In Fig.~\ref{fig:NCSM} we show some
results from Ref.~\cite{Navratil:2007we} for the spectra of $^{10}$B,
$^{11}$B, $^{12}$C and  $^{13}$C. We emphasize that the LECs $D$ and $E$
entering the N$^2$LO 3NF were determined in these calculations by the triton
binding energy and a global fit to selected properties of $^6$Li, $^{10}$B and
$^{12}$C. 
These studies clearly demonstrate that the chiral 3NF plays an important role
in the description of spectra and other properties of light nuclei. The
inclusion of the 3NF allows to considerably improve the agreement with the
data. Further results for light nuclei and the dilute neutron matter based on the
lattice formulation of chiral EFT are given in sections \ref{sec:fblatt} and \ref{sec:lattmany}.

\subsection{The role of the $\Delta$-isobar}
\label{sec:chiralEFTdelta}

The chiral expansion for the long-range part of the nuclear force discussed in
the previous section exhibits a somewhat unnatural convergence pattern in
certain cases such as e.~g.~for the central part of the $2\pi$-exchange
potential.  The origin of the 
unnaturally strong subleading contribution in this case potential can be
traced back to the large values of the dimension-two low-energy
constants (LECs) $c_{3,4}$ which are also responsible
for the numerical dominance of the subleading $3\pi$-exchange 
\cite{Kaiser:2001dm} and charge-symmetry breaking $2\pi$-exchange 2NF
\cite{Epelbaum:2005fd} over the corresponding leading contributions. 
The large values of these LECs are well understood in
terms of resonance saturation \cite{Bernard:1996gq}. In particular, 
the $\Delta$(1232) provides the dominant (significant)
contribution to $c_3$ ($c_4$). Given its low excitation energy,
$\Delta  \equiv m_\Delta - m = 293$~MeV, and strong 
coupling to the $\pi N$ system, the $\Delta$-isobar is known 
to play an important role in nuclear physics. 
One can, therefore, expect that the explicit inclusion of $\Delta$
in EFT will allow to resum a certain class of important contributions and 
improve the convergence as compared to the delta-less theory, provided
a proper power counting scheme such as the small scale expansion (SSE) 
\cite{Hemmert:1997ye} is employed, see also \cite{Jenkins:1991bs}. The SSE is a phenomenological extension
of chiral perturbation theory in which the small expansion parameter
includes external momenta,  
pion masses and the nucleon-delta mass splitting,
\beq
Q/ \Lambda \in \{ p/\Lambda, M_\pi/ \Lambda, \Delta/\Lambda \}~,
\eeq
i.e.~the delta-nucleon mass splitting is treated as $\Delta \equiv m_\Delta -
m \sim M_\pi$ rather
than $\Delta \sim \Lambda \gg M_\pi$. The improved convergence has been
explicitly demonstrated for pion-nucleon scattering where the description of the 
phase shifts at third order in the SSE comes out superior (inferior) to the
third (fourth) order chiral expansion in the delta-less theory
 \cite{Fettes:2000bb}. In the following, we will overview the additional
 contributions to the nuclear force which arise in the $\Delta$-full theory
as compared to the $\Delta$-less theory and discuss the implications for the
convergence of the low-momentum expansion.  Notice that in such a setting we
do not need to consider the $NN \to N\Delta$, $NN \to \Delta \Delta$ and
$N\Delta  \to \Delta \Delta$ transitions  which
would correspond to the coupled-channel approach. The adopted counting rules for the
nucleon mass and the delta-nucleon mass difference imply for the momentum
scale associated with real delta production $\sqrt{m \Delta} \sim \sqrt{m
  M_\pi} \sim \Lambda$. For typical external momenta (energies) of the
nucleons of the order $| \vec p \, | \sim M_\pi$ ($E_{\rm kin}  \sim
M_\pi^2/m$) we are interested in, the momenta associated  with real delta
production can be safely integrated out. We, therefore, only need to consider
the contributions to the nuclear force arising due to
virtual delta excitations.

The effective Lagrangian can be straightforwardly extended to include the
$\Delta$-degrees of freedom. To work out $\Delta$-contributions up to N$^2$LO,
the following additional terms in the heavy-baryon Lagrangian have to be taken
into account:
\begin{eqnarray}
\label{lagr}
\mathcal{L}_{\pi \Delta}^{(0)} &=& 
- \bar{T}_\mu^i \big[   i v \cdot D^{ij} - \Delta \delta^{ij} \big] 
g^{\mu \nu} T_\nu^j  + \ldots \,, \nn
\mathcal{L}_{\pi N \Delta}^{(0)} &=& h_A \, \bar T_\mu^i  P^{\mu \nu} \omega^i_{\nu } 
  N + \mbox{ h.c.} \,,
 \nn
\mathcal{L}_{\pi N \Delta}^{(1)} &=&  
 (b_3 + b_8 ) \, \bar T_\mu^i i P^{\mu \nu} \omega^i_{\nu \rho} v^\rho
  N + \mbox{ h.c.}  + \ldots \,,
\end{eqnarray}
where 
$T_\mu^i$ with $\mu$ ($i$) being the Lorentz (isospin) index 
denotes the large component of the delta field. 
Further, $D_\mu^{ij}$ refers to the chiral covariant derivative for the delta 
fields and $P_{\mu \nu}$ is the standard projector on the 3/2-components, 
$P_{\mu \nu} = g_{\mu \nu} - v_\mu v_\nu - 4S_\mu S_\nu /(1-d)$, 
with $d$ being the number of space-time dimensions. We also have $w_\alpha^i = 
\langle \tau^i u_\alpha \rangle /2$ and  $w_{\alpha\beta}^i =
\langle \tau^i [\partial_\alpha , u_\beta] \rangle / 2$. The only relevant LEC
in the lowest-order Lagrangian is the $\pi N \Delta$ axial coupling
$h_A$. At subleading order, the combination of $\pi N\Delta $ LECs $b_3 +b_8$
contributes. For more details
on the notation, the reader is referred to 
\cite{Hemmert:1997ye,Fettes:2000bb}, see also \cite{Bernard:2007zu} for a
recent review article and \cite{Pascalutsa:1998pw,Hacker:2005fh} for different
formulations.   
Finally, it should also be emphasized that 
the only possible derivative-less $NNN\Delta$ contact interaction 
\beq
\mathcal{L}_{N\Delta}^{(0)} \propto 
 ( \bar{T}_i^\mu N\bar{N}S_\mu\tau^i N + \mbox{h.~c.} )\,,
\eeq
vanishes due to the Pauli principle \cite{Epelbaum:2007sq}. 

The values of the LECs in the $\pi N$ Lagrangian are, clearly, different in
the $\Delta$-less  and
$\Delta$-full theories and can be naturally extracted from $\pi N$ scattering,
see \cite{Fettes:2000bb} for such a determination at the leading one-loop
level (i.e.~order $Q^3$). At subleading order, which is sufficient for our
purpose, the determination of 
$c_i$ from the $\pi N$ S- and P-wave threshold coefficients yields in the
delta-less theory \cite{Krebs:2007rh}
\begin{equation}
\label{ci_deltaless}
c_1 = -0.57, \;
c_2 = 2.84, \;  
c_3 = -3.87, \;  
c_4 = 2.89,
\end{equation}
where only central values are given and the units are GeV$^{-1}$. The above values are somewhat
smaller in magnitude than  the ones obtained at higher orders, see
e.g.~\cite{Fettes:1998ud}. Including the contributions form the
$\Delta$, one finds  
\beqa
\label{ci_deltafull}
&& c_1 = -0.57, \;   
c_2 = -0.25, \;  
c_3 = -0.79, \;  
c_4 = 1.33, \nn 
&& b_3 + b_8 = 1.40\, .
\eeqa
Notice that the LECs $c_{2,3,4}$ are strongly reduced in magnitude when the
$\Delta$-isobar is included. It should also be emphasized that the values 
of these LECs depend sensitively on the choice of $h_A$, which in the above
case was set to $h_A = 3 g_A/(2 \sqrt{2})$ from SU(4) (or large $N_c$). The results for the
threshold coefficients and the $2\pi$-exchange potential are, however, rather stable
\cite{Krebs:2007rh}. We also emphasize that the description of the P-wave threshold parameters 
improves significantly upon inclusion of the delta-isobar.   

We are now in the position to discuss the leading and subleading contributions
of the $\Delta$-isobar to the nuclear force. Since the appearance of a virtual
$\Delta$-isobar requires at least one loop, the corresponding contributions
first appear at NLO ($\nu = 2$). The relevant 2N and 3N diagrams can be
obtained from the ones of Figs.~\ref{fig:2NF} and \ref{fig:3NF} by replacing the nucleon
propagators by the ones of the $\Delta$-fields in all intermediate
states. We first discuss the 2NF. Similarly to the $\Delta$-less theory, the
additional contributions to 
the $1\pi$-exchange potential and contact interactions at both NLO and N$^2$LO
only lead to renormalization of various LECs. The $2\pi$-exchange diagrams were first 
discussed by Ord\'o\~nez et 
al.~\cite{Ordonez:1995rz} using time-ordered 
perturbation theory. These contributions were then calculated by Kaiser 
et al.~\cite{Kaiser:1998wa} using the Feynman graph technique. The corrections
at N$^2$LO have been worked out recently \cite{Krebs:2007rh}. We refrain
from showing here the resulting expressions which are rather involved and only
give the results for  the isovector 
tensor $2\pi$-exchange potential $W_T$, defined according to $V_{2N} = \fet \tau_1 \cdot \fet \tau_2 \, \vec \sigma_1 \cdot \vec q \, 
\sigma_2 \cdot \vec q \; W_T$, which may serve as a representative example: 
\begin{eqnarray}
W_T^{(2)} &=&  -\frac{h_A^2}{1296 \pi^2 F_\pi^4 \Delta }\;   
\Big[ 9 \pi g_A^2 \omega^2 \, A^{\tilde \Lambda}(q) + h_A^2 \big(
2L^{\tilde \Lambda}(q) \nn
&& {} + (4 \Delta^2 + \omega^2 ) D^{\tilde \Lambda}(q) \big)
\Big] \,, \nonumber
\eeqa
\beqa
W_T^{(3)} &=&  -\frac{h_A^2 \Delta}{648 \pi^2 F_\pi^4 }\;
\Big[ \big( 2 (b_3 + b_8) g_A (\omega^2 - 12 \Delta^2 ) \nn
&& {} - 9 c_4 (\omega^2 - 4
\Delta^2) \big) D^{\tilde \Lambda} (q)  \nn
&& {}+ 6 \big( 3 c_4 - 2 (b_3 + b_8) h_A \big)
L^{\tilde \Lambda} (q) \Big]\,.
\end{eqnarray}
Here, the new loop function $D^{\tilde \Lambda} (q)$ is defined via  
\begin{equation}
D^{\tilde \Lambda}(q) = \frac{1}{\Delta} \, \int_{2M_\pi}^{\tilde \Lambda}
\frac{d\mu}{\mu^2+q^2} \, \arctan \frac{\sqrt{\mu^2-4M_\pi^2}}{2\Delta}\,.
\end{equation}
The complete results for the $\Delta$-contributions can be found in
Refs.\cite{Kaiser:1998wa,Krebs:2007rh}. 
It is instructive  to verify the consistency between the $\Delta$-full and
$\Delta$-less theories which requires that the contributions due to intermediate
$\Delta$-excitations, expanded in powers of $1/\Delta$ can be absorbed
into a redefinition of the LECs in the $\Delta$-less theory. This is only
possible if the nonpolynomial (in momenta) terms up to N$^2$LO resulting from 
such an expansion have the same form as expressions in Eqs.~(\ref{VNLO},
\ref{VNNLO}). This indeed turns out to be the case: all expanded
nonpolynomial terms up to N$^2$LO are exactly reproduced by the shift in the
LECs $c_{3,4}$
\beq
\label{saturation}
c_3 = - 2 c_4 = -\frac{4 h_A^2}{9 \Delta }
\eeq 
in Eqs.~(\ref{VNLO}, \ref{VNNLO}). 

\begin{figure}[tb]
\begin{center}
\includegraphics[width=0.48\textwidth]{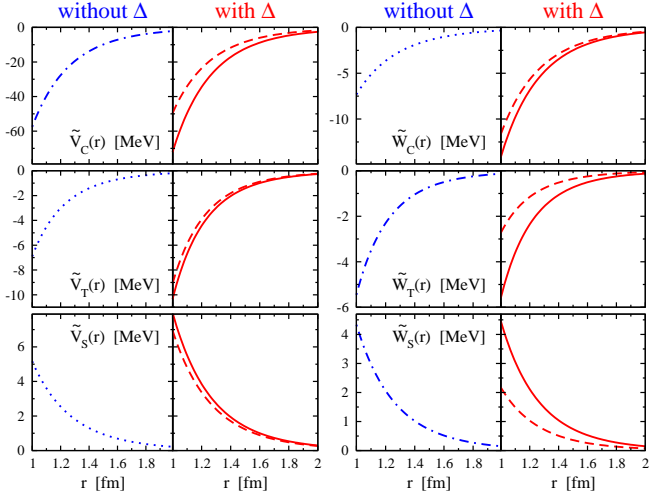}
\caption{Isoscalar (left panel) and isovector (right panel) components of the
  $2\pi$-exchange potential in coordinate space for $\tilde \Lambda = 700$~MeV. Dashed and
  solid (dotted and dashed-dotted) lines refer to
  the NLO and N$^2$LO results in the delta-full (delta-less) theory,
  respectively. There are no contributions to
  $\tilde V_C$ and $\tilde W_{T,S}$ ($\tilde V_{T,S}$ and $\tilde W_C$) at
      NLO (N$^2$LO) in the delta-less theory.}  \label{fig2} 
\end{center}
\end{figure}

\begin{figure}[tb]
\begin{center}
\includegraphics[width=0.48\textwidth]{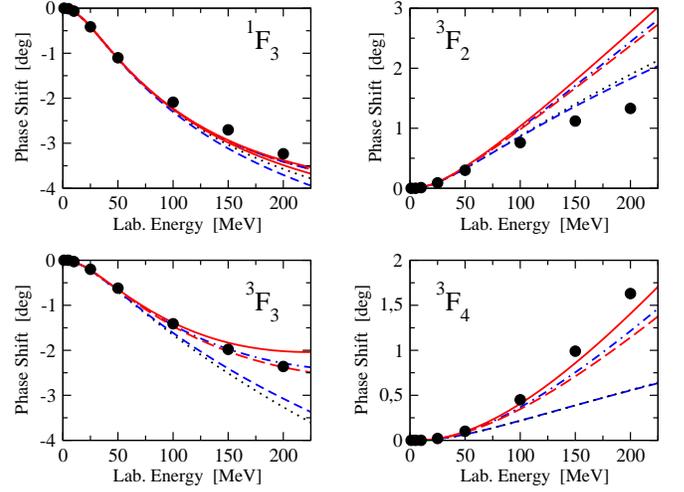}
\caption{F-wave NN phase shifts for $\tilde \Lambda = 700$~MeV. The dotted curve is the LO prediction, 
long-dashed (short-dashed) and solid (dashed-dotted) lines show the NLO and
N$^2$LO results with (without) the explicit $\Delta$-contributions. The filled circles depict
the results from the Nijmegen PWA \cite{Stoks:1993tb}.} \label{fig:pot_comp}
\end{center}
\end{figure}

To get more insight into the strength of various $2\pi$-exchange contributions in the
$\Delta$-full and $\Delta$-less theories, it is useful
to switch to coordinate space. The $2\pi$-exchange potential can then be written as
\beqa
\tilde V (r) &=& \tilde V_C +  \fet \tau_1 \cdot \fet \tau_2 \;  \tilde W_C 
+ \bigl[  \tilde V_S   +  \fet \tau_1 \cdot \fet \tau_2\; \tilde W_S \bigr]
\, \vec \sigma_1 \cdot \vec \sigma_2 \nn
&+& \bigl[  \tilde V_T   +  \fet \tau_1 \cdot \fet \tau_2\;  \tilde W_T  \bigr]
\, S_{12}\,,
\eeqa
where $S_{12} = 3 \vec \sigma_1 \cdot \hat r \; \vec \sigma_2 \cdot \hat r  - 
\vec \sigma_1 \cdot \vec \sigma_2$ is the tensor operator. The scalar
functions $\tilde V_i (r)$ and $\tilde W_i (r)$ are plotted in
Fig.~\ref{fig2} using the values for the LECs specified in
Eqs.~(\ref{ci_deltaless}) and (\ref{ci_deltafull}).   
As expected, one observes a more natural convergence pattern in the theory
with explicit deltas with the N$^2$LO contributions yielding typically
only modest corrections to the NLO result. This is, clearly, not the case
in the delta-less theory where the entire contributions to $\tilde V_C$ and
$\tilde W_{T,S}$ are generated at N$^2$LO. On the other hand, the N$^2$LO $2\pi$-exchange
potential in the delta-less theory provides a very good approximation
to the potential resulting at the same order in the delta-full theory. This
indicates that the saturation of the LECs $c_{3,4}$ is the most
important effect of the $\Delta$-isobar at the considered order. The results
for NN F- and other peripheral waves calculated using the Born approximation 
also clearly demonstrate the improved convergence in the theory with explicit
$\Delta$, see Fig.~\ref{fig:pot_comp}.   

As explained in section \ref{sec:chiralEFT}, the first nonvanishing
contributions to the 3NF appear in the $\Delta$-less theory at N$^2$LO. 
The situation is different in the $\Delta$-full theory where the first
3NF contribution is generated at NLO by the second graph in the first line of
Fig.~\ref{fig:3NF} with the intermediate nucleon propagator being replaced by
the one of the $\Delta$-field. In fact, the importance of the $\pi N$ $P_{33}$ partial
wave corresponding to the excitation of the $\Delta$ resonance in the 3NF has
been realized already 50  years by Fujita and Miyazawa
\cite{Fujita:1957zz}. The resulting expression for the 
$\Delta$-contribution to the $2\pi$-exchange 3NF
is exactly reproduced by the first term in Eq.~(\ref{leading}) if one uses the
$\Delta$-saturation values for the LECs $c_i$ from Eq.~(\ref{saturation}).
This, in fact, follows from the decoupling theorem and the fact
that the static $\Delta$ propagator is proportional to $\Delta^{-1}$.  
Notice that there are no short-range
3NFs with intermediate $\Delta$-excitation since the corresponding $NNN\Delta$
interaction is Pauli forbidden. Stated differently, the LECs $D$ and $E$ are
\emph{not} saturated by the $\Delta$-isobar. Surprisingly, one finds that there
are also no $\Delta$-contributions to the 3NF at N$^2$LO
\cite{Epelbaum:2007sq}. The $2\pi$-exchange 
diagrams with one insertion
of the subleading $\pi N \Delta$ vertex $\propto b_3 + b_8$ generate 
$1/m$-suppressed terms due to the time derivative entering
this vertex. Despite the fact that both the $\Delta$-full and $\Delta$-less theories
yield the same expressions for the $2\pi$-exchange 3NF at N$^2$LO, one should keep in mind
that the strengths of various terms are different. The extrapolation of the $\pi
  N$ amplitude from threshold, where the LECs are determined, to the
  kinematical region relevant for the 3NF is discussed by Pandharipande et 
al.~\cite{Pandharipande:2005sx} who claimed that both theories might yield 
sizeably different results if the expansion is truncated at low orders. Using
the values of the LECs from Eqs.~(\ref{ci_deltaless}) and (\ref{ci_deltafull}) one,
however, finds that the strengths of various terms differ from each other at
most by $\sim 7\%$ at N$^2$LO \cite{Epelbaum:2007sq}.  To conclude, the only effect of
including the $\Delta$-isobar as an explicit degree of freedom in the 3NF up to
N$^2$LO is the shift of the majour part of the $2\pi$-exchange contribution 
in Eq.~(\ref{leading}) from
N$^2$LO to NLO and some minor changes in the strengths of various terms in
this expression. 

We now discuss the role of the $\Delta$ for IB nuclear
forces. The observed unnatural convergence pattern for the CSB $2\pi$-exchange
2NF and CIB 3NF in the $\Delta$-less theory, see the discussion in section \ref{sec:chiralEFT},
is very similar to the one for isospin-conserving $2\pi$-exchange 2NF (in all
cases the large contributions are proportional to the LECs $c_{3,4}$) and
provides a strong motivation to explore the role of the $\Delta$ isobar in
this case. 

The leading IB $\Delta$-contributions to the $2\pi$-exchange 2NF result from 
the corresponding triangle, box and crossed-box diagrams with one insertion
of isospin-breaking pion, nucleon and delta mass shifts. The latter can be
deduced from the corresponding leading strong and electromagnetic Lagrangians
\cite{Epelbaum:2007sq} 
\beqa\label{massstr}
{\cal L}_{\pi \Delta ,\, {\rm IB}}^{(2)} &=&  - \bar{T}_i^\mu \,
c_5^\Delta \, \left(\chi_+ - \langle \chi_+\rangle \right) \,  \delta_{ij} \,
 g_{\mu\nu} \, T^\nu_j \,, \nn
{\cal L}_{\pi \Delta ,\, {\rm IB}}^{(3)} &=&  
- \bar{T}_i^\mu \, F_\pi^2\, \biggl[ 
  f_1^\Delta \, \delta_{ij} \, \langle Q_+^2 - Q_-^2\rangle
+ f_2^\Delta \, \delta_{ij} \, \langle Q_+ \rangle Q_+ \nn
&& {} + f_3^\Delta \, \delta_{ij} \, \langle Q_+ \rangle^2 
+ f_4^\Delta \, \langle \tau^i Q_+ \rangle\langle \tau^j Q_+ \rangle \nn
&& {}+ f_5^\Delta \, \langle \tau^i Q_- \rangle\langle \tau^j Q_- \rangle
\, \biggr] \, g_{\mu\nu} \, T^\nu_j + \ldots 
\eeqa
where $c_5^\Delta$ and $f_i^\Delta$ are the LECs and the ellipses in the last
line refer to strong terms which involve at least one pion field and are 
irrelevant for the following discussion. The masses of the physical delta
fields $( \Delta^{++}, \, \Delta^+, \, \Delta^0, \, \Delta^- )$ can be written
as
\beqa
&& m_{\Delta^{++}} = \tilde m_\Delta + \frac{\delta m_\Delta^1}{2} , \; \;
m_{\Delta^{+}} = \tilde m_\Delta + \frac{\delta m_\Delta^1}{6} 
+ \frac{\delta m_\Delta^2}{2}, \nn
&& m_{\Delta^{0}} = \tilde m_\Delta - \frac{\delta m_\Delta^1}{6} 
+ \frac{\delta m_\Delta^2}{2} , \; \;
m_{\Delta^{-}} = \tilde m_\Delta - \frac{ \delta m_\Delta^1}{2} ,  \nonumber
\eeqa
where $\delta m_\Delta^1$/$\delta m_\Delta^2$ denote the 
equidistant/non-equidistant splittings and the mass $\tilde m_\Delta$ contains an
isospin-invariant shift $\delta m_\Delta$ defined as $\tilde
m_\Delta = \krig{m}_\Delta + \delta m_\Delta$, with $\krig{m}_\Delta$ being
the delta mass in the chiral limit. The leading strong and electromagnetic
contributions to the splittings $\delta m_\Delta^{1,2}$ can be read off from
the Lagrangians in Eq.~(\ref{massstr}). While both strong and electromagnetic
terms contribute to the equidistant splitting  $\delta m_\Delta^{1}$, the
non-equidistant one at this order of of pure electromagnetic origin. In
Ref.~\cite{Epelbaum:2007sq}, the values for $\tilde m_\Delta$ and $\delta
m_\Delta^{1,2}$ were determined from the most recent particle data group values
for $m_{\Delta^{++}} = 1230.80 \pm 0.30$~MeV and $m_{\Delta^{0}} = 1233.45 \pm
0.35$~MeV \cite{PDG} together with the average mass $m_{\Delta} \equiv (m_{\Delta^{++}} +
  m_{\Delta^{+}} +m_{\Delta^{0}} +m_{\Delta^{-}}  )/4  = 1233$~MeV from
        Ref.~\cite{Arndt:2006bf} which leads to 
\beq
\label{splitt1}
\delta m_\Delta^{1} = -5.3 \pm 2.0 \mbox{ MeV} , \; \;  
\delta m_\Delta^{2} = -1.7 \pm 2.7 \mbox{ MeV}\,.
\eeq
If the quark model relation \cite{Rubinstein:1967aa} $m_{\Delta^{+}} -m_{\Delta^{0}} = m_p - m_n$ is
employed instead of using the average delta mass, the results change as
follows
\beq
\label{splitt2}
\delta m_\Delta^{1} = -3.9  \mbox{ MeV} , \; \;  
\delta m_\Delta^{2} = -0.3 \pm 0.3 \mbox{ MeV}\,.
\eeq
which is consistent with Eq.~(\ref{splitt1}). Notice that the values for $\delta
m_\Delta^{1,2}$ are of natural size. Indeed, based on  naive dimensional
analysis one expects $| \delta
m_\Delta^{1}  | \sim \epsilon M_\pi^2 / M_\rho \sim 8$~MeV
and  $| \delta
m_\Delta^{2}  | \sim e^2 M_\rho /(4 \pi)^2 \sim 0.5$~MeV.  For a related
discussion on the delta mass splittings in chiral EFT with 
a particular emphasis on their quark mass dependence the reader is referred to
Ref.\cite{Tiburzi:2005na}. 

\begin{figure}[tb]
\begin{center}
\includegraphics[width=0.4\textwidth]{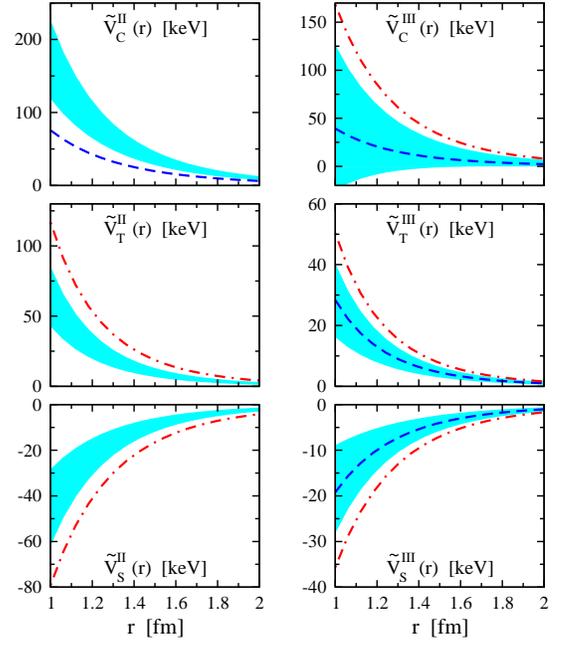}
\caption{Class-II (left panel) and class-III (right panel) $2\pi$-exchange
      potentials at N$^3$LO$\palka$ in the $\Delta$-full theory (shaded bands)
      compared to the results in the $\Delta$-less theory at N$^3$LO$\palka$  (dashed
      lines) and N$^4$LO$\palka$ (dashed-dotted lines). The bands arise from
      the variation of $\delta m_\Delta^1$ and $\delta m_\Delta^2$ according
      to Eq.~(\ref{splitt1}). Notice further that the leading
      (i.e.~N$^3$LO$\palka$) contributions to $\tilde V_{T,S}^{II}(r)$ and
      subleading (i.e.~N$^4$LO$\palka$) contributions to $\tilde V_{C}^{II} (r)$
      vanish in the $\Delta$-less theory. In all
      cases, the spectral function cutoff $\tilde 
      \Lambda = 700$~MeV is used. } 
\label{fig:pot_comp_iso}
\end{center}
\end{figure}

Having determined the values for the delta mass splittings, it is a
straightforward (but teddious) exercise to work out the leading 
$\Delta$-contributions to the IB $2\pi$-exchange potential. Notice that the
$\Delta$-contributions to the $1\pi$-exchange and contact potentials can be taken into
account by a redefinition of various LECs and will, therefore, not be
discussed. The explicit expressions for the $2\pi$-exchange contributions can
be found in  Ref.~\cite{Epelbaum:2008td}. In Fig.~\ref{fig:pot_comp_iso} we 
show the CIB and CSB central, tensor and spin-spin potentials in
coordinate space at N$^3$LO$\palka$ in the $\Delta$-full theory in comparison
with the  N$^3$LO$\palka$ and  N$^4$LO$\palka$  results in the  $\Delta$-less
theory. While in the $\Delta$-less theory, the leading and subleading class-II
$2\pi$-exchange potential arises entirely from the pion mass difference
$\delta M_\pi^2$, see the discussion in section \ref{sec:chiralEFT}, in the
$\Delta$-full theory one also finds contributions proportional to $\delta
m_\Delta^2$. Although these contributions are numerically small, they
provide a clear manifestation of effects which go beyond the 
subleading order in the $\Delta$-less theory. Furthermore, it is evident from
Fig.~\ref{fig:pot_comp_iso}  that the large
portion of the N$^4$LO$\palka$ CIB $2\pi$-exchange potential in the $\Delta$-less
theory is shifted to  N$^3$LO$\palka$ in the theory with explicit $\Delta$
degrees of freedom leading to a more natural convergence pattern. 
Similarly, a comparison of the corresponding CSB (i.e.~class-III) potentials
in two theories also indicates towards a more
natural convergence in the $\Delta$-full with the main part of the unnaturally
large subleading contribution in the $\Delta$-less theory being shifted to the lower order. 
Notice that the CSB $2\pi$-exchange potential at N$^3$LO$\palka$ in the
$\Delta$-full theory also receives
contributions from the delta splitting $\delta m_\Delta^1$ which are still
absent at N$^4$LO$\palka$ in the $\Delta$-less theory. For the central value,  
$\delta m_\Delta^1 = -5.3$~MeV, these contributions are numerically large and tend to
cancel the ones driven by the nucleon mass difference leading to a
significantly weaker resulting class-III $2\pi$-exchange potentials as compared to
the ones at subleading order in the $\Delta$-less theory. This can be viewed as
an indication that certain higher-order IB contributions still missing at
subleading order in the $\Delta$-less theory are unnaturally
large in the theory without explicit delta degrees of freedom. Last but not
least, effects from virtual $\Delta$-isobar excitation in the subleading $\pi
\gamma$- and $2\pi$-exchange potential induced by additional one-photon
exchange are considered in Ref.~\cite{Kaiser:2006ws}. 

Inclusion of the $\Delta$ as an explicit degree of freedom has also important
implications for the IB 3NF \cite{Epelbaum:2007sq}. As discussed in the
previous section, the 
strongest IB 3NF arises from taking into account the charge-to-neutral pion
mass difference in the $2\pi$-exchange 3N diagrams. In the $\Delta$-less
theory, the resulting charge-symmetry conserving 3NF, being formally
subleading (N$^4$LO$\palka$), is enhanced by the large values of the LECs
$c_{3,4}$. In the $\Delta$-full theory, the main part of this strong
$2\pi$-exchange contribution appears already at leading order (N$^3$LO$\palka$)
indicating a more natural convergence pattern. In addition to this obvious
effect, one obtains further $2\pi$-exchange contributions at N$^3$LO$\palka$ driven by the
delta and nucleon mass splittings $\delta m_\Delta^2$ (charge-symmetry conserving) and
$\delta m_\Delta^1$, $\delta m$  (charge-symmetry breaking). A close
inspection of the resulting expressions, which are all  proportional to
$\Delta^{-2}$, reveals that they are exactly reproduced in
the $\Delta$-less theory by the saturation of the sub-subleading 
isospin-conserving ($d_i$-terms in $\mathcal{L}_{\pi N}^{(2)}$) and IB $\pi
\pi NN$ vertices. Numerically, the strengths of these CSB 3NFs due to
intermediate delta excitation turn out to be rather small, 
$|\delta m_\Delta^1 - 3 \delta m_N |  g_A^2 h_A^2 M_\pi^6 /(432 \pi^2 F_\pi^4
\Delta^2 ) \sim 3~{\rm keV}$, which, however, is comparable to the typical size of the
remaining leading CSB 3NF \cite{Epelbaum:2004xf}, $g_A^4 M_\pi^4/(256 \pi^2 F_\pi^4) \sim 7$ keV. 


\subsection{Few-nucleon reactions involving pions}
\label{sec:further_scales}

\begin{figure}[tb]
\begin{center}
\includegraphics[width=0.4\textwidth]{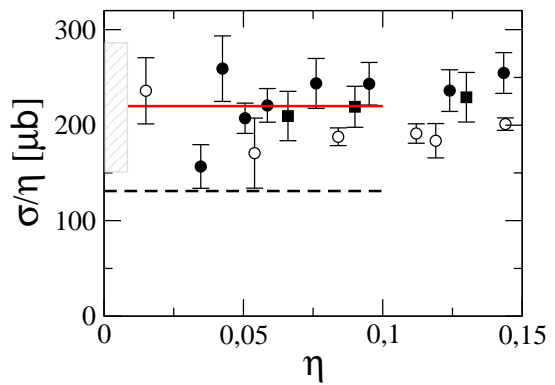}
\caption{LO (dashed line) and NLO (solid line) results for the total cross
  section for the reaction $pp \to d \pi^+$ in comparison with the data from 
\cite{Hutcheon:1991vt} (open circles), \cite{Heimberg:1996be} (filled circles)
and \cite{Drochner:1998ja}  (filled
squares). The hatched area gives the estimated uncertainty at NLO.  
Figure courtesy of C.~Hanhart.} 
\label{piprod}
\end{center}
\end{figure}

Few-nucleon reactions involving pions such as e.g.~$\pi d \to \pi d$ \cite{Beane:2002wk}, 
$\gamma d \to \pi^0 d$ \cite{Beane:1997iv,Krebs:2004ir}, $\pi \, ^3$He${} \to
\pi \, ^3$He \cite{Baru:2002cg}, $\pi^- d \to \gamma nn$ \cite{Gardestig:2005pp}, 
$\gamma d \to \pi^+ nn$ \cite{Lensky:2007zc} and $NN \to NN \pi$
\cite{Baru:2007cd}, where only some of the most 
recent references are given, provide another fascinating testing ground for the
chiral EFT framework. The calculations typically utilize the distorted-wave Born
approximation using transition operators derived in chiral EFT and employing
either phenomenological or chiral-EFT-based wave functions for the
few-nucleon states following Weinberg's original proposal
\cite{Weinberg:1992yk}.  
An important new ingredient in these applications is the appearance of the momentum
scale $p = \sqrt{m M_\pi}$ associated with real pion production which has
to be explicitly taken into account and requires an appropriate modification
of the power counting
\cite{Cohen:1995cc,Bernard:1998sz,daRocha:1999dm}. Such a modified
ordering scheme was proposed in \cite{Hanhart:2000gp} and applied in
\cite{Hanhart:2002bu} to calculate the pion production
operator in NN collisions at threshold at NLO.  Notice that the rather high
energies and momenta of the nucleons in the initial state require the
inclusion of the $\Delta$-isobar as an explicit degree of freedom. As a
characteristic feature of
the modified power counting scheme, one observes the appearance of 
half-integer powers of the small parameter $\chi = M_\pi/m$ in the 
expansion of the transition operators. One also finds that some pion loop
contributions are promoted to significantly lower orders compared to what is
expected from Weinberg's original power counting. An application of the modified power
counting to P-wave pion production in NN collisions up to N$^2$LO is carried
out in Ref.~\cite{Hanhart:2000gp}. At this order, only tree diagrams have to be
considered. The calculations showed
a satisfactory agreement with the data and also demonstrate the feasibility to
extract the LEC $D$ which enters the leading 3NF, see Eq.~(\ref{leading}),
from this reaction. Notice, however, that concerns have been raised in 
Ref.~\cite{Nakamura:2007vi} regarding the convergence of the chiral expansion in this
reaction. 
For S-wave pion production, one-loop diagrams already
start to contribute at NLO. As pointed out in Ref.~\cite{Lensky:2005jc}, it is
important to 
properly separate the truly irreducible contributions in the loop diagrams
from the reducible ones in order for the resulting pion production operator to
be renormalizable. Numerically, the NLO loop diagrams were found to provide an
important contribution to the cross section for $pp \to d
\pi^+$. Parametrizing the near-threshold cross section for this reaction as 
$\sigma = \alpha \eta + \mathcal{O} (\eta^3)$, where $\eta$ denotes the
outgoing pion momentum in units of its mass, the pion S-wave contribution at
LO and NLO
was found to be $\alpha^{\rm LO} = 131 \, \mu$b and $\alpha^{\rm NLO} = 220 \,
\mu$b, respectively~\cite{Lensky:2005jc}. The result at NLO agrees nicely with
various existing data sets, see Fig.~\ref{piprod}. 
A comprehensive review of meson production reactions in nucleon-nucleon
collisions can be found in Ref.~\cite{Hanhart:2003pg}.  

Pion production reactions in few-nucleon systems also proved useful to study
isospin-violating effects. Recent measurements of the forward-backward
assymmetry in the process $pn \to d \pi^0$ \cite{Opper:2003sb} and the total
cross  section in the
reaction $dd \to \alpha \pi^0$ \cite{Stephenson:2003dv} yielded a clear evidence of 
charge-symmetry breaking and serve as excellent testing ground to study
isospin violation in the nuclear force and the corresponding transition
operators. The first steps towards the theoretical understanding of these
reactions have been taken in Refs.~\cite{vanKolck:2000ip} and
\cite{Gardestig:2004hs,Nogga:2006cp}, respectively. Notice that the
appearance of the four-nucleon continuum states makes the theoretical analysis
of the process $dd \to \alpha \pi^0$ particularly challenging. We further
emphasize that new data on this reaction will be provided by WASA at
COSY \cite{Adam:2004ch}. Further details on these studies and related issues can be found in a
recent review article \cite{Miller:2006tv}. 

The role of the momentum scale $p = \sqrt{m M_\pi}$ and the related issue of
the nucleon recoil effects in reactions such e.g.~$\pi d$ scattering and pion photo- and
electroproduction off the deuteron was investigated in the context of chiral
EFT in Refs.~\cite{Baru:2004kw,Lensky:2005hb,Lensky:2006wd}. In particular, it
was realized that the importance of the recoil effects in a given
process is directly connected to the Pauli principle for the nucleons in the
intermediate states. Notice further that the reaction $\gamma d \to nn \pi^+$
\cite{Lensky:2007zc}
and the similar process $\pi^- d \to \gamma nn$ \cite{Gardestig:2005pp} were
proposed as a tool to 
extract the value of the neutron-neutron S-wave scattering length. For more
details on these and related reactions the reader is referred to
Ref.~\cite{Hanhart:2007ym}.

\subsection{Hyperon-nucleon \& hyperon-hyperon interactions}
\label{sec:YN}

The effects of strange quarks in nuclear matter can e.g. be tested
through the determination of the properties of so-called hyper-nuclei,
in which one (or two) nucleon(s) is (are) substituted by a hyperon
(hyperons). Such hyper-nuclei are produced by strangeness-exchange
reactions,  by associated strangeness production or by electroproduction
reactions at many labs world-wide, such as CERN, BNL, KEK, DA$\Phi$NE,
JLab, MAMI and GSI, see e.g. Ref.~\cite{Alberico:2001jb}.
More generally,  nuclear physics with strange quarks 
has a broad impact on contemporary physics  since it lies at the intersection 
of nuclear and elementary particle physics. Moreover, it has significant 
implications to the astrophysics of compact objects. Recent progress
in the field is reviewed in Ref.~\cite{BGM}.

The hyperon-nucleon (YN) interaction is at the heart of the hyper-nuclear
binding and thus a precise determination of its various components is
of utmost importance. Here, the situation is quite different compared
to the two-nucleon case. The data base on YN scattering is quite poor,
thus a partial wave analysis is not available and in any theoretical
approach one must directly compare to data. The poor status of our information
on the YN interaction is most clearly reflected in the present knowledge
of the $\Lambda N$ scattering lengths. E.g Ref.~\cite{Alexander:1969cx} 
gives for the singlet ($S$)  and triplet ($T$) scattering lenghts 
$a_s = -1.8_{-4.2}^{+2.3}\,{\rm fm}\,$,
$a_t = -1.8_{-0.8}^{+1.1}\,{\rm fm}\,$,
whereas in the six variants of the Nijmegen soft-core potential
model $a_s$ varies between $-2.5 \ldots -0.7\,$fm and $a_t$ between
$-2.2 \ldots -1.8\,$fm \cite{Rijken:1998yy}. In the
most modern version of the J\"ulich meson-exchange model
one finds $a_s \simeq -2.6\,$fm and $a_t \simeq
-1.7\,$fm~\cite{Haidenbauer:2005zh}. However, for the EFT approach
it is important to note that all these values are of natural size.
For a proposal to extract these
scattering lengths with high precision from production data, we
refer to Ref.~\cite{Gasparyan:2003cc}. Furthermore, since the masses
of the $\Lambda$ and the $\Sigma$ hyperons  are only about 75~MeV apart, 
the  coupling between the $\Lambda N$ and the $\Sigma N$ channels needs to
be taken into account. Moreover, for a sensible comparison with experimental
data,  it is preferable to solve the scattering equation in the particle
basis because then the Coulomb interaction in the charged channels can
be incorporated.

The hyperon-nucleon YN interaction has not been investigated using EFT as
extensively as the NN interaction. Hyperon and nucleon mass shifts in
nuclear matter, using chiral perturbation theory, have been studied 
in~\cite{Savage:1995kv}. These authors used a chiral interaction 
containing four-baryon contact
terms and pseudoscalar-meson exchanges. The hypertriton (a bound state
of a proton, a neutron and a $\Lambda$) and $\Lambda
d$ scattering were investigated in the framework of an EFT with contact 
interactions \cite{Hammer:2001ng}. Korpa et al. \cite{Korpa:2001au} performed a
next-to-leading order (NLO) EFT analysis of
YN scattering and hyperon mass shifts in nuclear matter. Their tree-level
amplitude contains four-baryon contact terms; pseudoscalar-meson exchanges
were not considered explicitly, but ${\rm SU(3)}$ breaking by meson masses
was modeled by incorporating dimension two terms coming from one-pion
exchange. The full scattering amplitude was calculated using the
Kaplan-Savage-Wise resummation scheme. The hyperon-nucleon scattering
data were described successfully for laboratory momenta below 200~MeV, using
12 free parameters. Some aspects of strong $\Lambda N$ scattering in effective
field theory and its relation to various formulations of lattice QCD are
discussed in  \cite{Beane:2003yx}. 

\begin{figure}[tb]
\includegraphics*[width=5.5cm]{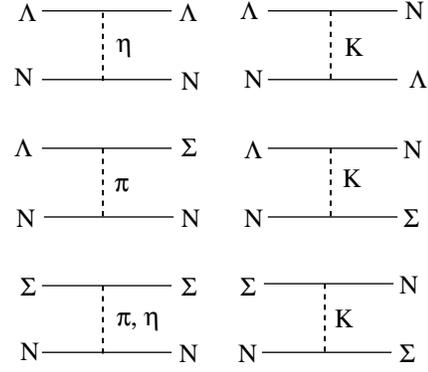}
\begin{center}
\caption{One-pseudoscalar-meson-exchange diagrams at LO for the
hyperon-nucleon interaction.
\label{fig:YNOBE}}
\vspace{-0.5cm}
\end{center}
\end{figure}

\begin{figure}[tb]
\includegraphics*[width=7.95cm]{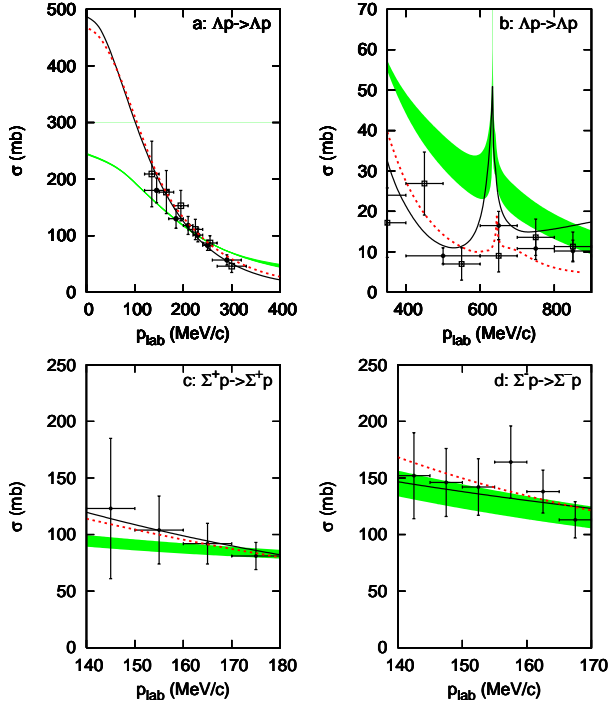}
\begin{center}
\caption{Total cross sections as a function of $p_{\rm lab}$ for
$\Lambda p \to \Lambda p$ and $\Sigma^\pm p \to \Sigma^\pm p$. The shaded
band is the LO EFT result for $\Lambda = 550 \ldots 700\,$MeV, the
dashed curve is the J\"ulich 04 model \protect\cite{Haidenbauer:2005zh},
and the solid curve is the Nijmegen NSC97f model \protect\cite{Rijken:1998yy}.
Note that the total cross sections for $\Sigma^\pm p \to \Sigma^\pm p$
are obtained by integrating the differential data in a limited
angular range, see Ref.\protect\cite{Eisele:1971mk}.
\label{fig:YNfit}}
\vspace{-0.5cm}
\end{center}
\end{figure}

Within the Weinberg counting scheme,
a detailed investigation of the YN interaction at  LO was presented 
in~\cite{Polinder:2006zh}. At LO, the YN potential is given by 
one-pseudoscalar-Goldstone-boson exchange diagrams, cf. 
Fig.~\ref{fig:YNOBE}, and contact interactons without derivatives.
\noindent The spin-space part of the one-pseudoscalar-meson-exchange potential 
resulting from the LO SU(3) effective chiral meson-baryon Lagrangian is, 
\beq
V^{B_1B_2\to B_1'B'_2} = -f_{B_1B'_1P}f_{B_2B'_2P}
\frac{\vec \sigma_1\cdot{\vec k} \, 
\vec \sigma_2\cdot{\vec k}}{{\vec k}^2+m^2_P}\ ,
\label{eq:YNOBE}
\eeq
where ${\vec k}$ is the momentum transfer, $P = \pi, K, \eta$, and the
$f_{B_1B'_1P}$, $f_{B_2B'_2P}$ are the appropriate coupling constants 
\begin{equation}
\begin{array}{rlrl}
f_{NN\pi}  = & f, & f_{NN\eta_8}  = & \frac{1}{\sqrt{3}}(4\alpha -1)f, \\ 
f_{\Lambda NK} = & -\frac{1}{\sqrt{3}}(1+2\alpha)f, &
f_{\Xi\Xi\pi}  = & -(1-2\alpha)f, \\  
f_{\Xi\Xi\eta_8}  = & -\frac{1}{\sqrt{3}}(1+2\alpha )f, & f_{\Xi\Lambda K} = & 
\frac{1}{\sqrt{3}}(4\alpha-1)f, \\
f_{\Lambda\Sigma\pi}  = & \frac{2}{\sqrt{3}}(1-\alpha)f, &
f_{\Sigma\Sigma\eta_8}  = & \frac{2}{\sqrt{3}}(1-\alpha )f, \\ 
f_{\Sigma NK} = & (1-2\alpha)f, & f_{\Sigma\Sigma\pi}  = & 2\alpha f, \\  
f_{\Lambda\Lambda\eta_8}  = & -\frac{2}{\sqrt{3}}(1-\alpha )f, & f_{\Xi\Sigma K} = & -f.
\end{array}
\label{eq:YNcoup}
\end{equation}
in terms of the coupling constant $f\equiv g_A/2F_\pi$ and 
the $F/(F+D)$-ratio $\alpha$  \cite{de Swart:1963gc}. The
corresponding isospin factors for the various channels 
multiplying the spin-space part of the potential in 
Eq.~(\ref{eq:YNOBE}) are tabulated in Tab.~\ref{tab:YNOBE}.
\begin{table}[tb]
\centering
\begin{tabular}{|c|r|r|r|r|}
\hline
Channel &Isospin &$\pi$ &$K$ &$\eta$\\
\hline
$NN\rightarrow NN$ &$0$ &$-3$ &$0$ &$1$ \\
                   &$1$ &$1$  &$0$ &$1$ \\
\hline
$\Lambda N\rightarrow \Lambda N$ &$\frac{1}{2}$ &$0$ &$1$ &$1$ \\
\hline
$\Lambda N\rightarrow \Sigma N$ &$\frac{1}{2}$ &$-\sqrt{3}$ &$-\sqrt{3}$ &$0$ \\
\hline
$\Sigma N\rightarrow \Sigma N$ &$\frac{1}{2}$ &$-2$ &$-1$ &$1$ \\
                               &$\frac{3}{2}$ &$1$ &$2$ &$1$ \\
\hline
\end{tabular}
\caption{The isospin factors for the various one--pseudoscalar-meson 
exchanges contributing to the hyperon-nucleon interaction.}
\label{tab:YNOBE}
\end{table}
\noindent It is important to stress that while the interaction potential
at LO is SU(3) symmetric, the kinematics of the various channels and the
masses of the exchanged mesons are to be taken at their physical values.
The LO contact terms for the octet baryon-baryon interactions, that are Hermitian 
and invariant under Lorentz transformations, are given by the SU(3)
invariants, 
\beqa
{\mathcal L}^1 &=& C^1_i \left<\bar{B}_a\bar{B}_b
\left(\Gamma_i B\right)_b\left(\Gamma_i B\right)_a\right>\ , \nonumber\\
{\mathcal L}^2 &=& C^2_i \left<\bar{B}_a\left(\Gamma_i B\right)_a\bar{B}_b
\left(\Gamma_i B\right)_b\right>\ , \nonumber \\
{\mathcal L}^3 &=& C^3_i \left<\bar{B}_a\left(\Gamma_i B\right)_a\right>
\left<\bar{B}_b\left(\Gamma_i B\right)_b\right>\  .
\label{eq:YYcont}
\eeqa
Here, $a$ and $b$ denote the Dirac indices of the particles, 
$B$ is the usual irreducible octet representation of SU(3) given by
\beqa
B&=&
\left(
\begin{array}{ccc}
\frac{\Sigma^0}{\sqrt{2}}+\frac{\Lambda}{\sqrt{6}} & \Sigma^+ & p \\
\Sigma^- & \frac{-\Sigma^0}{\sqrt{2}}+\frac{\Lambda}{\sqrt{6}} & n \\
-\Xi^- & \Xi^0 & -\frac{2\Lambda}{\sqrt{6}}
\end{array}
\right) \ ,
\label{eq:baryon}
\eeqa
and the brackets denote taking the trace in the three-dimensional 
flavor space. As an example, we display the resulting 
partial wave potentials  for $\Lambda N \to \Lambda N$
\beqa
V^{\Lambda\Lambda}_{1S0}&=&4\pi\biggl[\frac{1}{6}
\left(C^1_S-3C^1_T\right)+\frac{5}{3}\left(C^2_S-3C^2_T\right)\
\nonumber \\
&& \qquad 
+2\left(C^3_S-3C^3_T\right)\biggr] \nonumber \\
V^{\Lambda\Lambda}_{3S1}&=&4\pi\biggl[\frac{3}{2}
\left(C^1_S+C^1_T\right)+\left(C^2_S+C^2_T\right)\nonumber\\
&& \qquad 
+2\left(C^3_S+C^3_T\right)\biggr]~.
\label{eq:LNcont}
\eeqa
Similar expression for the isospin-1/2 and 3/2 $\Sigma N\to\Sigma N$ and the
$\Lambda N \to \Sigma N$ potentials are given in Ref.~\cite{Polinder:2006zh}. 
Note that only 5 of the $\{8\}\times\{8\}=\{27\}+\{10\}+\{10^*\}
+\{8\}_s+\{8\}_a+\{1\}$ representations are relevant for NN and YN
interactions,  since the $\{1\}$ occurs only in the $\Lambda\Lambda$, 
$\Xi N$ and $\Sigma\Sigma$ channels.  Equivalently, the six contact terms, 
$C_S^1$, $C_T^1$, $C_S^2$, $C_T^2$, $C_S^3$, $C_T^3$, enter the NN and YN 
potentials in only 5 different combinations. 
These 5 contact terms need to be determined by a fit to the experimental data.
The resulting chiral potential $V^{\rm LO} = V_{\rm OBE} + V_{\rm cont}$ in
the Lippmann-Schwinger equation is regulated with a regulator function
$f_\Lambda (p,p') = \exp(-(p^4+ {p'}^4)/\Lambda^4)$, where the cut-off
$\Lambda$ is varied between 550 and 700~MeV.  A fit to 35 low-energy data
(total cross sections from 
Refs.\cite{Alexander:1969cx,SechiZorn:1969hk,Eisele:1971mk,Engelmann:1966aa}
for $\Lambda p\to \Lambda p$, $\Sigma^-p \to \Lambda p$, $\Sigma^\pm p \to
\Sigma^\pm p$ and  $\Sigma^-p \to \Sigma^0 n$
with hyperon lab momenta between 110 and 300~MeV
and the inelastic capture ratio at rest \cite{de Swart:1962aa}) gives
a good descripton of the data, see Fig.~\ref{fig:YNfit}, with contact interactions
of natural size. 
\noindent
Note the strong cusp effect in $\Lambda p$ scattering at the
opening of the $\Sigma^+ n$ threshold at $p_{\rm lab} \simeq 600\,$MeV
(Fig.~\ref{fig:YNfit}b).
The chiral EFT also yields a correctly bound 
hypertriton~\cite{Haidenbauer:2007ra} if one fixes the relative strength of
the singlet and triplet S-waves accordingly. 
A $\Lambda p$ singlet scattering length of $-1.9\,$fm leads
to the correct binding energy. The corresponding triplet scattering length
is $a_t^{\Lambda p} = -1.2\,$fm and in the $\Sigma^+ p$ system, one finds
$a_s = -2.3\,$ and $a_t = -0.7\,$fm.  It is quite astonishing that with six
parameters only (5 LECs and the cut-off $\Lambda$) one achieves a quite
satisfatory desciption of the admittedly not very preciese YN scattering
data. Clearly, a NLO calculation should be performed and fits should
be done simultaneously to YN and NN data.
For a more detailed discussion of these 
results and a comparison to more conventional approaches to the YN
interaction, we refer the reader to Refs.~\cite{Polinder:2006zh,Haidenbauer:2007ra}.

The experimental situation on baryon-baryon scattering with $S=-2$, 
i.e in the YY and the $\Xi$N channels, is even poorer.
Only very recently doubly strange baryon-baryon scattering data at 
lower energies, below $p_{\rm lab}=0.8\,$GeV, were deduced for the first 
time \cite{Tamagawa:2001tk,Ahn:2005jz}. An upper limit of $24$ mb at 
$90\%$ confidence level was provided for elastic $\Xi^-p$ scattering, 
and for the $\Xi^-p\rightarrow \Lambda\Lambda$ cross section at 
$p_{\rm lab}=500\,$MeV a value of $4.3^{+6.3}_{-2.7}$ mb was 
reported \cite{Ahn:2005jz}. Within LO chiral EFT, baryon-baryon
scattering was analyzed by Polinder et al.~\cite{Polinder:2007mp}.
\begin{figure}[tb]
\includegraphics*[width=5.8cm]{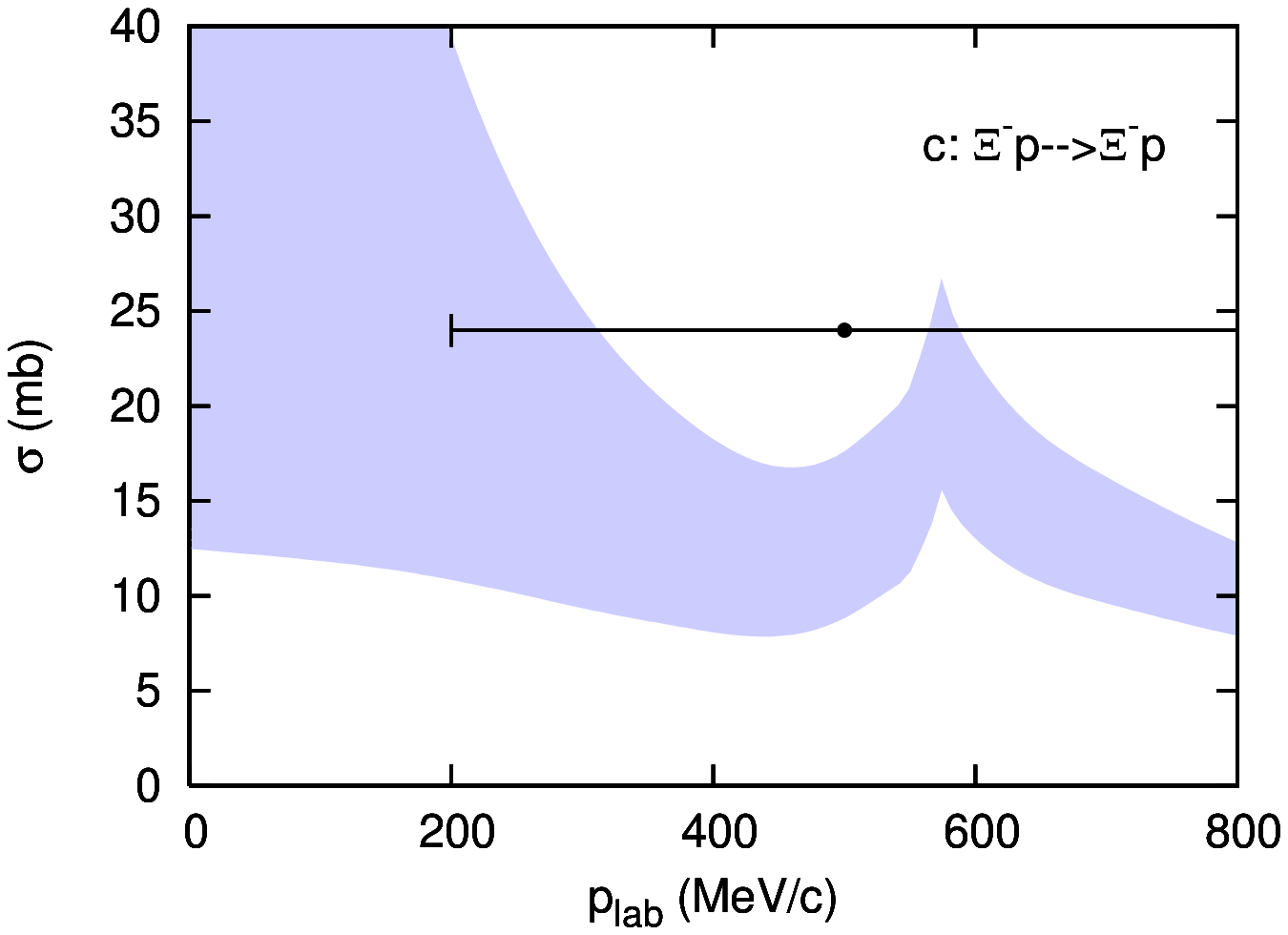}
\includegraphics*[width=5.8cm]{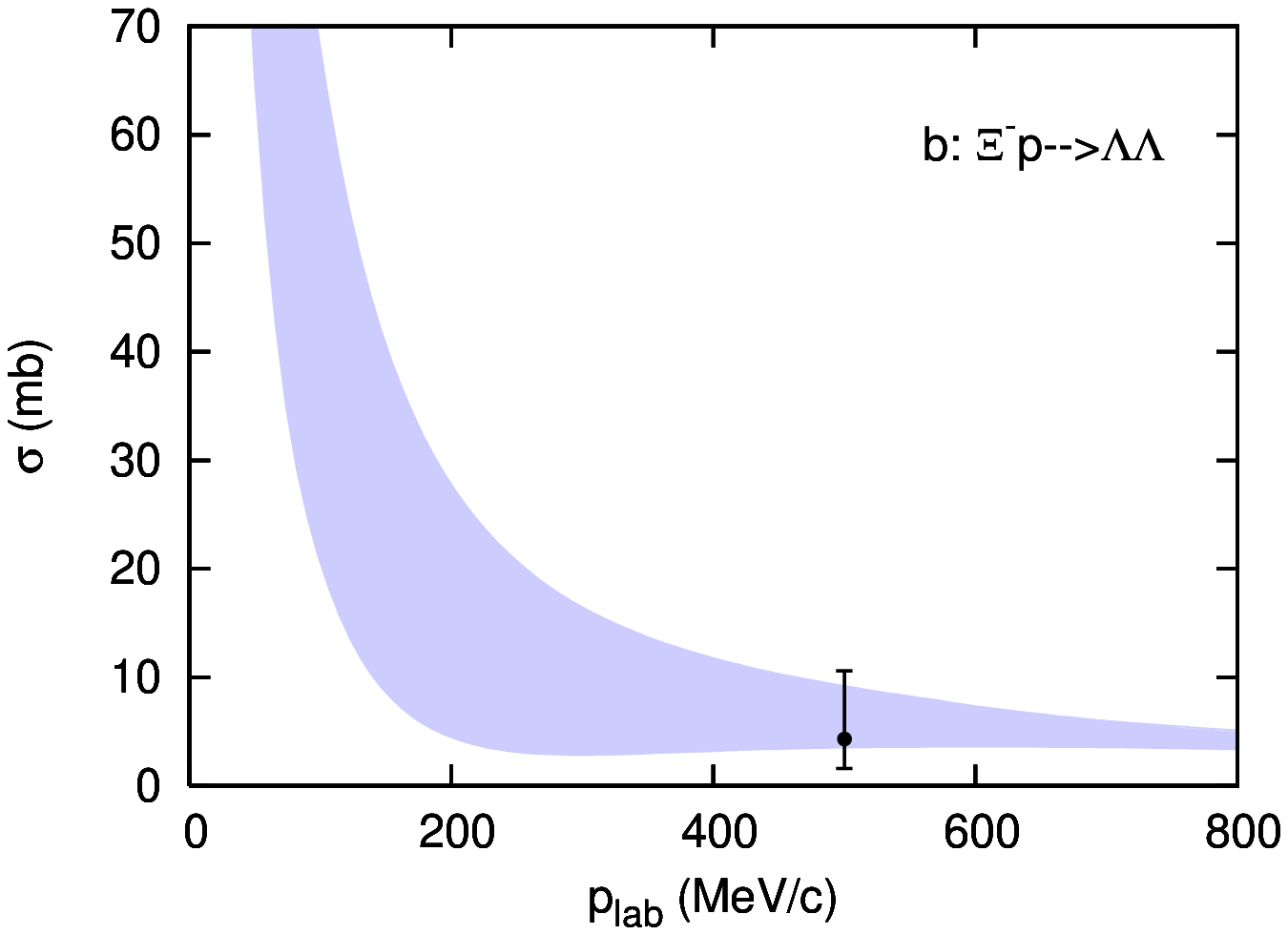}
\begin{center}
\caption{Total cross sections as a function of $p_{\rm lab}$ for
$\Xi^- p \to \Xi^- p$ (upper panel) and 
$\Xi^- p \to \Lambda\Lambda$. The shaded
band is the LO EFT result for $\Lambda = 600\,$MeV and
varying the LEC $C_{1S0}^{\Lambda\Lambda}$ of the additional singlet
contact term within natural bounds gievn a mildly attractive
interaction in the $^1S_0$ channel of the $\Lambda\Lambda$ interaction.
\label{fig:YY}}
\vspace{-0.5cm}
\end{center}
\end{figure}
\noindent
The contact terms and the couplings of the pseudoscalar mesons to the 
baryons are related via SU(3) symmetry to the $S=-1$ hyperon-nucleon
channels. There is one additional contact interaction whose strength
was varied within natural bounds in the $\Lambda \Lambda \to \Lambda \Lambda$
channel. This fixes its contribution in all other $S=-2$ baryon-baryon
channels because of SU(3) symmetry. As a constraint, the information deduced
from the recent candidate for ${}_{\Lambda\Lambda}^{\;\;\;6}{\rm He}$ with a
low binding energy \cite{Takahashi:2001nm}, the so-called Nagara event, 
that suggests that the $\Lambda\Lambda$ interaction 
should be only moderately attractive, was imposed. For a fixed cut-off, the prediction
of the $\Xi^- p \to \Xi^- p$ and the $\Xi^- p \to \Lambda\Lambda$ total
cross section in comparison to the available data are shown in
Fig.~\ref{fig:YY}. The resulting $\Lambda\Lambda$ scattering length in the $^1S_0$
channel is $a_s = -1.83 \ldots -1.38\,$fm. For comparison, in the
Nijmegen ESC04 model one finds in this channel  $a_s = -1.32\,$fm 
\cite{Rijken:2006kg} and in  the constituent quark model of
Fujiwara et al. \cite{Fujiwara:2006yh} one has  $a_s = -0.81\,$fm. 
Note also that this contact interaction does not contribute to certain
channels, so that at LO one can make parameter-free predictions for
$\Sigma^+\Sigma^+ \to \Sigma^+\Sigma^+$, $\Xi^0 p \to \Xi^0 p$ and
$\Xi^0 p \to \Sigma^+ \Lambda$. 
It is expected that in the coming years better-quality data on the fundamental 
$\Xi$N and YY interactions as well as much more information about the
physics  of hyper-nuclei will become available at the new facilities J-PARC
(Japan) and FAIR (Germany). The chiral EFT developed in
Ref.~\cite{Polinder:2007mp} 
can then be used  to analyze these upcoming data in a model-independent way.

\subsection{Nuclear lattice simulations}
\label{sec:fblatt}

Once the chiral nuclear forces are determined and the low-energy constants
appearing in the nuclear forces are fitted (in the two and three-nucleon
sector) one can make predictions in the four- and more-nucleon sectors
based on chiral EFT. However, the explicit numerical 
treatment of e.g. the 
Yakubowsky equations for more than four nucleons is a very
difficult task. One possible scheme to solve  the many-body problem is to put 
the chiral effective potential on the lattice and apply the powerful 
Monte-Carlo techniques which are already developed to high degree in lattice
QCD. One unique feature of the lattice effective field theory approach is the
ability to study in one formalism both few- and many-body systems as well as
zero- and nonzero-temperature phenomena. A large portion of the nuclear
phase diagram can be studied using exactly the same lattice action with
exactly the same operator coefficients. A second feature is the
computational advantage of many efficient Euclidean lattice methods developed
for lattice QCD and condensed matter applications.  This includes the use of
Markov chain Monte Carlo techniques, Hubbard-Stratonovich transformations, 
and non-local updating schemes such as a hybrid Monte Carlo. 
A third feature is the close theoretical link between nuclear lattice 
simulations and chiral effective field theory. One can write down the 
lattice Feynman rules and calculate lattice Feynman diagrams using precisely 
the same action used in the non-perturbative simulation. Since the lattice 
formalism is based on chiral effective field theory we have a systematic 
power counting expansion, an a priori estimate of errors for low-energy
scattering, and a clear theoretical connection to the underlying symmetries of QCD.
The first  studies combining lattice methods with effective field theory for
low-energy nuclear physics looked at infinite nuclear and neutron matter at
nonzero density and temperature (see Sec.~\ref{sec:lattmany}). Most of the
formalism for chiral EFT on the lattice was developed in~\cite{Lee:2004si}.
Nuclear lattice simulations were used to study the triton at leading-order in
pionless effective field theory with three-nucleon interactions
\cite{Borasoy:2005yc}. 

Let us briefly discuss the principles underlying such nuclear lattice
simulations (for a detailed discussion, we refer to Refs.~\cite{Lee:2004si,Borasoy:2006qn}).
In this framework, nucleons are represented 
as point-like Grassman-fields  and pions as point-like instantaneous
pseudoscalar fields. 
The lattice is defined by a volume $L^3 \times L_t$, with $L (L_t)$ the
spatial (temporal) size. The corresponding lattice spacings are called
$a$ and $a_t$, respectively. Typically, calculations are carried out using a lattice
length $L\simeq 20$~fm and the lattice spacing $a\simeq 2$~fm which corresponds
to the cutoff $\Lambda=\pi/a\simeq 300 \,{\rm MeV}$. At present, computational
resources prevent one from using smaller lattice sizes (larger UV cutoffs).
Still, the use of various forms of improved actions allows one to access the
systematic errors inherent in such simulations.

The basic quantity in nuclear lattice simulations is the correlation
function.  For $A$ nucleons in  Euclidean space it is defined by
\beq
Z_A(t)=\langle\Psi_A|\exp(-\tau H)|\Psi_A\rangle,
\eeq
where  $|\Psi_A\rangle$ refers to a Slater determinants for $A$ free 
nucleons, $H$ is the Hamiltonian of the system and $\tau$ the Euclidean time. The 
ground state energy of the $A$-nucleon system can be derived from the 
asymptotic behavior of the correlation function for large $\tau$,
\beq
E_A^0=-\lim_{\tau\rightarrow\infty}\frac{d}{d \tau}\ln Z_A(\tau).
\eeq
The expectation value of any normal ordered operator ${\cal O}$ can be derived in
a similar way by
\beqa
\langle\Psi_A^0|{\cal O}|\Psi_A^0\rangle &=&\lim_{\tau\leftarrow\infty}
\frac{Z_A^{{\cal O}}(\tau)}{Z_A(\tau)},  \\
Z_A^{{\cal O}}(t)&=&\langle\Psi_A|\exp(-\tau H/2){\cal O}\exp(-\tau H/2)|
\Psi_A\rangle,
\nonumber
\eeqa
where $|\Psi_A^0\rangle$ denote the ground states of 
$A$-nucleon system. It is convenient to describe NN contact
interactions by standard bilinear nucleon density operators using the
Hubbard-Stratonovich transformation. Using the relation
$\exp(\rho^2/2)\sim\int ds\, \exp(-s^2/2-s \,\rho)$
 one can
express terms quadratic in the nucleon density operator $\rho$ as terms 
linear in $\rho$ in the presence of auxiliary background fields (collectively
denoted by $s$). In this
representation, the full correlation function is related to the path integral
over pions and auxiliary fields,
\beqa
Z_A(t)&=& {\cal N}\int
Ds \prod_{I=1,2,3} D\pi_I Ds_I\exp(-S_{\pi\pi}-S_{ss})\nonumber\\
&\times& \!\!\! 
\langle\Psi_A|M^{(L_t-1)}(\pi_I,s,s_I)\cdots
M^{(0)}(\pi_I,s,s_I)|\Psi_A\rangle~.
\nonumber\\
\eeqa
Here $S_{\pi\pi}$ and $S_{ss}$ are free actions for pions and auxiliary fields
$s, s_I$ (where $s/s_I$ couples to the isospin-independent/dependent nucleon
bilinear), $I$ denotes isospin indices and ${\cal N}$ is an (irrelevant)
normalization constant. $M^{(n)}$ is a transfer matrix 
defined as an $n$'th step in the
temporal direction. We remark that the amplitude
$\langle\Psi_A|M^{(L_t-1)}(\pi_I,s_i)\cdots M^{(0)}(\pi_I,s_i)|\Psi_A\rangle$
is just a Slater determinant of single nucleon matrix elements 
${\cal M}_{i,j}$ with $i,j=1,\dots,A$.

To be specific, we give here the leading order action starting with the free
theory. The presentation here is somewhat sketchy. For an extensive discussion 
see~\cite{Borasoy:2006qn}. The free actions for the auxiliary fields and the pions are
\beqa
S_{ss}(s,s_I)&=&\frac{1}{2}\sum_{\vec{n}}s(\vec{n})^2
+\frac{1}{2}\sum_{I=1}^3\sum_{\vec{n}}s_I(\vec{n})^2~,\nonumber\\
S_{\pi\pi}(\pi_I)&=&\frac{\alpha_t}{2}\sum_{I=1}^3\sum_{\vec{n}}
\pi_I(\vec{n})(-\Delta+M_\pi^2)\pi_I(\vec{n})~, \nonumber\\
\eeqa
where $M_\pi$ is the physical pion mass and 
$\alpha_t=a_t/a$. 
For nucleons one may  use an
$O(a^4)$ improved free lattice Hamiltonian defined by
\beqa
H_{{\rm free}} &=& \frac{1}{m}\sum_{k=0}^3\sum_{\vec{n}_s,\hat{l}_s,i,j}f_k
\bigl[ a_{i,j}^\dagger(\vec{n}_s)(a_{i,j}(\vec{n}_s+k \hat{l}_s)
\nonumber\\
&&  \qquad\qquad + a_{i,j}(\vec{n}_s-k \hat{l}_s))\bigr], 
\eeqa
where the operators $a_{i,j}^\dagger(\vec{n}_s)$ and 
$a_{i,j}(\vec{n}_s)$ are the nucleon creation and
annihilation operators, $\vec{n}_s$ are spatial coordinates, $\hat{l}_s$ 
are spatial unit vectors,
the indices $i$ and $j$ represent spin and isospin indices, respectively, and
the coefficients $f_k$ are $f_{0,1,2,3}={49}/{2},-{3}/{4},{3}/{40},-{1}/{180}$.
To define the interactions one introduces  nucleon-density operators with
different spin/isospin polarizations
\beqa
\rho^{a^\dagger,a}(\vec{n}_s) &=&\sum_{i,j}a_{i,j}^\dagger(\vec{n}_s)
a_{i,j}(\vec{n}_s), \nonumber\\
\rho_I^{a^\dagger,a}(\vec{n}_s) &=&\sum_{i,j,j^\prime}
a_{i,j^\prime}^\dagger(\vec{n}_s)[\tau_I]_{j^\prime,j}a_{i,j}(\vec{n}_s),
\\
\rho_{I,S}^{a^\dagger,a}(\vec{n}_s) &=&\sum_{i,i^\prime,j,j^\prime}
a_{i^\prime,j^\prime}^\dagger(\vec{n}_s)[\sigma_S]_{i^\prime,i}
[\tau_I]_{j^\prime,j}a_{i,j}(\vec{n}_s).\nonumber
\eeqa
The transfer matrix for $n_t$-th step has, besides the free part, two important 
contributions:
\beqa
M^{(n_t)}&=&: \exp \biggl\{ -H_{{\rm free}}\alpha_t\nonumber\\
&-&\frac{g_A\alpha_t}{2 F_\pi}\sum_{S,I}\sum_{\vec{n}_s}\nabla_S
  \pi_I(\vec{n}_s,n_t)\rho_{S,I}^{a^\dagger,a}(\vec{n}_s)\nonumber\\
&+& \sqrt{-C\alpha_t}\sum_{\vec{n}_s} [s(\vec{n}_s,n_t)\rho^{a^\dagger,a}(\vec{n}_s)
\nonumber\\
&+&i\sqrt{C_I\alpha_t}\sum_{I}s_I(\vec{n}_s,n_t)\rho_I^{a^\dagger,a}(\vec{n}_s)]\biggr\}:.
\label{trasferMLO1}
\eeqa
Here $::$ denotes normal ordering. The first long-range contribution $\sim g_A$
includes the instantaneous pion-nucleon interaction and describes the
one-pion-exchange in the leading-order effective potential. The second
short-range contribution corresponds to the NN contact interactions. The
low-energy constants $C = C_S -2C_T$ and $C_I = -C_T$ (cf. sec.~\ref{sec:chiralEFT})  
have different signs, $C<0, C_I>0$. With these signs the pion-less theory can
be shown to have no sign-oscillations if
the number of protons and neutrons are equal and they stay pair-wise in
isospin-singlet states. In this case the multiplication with $\tau_2$ of the single-nucleon
matrix elements ${\cal M}$ from left and right is well defined and gives
$\tau_2{\cal M}\tau_2={\cal M}^*$.
For this reason, the determinant of ${\cal M}$ is real,
$\det{\cal M}^*=\det{\cal M}$.
Since $\tau_2$ is antisymmetric, the eigenvalues of ${\cal M}$ are doubly
degenerate. This leads to a positive Slater determinant~\cite{Chen:2003vy,Lee:2004ze}
\beq
\det{\cal M}\ge 0.
\eeq
The introduction of pions causes small sign oscillations which, however, are
not severe and appear to be suppressed.

\begin{figure}[tb]
\begin{center}
\includegraphics[width=0.49\textwidth,keepaspectratio,angle=0,clip]{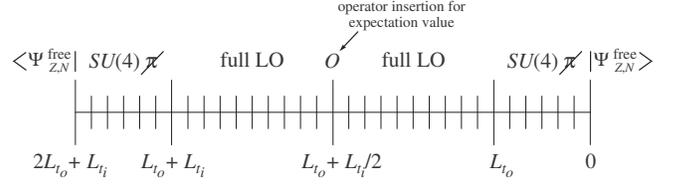}
\vspace{-0.15cm}
\caption[timesteps]{\label{timesteps} Overview of the various pieces of the
  transfer matrix calculation.}
\end{center}
\vspace{0.2cm}
\end{figure}

To perform simulations in a most efficient way one exploits the
approximate ${\rm SU}(4)$-Wigner~\cite{Wigner:1937zz} symmetry in the
NN system. The symmetry transformation is given by independent
rotations of the spin and isospin degrees of freedom.
\beq
\delta N=\alpha_{\mu\nu}\sigma^\mu\tau^\nu N \quad {\rm
  with} \quad\sigma^\mu=(1,\vec{\sigma})~, \quad \tau^\mu=(1,\vec{\tau}).
\eeq
One can show that in the limit where the NN S-wave scattering
lengths approach infinity the two-nucleon system becomes invariant under the
${\rm SU}(4)$-transformation~\cite{Mehen:1999qs}. 
The ${\rm SU}(4)$-breaking corrections come from the finite the
scattering length and higher order terms in the chiral expansion, 
these are of order ${\cal O}(( {1}/{a(^3S_1)}- {1}/{a(^1S_0)}),{q}/{\Lambda_\chi})$.
Since the NN scattering lengths
are very large, the ${\rm SU}(4)$-breaking corrections appear to be small. This
fact can be used to improve the performance of the lattice simulations. The
${\rm SU}(4)$ symmetric transfer matrix is given by
\beqa
M^{(n_t)} &=& :\exp\biggl[-H_{{\rm
      free}}\alpha_t \nonumber\\
&+&\sqrt{-C\alpha_t}\sum_{\vec{n}_s}s(\vec{n}_s,n_t)
\rho^{a^\dagger,a}(\vec{n}_s)\biggr]:.
\eeqa
In this case there are no sign oscillations for an 
even number of nucleons~\cite{Chen:2004rq}
and one only has one auxiliary field such that the simulations become much
cheaper. Although there is no positivity theorem for odd numbers of nucleons, 
sign oscillations seem to be suppressed also in systems with odd number of 
nucleons because it is only one particle away from an even system with no sign
oscillations.  Since the final result is close to the one
produced by a ${\rm SU}(4)$-symmetric simulation it pays to 
divide a simulations into three parts. To simulate the expectation value of
some observable one uses ${\rm SU}(4)$-symmetric transfer matrices in the first
and the last $L_{t_0}$ steps in order to filter the low-energy signal. After
this filtering, one starts the simulation with the complete (realistic) 
transfer matrices. A schematic overview of
the transfer matrix calculation is shown in Fig.~\ref{timesteps}.  
\noindent
Having set up the transfer matrix,  one utilizes the hybrid Monte-Carlo (HMC)
method~\cite{Duane:1987de} to update the field configurations. More
specifically, one introduces the conjugate fields $p_{\pi_I}$,
$p_s$, $p_{s_I}$ and uses molecular dynamics trajectories to generate new 
configurations for the fields $p_{\pi_I}$, $p_s$, $p_{s_I}$, $\pi_I$, $s$,
$s_I$ which keep the HMC Hamiltonian
\beqa
H_{{\rm HMC}} &=&
\frac{1}{2}\sum_{\vec{n}}\biggl(\sum_I\left[p_{\pi_I}^2(\vec{n})
+p_{s_I}^2(\vec{n})\right]+p_s^2(\vec{n})\biggr) \nonumber\\
&+& V(\pi_I,s,s_I),
\eeqa
constant, where the HMC potential is defined by
\beq
V(\pi_I,s,s_I)=S_{\pi\pi}+S_{ss}-\log|\det {\cal M}|.
\eeq
Upon completion of each molecular dynamics trajectory, an Metropolis 
accept or reject step for the new configuration according to the probability
distribution $\exp(-H_{{\rm HMC}})$ is applied. This process of molecular dynamics 
trajectory and Metropolis step is repeated many times.

Already at LO  promising results for binding energies, radii and density
correlations for the deuteron, triton and $^4$He are
obtained~\cite{Borasoy:2006qn}.  
On a $5^3$ lattice, the triton binding energy agrees with experiment 
within $5\%$ and the triton
root-mean-square radius is accurate to $30\%$. The binding energy for
$^4$He is within $25\%$ of the experimental value while the root-mean-square 
radius agrees within $10\%$. Note, however, that one has to overcome
a zero-range clustering instability that appears for 4 (or more particles) which
is mostly a combinatorial effect when more than two particles occupy the
same lattice site (for studies of this in other systems, see \cite{Lee:2005nm}
and references therein). To overcome this problem, one can e.g. smear out the contact
interactions with a Gaussian. Such terms improve the lattice
action and are formally of higher order (for more details, see
e.g. Ref.~\cite{Borasoy:2006qn}. At LO, one can also study the feasibility of 
simulations for light nuclei with more than four nucleons. It was observed 
that for $A\le 10$ the CPU time scales approximately linearly with the nucleon
number $A$.

\begin{figure}[tb]
\vspace{-0.8cm}
\includegraphics[width=0.55\textwidth,keepaspectratio,angle=0,clip]{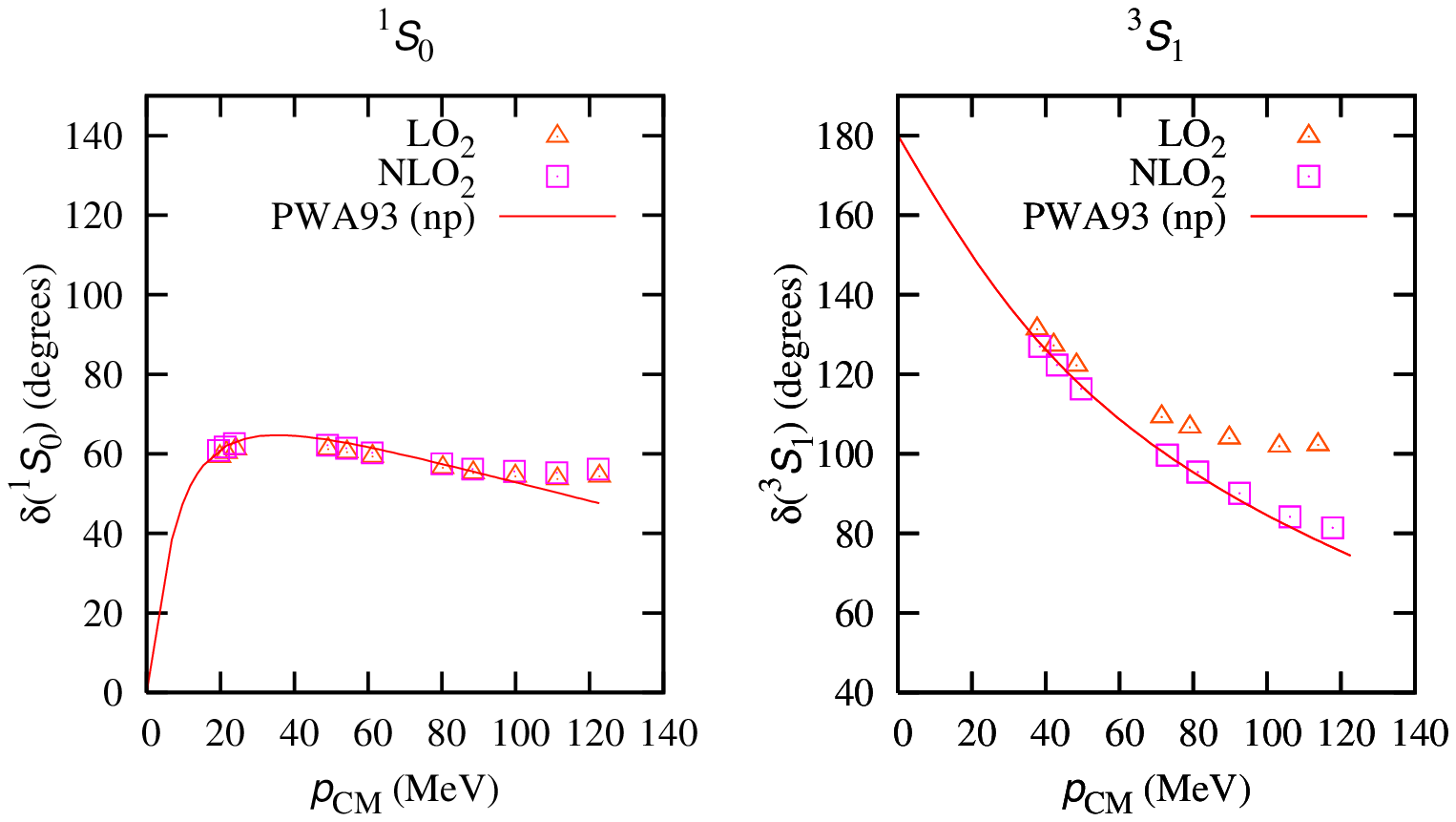}

\vspace{-0.8cm}
\includegraphics[width=0.55\textwidth,keepaspectratio,angle=0,clip]{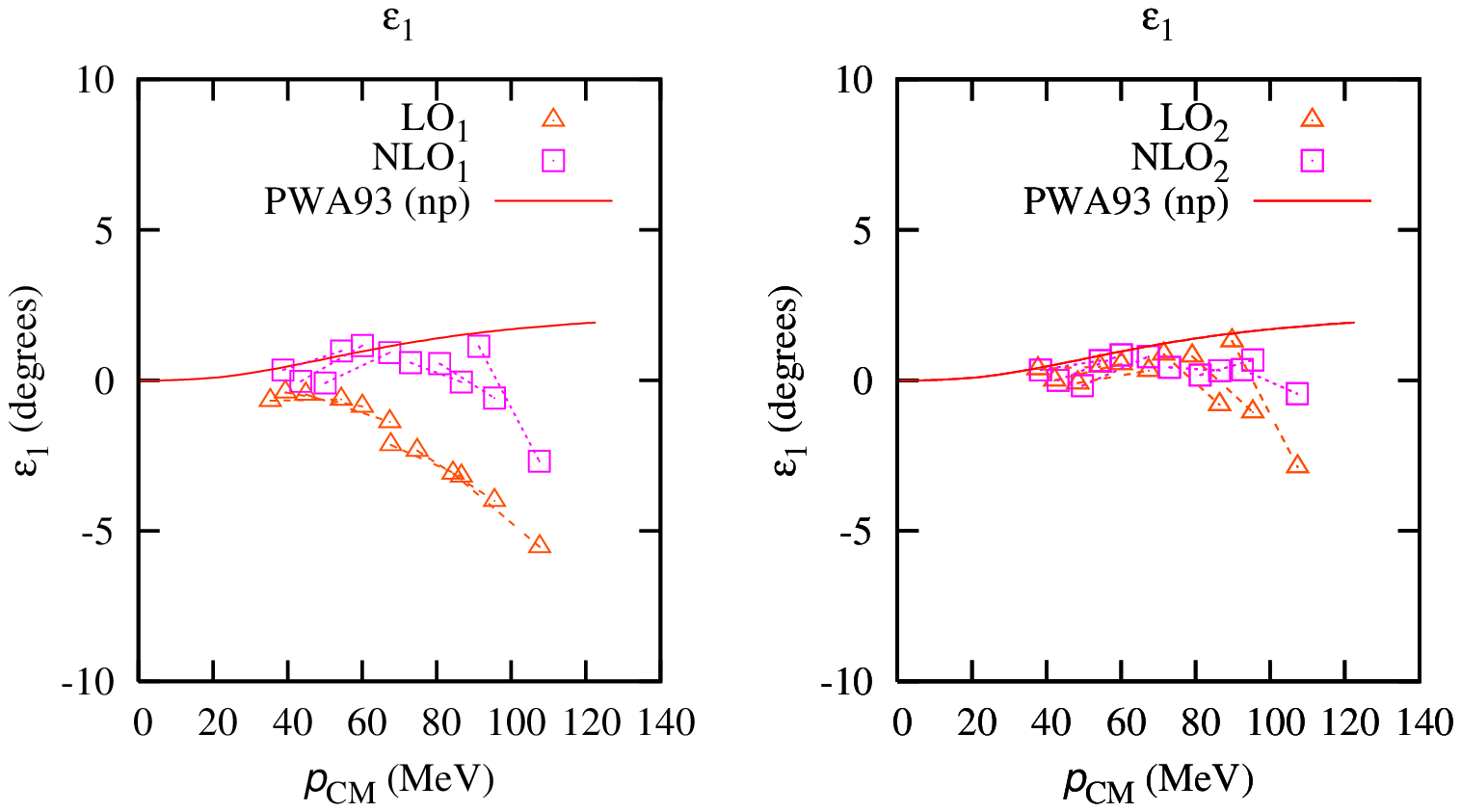}

\vspace{-0.2cm}
\caption[nlophaseshifts]{\label{nlophaseshifts} NN S-wave phase
  shifts and mixing angles versus
  center-of-mass momentum with actions ${\rm LO}_1$ and ${\rm LO}_2$.}
\end{figure}

At NLO there appear 9 LECs which can be fitted to the
Nijmegen NN scattering data and deuteron properties. 
Elastic scattering phase-shifts on the lattice are related by L\"uscher's 
formula to the energy levels of two-body
states in a finite large volume cubic box with periodic boundary
conditions (see Sec.~\ref{sec:lattice}). While this method is very useful 
at low momenta, it is not so useful for determining phase shifts on the
lattice at higher energies and
higher orbital angular momenta. Furthermore, spin-orbit coupling and
partial-wave mixing are difficult to measure accurately using L\"uscher's
method due to multiple-scattering artifacts produced by the periodic cubic
boundary conditions. In Ref.~\cite{Borasoy:2007vy} a more robust
approach to measure phase shifts for two nonrelativistic point particles
on the lattice using a spherical wall boundary was proposed. The basic idea 
is to impose a hard spherical wall boundary on the relative separation 
between the two interacting particles at some chosen radius. The reason for
this spherical wall is to remove copies of the two-particle interactions
due to the periodic boundaries on the lattice. This additional boundary condition 
allows for a direct extraction of the phase-shifts and mixing angles from the 
finite-volume spectrum. For more details, see~\cite{Borasoy:2007vy}.
Using the spherical wall method  the values of 9 LECs were determined by matching
three S-wave and four P-wave scattering data points, as well as deuteron binding
energy and quadrupole moment. In Fig.~\ref{nlophaseshifts} the NN S-wave phase-shifts and
the $^3S_1-^3D_1$mixing angle $\epsilon_1$ for two different actions, 
called ${\rm LO}_1$ and ${\rm LO}_2$,
are displayed. The action ${\rm LO}_1$ is the one given in
Eq.~(\ref{trasferMLO1}).  In the action  ${\rm LO}_2$ the contact interactions 
are smeared by a Gaussian. The two actions are identical at  leading order and 
differ only by higher-order terms, thus given an estimate of the higher order
corrections.
As can be seen from Fig.~\ref{nlophaseshifts}, the results of the lattice simulations
are in a good agreement with the partial wave results for momenta smaller than
$\sim 100$~MeV. Deviations between the two results for different actions appear
merely at larger momenta and are consistent with the expected higher order effects.
In the mean time, a novel action with spin-isopsin projected smearing has been
developed that gives a good description of the partial waves up to momenta
of the order of the pion mass.

\begin{figure}[tb]
\begin{center}
\includegraphics[width=0.45\textwidth,keepaspectratio,angle=0,clip]{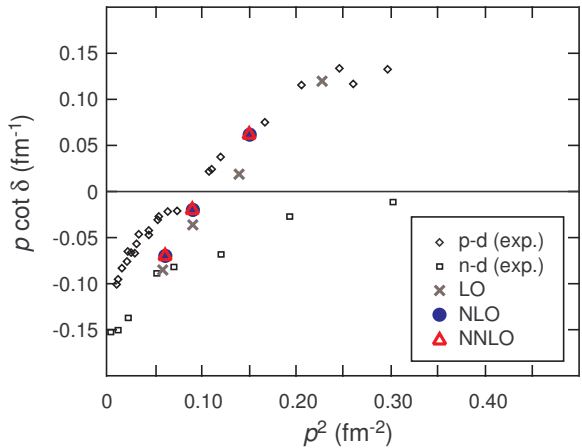}
\vspace{-0.15cm}
\caption[ndquartet]{\label{ndquartet} 
Neutron-deuteron scattering S-wave phase-shifts in the spin-$3/2$ quartet channel versus 
the square of relative momentum. The data for 
proton-deuteron and neutron-deuteron scattering are taken from Ref.~\cite{oers67nd}.
}
\end{center}
\vspace{-0.2cm}
\end{figure}

At N$^2$LO three-body forces start to show up which depend on two constants. 
These  LECs can be determined from a fit to neutron-deuteron scattering data in the
spin-$1/2$ doublet channel and the triton binding energy. These simulations
show a  very natural convergence pattern  with increasing chiral
order. For a  box length $\sim15$~fm the volume dependence already becomes
very small and the binding energy approaches its 
physical value. This is consistent with our expectation that the volume
dependence in nuclear lattice EFT simulations should become weak for 
$L\sim 20$~fm. In Fig.~\ref{ndquartet} the S-wave phase-shifts in the
spin-$3/2$ quartet  channel versus the square of relative momentum are shown. 
This channel was not taken into account in the fit procedure. Again one observes 
a very nice convergence with increasing chiral order. The predictions are 
located between the experimental data for proton-deuteron
and neutron-deuteron scattering data. Since  isospin-breaking was
not taken into account in the simulations the results are very satisfactory.  
At the same order, the (Coulomb-corrected) binding energy for $^4$He is 
overpredicted by  $5\%$, which is consistent with the expected theoretical 
accuracy of  these simulations.

The results of these studies demonstrate that lattice EFT is a promising 
tool for a quantitative description of light nuclei. In the future, 
it is planned to perform N$^2$LO Monte-Carlo simulations of light 
nuclei and probe neutron matter with larger number of neutrons in a box, see
also Sec.~\ref{sec:lattmany}. In addition, more detailed studies of finite
size effects and further improvements of the lattice action are called for.

\subsection{Quark mass dependence of nuclear forces and IR limit cycle in QCD}
\label{sec:chiralextrap}

The quark mass dependence of the chiral NN interaction was calculated
to next-to-leading order (NLO) in the chiral counting 
in Refs.~\cite{Beane:2002xf,Epelbaum:2002gb}.
At this order, the quark mass dependence is synonymous to
the pion mass dependence because of the Gell-Mann-Oakes-Renner relation:
$M_\pi^2 = -(m_u + m_d) \langle 0 | \bar{u} u | 0 \rangle/F_\pi^2\,,$
where $\langle 0 | \bar{u} u | 0 \rangle \approx (-290 \mbox{ MeV})^3$
is the quark condensate. In the following, we will therefore refer
only to the pion mass dependence which is more convenient for 
nuclear applications and treat the pion mass as a parameter that can be
varied by adjusting the values of the quark masses.
In the work of Refs.~\cite{Beane:2001bc,Beane:2002xf,Epelbaum:2002gb},
the pion mass dependence of the nucleon-nucleon scattering lengths in the 
$\trip$--$^3{\rm D}_1$ and $\sing$ channels
as well as the deuteron binding energy are calculated.
To next-to-leading order (NLO) in the chiral power counting,
the NN potential can be written as
\beq
V_{\rm NLO} = V^{\rm OPE} + V^{\rm TPE} + V^{\rm cont}\,,
\eeq
where $V^{\rm OPE}$, $ V^{\rm TPE}$, and  $V^{\rm cont}$
refer to the one-pion exchange, two-pion exchange, and contact potentials,
respectively. Explicit expressions can be found in 
Ref.~\cite{Epelbaum:2002gb} and section \ref{sec:chiralEFT}.  

In principle, the pion mass dependence of the
chiral NN potential is determined uniquely. 
However, the extrapolation away from the 
physical pion mass generates errors. The dominating source are the
constants  $\bar C_{S,T}$ and $\bar D_{S,T}$ in $V^{\rm cont}$
which give the corrections to the LO contact terms $\propto M_\pi^2$
and cannot be determined
independently from fits to data at the physical pion mass.
A smaller effect is due to the error in the LEC $\bar d_{16}$, which
governs the pion  mass dependence of $g_A$, 
from the chiral pion-nucleon Lagrangian which is enhanced
as one moves away from the physical pion mass. Both effects generate
increasing uncertainties as one extrapolates away from the physical point.

In the calculation of Ref.~\cite{Epelbaum:2002gb}, 
the central value $\bar d_{16}=-1.23$ GeV$^{-2}$ was used 
which is the average of three values given in Ref.~\cite{Fettes:2000fd}.
In addition, a variation of the LEC $\bar d_{16}$ in the range 
$\bar d_{16} = -0.91 \ldots -1.76$ GeV$^{-2}$ as given in  Ref.~\cite{Fettes:2000fd}  was
employed.  
The size of the two constants $\bar D_{S}$ and $\bar D_{T}$ was
constrained from naturalness arguments. It was argued that the 
corresponding dimensionless constants $F_\pi^2 \Lambda_\chi^2 \bar D_{S,T}$ 
can be expected to satisfy the bounds:
\beq
-3 \leq F_\pi^2 \Lambda_\chi^2 \bar D_{S,T} \leq 3\,,
\label{eq:STbound}
\eeq
where $\Lambda_\chi \simeq$ 1 GeV is the chiral symmetry breaking scale.
A more conservative error estimation was given in Ref.~\cite{Epelbaum:2002gk}. 
We note that Refs.~\cite{Beane:2001bc,Beane:2002xf} allowed for a larger 
variation of these LECs. However,
the bounds (\ref{eq:STbound}) are in agreement with resonance saturation 
estimates and similar relations are obeyed by the LECs whose values
are known \cite{Epelbaum:2001fm}.
For the constants $C_{S,T}$, e.g., this leads to
$C_{S}=-120.8$~GeV$^{-2}$ and $C_{T}=1.8$~GeV$^{-2}$ corresponding 
to the dimensionless coefficients $F_\pi^2 C_{S}$ = $-$1.03 and 
$F_\pi^2 C_{T}$ = 0.02, respectively. The unnaturally small value
of $F_\pi^2 C_{T}$ is a consequence of the approximate Wigner
SU(4) symmetry. 

The ranges from Eq.~(\ref{eq:STbound}) were
used to estimate the extrapolation errors of two-nucleon observables
like the deuteron binding energy and the spin-singlet and 
spin-triplet scattering
lengths in Ref.~\cite{Epelbaum:2002gb}. 
\begin{figure}[tb]
\centerline{\includegraphics*[width=7cm,angle=0,clip=true]{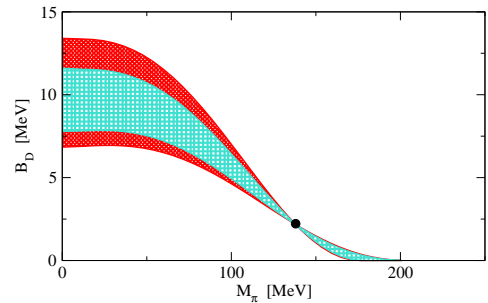}}
\caption{Deuteron binding energy as a function of the pion mass $M_\pi$.
The shaded areas correspond to the allowed values. The light shaded band
gives the uncertainty due to the unknown value of the LECs $\bar D_{\rm S,T}$
using the central value $\bar d_{16} -1.23$ GeV$^{-2}$. The dark shaded band
gives the uncertainty if, in addition to the variation of $\bar D_{\rm S,T}$,
the LEC $\bar d_{16}$ is varied in the range from $\bar d_{16} -0.91$
GeV$^{-2}$ to $\bar d_{16} -1.76$ GeV$^{-2}$ given in
Ref.~\cite{Fettes:2000fd}.   The heavy dot shows the binding energy for
the physical value of the pion mass. 
}
\label{fig:Edn}
\end{figure}
The resulting pion mass dependence of the deuteron 
binding energy is shown in  Fig.~\ref{fig:Edn}.
In the chiral limit the binding energy is
of natural size, $B_D \sim F_\pi^2/m \simeq 10$~MeV. Note, however, that 
in the calculation of \cite{Beane:2001bc,Beane:2002xf}
the assumed larger uncertainties in the LECs 
prevent one from making a definite statement about the binding
of the deuteron in the chiral limit. For pion masses above the physical value
the differences between the two calculations are considerably smaller.
The recent study of Mondejar and Soto seems to indicate that two-loop
diagrams generate a peculiar quark mass dependence of the contact 
interactions which are parametrically large \cite{Mondejar:2006yu}.
The influence of these effects on the quark mass dependence of, e.g., the 
deuteron binding energy remains to be worked out in detail.
\begin{figure}[tb]
\centerline{\includegraphics*[width=0.49\textwidth,angle=0,clip=true]{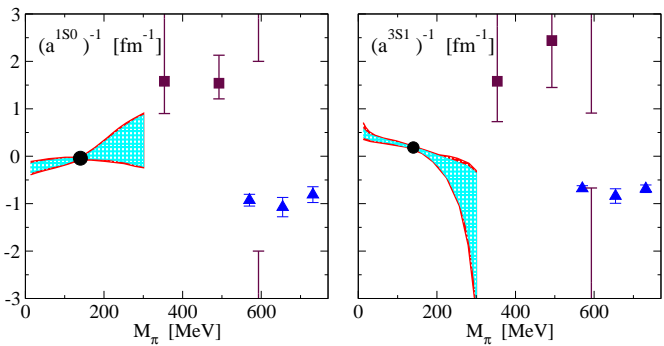}}  
\caption{Inverse of the S-wave scattering lengths in the  
spin-triplet and spin-singlet nucleon-nucleon channels 
as a function of the pion mass $M_\pi$. Filled triangles and rectangles
show the lattice calculations from
Refs.~\cite{Fukugita:1994na,Fukugita:1994ve} and \cite{Beane:2006mx},
respectively.  For remaining notation see Fig.~\ref{fig:Edn}. 
}
\label{fig:asat}
\end{figure}
In Fig.~\ref{fig:asat}, we show
the inverse scattering lengths in the spin-triplet and  spin-singlet 
channels  from Ref.~\cite{Epelbaum:2002gb}
together with some recent lattice results \cite{Beane:2006mx}.
However, the errors and pion masses are still too large to draw
any conclusions about the physical point.

Fig.~\ref{fig:asat} also shows that a scenario where 
both inverse scattering lengths vanish simultaneously
at a critical pion mass of about $200$ MeV
is possible. For pion masses below the critical value, the spin-triplet
scattering length would be positive and the deuteron would be bound. 
As the inverse spin-triplet scattering length decreases, 
the deuteron would becomes more and more
shallow and finally would become unbound at the critical mass. 
Above the critical pion mass the deuteron would exist as a shallow virtual 
state.  In the spin-singlet channel, the situation is reversed: the 
\lq\lq spin-singlet deuteron'' would be a virtual state below the 
critical pion mass and would become bound above. 
It is unlikely that this scenario of both inverse scattering lengths
vanishing simultaneously is realized in QCD at the physical
values of the quark masses. 
However, based on this behavior it was conjectured that 
one should be able to reach the critical point by varying  
the up- and down-quark masses $m_u$ and $m_d$ independently
because the spin-triplet and spin-singlet channels have different isospin
\cite{Braaten:2003eu}.
In this case, the triton would display the Efimov effect which
corresponds to the occurence of an infrared limit cycle in QCD.
It is evident that a complete investigation 
of this issue requires the inclusion
of isospin breaking corrections and therefore higher orders in the chiral
EFT. However, a number of studies have investigated the universal 
properties of the limit cycle by considering specific values of
$\bar D_{S}$ and $\bar D_{T}$.

In the exploratory study of
the three-nucleon system \cite{Braaten:2003eu}, the mean values of the 
error bands from \cite{Epelbaum:2002gb}
were used as input for the three-body calculations in the 
pionless EFT. 
Even though both scattering lengths were large for the mean values,
they did not become infinite at the same value of the pion mass and 
there was no exact limit cycle for this choice of parameters.
However, different sets of values
for $\bar D_{S}$ and $\bar D_{T}$ that lie within the bound 
given by Eq.~(\ref{eq:STbound}) and cause the spin-singlet and 
spin-triplet scattering lengths to become infinite at the same value of 
the pion mass can be found.

\begin{figure}[tb]
\centerline{\includegraphics*[width=8cm,angle=0,clip=true]{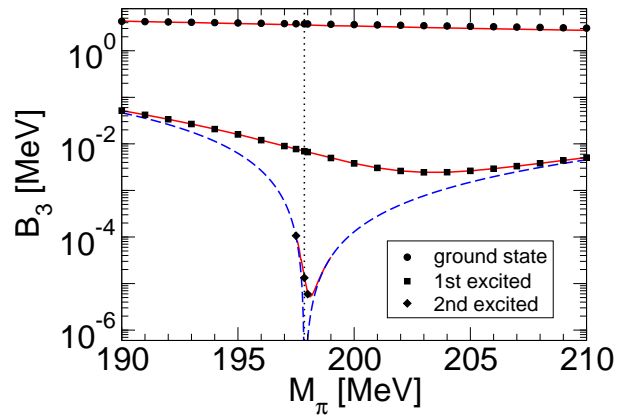}}
\caption{Binding energies $B_3$ of the triton ground and first two
excited states as a function of $M_\pi$.
The circles, squares, and diamonds give the chiral EFT result, while
the solid lines are calculations in the pionless theory.
The vertical dotted line indicates the critical pion mass $\mpic$
and the dashed lines are the bound state thresholds.
}
\label{fig:bind3}
\end{figure}

In Ref.~\cite{Epelbaum:2006jc}, the properties of the triton around
the critical pion mass were studied for one particular solution with 
a critical pion mass $\mpic=197.8577$ MeV. 
{}From the solution of the Faddeev equations, 
the binding energies of the triton and the first
two excited states in the vicinity of the limit cycle
were calculated for this scenario in chiral EFT.
The binding energies are given in Fig.~\ref{fig:bind3}
by the circles (ground state), squares (first excited state),
and diamonds (second excited state). The dashed lines indicate the
neutron-deuteron ($M_\pi \leq \mpic$) and 
neutron-spin-singlet-deuteron ($M_\pi \geq \mpic$) thresholds
where the three-body states become unstable. Directly at 
the critical mass, these thresholds coincide with the three-body
threshold and the triton has infinitely many excited states.
The solid lines are leading order calculations in the pionless theory using 
the pion mass dependence of the nucleon-nucleon scattering lengths
and one triton state from chiral EFT as input.
The chiral EFT results for the other triton states in the critical region 
are reproduced very well.
The binding energy of the triton ground state
varies only weakly over the whole range of pion masses and is about 
one half of the physical value at the critical point. The excited states are
influenced by the thresholds and vary much more strongly.

These studies were extended to N$^2$LO in the pionless EFT and 
neutron-deuteron scattering observables in Ref.~\cite{Hammer:2007kq}.
It was demonstrated that the higher order corrections in the vicinity
of the critical pion mass are small. 
\begin{figure}[tb]
\centerline{\includegraphics*[width=8cm,angle=0,clip=true]{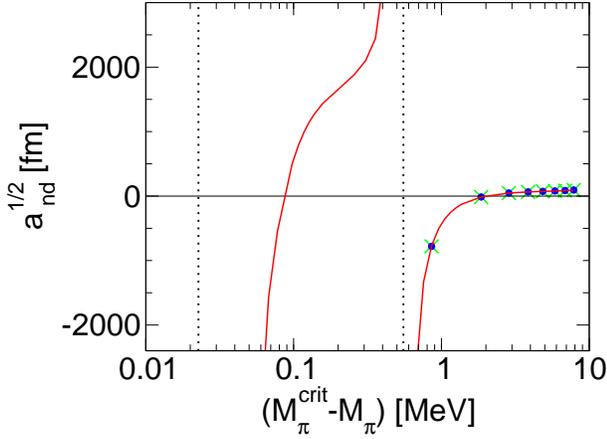}}
\caption{Doublet neutron-deuteron scattering length $a_{nd}^{1/2}$
in the critical region computed in the pionless EFT. The solid line gives 
the LO result, while the crosses and circles show the NLO and 
N$^2$LO results. The dotted lines indicate the 
pion masses at which $a_{nd}^{1/2}$ diverges.
}
\label{fig:aND_crit}
\end{figure}
This is illustrated in Fig.~\ref{fig:aND_crit}, where
we show the doublet scattering length $a_{nd}^{1/2}$ in the critical
region.  The solid line gives the LO result, while the crosses and
circles show the NLO and N$^2$LO results. The dotted lines indicate
the pion masses at which $a_{nd}^{1/2}$ diverges because the second
and third excited states of the triton appear at the neutron-deuteron
threshold.  These singularities in $a_{nd}^{1/2}(M_\pi)$ are a clear
signature that the limit cycle is approached in the critical
region.  Considering the specific case of
the doublet scattering length at
$M_\pi=190$~MeV, we have
\beq
a_{nd}^{1/2}(M_\pi=190~\text{MeV})=(93.18+0.80+0.14)~\text{fm}~.
\eeq
At this pion mass we are fairly far away from any singular
points in the function $a_{nd}^{1/2}(M_\pi)$. Consequently, the
pionless results follow a natural convergence pattern with the
expansion parameter $\gamma r_0$, which is $\approx 0.08$ at this 
value of $M_\pi$.
For three-body scattering observables in the critical region
no calculation in chiral EFT is available. The above example 
shows that both approaches complement each other.
If the pion mass dependence of one three-body observable is 
known, the pionless theory can be used to predict all other observables
with high precision and less computational effort. 
Figure \ref{fig:aND_crit} demonstrates clearly 
that the pionless theory converges rapidly in the critical region.

A final answer on the question of whether an infrared limit cycle can
be realized in QCD can only be given by solving QCD directly. In particular,
it would be very interesting to know whether this can be achieved
by approriately tuning the quark masses in a Lattice QCD simulation
\cite{Wilson:2004de}. Finally, we note that the dependence of nuclear 
binding on hadronic mass variation has also been studied based on the
Argonne potential \cite{Flambaum:2007mj}.

\section{Towards a Many-Body EFT for Nuclei}
\def\theequation{\arabic{section}.\arabic{equation}}
\setcounter{equation}{0}
\label{sec:many}

The EFT approach to the nuclear many-body problem is much
less established than the one for the forces and few-nucleon
systems. This is, on one hand, related to the apperance of
new scales like the Fermi momentum or induced by collective excitations,
and on the other hand to the computational problems related to 
solve the many-body problem. Thus, a variety of pathways are being
explored and here we can only give a brief overview about the
existing attempts and their status. For a pedagogical
review on the application of EFTs to finite density systems,
we refer to Furnstahl et al.~\cite{Furnstahl:2008df}.

\subsection{In-medium chiral perturbation theory}
\label{sec:CHPTrho}

Early attempts to formulate in-medium chiral perturbation theory 
where mostly triggered by the pioneering paper of Kaplan and
Nelson on kaon condensation~\cite{Nelson:1987dg}. Most of these
calculations were based on chiral Lagrangians at most 
bilinear in the nucleon fields and performed in the mean-field
approximation, $\bar N D N \to \rho_p {\rm Tr} D_{11} +
\rho_n {\rm Tr} D_{22}$ with $\rho_p (\rho_n)$ the proton 
(neutron) density, $D$ represents a generic differential operator
including the coupling to pions and external sources, the
trace runs over spinor indices and the subscripts run in
flavor space. Proceeding in this way, one keeps track
about the vacuum CHPT Lagrangians, but the chiral counting
in the medium is lost, as e.g. nucleon correlations are not
considered. The most elegant formulation of this approach 
based on the path-integral formulation is due to
Wirzba and collaboraotrs, see e.g.   
\cite{Thorsson:1995rj,Kirchbach:1996xy,Kirchbach:1997rk}.

\begin{figure}[tb]
\includegraphics*[width=5.5cm]{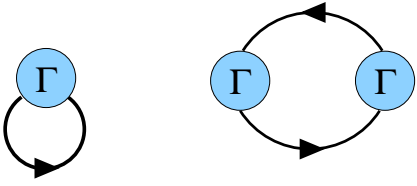}
\begin{center}
\caption{Generalized in-medium vertices of lowest order. 
The thick solid lines correspond to insertions of a 
Fermi-sea and each circle to the insertion of the
operator $\Gamma$ as defined in the text. 
\label{fig:genvertex}}
\vspace{-0.5cm}
\end{center}
\end{figure}

To go beyond the mean-field approximation, the in-medium
genrating functional for pions coupled to nucleons and
external sources was developed in  Refs.~\cite{Oller:2001sn,Meissner:2001gz}.
Leaving out multi-nucleon interactions, as systematic
in-medium CHPT can be developed by expanding  around the
nuclear matter ground state at asymptotic times and
intergrating out the nucleon fields in the path integral
representation, giving rise to the in-medium generating
functional (for the detailed derivation, we refer
to Ref.~\cite{Oller:2001sn})
\beqa
&& e^{i \tilde{{\cal Z}}[v,a,s,p]} = \int\! 
 [dU] \exp\Bigg\{i\!\int\! dx\, {\cal L}_{\pi\pi}\!
- i
\!\int\!\frac{d\vec{p}}{(2\pi)^3 2 E(p)}\nonumber\\ 
&\times& \!\!\!\!\int\! dx\,dy\, e^{ip(x-y)}\,
\hbox{Tr}\Bigg( A[I_4-D_0^{-1}A]^{-1}
\arrowvert_{(x,y)}
(\barr{p+m})\,n(p)\Bigg)
\nonumber \\ 
&+&\frac{1}{2}\!\int\!\!\! \frac{d\vec{p}}{(2\pi)^3 2E(p)}\!\int\! \!\! 
\frac{d\vec{q}}{(2\pi)^3 2 E(q)}
\!\int\! dx\,dx'\,dy\,dy'\,e^{ip(x-y)}\nonumber\\
&\times& \!\!\!e^{-iq(x'-y')}\, \hbox{Tr}\Bigg(
 A[I_4-D_0^{-1}A]^{-1}\arrowvert_{(x,x')}\, (\barr{q}+m)
n(q) \nonumber\\ 
&\times& \!\!\!A[I_4-D_0^{-1}A]^{-1}
\arrowvert_{(y',y)}(\barr{p}+m)\,n(p)\Bigg)+ \ldots\Bigg\}
\nonumber
\eeqa
\beqa
&& \,\,\, \equiv  \int\! 
 [dU] \exp\Bigg\{i\!\int\! dx\, {\tilde 
{\cal L}}_{\pi\pi}[U;v,a,s,p]\Bigg\}~,
\label{eq:fZ2}
\eeqa
where the operator $A$ is defined as the difference between the full
and the free Dirac operators, 
\beq
{\cal L}_{\bar\psi \psi} = \bar{\psi}(x) D(x) \psi(x) 
=  \bar{\psi}(x) (D_0(x)  -A(x)) \psi (x)~,
\eeq
with  $D_0 = i \gamma_\mu\partial^\mu -m$, while the diagonal 
flavor matrix $n(p)  = {\rm diag}(\theta(k_F^{(p)} - |\vec{p}|,
\theta(k_F^{(n)} - |\vec{p}|)$ parameterizes the upper cut-off 
of the three-momentum integrations in terms of the proton and neutron
Fermi momenta, respectively. Furthermore $I_4$ is the unit operator
in four dimension, $E(p)$ the on-shell energy of a nucleon with
mass $m$ and $v,a,s,p$ are vector, axial-vector, scalar and
pseudoscalar sources. The resulting in-medium effective 
Lagrangian $\tilde{\cal L}$
is given in terms of pions and external sources only and thus the
problem is reduced to that of vacuum CHPT, with the important
difference that the  $\tilde{\cal L}$ is {\sl non-covariant} as well
as {\sl non-local} (for a general analysis of the structure of
non-relativistic but local EFTs see~\cite{Leutwyler:1993gf}). 
In particular, we note the appearance of the  non-local vacuum 
vertex $\Gamma = -iA(I_4-D_0^{-1}A)^{-1}$ that generates a geometric 
series in terms of the local interaction operator $A$ and the 
free Dirac propagator, with $A$ itself being subject to the standard 
chiral expansion, $A= A^{(1)} + A^{(2)} + \ldots$,
see~\cite{Bernard:1992qa}. 
The generalized {\sl in-medium} vertices, cf Fig.~\ref{fig:genvertex}, 
consist of 
several non-local vacuum  vertices $\Gamma$  connected through the 
exchange of on-shell Fermi-sea states. These are the building blocks 
for the systematic expansion in small momenta, counting the Fermi 
momentum $k_F \sim 2M_\pi$ at  nuclear saturation as ${\cal O}(p)$.
\noindent
The in-medium chiral counting including the contributions
from nucleon propagators can now be given.  The choice of the
counting scheme depends on the energy flowing through the
nucleon lines, inducing a separate consideration of the
so-called {\sl standard} amd {\sl non-stantard} cases. Let
us consider first the former. Here, the energy flow is
of order $M_\pi \sim {\cal O}(p)$ and thus the nucleon propagator
counts as $D_0^{-1} \sim {\cal O}(p^{-1})$. The chiral dimension 
for a many-particle diagram with $L_\pi$ pion loops and $V_T$
vacuum and/or in-medium vertices of dimension $\delta_i$ is
\beq
\nu = 2L_\pi + 2 + \sum_{j=1}^{V_T} (\delta_j -2)~.
\eeq
Consequently, the lowest order in-medium contributions arise
at ${\cal O}(p^4)$ since the lowest order in-medium vertices have
dimesnion four due to the four-momentum Dirac delta-function attached
to any $\Gamma$ vertex. The first corrections at NLO arise
at ${\cal O}(p^5)$, which  should be contraested to the vacuum
case where LO (NLO) is  ${\cal O}(p^2) \, ({\cal O}(p^4))$. In
the absence of multi-nucleon interactions, the breakdown scale
is $\Lambda = \sqrt{6} \pi F_\pi \simeq 700\,$MeV for S-waves
and $\Lambda = \sqrt{6} \pi F_\pi/g_A \simeq 560\,$MeV for P-waves. However,
there is one subtlety with this power counting. Quite similar to
what happens in case of the TPE NN interaction, the energy flowing
into a nucleon line can vanish, so that the nucleon propagator
scales as ${\cal O} (p^{-2})$. To deal with this non-standard case, one
has to separately count the number of nucleon lines with energy
$E \leq k_F^2/2m$ and the normal lines with $E \sim M_\pi$. 
The explicit expression for the modified
counting index $\nu$ can be found in~\cite{Meissner:2001gz}.
In this case, the breakdown scale is $6\pi^2 F_\pi^2/2m \simeq
270\,$MeV for S-waves and further reduced by a factor of $1/g_A^2$
for P-wave interactions. Note that the so-defined in-medium CHPT 
not only encompasses but also transcends the so-called low-energy 
theorems of Refs.~\cite{Drukarev:1988kd,Furnstahl:1992pi,Birse:1994cz}.

We now discuss some results obtained in this scheme. 
The density dependence of the light quark condendates is given at NLO by
\beqa
 \label{qc}
 \langle \Omega|\bar{u}u|\Omega\rangle
&=&\langle \bar{u}u \rangle_{\rm vac}
 \left[1-\frac{2\sigma}{F^2_\pi M_\pi^2}\hat{\rho}
 +\frac{4 c_5}{F_\pi^2}\bar{\rho}\right]\nonumber~,\\
 \langle \Omega|\bar{d}d|\Omega\rangle
&=&\langle \bar{d}d \rangle_{\rm vac}
 \left[1-\frac{2\sigma}{F_\pi^2 M_\pi^2}\hat\rho
-\frac{4 c_5}{F_\pi^2}\bar\rho\right]~,
\eeqa
with $|\Omega\rangle$ the nuclear matter background, $\hat \rho
= (\rho_p + \rho_n)/2$,  $\bar \rho =(\rho_p - \rho_n)/2$ are the
isospin symmetric and asymmetric combinations of the proton and the
neutron densities, while $\sigma$ is the pion-nucleon sigma-term, 
$\sigma=-4c_1 M_\pi^2$ at ${\cal O}(p^2)$. The small isospin-breaking
contribution is given in terms of the LEC $c_5 = -(0.09\pm
0.01)\,$GeV. Furthermore, the subscript $\rm vac$ refers to the vacuum 
value of the corresponding quantity. Higher order corrections in
the density will be discussed in the next subsection. 

The propagation of pions in the medium can be analyzed by calculating
the spectral relations between the energy $\omega$ and the three-momentum
$\vec q$ for on-shell neutral and charged pions. For symmetric nuclear
matter (with density $\hat\rho$) and in the chiral limit one obtains 
the dispersion law
\beq
\omega^2 = \vec{q\,}^2 \left( 1 - \frac{4\hat\rho}{F_\pi^2}\, c_2\right)~.
\eeq
Since the in-medium pion velocity $\tilde v = d\omega / d|\vec{q}\,|=
1-2\hat\rho c_2/F_\pi^2$ must be smaller than the velcoty of light
\cite{Leutwyler:1993gf,Pisarski:1996mt}, this imposes the constraint 
$c_2 \geq 0$, which is satisfied by the actual value of this LEC.
It was also established in~\cite{Meissner:2001gz} that for standard
values of the LEC $c_3$, chiral symmetry can account for the observed
mass shift of the negatively charged pion in deeply bound pionic
states in $^{207}$Pb \cite{Gilg:1999qa,Itahashi:1999qb}. At NLO,
one obtains $\Delta M_\pi = 18\pm 5\,$MeV which is compatible with the
experimental result, $\Delta M_\pi = 23\ldots 27\,$MeV. Of interest 
is also splitting of the temporal and space-like components of the
pion decay constant, which read at NLO in symmetric nuclear matter
\beqa
F_t = F_\pi \left\{ 1 - \frac{\hat\rho}{\rho_0} (0.26\pm 0.04)\right\}~,
\nonumber\\
F_s = F_\pi \left\{ 1 - \frac{\hat\rho}{\rho_0} (1.23\pm 0.07)\right\}~,
\eeqa
with $\rho_0$ the nuclear matter density. The ratio $F_s/F_t = \tilde{v}^2
< 1$ is consistent with the discussion about the in-medium pion
velocity. One can also show that the corrections at ${\cal O}(p^5)$
do not spoil the validity of the Gell-Mann--Oakes--Renner relation,
see also~\cite{Thorsson:1995rj}, in particular both $F_t$ and the quark
condensate decrease with increasing density. For a more detailed
discussion of 2-, 3- and 4-point functions in the medium, we
reder to Ref.~\cite{Meissner:2001gz}.

The missing ingedrient in these calculations are the effects of
multi-nucleon interactions. It has recently been shown how
these can be included in the path integral formulation~\cite{Oller,LMOW}. 
For that, one introduces heavy fields $H$ that couple to nucleon
bilinears in appropriate spin-isospin combinations, 
$\tilde{{\cal L}} [U;v,a,s,p]\to \tilde{{\cal L}} [U;H;,a,s,p]$,
and letting the mass of the $H$-fields tend to infinity, cf. 
Fig.~\ref{fig:resosat}. In that way, one can formally integrate
out the multi-fermion interactions from the generating functional.

Further progress has been made by Girlanda et al. in 
Ref.~\cite{Girlanda:2003cq}. They developed a generalization of
in-medium CHPT for finite systems. This provides a framework
to study pion-nuclear bound states, for which the finite volume
and the surface of the nucleus are important ingredients. The
corresponding chiral counting is applied to the underlying pion-nucleon
interactions and also to the relevant nuclear matrix elements. The
central object of this approach are the Greens function in the
presence of a nucleus, $G_A(X\to Y)$ that describe the general process
$A + X \to A +Y$, where $X,Y$ represent some number of external pions
and photons and $A$ is a nucleus made of $A$ nucleons. The presence of the 
nucleus is parameterized in terms of proton and neutron distribution functions
that are taken from phenomenology. In the limit of uniform desnity,
this approach reduce to the in-medium CHPT described above. As an
example, the pion-nucleus optical potential  is calculated at NLO,
\beqa\label{eq:pinucpot}
U(E;\vec{q}', \vec{q}) &=& \!\!\!\! \int\!\! d^3\vec{x} 
{\rm e}^{-i(\vec{q}'-\vec{q})\cdot \vec{x}} \left[ \tilde{U}
(E;\vec{q}', \vec{q}, \vec{x}) + {\cal O}(p^6)\right],\nonumber\\
\tilde{U}(E;\vec{q}', \vec{q},\vec{x}) &=& - \!\!\int\!\! d^3\vec{r}
\frac{e^2 E}{4\pi |\vec{x}-\vec{r}|} \, 2\rho_p(\vec{r})
+ \ldots 
\eeqa
where $E$ is the pion energy, $\vec{q}', \vec{q}$ are the out-going and the
in-comimg three-momenta, in order, and $\rho_p$ is the proton charge density. 
The ellipsis in Eq.~(\ref{eq:pinucpot}) stand for the contributions from 
the hard virtual photons and from the strong interaction,
for details see~\cite{Girlanda:2003cq}. As particularly stressed
in~\cite{Girlanda:2004qa}, this approach allows one to identify 
unambiguously the nuclear finite size effects and to disentangle the 
S-, P - and D-wave contributions to the optical potential without invoking 
the local density approximation. For a more detailed discussion concerning
also the comparison with more traditional approaches to pion-nuclues
physics, we refer to Ref.~\cite{Girlanda:2003cq}. 

\subsection{Perturbative chiral nuclear dynamics}
\label{sec:TUMrho}

\begin{figure}[tb]
\includegraphics*[width=5.5cm]{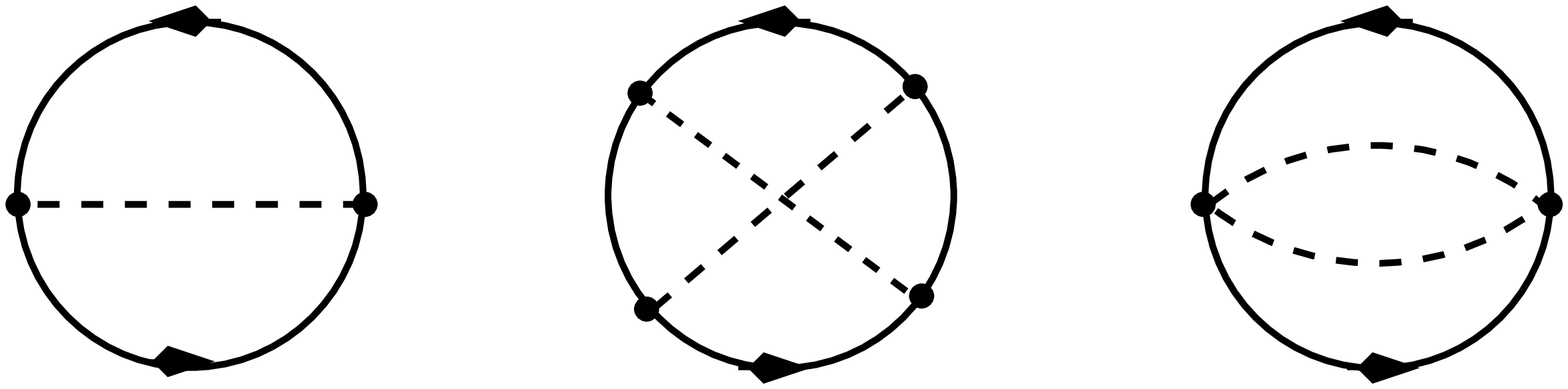}

\vspace{0.2cm}

\includegraphics*[width=5.5cm]{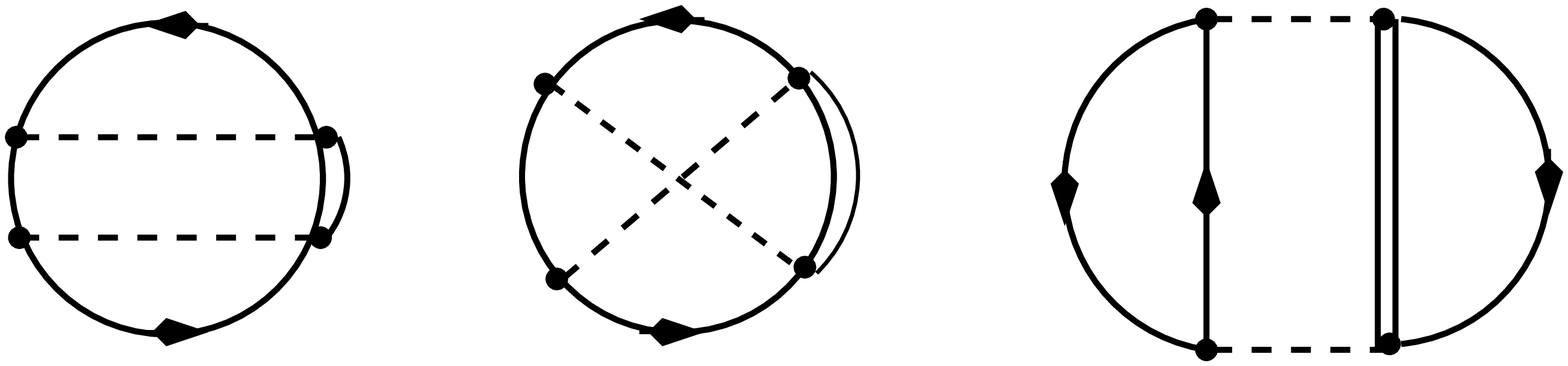}
\begin{center}
\caption{Upper panel: one- and two pion exchange
diagrams contributing to the energy per particle
at two and three loops. Solid and dashed lines 
denote nucleons and pions, respectively.
Lower panel: Three-body diagrams related to $2\pi$-exchange
with single delta excitations (double lines). 
These represent interactions
between three nucleons in the Fermi sea.
\label{fig:TUMdiag}}
\vspace{-0.5cm}
\end{center}
\end{figure}

A somewhat different path, that has turned out to be very
successfull phenomenologically, has been taken by the
Munich group \cite{Kaiser:2001jx,Kaiser:2001ra,Fritsch:2004nx} (for 
a related work, see Ref.~\cite{Lutz:1999vc}).  Its key element is the
separation of long- and short-distance dynamics. The ordering scheme
counts pion massses and momentum, the Fermi momentum and the nucleon-delta
mass splitting as small quantities~\footnote{We unify here the two formulations
with and without explicit deltas presented by this group.}, motivated by 
the fact that at nuclear matter density $k_F \simeq
2M_\pi$. Therefore, pions must be included and propagation effects of the
delta will be resolved. The long-distance physics is calculated perturbatively
including one- and two pion exchange Hartree and Fock graphs, see
Fig.\ref{fig:TUMdiag} for some typical diagrams contributing to the energy per 
particle in nuclear matter. The short-distance dynamics is also treated
perturbatively, either by fine-tuning of an UV cutoff or adjusting the
parameters that appear at a given observable at a given order. This differs
from the treatment of the contact interactions in free NN scattering 
\noindent
Within this scheme, one can reproduce the empirical saturation point of
nuclear matter by adjusting (fine-tuning) one parameter related to the
short-distance dynamics. This corresponds to a novel mechanism for
nuclear matter saturation due to the repulsive contribution to the
energy per particle generated by Pauli-blocking in second order (iterated)
one-pion exchange. This can be better understood by considering the 
realistic parameterization of the energy per particle in isospin symmetric
nuclear matter,
\beq
\bar{E}(k_F) = \frac{3k_F^2}{10m} - \alpha \frac{k_F^3}{m^2} +
\beta \frac{k_F^4}{m^3}~,
\eeq
where $\alpha, \beta$ are dimensionless parameters. If $\alpha, \beta <0$,
a saturation minimum will be obtained. In the deltaless
theory, one obtains exactly such a form in the chiral limit and the parameters 
can be given in closed form. In fact, the experssion for $\beta$ is
parameter-free. Fine-tuning the 
short-distance contribution to $\alpha$, one obtains the proper binding
energy of nuclear matter. Qualitatively, this picture is not changed
when the effects related to the finite pion mass and the delta-excitation
are included. Furthermore, one obtains as a by-product a realistic value of
the nuclear matter compressibility.

This approach has been applied and extended in varoius ways. Spin-orbit 
interactions in nuclei and hyper-nuclei were considered in
\cite{Kaiser:2002yg,Kaiser:2003ux,Kaiser:2004fe,Kaiser:2008gq}. 
This led to a nice explanation of the very strong spin-orbit interactions 
in ordinary nuclei contrasted to
the remarkably weak spin-orbit splitting in $\Lambda$ hyper-nuclei. Corrections
to the in-medium chiral condensate beyond the linear density approximation
were calculated in Refs.~\cite{Kaiser:2007nv,Kaiser:2008qu}. Further, 
a systematic analysis of the nuclear energy density functional based on 
a unfication of chiral pion nuclear dynamics with strong scalar and
vector mean fields was performed and applied to the the properties of
nuclear matter and finite nuclei, see 
Refs.~\cite{Kaiser:2002jz,Finelli:2002na,Finelli:2003fk,Finelli:2007wm}.
For more details on these interesting calculations, the reader is referred
to the original articles.

\subsection{EFT for halo nuclei}

A special class of nuclear systems exhibiting universal behavior
are {\it halo nuclei}. 
A halo nucleus consists of a tightly bound core 
surrounded by one or more loosely bound valence nucleons.
The valence nucleons are characterized by a very low
separation energy compared to those in the core. 
As a consequence, the radius of the halo nucleus is large 
compared to the radius of the core.  A trivial example is the deuteron, 
which can be considered a 2-body halo nucleus. The root mean square radius of
the deuteron ${\langle r^2 \rangle}^{1/2}\approx 3$ fm is about  
four times larger than the size of the constituent nucleons. 
Halo nuclei with two valence nucleons are particularly interesting 
examples of 3-body systems.  If none of the 
2-body subsystems are bound, they are called {\it Borromean} halo nuclei. 
This name is derived from the heraldic symbol of the Borromeo family 
of Italy, which consists of three rings interlocked in such way that 
if any one of the rings is removed the other two separate.
The most
carefully studied Borromean halo nuclei are $^6$He and $^{11}$Li, 
which have two weakly bound valence neutrons \cite{Zhukov-93, Jensen-04}. 
In the case of $^6$He, the core is a $^4$He nucleus, 
which is also known as the $\alpha$ particle.
The two-neutron separation energy for $^6$He is about 1 MeV,
small compared to the binding energy of the $\alpha$ particle
which is about 28 MeV.   The neutron-$\alpha$ ($n\alpha$) system 
has no bound states and the $^6$He nucleus is therefore Borromean.
There is, however, a strong P-wave 
resonance in the $J=3/2$ channel of $n \alpha$ scattering
which is sometimes referred to as $^5$He.  This resonance is
responsible for the binding of $^6$He. Thus $^6$He
can be interpreted as a bound state of an $\alpha$-particle
and two neutrons, both of which are in 
$P_{3/2}$ configurations.

Because of the separation of scales in halo nuclei, they can be described by
extensions of the pionless EFT. 
One can assume the core to be structureless and treats 
the nucleus as a few-body system of the core and the valence nucleons.
Corrections from the structure of the core appear in higher orders and
can be included in perturbation theory. Cluster models of halo nuclei
then appear as leading order approximations in the EFT. 
A new facet is the appearance of resonances as in the neutron-alpha system
which leads to a more complicated singularity structure 
and renormalization compared to the few-nucleon system discussed above
\cite{Bertulani:2002sz}.

The first application of effective field theory methods to halo nuclei 
was carried out in Refs.~\cite{Bertulani:2002sz,Bedaque:2003wa}, where the 
$n\alpha$ system (``$^5$He'') was considered. It was found that 
for resonant P-wave interactions both the scattering length and effective 
range have to be resummed at leading order. At threshold, however, only 
one combination of coupling constants is fine-tuned and the EFT becomes
perturbative. More recent studies have 
focused on the consistent inclusion of the Coulomb interation in two-body 
halo nuclei such as the $p\alpha$ and $\alpha\alpha$ 
systems \cite{Higa:2008rx,Higa:2008dn}. 
In particular, the $\alpha\alpha$ system shows
a surprising amount of fine-tuning between the strong and electromagnetic
interaction. It can be understood in an expansion 
around the limit where, when electromagnetic interactions are turned off, 
the $^8$Be ground state is exactly at threshold and exhibits conformal 
invariance. In this scenario, the Hoyle state in $^{12}$C would appear
as a remnant of an excited Efimov state \cite{Efimov:1981aa}.
In order to better understand the modification of the Efimov spectrum
and limit cycles by long-range interactions such as the Coulomb interaction, 
a one dimensional inverse square potential supplemented with a 
Coulomb interaction was investigated in \cite{Hammer:2008ra}.
The results indicate that the counterterm required to renormalize the 
inverse square potential alone is sufficient to renormalize the full 
problem. However, the breaking of the discrete scale invariance through 
the Coulomb interaction leads to a modified bound state spectrum. The 
shallow bound states are strongly influenced by the Coulomb interaction 
while the deep bound states are dominated by the inverse square potential.

\begin{figure}[tb]
	\centering
		\includegraphics*[width=8cm,angle=0]{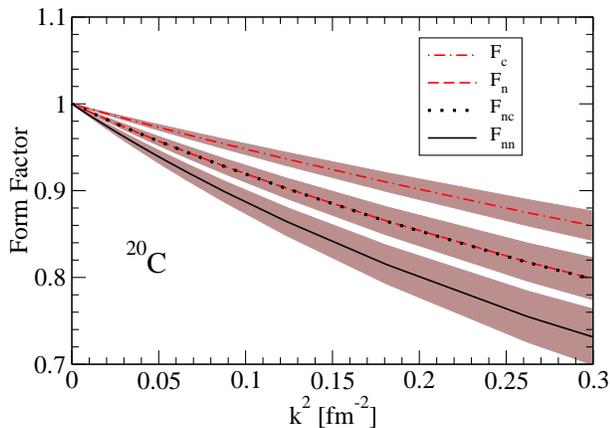}
	\caption{The various one- and two-body matter density form factors with leading order error bands 
	for the ground state of $^{20}$C in the low-energy 
	region: ${\mathcal F}_{nn}(k^2)$ black solid line; ${\mathcal F}_{nc}(k^2)$ black dotted line; 
	${\mathcal F}_{n}(k^2)$ lighter (red) dashed line; ${\mathcal F}_{c}(k^2)$ lighter (red) dot-dashed line.}
	\label{fig:FF20C}
\end{figure}

Three-body halo nuclei composed of a core and two valence neutrons
are of particular interest due to the possibility of these systems 
to display the Efimov effect \cite{Efimov-70}. Since the scattering length 
can not easily be varied in halo nuclei, one has to look for excited 
states. Such studies have previously been carried out in 
cluster models and the renormalized zero-range model
\cite{Federov:1994cf,Amorim:1997mq,Mazumdar:2000dg}.
A comprehensive study of S-wave halo nuclei in EFT including
structure calculations with error estimates was recently carried out 
in Ref.~\cite{Canham:2008jd}.
Currently, the only possible candidate for an excited Efimov state is $^{20}$C,
which consists of a core nucleus with spin and parity quantum numbers
$J^P=0^+$ and two valence neutrons.  The nucleus $^{19}$C 
is expected to have a $\frac{1}{2}^+$ state near threshold,
implying a shallow neutron-core bound state and therefore a large 
neutron-core scattering length. The value 
of the $^{19}$C energy, however, is not known well enough to 
make a definite statement about the appearance of an excited state
in $^{20}$C. The matter form factors of halo nuclei can also be calculated
in the halo EFT. As an example, we show the 
various one- and two-body matter density form factors with 
leading order error bands for the ground state of $^{20}$C at low
momentum transfers in Fig.~\ref{fig:FF20C}:
${\mathcal F}_{nn}(k^2)$, ${\mathcal F}_{nc}(k^2)$, 
${\mathcal F}_{n}(k^2)$, and ${\mathcal F}_{c}(k^2)$.
A definition of the form factors can be found in \cite{Canham:2008jd}.
The theory breaks down for momentum transfers of the order of the pion-mass 
squared ($k^2\approx 0.5$ fm$^{-2}$) where the 
one-pion exchange interaction cannot be approximated by short-range 
contact interactions anymore.
{}From the slope of the matter form factors one can extract the 
corresponding radii:
\beq
{\mathcal F}(k^2) =  1 
- {1 \over 6} k^2 \left\langle r^2 \right\rangle + \ldots\, .
\label{FFexpand}
\eeq
Information on these radii has been extracted from experiment for
some halo nuclei.
For the neutron-neutron radius of the Borromean halo nucleus $^{14}$Be
for example, the leading order halo EFT result is
$\sqrt{\langle r_{nn}^2\rangle}=4.1 \pm 0.5$~fm.
The value $\sqrt{\langle r_{nn}^2\rangle_{exp}}
=5.4 \pm 1.0$~fm was obtained from 
three-body correlations in the dissociation of $^{14}$Be using
a technique based on intensity interferometry and Dalitz plots 
\cite{Marques:2001pe}.  Within the errors there is good agreement
between both values. However, one should also keep in mind that there 
is some model dependence in the experimental result.
Results for further halo nuclei are given in Ref.~\cite{Canham:2008jd}.
A few recent studies have also investigated scattering observables.
In particular, in Refs.~\cite{Yamashita:2007ej,Mazumdar:2006tn}
the trajectory of the possible $^{20}$C excited state was extended
into the scattering region in order to find a
resonance in $n$-$^{19}$C scattering.

The simplest strange halo nucleus is the hypertriton, a 3-body
bound state of a proton, neutron, and a strange particle called the $\Lambda$. 
The total binding energy is only about 2.4 MeV. 
The hypertriton is not Borromean, because the proton-neutron subsystem 
has a bound state, the deuteron. 
The separation energy for the $\Lambda$, $E_\Lambda = 0.13 \pm 0.05$ MeV, 
is tiny compared to the binding energy $B_D = 2.224$ MeV of the 
deuteron.  The hypertriton can therefore also be considered a 2-body 
halo nucleus.
It has been studied in both 2-body and 3-body approaches
\cite{Cobis:1996ru,Fedorov:2001wj,Congl92}.
A study of the hypertriton in the halo EFT was carried out in
Ref.~\cite{Hammer:2001ng}.
An important feature of the halo EFT is the possibility to quantify 
theoretical errors through error bands. Calculations can be improved
systematically through the inclusion of higher order terms.

Another interesting application of this effective theory will be the 
study of Coulomb excitation data from existing and future facilities 
with exotic beams (such as FAIR and FRIB).
In these experiments a nuclear beam scatters 
off the Coulomb field of a heavy
nucleus. Such processes can populate excited states of the projectile
which subsequently decay, leading to its ``Coulomb dissociation''
\cite{Bert88}. Effective theories offer a systematic
framework for a full quantum-mechanical treatment of these reactions.
In summary, with new improved experimental data for these weakly bound 
nuclei, much more knowledge can be obtained about the structure of these 
interesting systems as well as discovering 
whether they show universal behavior and excited Efimov states.

\subsection{$\fet{V_{{\rm low\: k}}}$ potentials: construction and 
applications} 

\begin{figure*}[tb]
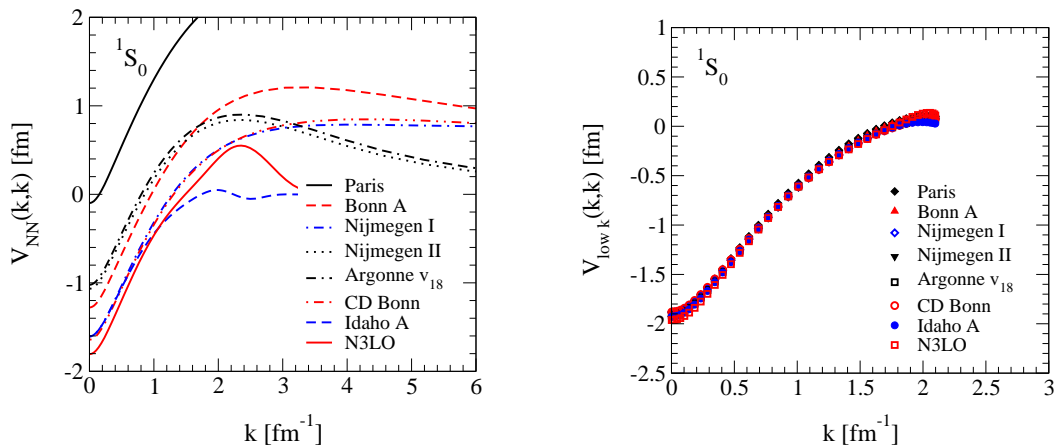

\centerline{
\includegraphics*[width=6.3cm,angle=0]{vnn_1s0_before_v2.eps}\qquad
\qquad
\includegraphics*[width=6.3cm,angle=0]{vlowk_1s0_after.eps}}
\caption{The diagonal matrix elements of the nucleon-nucleon
potential in the  $^1$S$_0$ partial wave. Left panel: Various realistic
nucleon-nucleon potentials and the chiral N$^3$LO potential by 
EM \cite{Entem:2003ft}. 
Right panel: Same potentials evolved down
to a cutoff $\Lambda = 2.2 $ fm$^{-1}$.
(Figure courtesy of A. Schwenk.)}
	\label{fig:vlowk}
\end{figure*}

Nuclear interactions, like all interactions, depend on the resolution scale.
For a momentum cutoff $\Lambda$, only details of the interactions at 
distances larger than $1/\Lambda$ can be resolved. The 
interaction potential $V$ 
consists of 2-, 3-, and higher-body terms and can be written as
\beq
V(\Lambda)= V_2(\Lambda)+V_3(\Lambda)+V_4(\Lambda)+\ldots\,.
\eeq
While $V$ depends on $\Lambda$, observables are independent of $\Lambda$.
This property can
be used to construct so-called low-momentum potentials with a 
cutoff $\Lambda' < \Lambda$ that describe
low-energy physics in terms of low-energy degrees of freedom only.
Various methods for constructing such low-momentum potentials
are known. (See, e.g. Ref.~\cite{Bogner:2003wn} for a review.)
The first construction of a low-momentum potential 
was carried out in \cite{Epelbaum:1998hg,Epelbaum:1998na} 
based on the Okubo
method \cite{Okubo:1954aa,Fukuda:1954aa}. 
Bogner and collaborators have pushed this idea further and 
constructed low-momentum potentials for various realistic nucleon-nucleon
interactions using renormalization group (RG) techniques.
They showed that these potentials all converge to the
same universal $V_{\rm low\: k}$ if the cutoff is lowered to low enough values
($\Lambda \simlt 2$ fm$^{-1}$). This is illustrated in Fig.~\ref{fig:vlowk}
for the $^1$S$_0$ partial wave.
The left panel shows various realistic
nucleon-nucleon potentials and the chiral N$^3$LO potential by 
EM \cite{Entem:2003ft}. 
The right panel shows the same potentials evolved 
down to a cutoff $\Lambda = 2.2 $ fm$^{-1}$. At this cutoff all
potentials have collapsed to the same universal curve.

As the resolution scale is lowered the
physics previously present in high-momentum modes now appears 
in many-body forces that are generated through the RG transformation.
The low-momentum potential constructed this way is phase-equivalent 
by construction. In most early calculations, the many-body forces 
generated by the RG have been neglected for simplicity. In this case
phase equivalence only holds in the 2-body subspace. Recent 
advances based on similarity RG techniques, however,
suggest that these limitations can be overcome soon 
\cite{Bogner:2006pc,Jurgenson:2008jp}.
Low-momentum potentials have also
been constructed for hyperon-nucleon interactions
\cite{Schaefer:2005fi,Dapo:2008qv}. Here the various 
realistic potentials are less constrained by data and the $V_{\rm low\: k}$
interactions show only convergence in some channels.

In Ref.~\cite{Nogga:2004ab}, the three- and four-nucleon systems
were studied using the $V_{\rm low\: k}$
potential supplemented by the leading order chiral 3N forces.
This procedure was motivated by the expectation that the many body-forces
have to reduce to the leading order chiral forces at low enough momentum.
The free parameters in the chiral 3N force were then fitted to 
experiment for each value of the cutoff. If the  chiral 3N force
is left out, the cutoff variation generates the Tjon line. Since the
RG evolution leaves the two-body observables unchanged by construction,
the variation can only go along the Tjon line.
This is in conjunction with the findings in the pionless EFT
\cite{Platter:2004zs}. For a study of potential problems with using
inconsistent 2N and 3N low-momentum potentials in scattering calculations
see Ref.~\cite{Fujii:2004dd}. 

Because of the low cutoff,
the low-momentum potential has advantages in nuclear structure
calculations where smaller model spaces are desirable because
of the computational effort involved. For a summary
of recent applications, see Ref.~\cite{Schwenk:2008su}.
As the RG evolution shifts contributions between the potential
and the integrals over intermediate states in loop integrals
which are restricted by $\Lambda$, the RG transformation can eliminate
sources of non-perturbative behavior such as strong short-range 
repulsion or tensor-forces \cite{Bogner:2005sn,Bogner:2006tw}.
This suggests that perturbative nuclear matter calculations 
are possible. At these low resolution scales, nuclear matter
saturation would be largely driven by three-body forces. Moreover,
perturbative nuclear matter calculations would also 
provide a solid basis for the construction of a universal 
density functional for nuclei with controlled errors 
\cite{Furnstahl:2007xm,Scidac:2007aa}.

\subsection{Lattice simulations of many-nucleon systems}
\label{sec:lattmany}

Nuclear matter studies utilizing lattice simulations were pioneered
by Brockmann and Frank~\cite{Brockmann:1992in}, who calculated the
quantum corrections to the Walecka model~\cite{Serot:1984ey} and
by M\"uller et al.~\cite{Muller:1999cp}, who investigated nuclear
matter properties utilizing a Hamiltonian that accomodates on-site
and next-to-neighbour parts of the central, spin- and isospin-exchange
nucleon-nucleon interactions. 
The first connection between chiral EFT and the
properties of nucleon and neutron matter using Monte-Carlo
methods was done in the groundbreaking work of Borasoy, Lee
and Sch\"afer~\cite{Lee:2004si}. They laid out the framework for
nuclear lattice simulations with chiral EFT and presented leading order
results for hot neutron matter at temperatures $T = 20 - 40\,$K and densities
below twice the nuclear matter density. Neutron matter in a periodic box 
based on the lattice  representation of the chiral NLO potential 
(see sect.~\ref{sec:fblatt}) was performed in Ref.~\cite{Borasoy:2007vk},
probing the 
density range from 2\% to 8\% of normal nuclear matter density. Dilute
neutron matter is a particularly good testing ground for  chiral EFT
applied to many-nucleon systems because of the Pauli suppression of
three-body forces. Furthermore, neutron matter at $k_F \sim 80\,$MeV,
with $k_F = (3\pi^2 N)^{1/3}/L$ (for $N$ neutrons in a box of volume $L^3$),
is close to the so-called unitary limit. In this limit, the scattering
length is infinite and the range of the interaction is zero, so that
the scattering amplitude takes its largest possible value (as given
by unitarity). In this limit,
the only dimensionful parameter describing the ground state of the 
many-fermion system is the
particle density. Thus, the ground state energy $E_0$ of the system
obeys the simple relation
\beq
E_0 = \xi \, E_0^{\rm free}~,
\eeq
where $\xi$ is a dimensionless measurable constant and $ E_0^{\rm free}$
the ground state energy of a free Fermi gas. Due to its universal nature,
the unitary limit can be studied in ultracold atomic systems like
$^6$Li or $^{40}$Ka utilizing Feshbach-resonance techniques. 
Recently measured values for $\xi$ scatter considerably and have 
sizeable error bars (for a  review, see \cite{Giorgini08}). 
There also have been numerous calculations of 
$\xi$ employing very different many-body techniques, see 
Ref.~\cite{Furnstahl:2008df} for a recent review.

\begin{figure}[tb]
\includegraphics*[width=7.5cm]{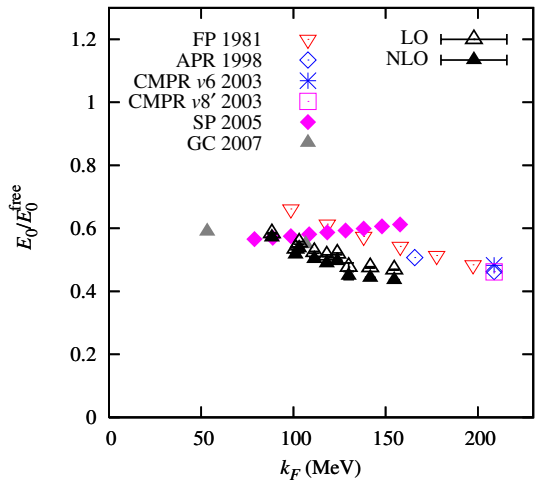}
\begin{center}
\caption{Results for $E_0/E_0^{\rm free}$ versus the Fermi
momentum $k_F$ at LO (open triangles) and NLO (filled triangles).
For comparison, we also display the results of 
FP~1981~\cite{Friedman:1981qw}, APR~1998~\cite{Akmal:1998cf},
CMPR~$v6$ and $v8^\prime$~2003~\cite{Carlson:2003wm},
SP~2005~\cite{Schwenk:2005ka} and GC~2007~\cite{Gezerlis:2007fs}.
\label{fig:xsi}}
\vspace{-0.5cm}
\end{center}
\end{figure}

Recent EFT simulations at LO and NLO with an
improved action for 8, 12 and 16 neutrons boxes of length $L = 10, 12$
and $14\,$fm are shown in Fig.~\ref{fig:xsi} \cite{EKLM}. The chiral EFT
\noindent
results are consistent with most earlier calculations based on different
methods (note the markedly different slope obtained in~\cite{Schwenk:2005ka}).
A good fit to the lattice data is obtained by (the
structure of the correction terms is discussed in detail in 
Ref.~\cite{Borasoy:2007vk})
\beq
\frac{E_0}{E_0^{{\rm free}}}\simeq \xi-\frac{\xi_1}{k_F\,a}+
0.16\,\, k_F\,r_0-(0.51{\rm fm}^3) k_F^3~,
\eeq
with  $\xi\simeq 0.31$ and $\xi_1\simeq 0.81$. This is consistent
with the Monte Carlo studies of many-fermion systems in 
Ref.~\cite{Lee:2008xsa} but smaller than the value for $\xi$ 
obtained in fixed-node Greens function Monte Carlo calculations, 
see e.g.~\cite{Carlson:2003wm}. This suggests that the upper 
bound on the ground state energy in that type of approach
might be lowered further by a more optimal fermionic nodal
surface. Clearly, such numerical simulations of many-nucleon
systems have become a valuable tool to further constrain the
nuclear equation of state at moderate densities and lead to
further insight into the physics of strongly-coupled many-body systems. 
In particular, they provide another nice link between nuclear and
atomic physics as already discussed in sec.~\ref{sec:EFTfoundations}.

\section{Summary and perspectives}
\def\theequation{\arabic{section}.\arabic{equation}}
\setcounter{equation}{0}
\label{sec:pers}

In this review, we have described the theory  
that has emerged by applying effective field theory methods to the
nuclear force problem. This method allows for a systematic derivation 
of nuclear forces with a direct connection to QCD via its symmetries.
The review focused on the derivation of the forces and their
application in the few-nucleon problem where most work has been carried
out so far. However, there are many  frontiers where future 
work is required. These include a better understanding of 
nonperturbative renormalization and improved renormalization schemes,
the consistent inclusion of electroweak currents, and the development
of a consistent EFT for the nuclear many-body problem. 
The application of new techniques and
advanced calculational methods for the many-body problem will be decisive 
to achieve the latter. Promising approaches include the renormalization
group, nuclear lattice calculations, coupled cluster approaches,
the no-core shell model, and density functional theory.
For very low-energy processes, these approaches can be complemented by 
the pionless or halo EFT which is an ideal tool to unravel universal
properties and establish connections to other fields of physics. 

\section*{Acknowledgments}

We thank all our collaborators for sharing their insights into
the topics discussed here. The work of E.E. was supported 
in parts by funds provided from the Helmholtz Association 
to the young investigator group ``Few-Nucleon Systems in 
Chiral Effective Field Theory'' (grant  VH-NG-222).
This work was further supported by the BMBF under contract No.
06BN411, by the DFG (SFB/TR 16 ``Subnuclear Structure
of Matter''), by the Helmholtz Association through funds provided 
to the virtual institute ``Spin and strong QCD'' (VH-VI-231),
and by the EU Integrated Infrastructure Initiative Hadron
Physics Project under contract number RII3-CT-2004-506078.



\begin{thebibliography}{99}


\bibitem{Peccei:2006as}
  R.~D.~Peccei,
  Lect.\ Notes Phys.\  {\bf 741}, 3 (2008).

\bibitem{Vafa:1983tf}
  C.~Vafa and E.~Witten,
  Nucl.\ Phys.\ B {\bf 234}, 173 (1984).

\bibitem{Giusti:2007cn}
  L.~Giusti and S.~Necco,
  JHEP {\bf 0704}, 090 (2007).


\bibitem{Banks:1979yr}
  T.~Banks and A.~Casher,
  Nucl.\ Phys.\ B {\bf 169}, 103 (1980).

\bibitem{Leutwyler:1992yt}
  H.~Leutwyler and A.~Smilga,
  Phys.\ Rev.\ D {\bf 46} (1992) 5607.

\bibitem{Stern:1998dy}
  J.~Stern,
  arXiv:hep-ph/9801282.

\bibitem{Goldstone:1961eq}
  J.~Goldstone,
  Nuovo Cim.\  {\bf 19}, 154 (1961).

\bibitem{Goldstone:1962es}
  J.~Goldstone, A.~Salam and S.~Weinberg,
  Phys.\ Rev.\  {\bf 127}, 965 (1962).

\bibitem{Brown:1970th}
  G.~E.~Brown,
  Comments Nucl.\ Part.\ Phys.\  {\bf 4}, 140 (1970).


\bibitem{Friar:1995dt}
  J.~L.~Friar, D.~G.~Madland and B.~W.~Lynn,
  Phys.\ Rev.\  C {\bf 53}, 3085 (1996).

\bibitem{Friar:1996zw}
  J.~L.~Friar,
  Few Body Syst.\  {\bf 22}, 161 (1997).

\bibitem{Kaiser:2006tu}
  N.~Kaiser, M.~Muhlbauer and W.~Weise,
  Eur.\ Phys.\ J.\  A {\bf 31}, 53 (2007).

\bibitem{Delfino:2007zu}
  A.~Delfino, T.~Frederico, V.~S.~Timoteo and L.~Tomio,
  Phys.\ Lett.\  B {\bf 634}, 185 (2006).


\bibitem{Yukawa:1935xg}
  H.~Yukawa,
  Proc.\ Phys.\ Math.\ Soc.\ Jap.\  {\bf 17}, 48 (1935).

\bibitem{Cottingham:1973wt}
  W.~N.~Cottingham, M.~Lacombe, B.~Loiseau, J.~M.~Richard and R.~Vinh Mau,
  Phys.\ Rev.\  D {\bf 8}, 800 (1973).

\bibitem{Jackson:1975be}
  A.~D.~Jackson, D.~O.~Riska and B.~Verwest,
  Nucl.\ Phys.\  A {\bf 249}, 397 (1975).


\bibitem{Machleidt:2001rw} 
  R.~Machleidt and I.~Slaus, 
  J.\ Phys.\ G {\bf 27}, R69 (2001).

\bibitem{Machleidt:2000ge}
  R.~Machleidt,
  Phys.\ Rev.\  C {\bf 63}, 024001 (2001).

\bibitem{Stoks:1994wp}
  V.~G.~J.~Stoks, R.~A.~M.~Klomp, C.~P.~F.~Terheggen and J.~J.~de Swart,
  Phys.\ Rev.\  C {\bf 49}, 2950 (1994).

\bibitem{Wiringa:1994wb}
  R.~B.~Wiringa, V.~G.~J.~Stoks and R.~Schiavilla,
  Phys.\ Rev.\  C {\bf 51}, 38 (1995).

\bibitem{Pieper:2001mp}
  S.~C.~Pieper and R.~B.~Wiringa,
  Ann.\ Rev.\ Nucl.\ Part.\ Sci.\  {\bf 51}, 53 (2001).

\bibitem{Gloeckle:1995jg}
  W.~Gl\"ockle, H.~Wita{\l}a, D.~Huber, H.~Kamada and J.~Golak,
  Phys.\ Rept.\  {\bf 274}, 107 (1996).

\bibitem{Fujita:1957zz}
  J.~Fujita and H.~Miyazawa,
  Prog.\ Theor.\ Phys.\  {\bf 17}, 360 (1957).

\bibitem{TM1}
B.~H.~J.~McKellar and R.~Rajaraman,
Phys. Rev. Lett. {\bf 21}, 450 (1968).

\bibitem{Coon:1974vc}
  S.~A.~Coon, M.~D.~Scadron and B.~R.~Barrett,
  Nucl.\ Phys.\  A {\bf 242}, 467 (1975).

\bibitem{Coelho:1984hk}
  H.~T.~Coelho, T.~K.~Das and M.~R.~Robilotta,
  Phys.\ Rev.\  C {\bf 28}, 1812 (1983).

\bibitem{Pudliner:1997ck}
  B.~S.~Pudliner, V.~R.~Pandharipande, J.~Carlson, S.~C.~Pieper and R.~B.~Wiringa,
  Phys.\ Rev.\  C {\bf 56}, 1720 (1997).

\bibitem{Pieper:2001ap}
  S.~C.~Pieper, V.~R.~Pandharipande, R.~B.~Wiringa and J.~Carlson,
  Phys.\ Rev.\  C {\bf 64}, 014001 (2001).


\bibitem{Georgi:1994qn}
  H.~Georgi,
  Ann.\ Rev.\ Nucl.\ Part.\ Sci.\  {\bf 43}, 209 (1993).

\bibitem{Manohar:1996cq}
  A.~V.~Manohar,
  arXiv:hep-ph/9606222.

\bibitem{Burgess:2007pt}
  C.~P.~Burgess,
  Ann.\ Rev.\ Nucl.\ Part.\ Sci.\  {\bf 57}, 329 (2007).

\bibitem{Wilson-83}
K.G.~Wilson,
Rev.\ Mod.\ Phys.\ {\bf 55}, 583 (1983).

\bibitem{Lepage-89}
G.P.~Lepage, ``What is Renormalization?,''
in {\it From Actions to Answers} (TASI-89), edited by
        T.~DeGrand and D.~Toussaint (World Scientific, Singapore, 1989)
[arXiv:hep-ph/0506330].


\bibitem{Weinberg:1978kz} 
  S.~Weinberg, 
  Physica A {\bf 96}, 327 (1979). 

\bibitem{Gasser:1983yg} 
  J.~Gasser and H.~Leutwyler, 
  Annals Phys.\  {\bf 158}, 142 (1984). 
 
\bibitem{Gasser:1984gg} 
  J.~Gasser and H.~Leutwyler, 
  Nucl.\ Phys.\  B {\bf 250}, 465 (1985). 
 
\bibitem{Bernard:2006gx}
  V.~Bernard and U.-G.~Mei{\ss}ner, 
  Ann.\ Rev.\ Nucl.\ Part.\ Sci.\  {\bf 57}, 33 (2007).

\bibitem{Bernard:2007zu}
  V.~Bernard,
  Prog.\ Part.\ Nucl.\ Phys.\  {\bf 60}, 82 (2008).

\bibitem{Kaplan:1998tg}
  D.~B.~Kaplan, M.~J.~Savage and M.~B.~Wise,
  Phys.\ Lett.\  B {\bf 424}, 390 (1998).

\bibitem{Kaplan:1998we}
  D.~B.~Kaplan, M.~J.~Savage and M.~B.~Wise,
  Nucl.\ Phys.\  B {\bf 534}, 329 (1998).

\bibitem{Braaten:2004rn} 
  E.~Braaten and H.-W.~Hammer, 
  Phys.\ Rept.\  {\bf 428}, 259 (2006).


\bibitem{Fukugita:1994na}
  M.~Fukugita, Y.~Kuramashi, H.~Mino, M.~Okawa and A.~Ukawa,
  Phys.\ Rev.\ Lett.\  {\bf 73}, 2176 (1994).

\bibitem{Fukugita:1994ve}
  M.~Fukugita, Y.~Kuramashi, M.~Okawa, H.~Mino and A.~Ukawa,
  Phys.\ Rev.\  D {\bf 52}, 3003 (1995).

\bibitem{Hamber:1983vu}
  H.~W.~Hamber, E.~Marinari, G.~Parisi and C.~Rebbi,
  Nucl.\ Phys.\  B {\bf 225}, 475 (1983).

\bibitem{Luscher:1986pf}
  M.~L\"uscher,
  Commun.\ Math.\ Phys.\  {\bf 105}, 153 (1986).

\bibitem{Luscher:1990ux}
  M.~L\"uscher,
  Nucl.\ Phys.\  B {\bf 354}, 531 (1991).


\bibitem{Beane:2003da}
  S.~R.~Beane, P.~F.~Bedaque, A.~Parreno and M.~J.~Savage,
  Phys.\ Lett.\  B {\bf 585}, 106 (2004).

\bibitem{Bedaque:2002mn}
  P.~F.~Bedaque and U.~van Kolck,
  Ann.\ Rev.\ Nucl.\ Part.\ Sci.\  {\bf 52}, 339 (2002).

\bibitem{Beane:2006mx}
  S.~R.~Beane, P.~F.~Bedaque, K.~Orginos and M.~J.~Savage,
  Phys.\ Rev.\ Lett.\  {\bf 97}, 012001 (2006).

\bibitem{Beane:2003yx}
  S.~R.~Beane, P.~F.~Bedaque, A.~Parreno and M.~J.~Savage,
  Nucl.\ Phys.\  A {\bf 747}, 55 (2005).

\bibitem{Beane:2006gf}
  S.~R.~Beane, P.~F.~Bedaque, T.~C.~Luu, K.~Orginos, E.~Pallante, A.~Parreno and M.~J.~Savage
                  [NPLQCD Collaboration],
  Nucl.\ Phys.\  A {\bf 794}, 62 (2007).

\bibitem{Beane:2008dv}
  S.~R.~Beane, K.~Orginos and M.~J.~Savage,
  Int.\ J.\ Mod.\ Phys.\ E {\bf 17}, 1157 (2008).


\bibitem{Ishii:2006ec}
  N.~Ishii, S.~Aoki and T.~Hatsuda,
  Phys.\ Rev.\ Lett.\  {\bf 99}, 022001 (2007).

\bibitem{Aoki:2005uf}
  S.~Aoki {\it et al.}  [CP-PACS Collaboration],
  Phys.\ Rev.\  D {\bf 71}, 094504 (2005).

\bibitem{Ishii:2007xz}
  N.~Ishii, S.~Aoki and T.~Hatsuda,
  PoS {\bf LAT2007}, 146 (2007).

\bibitem{Wilczek:2007}
F.~Wilczek, Nature {\bf 445}, 156 (2007).

\bibitem{Nemura:2008sp}
  H.~Nemura, N.~Ishii, S.~Aoki and T.~Hatsuda,
  arXiv:0806.1094 [nucl-th].

\bibitem{Latt08conf}
Lattice 2008 conference website:\\ http://conferences.jlab.org/lattice2008/\,.

\bibitem{SRIVASTAVA75}M.K.\ Srivastava and D.W.L.\ Sprung,
  Adv.\ Nucl.\ Phys.\ {\bf 8}, 121 (1975).
 
\bibitem{Haag58} 
R. Haag, Phys. Rev. {\bf 112}, 669 (1958). 
 

\bibitem{COLEMAN69}S.~Coleman, J.~Wess, and B.~Zumino, Phys.\ Rev.\
    {\bf 177}, 2239 (1969).

\bibitem{POLITZER80}H.D. Politzer, Nucl.\ Phys.\ B {\bf 172}, 349 (1980).

\bibitem{GEORGI91}H.~Georgi, Nucl.\ Phys.\ B {\bf 361}, 339 (1991).

\bibitem{KILIAN94}W.~Kilian and T.~Ohl, Phys.\ Rev.\ D {\bf 50}, 4649 (1994).

\bibitem{SCHERER95a}S. Scherer and H.W. Fearing, Phys.\ Rev.\ D {\bf 52},
6445 (1995).

\bibitem{ARZT95}C.~Arzt, Phys.\ Lett.\ B {\bf 342}, 189 (1995).

\bibitem{FEARING98}H.W.\ Fearing, Phys.\ Rev.\ Lett.\ {\bf 81}, 758 (1998).

\bibitem{FEARING99}H.W.\ Fearing and S.~Scherer, Phys.\ Rev.\ C {\bf 62},
034003 (2000).

\bibitem{SCHERER95b}S.\ Scherer and H.W.\ Fearing, Phys.\ Rev.\ C {\bf 51},
359 (1995).

\bibitem{FHK99}J.L.\ Friar, D.\ H{\"u}ber, and U.\ van Kolck,
      Phys.\ Rev.\ C {\bf 59}, 53 (1999).

\bibitem{CFM96}T.D.\ Cohen, J.L.\ Friar, G.A.\ Miller, and U.\ van Kolck,
      Phys.\ Rev.\ C {\bf 53}, 2661 (1996).

%
\bibitem{KSW99}D.~B.\ Kaplan, M.~J.\ Savage, and M.~B.\ Wise,
        Phys.\ Rev.\ C {\bf 59}, 617 (1999).

\bibitem{ARNAN73}I.R.\ Afnan and F.J.D.\ Serduke, Phys.\ Lett.\ {\bf 44B},
143 (1973).
%
\bibitem{POLYZOU90}W.N.\ Polyzou and W.~Gl\"ockle, Few Body Syst.\
 {\bf 9}, 97 (1990).

\bibitem{AMGHAR95}A. Amghar and B.~Desplanques,
 Nucl.\ Phys.\ A {\bf 585}, 657 (1995).
%

\bibitem{Furnstahl:2000we} 
  R.~J.~Furnstahl, H.-W.~Hammer and N.~Tirfessa, 
  Nucl.\ Phys.\  A {\bf 689}, 846 (2001). 


\bibitem{Hammer:2000xg}
  H.-W.~Hammer and R.~J.~Furnstahl,
  Nucl.\ Phys.\  A {\bf 678}, 277 (2000).

\bibitem{COESTER70}F.~Coester, S.~Cohen, B.~Day, and C.M.\ Vincent,
    Phys.\ Rev.\ C {\bf 1}, 769 (1970).


\bibitem{Furnstahl:2001xq}
  R.~J.~Furnstahl and H.-W.~Hammer,
  Phys.\ Lett.\  B {\bf 531}, 203 (2002).

\bibitem{Jaminon:1990zz}
  M.~Jaminon and C.~Mahaux,
  Phys.\ Rev.\  C {\bf 41}, 697 (1990).






\bibitem{Mohr:2005pv}
  R.~F.~Mohr, R.~J.~Furnstahl, R.~J.~Perry, K.~G.~Wilson and H.-W.~Hammer,
  Annals Phys.\  {\bf 321}, 225 (2006).

\bibitem{Braaten:2003eu} 
  E.~Braaten and H.-W.~Hammer, 
  Phys.\ Rev.\ Lett.\  {\bf 91}, 102002 (2003). 
 
\bibitem{Efimov-70}
V. Efimov, 
Phy. Lett. {\bf 33B} (1970) 563.

\bibitem{Braaten:2006vd}
  E.~Braaten and H.-W.~Hammer,
  Annals Phys.\  {\bf 322}, 120 (2007).

\bibitem{Gegelia:1998xr}
  J.~Gegelia,
  arXiv:nucl-th/9802038.

\bibitem{vanKolck:1998bw}
  U.~van Kolck,
  Nucl.\ Phys.\  A {\bf 645}, 273 (1999).


\bibitem{Beane:2000fx}
  S.~R.~Beane, P.~F.~Bedaque, W.~C.~Haxton, D.~R.~Phillips and M.~J.~Savage,
  arXiv:nucl-th/0008064.


\bibitem{Christlmeier:2008ye}
  S.~Christlmeier and H.~W.~Grie{\ss}hammer,
  Phys.\ Rev.\  C {\bf 77}, 064001 (2008).

\bibitem{Ryezayeva:2008zz}
  N.~Ryezayeva {\it et al.},
  Phys.\ Rev.\ Lett.\  {\bf 100}, 172501 (2008).

\bibitem{Ando:2007in}
  S.~I.~Ando,
  Eur.\ Phys.\ J.\  A {\bf 33}, 185 (2007).


\bibitem{Kaplan:1996nv}
  D.~B.~Kaplan,
  Nucl.\ Phys.\  B {\bf 494}, 471 (1997).


\bibitem{Bedaque:1999ve}
  P.~F.~Bedaque, H.-W.~Hammer and U.~van Kolck,
  Nucl.\ Phys.\  A {\bf 676}, 357 (2000).

\bibitem{Phillips:1999hh}
  D.~R.~Phillips, G.~Rupak and M.~J.~Savage,
  Phys.\ Lett.\  B {\bf 473}, 209 (2000).

\bibitem{Skorniakov:1957aa}
G.V.~Skorniakov and K.A.~Ter-Martirosian,
Sov.\ Phys.\ JETP {\bf 4}, 648 (1957) 
        [J.\ Exptl.\ Theoret.\ Phys.\ (U.S.S.R.) {\bf 31}, 775 (1956)].

\bibitem{Danilov:1961aa}
G.S.~Danilov, 
Sov.\ Phys.\ JETP {\bf 13}, 349 (1961)
        [J.\ Exptl.\ Theoret.\ Phys.\ (U.S.S.R.) {\bf 40}, 498 (1961)].


\bibitem{Bedaque:1998kg}
  P.~F.~Bedaque, H.-W.~Hammer and U.~van Kolck,
  Phys.\ Rev.\ Lett.\  {\bf 82}, 463 (1999).

\bibitem{Bedaque:1998km}
  P.~F.~Bedaque, H.-W.~Hammer and U.~van Kolck,
  Nucl.\ Phys.\  A {\bf 646}, 444 (1999).

\bibitem{Kharchenko:1973aa}
V.F.~Kharchenko,
Sov.\ J.\ Nucl.\ Phys.\ {\bf 16}, 173 (1973)
        [Yad.\ Fiz.\ {\bf 16}, 310 (1972)].


\bibitem{Hammer:2000nf}
  H.-W.~Hammer and T.~Mehen,
  Nucl.\ Phys.\  A {\bf 690}, 535 (2001).

\bibitem{Afnan:2003bs}
  I.~R.~Afnan and D.~R.~Phillips,
  Phys.\ Rev.\  C {\bf 69}, 034010 (2004).

\bibitem{Platter:2008cx}
  L.~Platter, C.~Ji and D.~R.~Phillips,
  arXiv:0808.1230 [cond-mat.other].

\bibitem{Bedaque:2002yg}
  P.~F.~Bedaque, G.~Rupak, H.~W.~Grie{\ss}hammer and H.-W.~Hammer,
  Nucl.\ Phys.\  A {\bf 714}, 589 (2003).

\bibitem{Bedaque:1997qi}
  P.~F.~Bedaque and U.~van Kolck,
  Phys.\ Lett.\  B {\bf 428}, 221 (1998).

\bibitem{Bedaque:1998mb}
  P.~F.~Bedaque, H.-W.~Hammer and U.~van Kolck,
  Phys.\ Rev.\  C {\bf 58}, 641 (1998).


\bibitem{Phillips68}
A.C.~Phillips, 
 Nucl.\ Phys.\ A {\bf 107}, 209 (1968).


\bibitem {Efimov:1991aa}
V.~Efimov, 
Phys.\ Rev.\ C {\bf 44}, 2303 (1991).


\bibitem{Griesshammer:2004pe}
  H.~W.~Grie{\ss}hammer,
  Nucl.\ Phys.\  A {\bf 744}, 192 (2004).

\bibitem{Platter:2006ev}
  L.~Platter and D.~R.~Phillips,
  Few Body Syst.\  {\bf 40}, 35 (2006).

\bibitem{Platter:2006ad}
  L.~Platter,
  Phys.\ Rev.\  C {\bf 74}, 037001 (2006).


\bibitem{Gabbiani:1999yv}
  F.~Gabbiani, P.~F.~Bedaque and H.~W.~Grie{\ss}hammer,
  Nucl.\ Phys.\  A {\bf 675}, 601 (2000).

\bibitem{Griesshammer:2005ga}
  H.~W.~Grie{\ss}hammer,
  Nucl.\ Phys.\  A {\bf 760}, 110 (2005).


\bibitem{Barford:2004fz}
  T.~Barford and M.~C.~Birse,
  J.\ Phys.\ A  {\bf 38}, 697 (2005).

\bibitem{Birse:2008wt}
  M.~C.~Birse,
  Phys.\ Rev.\  C {\bf 77}, 047001 (2008).

\bibitem{Ando:2008jb}
  S.~I.~Ando and M.~C.~Birse,
  Phys.\ Rev.\  C {\bf 78}, 024004 (2008).


\bibitem{Efimov:1981aa}
V.~Efimov,
Nucl.\ Phys.\ A {\bf 362}, 45 (1981); 
Erratum ibid. {\bf 378}, 581(E) (1982); 

\bibitem{Platter:2005sj}
  L.~Platter and H.-W.~Hammer,
  Nucl.\ Phys.\  A {\bf 766}, 132 (2006).

\bibitem{Sadeghi:2006aa}
  H.~Sadeghi, S.~Bayegan and H.~W.~Grie{\ss}hammer,
  Phys.\ Lett.\ B {\bf 643}, 263 (2006).

\bibitem{Sadeghi:2005aa}
  H.~Sadeghi and S.~Bayegan,
 Nucl.\ Phys.\ A {\bf 753}, 291 (2005). 


\bibitem{Platter:2004qn}
  L.~Platter, H.-W.~Hammer and U.-G.~Mei\ss ner,
  Phys.\ Rev.\  A {\bf 70}, 052101 (2004).

\bibitem{Platter:2004zs}
  L.~Platter, H.-W.~Hammer and U.-G.~Mei\ss ner,
  Phys.\ Lett.\  B {\bf 607}, 254 (2005).


\bibitem{Nogga:2000uu}
  A.~Nogga, H.~Kamada and W.~Gl\"ockle,
  Phys.\ Rev.\ Lett.\  {\bf 85}, 944 (2000).


\bibitem{Epelbaum:2000mx}
  E.~Epelbaum, H.~Kamada, A.~Nogga, H.~Wita{\l}a, W.~Gl\"ockle and U.-G.~Mei\ss ner,
  Phys.\ Rev.\ Lett.\  {\bf 86}, 4787 (2001).

\bibitem{Epelbaum:2002vt} 
  E.~Epelbaum, A.~Nogga, W.~Gl\"ockle, H.~Kamada, U.-G.~Mei{\ss}ner and H.~Wita{\l}a, 
  Phys.\ Rev.\  C {\bf 66}, 064001 (2002).

\bibitem{Nogga:2004ab}
  A.~Nogga, S.~K.~Bogner and A.~Schwenk,
  Phys.\ Rev.\  C {\bf 70}, 061002 (2004).

\bibitem{Hammer:2006ct}
  H.-W.~Hammer and L.~Platter,
  Eur.\ Phys.\ J.\  A {\bf 32}, 113 (2007).

\bibitem{Yama06}
M.T.~Yamashita, L.~Tomio, A.~Delfino, T.~Frederico,
Europhys.\ Lett.\ {\bf 75}, 555 (2006).

\bibitem{vStech08}
J.~von Stecher, J.P.~D'Incao, C.H.~Greene
arXiv:0810.3876.


\bibitem{Stetcu:2006ey}
  I.~Stetcu, B.~R.~Barrett and U.~van Kolck,
  Phys.\ Lett.\  B {\bf 653}, 358 (2007).

\bibitem{Stetcu:2007ms}
  I.~Stetcu, B.~R.~Barrett, U.~van Kolck and J.~P.~Vary,
  Phys.\ Rev.\  A {\bf 76}, 063613 (2007).


\bibitem{Weinberg:1990rz}
  S.~Weinberg,
  Phys.\ Lett.\  B {\bf 251}, 288 (1990).
 
\bibitem{Weinberg:1991um}
  S.~Weinberg,
  Nucl.\ Phys.\  B {\bf 363}, 3 (1991).


\bibitem{Cohen:1998jr}
  T.~D.~Cohen and J.~M.~Hansen,
  Phys.\ Rev.\  C {\bf 59}, 13 (1999).

\bibitem{Cohen:1999iaa}
  T.~D.~Cohen and J.~M.~Hansen,
  Phys.\ Rev.\  C {\bf 59}, 3047 (1999).

\bibitem{Fleming:1999ee}
  S.~Fleming, T.~Mehen and I.~W.~Stewart,
  Nucl.\ Phys.\  A {\bf 677}, 313 (2000).

\bibitem{Lutz:1999yr}
  M.~Lutz,
  Nucl.\ Phys.\  A {\bf 677}, 241 (2000).

\bibitem{Oller:2003px}
  J.~A.~Oller,
  Nucl.\ Phys.\  A {\bf 725}, 85 (2003).

\bibitem{Soto:2007pg}
  J.~Soto and J.~Tarrus,
  Phys.\ Rev.\  C {\bf 78}, 024003 (2008).

\bibitem{Mondejar:2006yu}
  J.~Mondejar and J.~Soto,
  Eur.\ Phys.\ J.\  A {\bf 32}, 77 (2007).

\bibitem{Nogga:2005hy}
  A.~Nogga, R.~G.~E.~Timmermans and U.~van Kolck,
  Phys.\ Rev.\  C {\bf 72}, 054006 (2005).

\bibitem{Birse:2005um}
  M.~C.~Birse,
  Phys.\ Rev.\  C {\bf 74}, 014003 (2006).

\bibitem{Phillips:1959aa}
R.~J.~N. Phillips,
\newblock Reports on Progress in Physics {\bf XXII}, 562 (1959).


\bibitem{Kaiser:1997mw}
  N.~Kaiser, R.~Brockmann and W.~Weise,
  Nucl.\ Phys.\  A {\bf 625}, 758 (1997).

\bibitem{Epelbaum:1998ka}
  E.~Epelbaum, W.~Gl\"ockle and U.-G.~Mei\ss ner,
  Nucl.\ Phys.\  A {\bf 637}, 107 (1998).

\bibitem{Jenkins:1990jv}
  E.~E.~Jenkins and A.~V.~Manohar,
  Phys.\ Lett.\  B {\bf 255} (1991) 558.

\bibitem{Bernard:1992qa}
  V.~Bernard, N.~Kaiser, J.~Kambor and U.-G.~Mei\ss ner,
  Nucl.\ Phys.\  B {\bf 388}, 315 (1992).

\bibitem{Fettes:2000gb}
  N.~Fettes, U.-G.~Mei\ss ner, M.~Mojzis and S.~Steininger,
  Annals Phys.\  {\bf 283}, 273 (2000)
  [Erratum-ibid.\  {\bf 288}, 249 (2001)].

\bibitem{Gasser:1987rb}
  J.~Gasser, M.~E.~Sainio and A.~Svarc,
  Nucl.\ Phys.\  B {\bf 307}, 779 (1988).

\bibitem{Epelbaum:2000kv}
E.~Epelbaum,
\newblock {\em The nucleon nucleon interaction in a chiral effective field
  theory},
\newblock PhD thesis, Ruhr-University Bochum, Germany, 2000,
\newblock JUL-3803.

\bibitem{Ordonez:1993tn}
  C.~Ordonez, L.~Ray and U.~van Kolck,
  Phys.\ Rev.\ Lett.\  {\bf 72}, 1982 (1994).

\bibitem{Ordonez:1995rz}
  C.~Ordonez, L.~Ray and U.~van Kolck,
  Phys.\ Rev.\  C {\bf 53}, 2086 (1996).

\bibitem{Friar:1994zz}
  J.~L.~Friar and S.~A.~Coon,
  Phys.\ Rev.\  C {\bf 49}, 1272 (1994).

\bibitem{Friar:1977xh}
  J.~L.~Friar,
  Annals Phys.\  {\bf 104}, 380 (1977).


\bibitem{Epelbaum:1999dj}
  E.~Epelbaum, W.~Gl\"ockle and U.-G.~Mei\ss ner,
  Nucl.\ Phys.\  A {\bf 671}, 295 (2000).

\bibitem{Epelbaum:2002gb}
  E.~Epelbaum, U.-G.~Mei\ss ner and W.~Gl\"ockle,
  Nucl.\ Phys.\  A {\bf 714}, 535 (2003).

\bibitem{Epelbaum:2003gr}
  E.~Epelbaum, W.~Gl\"ockle and U.-G.~Mei\ss ner,
  Eur.\ Phys.\ J.\  A {\bf 19}, 125 (2004).

\bibitem{Epelbaum:2003xx}
  E.~Epelbaum, W.~Gl\"ockle and U.-G.~Mei\ss ner,
  Eur.\ Phys.\ J.\  A {\bf 19}, 401 (2004).


\bibitem{Ordonez:1992xp}
  C.~Ordonez and U.~van Kolck,
  Phys.\ Lett.\  B {\bf 291}, 459 (1992).

\bibitem{Coon:1986kq}
  S.~A.~Coon and J.~L.~Friar,
  Phys.\ Rev.\  C {\bf 34}, 1060 (1986).

\bibitem{Eden:1996ey}
  J.~A.~Eden and M.~F.~Gari,
  Phys.\ Rev.\  C {\bf 53}, 1510 (1996).


\bibitem{Bernard:1996gq}
  V.~Bernard, N.~Kaiser and U.-G.~Mei\ss ner,
  Nucl.\ Phys.\  A {\bf 615}, 483 (1997).

\bibitem{Robilotta:2000py}
  M.~R.~Robilotta,
  Phys.\ Rev.\  C {\bf 63}, 044004 (2001).

\bibitem{Becher:1999he}
  T.~Becher and H.~Leutwyler,
  Eur.\ Phys.\ J.\  C {\bf 9}, 643 (1999).

\bibitem{Epelbaum:2005pn}
  E.~Epelbaum,
  Prog.\ Part.\ Nucl.\ Phys.\  {\bf 57}, 654 (2006).


\bibitem{Rentmeester:1999vw}
  M.~C.~M.~Rentmeester, R.~G.~E.~Timmermans, J.~L.~Friar and J.~J.~de Swart,
  Phys.\ Rev.\ Lett.\  {\bf 82}, 4992 (1999).

\bibitem{Rentmeester:2003mf}
  M.~C.~M.~Rentmeester, R.~G.~E.~Timmermans and J.~J.~de Swart,
  Phys.\ Rev.\  C {\bf 67}, 044001 (2003).

\bibitem{Entem:2003cs}
  D.~R.~Entem and R.~Machleidt,
  arXiv:nucl-th/0303017.

\bibitem{Birse:2003nz}
  M.~C.~Birse and J.~A.~McGovern,
  Phys.\ Rev.\  C {\bf 70}, 054002 (2004).

\bibitem{Birse:2007sx}
  M.~C.~Birse,
  Phys.\ Rev.\  C {\bf 76}, 034002 (2007).

\bibitem{vanKolck:1994yi}
  U.~van Kolck,
  Phys.\ Rev.\  C {\bf 49}, 2932 (1994).

\bibitem{Friar:1999sj}
  J.~L.~Friar,
  Phys.\ Rev.\  C {\bf 60}, 034002 (1999).


\bibitem{Kaiser:2001pc}
  N.~Kaiser,
  Phys.\ Rev.\  C {\bf 64}, 057001 (2001).

\bibitem{Entem:2002sf}
  D.~R.~Entem and R.~Machleidt,
  Phys.\ Rev.\  C {\bf 66}, 014002 (2002).

\bibitem{Kaiser:2001at}
  N.~Kaiser,
  Phys.\ Rev.\  C {\bf 65}, 017001 (2002).

\bibitem{Higa:2003jk}
  R.~Higa and M.~R.~Robilotta,
  Phys.\ Rev.\  C {\bf 68}, 024004 (2003).

\bibitem{Higa:2003sz}
  R.~Higa, M.~R.~Robilotta and C.~A.~da Rocha,
  Phys.\ Rev.\  C {\bf 69}, 034009 (2004).

\bibitem{Higa:2005ip}
  R.~Higa, M.~R.~Robilotta and C.~A.~da Rocha,
  arXiv:nucl-th/0501076.

\bibitem{Kaiser:1999ff}
  N.~Kaiser,
  Phys.\ Rev.\  C {\bf 61} (2000) 014003.

\bibitem{Kaiser:1999jg}
  N.~Kaiser,
  Phys.\ Rev.\  C {\bf 62} (2000) 024001.

\bibitem{Pupin:1999ba}
  J.~C.~Pupin and M.~R.~Robilotta,
  Phys.\ Rev.\  C {\bf 60}, 014003 (1999).

\bibitem{Kaiser:2001dm}
  N.~Kaiser,
  Phys.\ Rev.\  C {\bf 63}, 044010 (2001).

\bibitem{Bernard:2007sp}
  V.~Bernard, E.~Epelbaum, H.~Krebs and U.-G.~Mei\ss ner,
  Phys.\ Rev.\  C {\bf 77}, 064004 (2008).

\bibitem{Ishikawa:2007zz}
  S.~Ishikawa and M.~R.~Robilotta,
  Phys.\ Rev.\  C {\bf 76}, 014006 (2007).


\bibitem{Bernard:1995gx}
  V.~Bernard, N.~Kaiser and U.-G.~Mei\ss ner,
  Nucl.\ Phys.\  B {\bf 457}, 147 (1995).

\bibitem{Fettes:1998ud}
  N.~Fettes, U.-G.~Mei\ss ner and S.~Steininger,
  Nucl.\ Phys.\  A {\bf 640}, 199 (1998).

\bibitem{Buettiker:1999ap}
  P.~Buettiker and U.-G.~Mei\ss ner,
  Nucl.\ Phys.\  A {\bf 668}, 97 (2000).

\bibitem{Epelbaum:2007us}
  E.~Epelbaum,
  Eur.\ Phys.\ J.\  A {\bf 34}, 197 (2007).

\bibitem{Belitsky:2002ni}
  A.~V.~Belitsky and T.~D.~Cohen,
  Phys.\ Rev.\  C {\bf 65}, 064008 (2002).

\bibitem{Cohen:2002qn}
  T.~D.~Cohen and B.~A.~Gelman,
  Phys.\ Lett.\  B {\bf 540}, 227 (2002).

\bibitem{Cohen:2002im}
  T.~D.~Cohen,
  Phys.\ Rev.\  C {\bf 66}, 064003 (2002).


\bibitem{Fujita:1962}
  J.-I.~Fujita, M.~Kawai and M.~Tanifuji,
  Nucl.\ Phys.\  {\bf 29}, 252 (1962).

\bibitem{Robilotta:2006xq}
  M.~R.~Robilotta,
  Phys.\ Rev.\  C {\bf 74}, 044002 (2006)
  [Erratum-ibid.\  C {\bf 74}, 059902 (2006)].

\bibitem{Epelbaum:2006eu}
  E.~Epelbaum,
  Phys.\ Lett.\  B {\bf 639}, 456 (2006).

\bibitem{Mcmanus:1980ep}
  H.~McManus and D.~O.~Riska,
  Phys.\ Lett.\  B {\bf 92}, 29 (1980).

\bibitem{Robilotta:1985gv}
  M.~R.~Robilotta,
  Phys.\ Rev.\  C {\bf 31}, 974 (1985).

\bibitem{Rozpedzik:2006yi}
  D.~Rozpedzik {\it et al.},
  Acta Phys.\ Polon.\  B {\bf 37}, 2889 (2006).

\bibitem{Miller:1990iz}
  G.~A.~Miller, B.~M.~K.~Nefkens and I.~Slaus,
  Phys.\ Rept.\  {\bf 194}, 1 (1990).

\bibitem{Miller:1994zh}
  G.~A.~Miller and W.~T.~H.~Van Oers,
  arXiv:nucl-th/9409013.


\bibitem{Howell:1998aa}
  C.~R.~Howell, et al.,
  %
  Phys.\ Lett.\  B {\bf 444}, 252 (1998).

\bibitem{GonzalezTrotter:1999zz}
  D.~E.~Gonzalez Trotter {\it et al.},
  Phys.\ Rev.\ Lett.\  {\bf 83}, 3788 (1999).

\bibitem{Huhn:2000aa}
  V.~Huhn {\it et al.},
  %
  Phys.\ Rev.\ Lett.\  {\bf 85}, 1190 (2000).

\bibitem{Howell:2008dt}
  C.~R.~Howell,
  arXiv:0805.1177 [nucl-ex].

\bibitem{Leutwyler:1996qg}
  H.~Leutwyler,
  Phys.\ Lett.\  B {\bf 378}, 313 (1996).


\bibitem{Urech:1994hd}
  R.~Urech,
  Nucl.\ Phys.\  B {\bf 433}, 234 (1995).

\bibitem{Neufeld:1995mu}
  H.~Neufeld and H.~Rupertsberger,
  Z.\ Phys.\  C {\bf 71}, 131 (1996).

\bibitem{Meissner:1997fa}
  U.-G.~Mei\ss ner, G.~M\"uller and S.~Steininger,
  Phys.\ Lett.\  B {\bf 406}, 154 (1997)
  [Erratum-ibid.\  B {\bf 407}, 454 (1997)].

\bibitem{Knecht:1997jw}
  M.~Knecht and R.~Urech,
  Nucl.\ Phys.\  B {\bf 519}, 329 (1998).

\bibitem{Meissner:1997ii}
  U.-G.~Mei\ss ner and S.~Steininger,
  Phys.\ Lett.\  B {\bf 419}, 403 (1998).

\bibitem{Muller:1999ww}
  G.~M\"uller and U.-G.~Mei\ss ner,
  Nucl.\ Phys.\  B {\bf 556}, 265 (1999).

\bibitem{Gasser:2002am}
  J.~Gasser, M.~A.~Ivanov, E.~Lipartia, M.~Mojzis and A.~Rusetsky,
  Eur.\ Phys.\ J.\  C {\bf 26}, 13 (2002).

\bibitem{Walzl:2000cx}
  M.~Walzl, U.-G.~Mei\ss ner and E.~Epelbaum,
  Nucl.\ Phys.\  A {\bf 693}, 663 (2001).

\bibitem{Epelbaum:2004xf}
  E.~Epelbaum, U.-G.~Mei\ss ner and J.~E.~Palomar,
  Phys.\ Rev.\  C {\bf 71}, 024001 (2005).

\bibitem{Epelbaum:2005fd}
  E.~Epelbaum and U.-G.~Mei\ss ner,
  Phys.\ Rev.\  C {\bf 72}, 044001 (2005).

\bibitem{VanKolck:1993ee}
  U.~L.~Van Kolck,
  Soft Physics: Applications Of Effective Chiral Lagrangians To Nuclear
  Physics And Quark Models, PhD thesis, University of Texas, Austin, USA,
  1993, UNI-94-01021.

\bibitem{vanKolck:1996rm}
  U.~van Kolck, J.~L.~Friar and J.~T.~Goldman,
  Phys.\ Lett.\  B {\bf 371}, 169 (1996).

\bibitem{Friar:1999zr}
  J.~L.~Friar and U.~van Kolck,
  Phys.\ Rev.\  C {\bf 60}, 034006 (1999).

\bibitem{Friar:2003yv}
  J.~L.~Friar, U.~van Kolck, G.~L.~Payne and S.~A.~Coon,
  Phys.\ Rev.\  C {\bf 68}, 024003 (2003).

\bibitem{Friar:2004ca}
  J.~L.~Friar, U.~van Kolck, M.~C.~M.~Rentmeester and R.~G.~E.~Timmermans,
  Phys.\ Rev.\  C {\bf 70}, 044001 (2004).

\bibitem{Friar:2004rg}
  J.~L.~Friar, G.~L.~Payne and U.~van Kolck,
  Phys.\ Rev.\  C {\bf 71}, 024003 (2005).

\bibitem{Gasser:1982ap}
  J.~Gasser and H.~Leutwyler,
  Phys.\ Rept.\  {\bf 87}, 77 (1982).

\bibitem{Beane:2006fk}
  S.~R.~Beane, K.~Orginos and M.~J.~Savage,
  Nucl.\ Phys.\  B {\bf 768}, 38 (2007).


\bibitem{Henley:1979ig}
  E.~M.~Henley and G.~A.~Miller,
{\it  In *Rho M, Wilkinson D: Mesons In Nuclei, Vol.I*, Amsterdam 1979, 405-434}

\bibitem{Austen:1983te}
  G.~J.~M.~Austen and J.~J.~de Swart,
  Phys.\ Rev.\ Lett.\  {\bf 50}, 2039 (1983).

\bibitem{Kaiser:2006ws}
  N.~Kaiser,
  Phys.\ Rev.\  C {\bf 73}, 044001 (2006).

\bibitem{Ueling:1935aa}
  E.A.~Uehling,
  Phys.\ Rev.\  {\bf 48}, 55 (1935).


\bibitem{Durand:1957zz}
  L.~Durand,
  Phys.\ Rev.\  {\bf 108}, 1597 (1957).

\bibitem{Stoks:1990bb}
V.G. Stoks,
  The magnetic moment interaction in NN phase shift analysis,
  PhD thesis, University Nijmegen, The Netherlands, 1990.

\bibitem{Kievsky:2003eu}
  A.~Kievsky, M.~Viviani and L.~E.~Marcucci,
  Phys.\ Rev.\  C {\bf 69}, 014002 (2004).

\bibitem{Witaa:2003gs}
  H.~Wita{\l}a, J.~Golak, R.~Skibinski, C.~R.~Howell and W.~Tornow,
  Phys.\ Rev.\  C {\bf 67}, 064002 (2003).

\bibitem{Rupak:2001ci}
  G.~Rupak and X.~W.~Kong,
  Nucl.\ Phys.\  A {\bf 717}, 73 (2003).

\bibitem{deSwart:1997ep}
  J.~J.~de Swart, M.~C.~M.~Rentmeester and R.~G.~E.~Timmermans,
  PiN Newslett.\  {\bf 13}, 96 (1997).

\bibitem{Cheung:1979ma}
  C.~Y.~Cheung, E.~M.~Henley and G.~A.~Miller,
  Nucl.\ Phys.\  A {\bf 348}, 365 (1980).

\bibitem{vanKolck:1997fu}
  U.~van Kolck, M.~C.~M.~Rentmeester, J.~L.~Friar, J.~T.~Goldman and J.~J.~de Swart,
  Phys.\ Rev.\ Lett.\  {\bf 80}, 4386 (1998).


\bibitem{Coon:1995qh}
  S.~A.~Coon and J.~A.~Niskanen,
  Phys.\ Rev.\  C {\bf 53}, 1154 (1996).

\bibitem{Niskanen:2001aj}
  J.~A.~Niskanen,
  Phys.\ Rev.\  C {\bf 65}, 037001 (2002).

\bibitem{Meissner:2005ne}
  U.-G.~Mei\ss ner, U.~Raha and A.~Rusetsky,
  Phys.\ Lett.\  B {\bf 639}, 478 (2006).

\bibitem{Kaiser:2006ck}
  N.~Kaiser,
  Phys.\ Rev.\  C {\bf 73}, 064003 (2006).

\bibitem{Kaiser:2006na}
  N.~Kaiser,
  Phys.\ Rev.\  C {\bf 74}, 067001 (2006).


\bibitem{Beane:1997pk}
  S.~R.~Beane, T.~D.~Cohen and D.~R.~Phillips,
  Nucl.\ Phys.\  A {\bf 632}, 445 (1998).

\bibitem{Phillips:1997xu}
  D.~R.~Phillips, S.~R.~Beane and T.~D.~Cohen,
  Annals Phys.\  {\bf 263}, 255 (1998).

\bibitem{Birse:1998dk}
  M.~C.~Birse, J.~A.~McGovern and K.~G.~Richardson,
  Phys.\ Lett.\  B {\bf 464}, 169 (1999).

\bibitem{Phillips:1998uy}
  D.~R.~Phillips, S.~R.~Beane and M.~C.~Birse,
  J.\ Phys.\ A  {\bf 32}, 3397 (1999).


\bibitem{Gegelia:1998iu}
  J.~Gegelia,
  J.\ Phys.\ G {\bf 25}, 1681 (1999).

\bibitem{Yang:2004ss}
  J.~F.~Yang and J.~H.~Huang,
  Phys.\ Rev.\  C {\bf 71}, 034001 (2005)
  [Erratum-ibid.\  C {\bf 71}, 069901 (2005)].

\bibitem{Harada:2006cw}
  K.~Harada and H.~Kubo,
  Nucl.\ Phys.\  B {\bf 758}, 304 (2006).

\bibitem{Harada:2007ua}
  K.~Harada, H.~Kubo and A.~Ninomiya,
  arXiv:nucl-th/0702074.

\bibitem{Lepage:1997}
G.P. Lepage,
arXiv:nucl-th/9706029.

\bibitem{Lepage:2000}
G.P. Lepage,
{\it How to renormalize the Schr\"odinger equation}, talk given at
  the {INT} program {Effective Field Theories and Effective Interactions, INT,
  Seattle, USA, June 25-August 2, 2000}.

\bibitem{Cohen:1998bv}
  T.~D.~Cohen and J.~M.~Hansen,
  Phys.\ Lett.\  B {\bf 440}, 233 (1998).

\bibitem{Frederico:1999ps}
  T.~Frederico, V.~S.~Timoteo and L.~Tomio,
  Nucl.\ Phys.\  A {\bf 653}, 209 (1999).

\bibitem{Phillips:1999bf}
  D.~R.~Phillips, I.~R.~Afnan and A.~G.~Henry-Edwards,
  Phys.\ Rev.\  C {\bf 61}, 044002 (2000).

\bibitem{Gegelia:2001ev}
  J.~Gegelia and G.~Japaridze,
  Phys.\ Lett.\  B {\bf 517}, 476 (2001).

\bibitem{PavonValderrama:2003np}
  M.~Pavon Valderrama and E.~Ruiz Arriola,
  Phys.\ Lett.\  B {\bf 580}, 149 (2004).

\bibitem{PavonValderrama:2004nb}
  M.~Pavon Valderrama and E.~Ruiz Arriola,
  Phys.\ Rev.\  C {\bf 70}, 044006 (2004).

\bibitem{Gegelia:2004pz}
  J.~Gegelia and S.~Scherer,
  Int.\ J.\ Mod.\ Phys.\  A {\bf 21}, 1079 (2006).

\bibitem{PavonValderrama:2005gu}
  M.~Pavon Valderrama and E.~Ruiz Arriola,
  Phys.\ Rev.\  C {\bf 72}, 054002 (2005).

\bibitem{PavonValderrama:2005wv}
  M.~Pavon Valderrama and E.~R.~Arriola,
  Phys.\ Rev.\  C {\bf 74}, 054001 (2006).

\bibitem{PavonValderrama:2005uj}
  M.~Pavon Valderrama and E.~Ruiz Arriola,
  Phys.\ Rev.\  C {\bf 74}, 064004 (2006)
  [Erratum-ibid.\  C {\bf 75}, 059905 (2007)].

\bibitem{Epelbaum:2006pt}
  E.~Epelbaum and U.-G.~Mei{\ss}ner,
  arXiv:nucl-th/0609037.

\bibitem{Djukanovic:2006mc}
  D.~Djukanovic, J.~Gegelia, S.~Scherer and M.~R.~Schindler,
  Few Body Syst.\  {\bf 41}, 141 (2007).

\bibitem{Higa:2007gz}
  R.~Higa, M.~Pavon Valderrama and E.~Ruiz Arriola,
  Phys.\ Rev.\  C {\bf 77}, 034003 (2008).

\bibitem{Yang:2007hb}
  C.~J.~Yang, C.~Elster and D.~R.~Phillips,
  Phys.\ Rev.\  C {\bf 77}, 014002 (2008).

\bibitem{Long:2007vp}
  B.~Long and U.~van Kolck,
  Annals Phys.\  {\bf 323}, 1304 (2008).

\bibitem{Entem:2007jg}
  D.~R.~Entem, E.~Ruiz Arriola, M.~Pavon Valderrama and R.~Machleidt,
  Phys.\ Rev.\  C {\bf 77}, 044006 (2008).

\bibitem{Valderrama:2008kj}
  M.~P.~Valderrama and E.~R.~Arriola,
  arXiv:0809.3186 [nucl-th].

\bibitem{Shukla:2008sp}
  D.~Shukla, D.~R.~Phillips and E.~Mortenson,
  J.\ Phys.\ G {\bf 35}, 115009 (2008).

\bibitem{Beane:2001bc}
S.~R.~Beane, P.~F.~Bedaque, M.~J.~Savage, and U.~van Kolck,
Nucl.\ Phys.\ A {\bf 700}, 377 (2002).

\bibitem{Beane:2000wh}
  S.~R.~Beane, P.~F.~Bedaque, L.~Childress, A.~Kryjevski, J.~McGuire and U.~v.~Kolck,
  Phys.\ Rev.\  A {\bf 64}, 042103 (2001).

\bibitem{Bawin:2003dm}
  M.~Bawin and S.~A.~Coon,
  Phys.\ Rev.\  A {\bf 67}, 042712 (2003).

\bibitem{Braaten:2004pg}
  E.~Braaten and D.~Phillips,
 Phys.\ Rev.\ A {\bf 70}, 052111 (2004).

\bibitem{Hammer:2005sa}
  H.-W.~Hammer and B.~G.~Swingle,
  Annals Phys.\  {\bf 321}, 306 (2006).

\bibitem{Entem:2003ft}
  D.~R.~Entem and R.~Machleidt,
  Phys.\ Rev.\  C {\bf 68}, 041001 (2003).

\bibitem{Epelbaum:2004fk}
  E.~Epelbaum, W.~Gl\"ockle and U.-G.~Mei\ss ner,
  Nucl.\ Phys.\  A {\bf 747}, 362 (2005).

\bibitem{Stoks:1993tb}
V.G.J. Stoks et~al.,
Phys.\ Rev.\ C {\bf 48}, 792 (1993).

\bibitem{NNonline}
NN-Online program, M.~C.~M.~Rentmeester et al., http://nn-online.org.

\bibitem{SAID}
SAID on-line program, R.~A.Arndt et al., http://gwdac.phys.gwu.edu.

\bibitem{Witala:2004pv}
  H.~Wita{\l}a, J.~Golak, W.~Gl\"ockle and H.~Kamada,
  Phys.\ Rev.\  C {\bf 71}, 054001 (2005).

\bibitem{Lin:2007kg}
  T.~Lin, C.~Elster, W.~N.~Polyzou and W.~Gl\"ockle,
  Phys.\ Lett.\  B {\bf 660}, 345 (2008).

\bibitem{Lin:2008sy}
  T.~Lin, C.~Elster, W.~N.~Polyzou, H.~Wita{\l}a and W.~Gl\"ockle,
  Phys.\ Rev.\  C {\bf 78}, 024002 (2008).

\bibitem{Kamada:1999wy}
  H.~Kamada and W.~Gl\"ockle,
  Phys.\ Rev.\ Lett.\  {\bf 80}, 2547 (1998).

\bibitem{Epelbaum:2001fm}
  E.~Epelbaum, U.-G.~Mei\ss ner, W.~Gl\"ockle, and C.~Elster,
  Phys.\ Rev.\ C {\bf 65}, 044001 (2002).

\bibitem{Nogga:2005hp}
  A.~Nogga, P.~Navratil, B.~R.~Barrett and J.~P.~Vary,
  Phys.\ Rev.\  C {\bf 73}, 064002 (2006).

\bibitem{Navratil:2007we}
  P.~Navratil, V.~G.~Gueorguiev, J.~P.~Vary, W.~E.~Ormand and A.~Nogga,
  Phys.\ Rev.\ Lett.\  {\bf 99}, 042501 (2007).

\bibitem{Hanhart:2000gp}
  C.~Hanhart, U.~van Kolck and G.~A.~Miller,
  Phys.\ Rev.\ Lett.\  {\bf 85}, 2905 (2000).

\bibitem{Nakamura:2007vi}
  S.~X.~Nakamura,
  Phys.\ Rev.\  C {\bf 77}, 054001 (2008).


\bibitem{Epelbaum:2002ji}
  E.~Epelbaum, A.~Nogga, W.~Gl\"ockle, H.~Kamada, U.-G.~Mei\ss ner and H.~Wita{\l}a,
  Eur.\ Phys.\ J.\  A {\bf 15}, 543 (2002).


\bibitem{Ermisch:2003zq}
  K.~Ermisch {\it et al.},
  Phys.\ Rev.\  C {\bf 68}, 051001 (2003).

\bibitem{Duweke:2004xv}
  C.~Duweke {\it et al.},
  Phys.\ Rev.\  C {\bf 71}, 054003 (2005).

\bibitem{Ermisch:2005kf}
  K.~Ermisch {\it et al.},
  Phys.\ Rev.\  C {\bf 71}, 064004 (2005).

\bibitem{Kistryn:2005fi}
  S.~Kistryn {\it et al.},
  Phys.\ Rev.\  C {\bf 72}, 044006 (2005).

\bibitem{Witala:2006nn}
  H.~Wita{\l}a {\it et al.},
  Phys.\ Rev.\  C {\bf 73}, 044004 (2006).

\bibitem{Biegun:2006zc}
  A.~Biegun {\it et al.},
  Acta Phys.\ Polon.\  B {\bf 37}, 213 (2006).

\bibitem{Ley:2006hu}
  J.~Ley {\it et al.},
  Phys.\ Rev.\  C {\bf 73}, 064001 (2006).

\bibitem{Stephan:2007zza}
  E.~Stephan {\it et al.},
  Phys.\ Rev.\  C {\bf 76}, 057001 (2007).

\bibitem{Huber:1996td}
  D.~Huber, H.~Wita{\l}a, A.~Nogga, W.~Gl\"ockle and H.~Kamada,
  Few Body Syst.\  {\bf 22}, 107 (1997).


\bibitem{Kamada:2008zz}
  H.~Kamada {\it et al.},
  AIP Conf.\ Proc.\  {\bf 1011}, 59 (2008).


\bibitem{Deltuva:2005cc}
  A.~Deltuva, A.~C.~Fonseca and P.~U.~Sauer,
  Phys.\ Rev.\  C {\bf 72}, 054004 (2005)
  [Erratum-ibid.\  C {\bf 72}, 059903 (2005)].

\bibitem{Deltuva:2005zz}
  A.~Deltuva, A.~C.~Fonseca and P.~U.~Sauer,
  Phys.\ Rev.\ Lett.\  {\bf 95}, 092301 (2005).

\bibitem{Howell:1987aa}
C.~R. Howell {\em et~al.},
Few-Body Systems {\bf 2}, 19 (1987).

\bibitem{Sagara:1994zz}
  K.~Sagara, H.~Oguri, S.~Shimizu, K.~Maeda, H.~Nakamura, T.~Nakashima and S.~Morinobu,
  Phys.\ Rev.\  C {\bf 50}, 576 (1994).

\bibitem{Rauprich:1988aa}
G.~Rauprich {\em et~al.},
Few-Body Systems {\bf 5}, 67 (1988).

\bibitem{Sperisen:1984aa}
F.~Sperisen {\em et~al.},
Nucl. Phys. {\bf A422}, 81 (1984).

\bibitem{Witala:1993aa}
H.~Wita{\l}a {\em et~al.},
Few-Body Systems  {\bf 15}, 67 (1993).

\bibitem{Glombik:1995aa}
A.~Glombik {\em et~al.},
AIP Conference Proc. {\bf 334}, 486 (1995).

\bibitem{Kretschmer:1995aa}
W.~Kretschmer {\em et~al.},
AIP Conference Proc. {\bf 339}, 335 (1995).

\bibitem{Strate:1989aa}
J.~Strate {\em et~al.},
Nucl. Phys. A {\bf 501}, 51 (1989).


\bibitem{Setze:1996aa}
H.R.~Setze {\em et~al.},
Phys. Lett. B {\bf 388}, 229 (1996).

\bibitem{Rauprich:1991aa}
G.~Rauprich {\em et~al.},
Nucl. Phys. A {\bf 535}, 313 (1991).

\bibitem{Coon:2001pv}
  S.~A.~Coon and H.~K.~Han,
  Few Body Syst.\  {\bf 30}, 131 (2001).

\bibitem{Fisher:2006pm}
  B.~M.~Fisher {\it et al.},
  Phys.\ Rev.\  C {\bf 74}, 034001 (2006).

\bibitem{Deltuva:2007xv}
  A.~Deltuva and A.~C.~Fonseca,
  Phys.\ Rev.\ Lett.\  {\bf 98}, 162502 (2007).

\bibitem{Deltuva:2007xw}
  A.~Deltuva and A.~C.~Fonseca,
  Phys.\ Rev.\  C {\bf 76}, 021001 (2007).

\bibitem{Lazauskas:2004uq}
  R.~Lazauskas, J.~Carbonell, A.~C.~Fonseca, M.~Viviani, A.~Kievsky and S.~Rosati,
  Phys.\ Rev.\  C {\bf 71}, 034004 (2005).

\bibitem{Quaglioni:2008sm}
  S.~Quaglioni and P.~Navratil,
  arXiv:0804.1560 [nucl-th].


\bibitem{Navratil:2007aj}
  P.~Navratil, V.~G.~Gueorguiev, J.~P.~Vary, W.~E.~Ormand, A.~Nogga and S.~Quaglioni,
  arXiv:0712.1207 [nucl-th].

\bibitem{Hemmert:1997ye}
  T.~R.~Hemmert, B.~R.~Holstein and J.~Kambor,
  J.\ Phys.\ G {\bf 24}, 1831 (1998).

\bibitem{Jenkins:1991bs}
  E.~E.~Jenkins and A.~V.~Manohar,
  Phys.\ Lett.\  B {\bf 281}, 336 (1992).

\bibitem{Pascalutsa:1998pw}
  V.~Pascalutsa,
  Phys.\ Rev.\  D {\bf 58}, 096002 (1998).

\bibitem{Hacker:2005fh}
  C.~Hacker, N.~Wies, J.~Gegelia and S.~Scherer,
  Phys.\ Rev.\  C {\bf 72}, 055203 (2005).


\bibitem{Fettes:2000bb}
  N.~Fettes and U.-G.~Mei\ss ner,
  Nucl.\ Phys.\  A {\bf 679}, 629 (2001).


\bibitem{Epelbaum:2007sq}
  E.~Epelbaum, H.~Krebs and U.-G.~Mei\ss ner,
  Nucl.\ Phys.\  A {\bf 806}, 65 (2008).

\bibitem{Krebs:2007rh}
  H.~Krebs, E.~Epelbaum and U.-G.~Mei\ss ner,
  Eur.\ Phys.\ J.\  A {\bf 32}, 127 (2007).

\bibitem{Kaiser:1998wa}
  N.~Kaiser, S.~Gerstendorfer and W.~Weise,
  Nucl.\ Phys.\  A {\bf 637}, 395 (1998).

\bibitem{Pandharipande:2005sx}
  V.~R.~Pandharipande, D.~R.~Phillips and U.~van Kolck,
  Phys.\ Rev.\  C {\bf 71}, 064002 (2005).

\bibitem{PDG}
Particle Data Group, see the website http://pdg.lbl.gov/.

\bibitem{Arndt:2006bf}
  R.~A.~Arndt, W.~J.~Briscoe, I.~I.~Strakovsky and R.~L.~Workman,
  Phys.\ Rev.\  C {\bf 74}, 045205 (2006).

\bibitem{Rubinstein:1967aa}
H.~R. Rubinstein, F.~Scheck, and R.~H. Sokolov,
 Phys. Rev. {\bf 154}, 1608 (1967).

\bibitem{Tiburzi:2005na}
 B.~C.~Tiburzi and A.~Walker-Loud,
 Nucl.\ Phys.\  A {\bf 764}, 274 (2006).

\bibitem{Epelbaum:2008td}
  E.~Epelbaum, H.~Krebs and U.-G.~Mei\ss ner,
  Phys.\ Rev.\  C {\bf 77}, 034004 (2008).

\bibitem{Beane:2002wk}
  S.~R.~Beane, V.~Bernard, E.~Epelbaum, U.-G.~Mei{\ss}ner and D.~R.~Phillips,
  Nucl.\ Phys.\  A {\bf 720}, 399 (2003).

\bibitem{Beane:1997iv}
  S.~R.~Beane, V.~Bernard, T.~S.~H.~Lee, U.-G.~Mei{\ss}ner and U.~van Kolck,
  Nucl.\ Phys.\  A {\bf 618}, 381 (1997).

\bibitem{Krebs:2004ir}
  H.~Krebs, V.~Bernard and U.-G.~Mei{\ss}ner,
  Eur.\ Phys.\ J.\  A {\bf 22}, 503 (2004).

\bibitem{Baru:2002cg}
  V.~Baru, J.~Haidenbauer, C.~Hanhart and J.~A.~Niskanen,
  Eur.\ Phys.\ J.\  A {\bf 16}, 437 (2003).

\bibitem{Gardestig:2005pp}
  A.~Gardestig and D.~R.~Phillips,
  Phys.\ Rev.\  C {\bf 73}, 014002 (2006).

\bibitem{Lensky:2007zc}
  V.~Lensky, V.~Baru, E.~Epelbaum, C.~Hanhart, J.~Haidenbauer, A.~E.~Kudryavtsev and U.-G.~Mei{\ss}ner,
  Eur.\ Phys.\ J.\  A {\bf 33}, 339 (2007).

\bibitem{Baru:2007cd}
  V.~Baru, J.~Haidenbauer, C.~Hanhart, A.~E.~Kudryavtsev, V.~Lensky and U.-G.~Mei{\ss}ner,
  [arXiv:0711.2748 [nucl-th]].

\bibitem{Weinberg:1992yk}
  S.~Weinberg,
  Phys.\ Lett.\  B {\bf 295}, 114 (1992).

\bibitem{Cohen:1995cc}
  T.~D.~Cohen, J.~L.~Friar, G.~A.~Miller and U.~van Kolck,
  Phys.\ Rev.\  C {\bf 53}, 2661 (1996).

\bibitem{Bernard:1998sz}
  V.~Bernard, N.~Kaiser and U.-G.~Mei{\ss}ner,
  Eur.\ Phys.\ J.\  A {\bf 4}, 259 (1999).

\bibitem{daRocha:1999dm}
  C.~da Rocha, G.~Miller and U.~van Kolck,
  Phys.\ Rev.\  C {\bf 61}, 034613 (2000).

\bibitem{Hanhart:2002bu}
  C.~Hanhart and N.~Kaiser,
  Phys.\ Rev.\  C {\bf 66}, 054005 (2002).

\bibitem{Lensky:2005jc}
  V.~Lensky, V.~Baru, J.~Haidenbauer, C.~Hanhart, A.~E.~Kudryavtsev and U.-G.~Mei{\ss}ner,
  Eur.\ Phys.\ J.\  A {\bf 27}, 37 (2006).

\bibitem{Hutcheon:1991vt}
  D.~A.~Hutcheon {\it et al.},
  Nucl.\ Phys.\  A {\bf 535}, 618 (1991).

\bibitem{Heimberg:1996be}
  P.~Heimberg {\it et al.},
  Phys.\ Rev.\ Lett.\  {\bf 77}, 1012 (1996).

\bibitem{Drochner:1998ja}
  M.~Drochner {\it et al.}  [GEM Collaboration],
  Nucl.\ Phys.\  A {\bf 643}, 55 (1998).

\bibitem{Hanhart:2003pg}
  C.~Hanhart,
  Phys.\ Rept.\  {\bf 397}, 155 (2004).


\bibitem{Opper:2003sb}
  A.~K.~Opper {\it et al.},
  Phys.\ Rev.\ Lett.\  {\bf 91}, 212302 (2003).

\bibitem{Stephenson:2003dv}
  E.~J.~Stephenson {\it et al.},
  Phys.\ Rev.\ Lett.\  {\bf 91}, 142302 (2003).

\bibitem{vanKolck:2000ip}
  U.~van Kolck, J.~A.~Niskanen and G.~A.~Miller,
  Phys.\ Lett.\  B {\bf 493}, 65 (2000).

\bibitem{Gardestig:2004hs}
  A.~Gardestig {\it et al.},
  Phys.\ Rev.\  C {\bf 69}, 044606 (2004).

\bibitem{Nogga:2006cp}
  A.~Nogga {\it et al.},
  Phys.\ Lett.\  B {\bf 639}, 465 (2006).

\bibitem{Adam:2004ch}
  H.~H.~Adam {\it et al.}  [WASA-at-COSY Collaboration],
  arXiv:nucl-ex/0411038.

\bibitem{Miller:2006tv}
  G.~A.~Miller, A.~K.~Opper and E.~J.~Stephenson,
  Ann.\ Rev.\ Nucl.\ Part.\ Sci.\  {\bf 56}, 253 (2006).

\bibitem{Baru:2004kw}
  V.~Baru, C.~Hanhart, A.~E.~Kudryavtsev and U.-G.~Mei{\ss}ner,
  Phys.\ Lett.\  B {\bf 589}, 118 (2004).

\bibitem{Lensky:2005hb}
  V.~Lensky, V.~Baru, J.~Haidenbauer, C.~Hanhart, A.~E.~Kudryavtsev and U.-G.~Mei{\ss}ner,
  Eur.\ Phys.\ J.\  A {\bf 26}, 107 (2005).

\bibitem{Lensky:2006wd}
  V.~Lensky, V.~Baru, J.~Haidenbauer, C.~Hanhart, A.~E.~Kudryavtsev and U.-G.~Mei{\ss}ner,
  Phys.\ Lett.\  B {\bf 648}, 46 (2007).

\bibitem{Hanhart:2007ym}
  C.~Hanhart,
  arXiv:nucl-th/0703028.

\bibitem{Alberico:2001jb}
  W.~M.~Alberico and G.~Garbarino,
  Phys.\ Rept.\  {\bf 369}, 1 (2002).

\bibitem{BGM}
Bydzovsky, P.; Gal, A.; Mares, J. (Eds.)
Lecture Notes in Physics  {\bf  724} 2007
(Springer, Heidelberg). 


\bibitem{Alexander:1969cx}
  G.~Alexander, U.~Karshon, A.~Shapira, G.~Yekutieli, R.~Engelmann, H.~Filthuth and W.~Lughofer,
  Phys.\ Rev.\  {\bf 173}, 1452 (1968).

\bibitem{Rijken:1998yy}
  T.~A.~Rijken, V.~G.~J.~Stoks and Y.~Yamamoto,
  Phys.\ Rev.\  C {\bf 59}, 21 (1999).

\bibitem{Haidenbauer:2005zh}
  J.~Haidenbauer and U.-G.~Mei{\ss}ner,
  Phys.\ Rev.\  C {\bf 72}, 044005 (2005).

\bibitem{Gasparyan:2003cc}
  A.~Gasparyan, J.~Haidenbauer, C.~Hanhart and J.~Speth,
  Phys.\ Rev.\  C {\bf 69} (2004) 034006.

\bibitem{Savage:1995kv}
  M.~J.~Savage and M.~B.~Wise,
  Phys.\ Rev.\  D {\bf 53}, 349 (1996).

\bibitem{Hammer:2001ng}
  H.-W.~Hammer,
  Nucl.\ Phys.\  A {\bf 705}, 173 (2002).

\bibitem{Korpa:2001au}
  C.~L.~Korpa, A.~E.~L.~Dieperink and R.~G.~E.~Timmermans,
  Phys.\ Rev.\  C {\bf 65}, 015208 (2002).

\bibitem{Polinder:2006zh}
  H.~Polinder, J.~Haidenbauer and U.-G.~Mei{\ss}ner,
  Nucl.\ Phys.\  A {\bf 779}, 244 (2006).

\bibitem{de Swart:1963gc}
  J.~J.~de Swart,
  Rev.\ Mod.\ Phys.\  {\bf 35}, 916 (1963).

\bibitem{SechiZorn:1969hk}
  B.~Sechi-Zorn, B.~Kehoe, J.~Twitty and R.~A.~Burnstein,
  Phys.\ Rev.\  {\bf 175}, 1735 (1968).

\bibitem{Eisele:1971mk}
  F.~Eisele, H.~Filthuth, W.~Foehlisch, V.~Hepp and G.~Zech,
  Phys.\ Lett.\  B {\bf 37}, 204 (1971).

\bibitem{Engelmann:1966aa}
  R.~Engelmann, H.~Filthuth, V.~Hepp and E.~Kluge,
  Phys.\ Lett.\  {\bf 21}, 587 (1966).

\bibitem{de Swart:1962aa}
  J.~J.~de Swart and C.~Dullemond,
  Ann.\ Phys.\  {\bf 19}, 485 (1962).


\bibitem{Haidenbauer:2007ra}
  J.~Haidenbauer, U.-G.~Mei{\ss}ner, A.~Nogga and H.~Polinder,
  Lect.\ Notes Phys.\  {\bf 724}, 113 (2007).

\bibitem{Tamagawa:2001tk}
  T.~Tamagawa {\it et al.},
  Nucl.\ Phys.\  A {\bf 691}, 234 (2001).

\bibitem{Ahn:2005jz}
  J.~K.~Ahn {\it et al.},
  Phys.\ Lett.\  B {\bf 633}, 214 (2006).

\bibitem{Polinder:2007mp}
  H.~Polinder, J.~Haidenbauer and U.-G.~Mei\ss ner,
  Phys.\ Lett.\  B {\bf 653}, 29 (2007).


\bibitem{Takahashi:2001nm}
  H.~Takahashi {\it et al.},
  Phys.\ Rev.\ Lett.\  {\bf 87} (2001) 212502.

\bibitem{Rijken:2006kg}
  T.~A.~Rijken and Y.~Yamamoto,
  arXiv:nucl-th/0608074.

\bibitem{Fujiwara:2006yh}
  Y.~Fujiwara, Y.~Suzuki and C.~Nakamoto,
  Prog.\ Part.\ Nucl.\ Phys.\  {\bf 58} (2007) 439.


\bibitem{Lee:2004si}
  D.~Lee, B.~Borasoy and T.~Sch\"afer,
  Phys.\ Rev.\  C {\bf 70}, 014007 (2004).

\bibitem{Borasoy:2005yc}
  B.~Borasoy, H.~Krebs, D.~Lee and U.-G.~Mei{\ss}ner,
  Nucl.\ Phys.\  A {\bf 768}, 179 (2006).

\bibitem{Borasoy:2006qn}
  B.~Borasoy, E.~Epelbaum, H.~Krebs, D.~Lee and U.-G.~Mei\ss ner,
  Eur.\ Phys.\ J.\  A {\bf 31}, 105 (2007).

\bibitem{Chen:2003vy}
  J.~W.~N.~Chen and D.~B.~Kaplan,
  Phys.\ Rev.\ Lett.\  {\bf 92}, 257002 (2004).

\bibitem{Lee:2004ze}
  D.~Lee,
  Phys.\ Rev.\  C {\bf 70}, 064002 (2004).

\bibitem{Wigner:1937zz}
  E.~Wigner,
  Phys.\ Rev.\  {\bf 51}, 947 (1937).

\bibitem{Mehen:1999qs}
  T.~Mehen, I.~W.~Stewart and M.~B.~Wise,
  Phys.\ Rev.\ Lett.\  {\bf 83}, 931 (1999).

\bibitem{Chen:2004rq}
  J.~W.~Chen, D.~Lee and T.~Schafer,
  Phys.\ Rev.\ Lett.\  {\bf 93}, 242302 (2004).

\bibitem{Duane:1987de}
  S.~Duane, A.~D.~Kennedy, B.~J.~Pendleton and D.~Roweth,
  Phys.\ Lett.\  B {\bf 195}, 216 (1987).

\bibitem{Lee:2005nm}
  D.~Lee,
  Phys.\ Rev.\  A {\bf 73}, 063204 (2006).

\bibitem{Borasoy:2007vy}
  B.~Borasoy, E.~Epelbaum, H.~Krebs, D.~Lee and U.-G.~Mei\ss ner,
  Eur.\ Phys.\ J.\  A {\bf 34}, 185 (2007).

\bibitem{oers67nd}
W. T. H. van Oers, J. D. Seagrave, Phys.\ Lett.\ B {\bf 24}, 562 (1967). 


\bibitem{Beane:2002xf}
S.~R.~Beane and M.~J.~Savage,
        Nucl.\ Phys.\ A {\bf 717}, 91 (2003);
Nucl.\ Phys.\ A {\bf 713}, 148 (2003).

\bibitem{Fettes:2000fd}
  N.~Fettes, 
  Pion--nucleon physics in chiral perturbation theory, Ph.D.~Thesis, 
  Universit\"at Bonn, Germany, 2000, JUL-3814.

\bibitem{Epelbaum:2002gk}
  E.~Epelbaum, U.-G.~Mei{\ss}ner and W.~Gl\"ockle,
  arXiv:nucl-th/0208040.

\bibitem{Epelbaum:2006jc}
  E.~Epelbaum, H.-W.~Hammer, U.-G.~Mei\ss ner and A.~Nogga,
  Eur.\ Phys.\ J.\  C {\bf 48}, 169 (2006).

\bibitem{Hammer:2007kq}
  H.-W.~Hammer, D.~R.~Phillips and L.~Platter,
  Eur.\ Phys.\ J.\  A {\bf 32}, 335 (2007).

\bibitem{Wilson:2004de}
  K.~G.~Wilson,
  Nucl.\ Phys.\ Proc.\ Suppl.\  {\bf 140}, 3 (2005).

\bibitem{Flambaum:2007mj}
  V.~V.~Flambaum and R.~B.~Wiringa,
  Phys.\ Rev.\  C {\bf 76}, 054002 (2007).


\bibitem{Furnstahl:2008df}
  R.~J.~Furnstahl, G.~Rupak and T.~Schafer,
  Ann.\ Rev.\ Nucl.\ Part.\ Sci.\  {\bf 58}, 1 (2008).

\bibitem{Nelson:1987dg}
  A.~E.~Nelson and D.~B.~Kaplan,
  Phys.\ Lett.\  B {\bf 192} (1987) 193.

\bibitem{Thorsson:1995rj}
  V.~Thorsson and A.~Wirzba,
  Nucl.\ Phys.\  A {\bf 589}, 633 (1995).

\bibitem{Kirchbach:1996xy}
  M.~Kirchbach and A.~Wirzba,
  Nucl.\ Phys.\  A {\bf 604}, 395 (1996).

\bibitem{Kirchbach:1997rk}
  M.~Kirchbach and A.~Wirzba,
  Nucl.\ Phys.\  A {\bf 616}, 648 (1997).

\bibitem{Oller:2001sn}
  J.~A.~Oller,
  Phys.\ Rev.\  C {\bf 65}, 025204 (2002).

\bibitem{Meissner:2001gz} 
  U.-G.~Mei{\ss}ner, J.~A.~Oller and A.~Wirzba, 
  Annals Phys.\  {\bf 297}, 27 (2002).

\bibitem{Leutwyler:1993gf}
  H.~Leutwyler,
  Phys.\ Rev.\  D {\bf 49}, 3033 (1994).


\bibitem{Drukarev:1988kd}
  E.~G.~Drukarev and E.~M.~Levin,
  Nucl.\ Phys.\  A {\bf 511}, 679 (1990)
  [Erratum-ibid.\  A {\bf 516}, 715 (1990)].

\bibitem{Furnstahl:1992pi}
  R.~J.~Furnstahl, D.~K.~Griegel and T.~D.~Cohen,
  Phys.\ Rev.\  C {\bf 46}, 1507 (1992).

\bibitem{Birse:1994cz}
  M.~C.~Birse,
  J.\ Phys.\ G {\bf 20}, 1537 (1994).

\bibitem{Pisarski:1996mt}
  R.~D.~Pisarski and M.~Tytgat,
  Phys.\ Rev.\  D {\bf 54}, 2989 (1996).

\bibitem{Gilg:1999qa}
  H.~Gilg {\it et al.},
  Phys.\ Rev.\  C {\bf 62}, 025201 (2000).

\bibitem{Itahashi:1999qb}
  K.~Itahashi {\it et al.},
  Phys.\ Rev.\  C {\bf 62}, 025202 (2000).

\bibitem{Oller} J.A.~Oller, private communication

\bibitem{LMOW} A.~Lacour, U.-G.~Mei{\ss}ner, J.A.~Oller, A.~Wirzba,
in preparation.

\bibitem{Girlanda:2003cq} 
  L.~Girlanda, A.~Rusetsky and W.~Weise, 
  Annals Phys.\  {\bf 312}, 92 (2004).

\bibitem{Girlanda:2004qa}
  L.~Girlanda, A.~Rusetsky and W.~Weise,
  Nucl.\ Phys.\  A {\bf 755}, 653 (2005).




\bibitem{Kaiser:2001jx}
  N.~Kaiser, S.~Fritsch and W.~Weise,
  Nucl.\ Phys.\  A {\bf 697}, 255 (2002).

\bibitem{Fritsch:2004nx}
  S.~Fritsch, N.~Kaiser and W.~Weise,
  Nucl.\ Phys.\  A {\bf 750}, 259 (2005).

\bibitem{Kaiser:2001ra}
  N.~Kaiser, S.~Fritsch and W.~Weise,
  Nucl.\ Phys.\  A {\bf 700}, 343 (2002).

\bibitem{Lutz:1999vc}
  M.~Lutz, B.~Friman and C.~Appel,
  Phys.\ Lett.\  B {\bf 474}, 7 (2000).

\bibitem{Kaiser:2002yg}
  N.~Kaiser,
  Nucl.\ Phys.\  A {\bf 709}, 251 (2002).

\bibitem{Kaiser:2003ux}
  N.~Kaiser,
  Phys.\ Rev.\  C {\bf 68}, 054001 (2003).

\bibitem{Kaiser:2004fe}
  N.~Kaiser and W.~Weise,
  Phys.\ Rev.\  C {\bf 71}, 015203 (2005).

\bibitem{Kaiser:2008gq}
  N.~Kaiser and W.~Weise,
  Nucl.\ Phys.\  A {\bf 804}, 60 (2008).

\bibitem{Kaiser:2007nv}
  N.~Kaiser, P.~de Homont and W.~Weise,
  Phys.\ Rev.\  C {\bf 77}, 025204 (2008).

\bibitem{Kaiser:2008qu}
  N.~Kaiser and W.~Weise,
  arXiv:0808.0856 [nucl-th].

\bibitem{Kaiser:2002jz}
  N.~Kaiser, S.~Fritsch and W.~Weise,
  Nucl.\ Phys.\  A {\bf 724}, 47 (2003).

\bibitem{Finelli:2002na}
  P.~Finelli, N.~Kaiser, D.~Vretenar and W.~Weise,
  Eur.\ Phys.\ J.\  A {\bf 17}, 573 (2003).

\bibitem{Finelli:2003fk}
  P.~Finelli, N.~Kaiser, D.~Vretenar and W.~Weise,
  Nucl.\ Phys.\  A {\bf 735}, 449 (2004).

\bibitem{Finelli:2007wm}
  P.~Finelli, N.~Kaiser, D.~Vretenar and W.~Weise,
  Phys.\ Lett.\  B {\bf 658}, 90 (2007).



\bibitem{Zhukov-93}
M.V. Zhukov, B.V. Danilin, D.V. Fedorov, J.M. Bang, I.J. Thompson, 
and J.S. Vaagen, 
Phys. Rep. {\bf 231}, 151 (1993).

\bibitem{Jensen-04}
A.S. Jensen, K. Riisager, D.V. Fedorov, and E. Garrido, 
Rev. Mod. Phys. {\bf 76}, 215 (2004).

\bibitem{Bertulani:2002sz} 
  C.~A.~Bertulani, H.-W.~Hammer and U.~Van Kolck, 
  Nucl.\ Phys.\  A {\bf 712}, 37 (2002).
 
\bibitem{Bedaque:2003wa}
  P.~F.~Bedaque, H.-W.~Hammer and U.~van Kolck,
  Phys.\ Lett.\  B {\bf 569}, 159 (2003).


\bibitem{Higa:2008rx}
  R.~Higa,
  arXiv:0809.5157 [nucl-th].

\bibitem{Higa:2008dn}
  R.~Higa, H.-W.~Hammer and U.~van Kolck,
  Nucl.\ Phys.\  A {\bf 809}, 171 (2008).

\bibitem{Hammer:2008ra}
  H.-W.~Hammer and R.~Higa,
Eur.\ Phys.\ J.\  A {\bf 37}, 193 (2008).


\bibitem{Federov:1994cf}
  D.~V.~Federov, A.~S.~Jensen and K.~Riisager,
  Phys.\ Rev.\ Lett.\  {\bf 73}, 2817 (1994).


\bibitem{Amorim:1997mq}
  A.~E.~A.~Amorim, T.~Frederico and L.~Tomio,
  Phys.\ Rev.\  C {\bf 56}, R2378 (1997).

\bibitem{Mazumdar:2000dg}
  I.~Mazumdar, V.~Arora and V.~S.~Bhasin,
  Phys.\ Rev.\  C {\bf 61}, 051303 (2000).

\bibitem{Canham:2008jd}
  D.~L.~Canham and H.~W.~Hammer,
Eur.\ Phys.\ J.\  A {\bf 37}, 367 (2008).

\bibitem{Marques:2001pe}
  F.~M.~Marques {\it et al.},
  Phys.\ Rev.\  C {\bf 64}, 061301 (2001).

\bibitem{Yamashita:2007ej}
  M.~T.~Yamashita, T.~Frederico and L.~Tomio,
  Phys.\ Lett.\ B {\bf 660}, 339 (2008).


\bibitem{Mazumdar:2006tn}
  I.~Mazumdar, A.~R.~P.~Rau and V.~S.~Bhasin,
  Phys.\ Rev.\ Lett.\  {\bf 97}, 062503 (2006).


\bibitem{Cobis:1996ru}
A.~Cobis, A.S.~Jensen and D.V.~Fedorov,
J.\ Phys.\ G {\bf 23}, 401 (1997).

\bibitem{Fedorov:2001wj}
D.V.~Fedorov and A.S.~Jensen,
Nucl.\ Phys.\ A {\bf 697}, 783 (2002).

\bibitem{Congl92}
J.G.~Congleton,
J.\ Phys.\ G {\bf 18}, 339 (1992).

\bibitem{Bert88}
C.A.~Bertulani and G.~Baur, Phys.\ Rep.\ {\bf 163}, 299 (1988).

\bibitem{Bogner:2003wn} 
  S.~K.~Bogner, T.~T.~S.~Kuo and A.~Schwenk, 
  Phys.\ Rept.\  {\bf 386}, 1 (2003).

\bibitem{Epelbaum:1998hg}
  E.~Epelbaum, W.~Gl\"ockle and U.-G.~Mei\ss ner,
  Phys.\ Lett.\  B {\bf 439}, 1 (1998).

\bibitem{Epelbaum:1998na} 
  E.~Epelbaum, W.~Gl\"ockle, A.~Kr\"uger and U.-G.~Mei{\ss}ner, 
  Nucl.\ Phys.\  A {\bf 645}, 413 (1999).
 
\bibitem{Okubo:1954aa}
S.~Okubo, Prog.\ Theor.\ Phys.\ {\bf 12}, 603 (1954).

\bibitem{Fukuda:1954aa}
N.~Fukuda, K.~Sawada, and M.~Taketani, Prog.\ Theor.\ Phys.\ {\bf 12}, 
156 (1954).
\bibitem{Bogner:2006pc} 
  S.~K.~Bogner, R.~J.~Furnstahl and R.~J.~Perry, 
  Phys.\ Rev.\ C {\bf 75}, 061001 (2007).
 
\bibitem{Jurgenson:2008jp}
  E.~D.~Jurgenson and R.~J.~Furnstahl,
  arXiv:0809.4199 [nucl-th].

\bibitem{Schaefer:2005fi} 
  B.~J.~Schaefer, M.~Wagner, J.~Wambach, T.~T.~S.~Kuo and G.~E.~Brown, 
  Phys.\ Rev.\  C {\bf 73}, 011001 (2006).
 

\bibitem{Dapo:2008qv}
  H.~Dapo, B.~J.~Schaefer and J.~Wambach,
  Eur.\ Phys.\ J.\  A {\bf 36}, 101 (2008).

\bibitem{Fujii:2004dd}
  S.~Fujii, E.~Epelbaum, H.~Kamada, R.~Okamoto, K.~Suzuki and W.~Gl\"ockle,
  Phys.\ Rev.\  C {\bf 70}, 024003 (2004).

\bibitem{Schwenk:2008su}
  A.~Schwenk and J.~D.~Holt,
  AIP Conf.\ Proc.\  {\bf 1011}, 159 (2008).

\bibitem{Bogner:2005sn}
  S.~K.~Bogner, A.~Schwenk, R.~J.~Furnstahl and A.~Nogga,
  Nucl.\ Phys.\  A {\bf 763}, 59 (2005).


\bibitem{Bogner:2006tw}
  S.~K.~Bogner, R.~J.~Furnstahl, S.~Ramanan and A.~Schwenk,
  Nucl.\ Phys.\  A {\bf 773}, 203 (2006).

\bibitem{Furnstahl:2007xm} 
  R.~J.~Furnstahl, 
  arXiv:nucl-th/0702040. 
 
\bibitem{Scidac:2007aa}
G.F. Bertsch, D.J.~Dean, and W.~Nazarewicz,
SciDAC Review {\bf 6}, 42 (2007).


\bibitem{Brockmann:1992in}
  R.~Brockmann and J.~Frank,
  Phys.\ Rev.\ Lett.\  {\bf 68}, 1830 (1992).

\bibitem{Serot:1984ey}
  B.~D.~Serot and J.~D.~Walecka,
  Adv.\ Nucl.\ Phys.\  {\bf 16}, 1 (1986).

\bibitem{Muller:1999cp}
  H.~M.~Muller, S.~E.~Koonin, R.~Seki and U.~van Kolck,
  Phys.\ Rev.\  C {\bf 61}, 044320 (2000).

\bibitem{Borasoy:2007vk}
  B.~Borasoy, E.~Epelbaum, H.~Krebs, D.~Lee and U.-G.~Mei{\ss}ner,
  Eur.\ Phys.\ J.\  A {\bf 35}, 357 (2008).

\bibitem{Giorgini08}
S.~Giorgini, L.P.~Pitaevskii, and S.~Stringari,
Rev.\ Mod.\ Phys.\  {\bf 80}, 1215 (2008). 

\bibitem{EKLM}
 E.~Epelbaum, H.~Krebs, D.~Lee and U.-G.~Mei{\ss}ner, forthcoming.



\bibitem{Friedman:1981qw}
  B.~Friedman and V.~R.~Pandharipande,
  Nucl.\ Phys.\  A {\bf 361} (1981) 502.

\bibitem{Akmal:1998cf}
  A.~Akmal, V.~R.~Pandharipande and D.~G.~Ravenhall,
  Phys.\ Rev.\  C {\bf 58}, 1804 (1998).

\bibitem{Carlson:2003wm}
  J.~Carlson, J.~.~J.~Morales, V.~R.~Pandharipande and D.~G.~Ravenhall,
  Phys.\ Rev.\  C {\bf 68}, 025802 (2003).

\bibitem{Schwenk:2005ka}
  A.~Schwenk and C.~J.~Pethick,
  Phys.\ Rev.\ Lett.\  {\bf 95}, 160401 (2005).

\bibitem{Gezerlis:2007fs}
  A.~Gezerlis and J.~Carlson,
  Phys.\ Rev.\  C {\bf 77}, 032801 (2008).

\bibitem{Lee:2008xsa}
  D.~Lee,
  Phys.\ Rev.\  C {\bf 78}, 024001 (2008).






\end{thebibliography}
\end{document}